\documentclass{article}

\pdfoutput=1

\usepackage{jheppub}
\usepackage{float}
\usepackage{ulem}
\usepackage{mathrsfs}
\usepackage{natbib}
\usepackage{tikz}
\usepackage{setspace}
\usepackage{amsfonts}
\usepackage{amssymb}
\usepackage{amsmath}
\usepackage{esint}                   
\usepackage{epsfig}
\usepackage{latexsym}
\usepackage{color}
\usepackage{graphicx}
\usepackage{hyperref}
\usepackage{ulem}

\usepackage{mathtools}

\definecolor{myDeepPurple}{RGB}{95, 5, 120}   
\definecolor{myDeepBlue}{RGB}{15, 35, 110}    
\definecolor{myDeepOrange}{RGB}{170, 60, 0}   
\hypersetup{
    colorlinks = true,
    citecolor  = myDeepPurple,
    linkcolor  = myDeepBlue,
    urlcolor   = myDeepOrange
}

\usepackage{nicefrac}
\usepackage{inputenc}
\usepackage{pstricks}
\usepackage{slashed}
\usepackage{multirow}
\usepackage{enumerate}
\usepackage{subcaption,comment}

\def\be{\begin{equation}}
\def\ee{\end{equation}}
\def\ba{\begin{eqnarray}}
\def\ea{\end{eqnarray}}

\def\r{\rho}
\def\a{\alpha}

\def\b{\beta}

\def\d{\delta}

\def\th{\theta}

\def\m{\mu}

\def\om{\omega}

\def\vphi{\varphi}

\def\cA{{\cal A}}

\def\cE{{\cal E}}
\def\cJ{{\cal J}}

\def\no{\noindent}

\def\IR{\relax{\rm I\kern-.18em R}}

\def\inv{^{\raise.0ex\hbox{${\scriptscriptstyle -}$}\kern-.05em 1}}

\allowdisplaybreaks

\begin{document}
\hfill HU-EP-25/38

\title{Supersymmetric AdS Solitons, Coulomb Branch Flows and Twisted Compactifications}

\author[a]{Dimitrios Chatzis,}
\author[a]{Madison Hammond, }
\author[b]{Georgios Itsios, }
\author[a]{Carlos Nunez,  }
\author[c]{ and 
Dimitrios Zoakos}
\affiliation[a]{Centre for Quantum Fields and Gravity, Department of Physics, Swansea University, Swansea SA2 8PP, United Kingdom}
\affiliation[b]{Institut f\"{u}r Physik, Humboldt-Universit\"{a}t zu Berlin,\\
IRIS Geb\"{a}ude, Zum Gro{\ss}en Windkanal 2, 12489 Berlin, Germany}
\affiliation[c]{Department of Physics, University of Patras, 26504 Patras, Greece}

\emailAdd{dchatzis@proton.me}
\emailAdd{m.hammond.2412736@swansea.ac.uk}
\emailAdd{georgios.itsios@physik.hu-berlin.de}
\emailAdd{c.nunez@swansea.ac.uk}
\emailAdd{dzoakos@upatras.gr}
\abstract{This work, which accompanies \cite{Chatzis:2025dnu}, is about constructing smooth solutions in type II and eleven dimensional supergravity which describe supersymmetry preserving RG flows from four-dimensional SCFTs in the UV to  three-dimensional SQFTs in the IR, through holography. We show that all the different UV fixed points flow to theories which confine external quarks and have a mass gap. We proceed by presenting extended calculations of a plethora of observables and analyse the dual field theories in great detail. This includes a boundary analysis and application of holographic renormalization methods in the simplest case of the type IIB solution. Many of the observables computed here have a universal behaviour: they factorize into two parts, one of which includes information about the UV SCFTs, and the other describing the dynamics of the RG flow, which is the same regardless of the UV fixed point.
}

\maketitle

\flushbottom



\section{Introduction and organization of the paper}
In the ongoing quest to deepen our understanding of the strongly coupled dynamics of $\mathrm{QFTs}$, particularly those that exhibit features akin to real-world $\mathrm{QCD}_4$, the gauge/gravity duality has proven to be one of the most powerful theoretical tools since its introduction by Maldacena in \cite{Maldacena:1997re}. The duality provides a means of studying non-perturbative aspects of gauge theories through a classical gravitational description, enabling the investigation of phenomena such as confinement and phase structure, chiral symmetry breaking, the computation of form factors, and the spectra of mesons, hadrons, and glueballs, among many others. The gravitational systems that holographically capture these features arise as extensions of the original duality between type II string theory on $\mathrm{AdS}_5\times\mathbb{S}^5$ and ${\cal N}=4$ $\mathrm{SYM}$ on $\mathbb{R}^{1,3}$ \cite{Klebanov:1998hh,Girardello:1999hj, Polchinski:2000uf, Klebanov:2000hb}.

\vspace{1em}

Over the years, a substantial body of work has been devoted to extending and generalizing the original duality, thereby broadening our understanding of strongly coupled gauge dynamics reminiscent of those encountered in $\mathrm{QCD}_4$. One class of such developments involves solutions with branes wrapped on internal cycles \cite{Maldacena:2000yy,Atiyah:2000zz,Edelstein:2001pu, Nunez:2001pt}, while another considers D-brane configurations at the tip of deformed Calabi–Yau singularities \cite{Klebanov:1998hh,Klebanov:2000hb,Klebanov:2000nc}. These constructions have been remarkably successful in reproducing several desirable features, including the ability to describe a broad range of ${\cal N}=1,2$ four-dimensional quiver gauge theories with bifundamental matter, as well as systems with reduced or broken supersymmetry \cite{Maldacena:2000yy,Klebanov:2000hb, Gauntlett:2001ps, Bigazzi:2001aj, Bigazzi:2002gyi, Bertolini:2003iv}. 

Nevertheless, the conventional approaches to constructing backgrounds dual to confining $\mathrm{QFTs}$ often face challenges related to their $\mathrm{UV}$ behaviour. In the dual field theories of these models, the number of degrees of freedom tends to grow without bound as one approaches the $\mathrm{UV}$, corresponding to an infinite sequence of Seiberg dualities \cite{Klebanov:2000hb}. This behaviour complicates the application of holographic renormalization techniques developed in subsequent works \cite{deHaro:2000vlm,Bianchi:2001de,Papadimitriou:2004ap}, making it technically demanding to extract precise information about operators and their correlation functions. Recently, this issue has been addressed in \cite{Aramini:2025twg}, where the introduction of orientifold planes was shown to regulate the $\mathrm{UV}$ dynamics, leading to the emergence of a fixed point that effectively terminates the duality cascade with a finite number of degrees of freedom.

\vspace{1em}

An alternative approach to constructing confining holographic backgrounds is provided by the so-called AdS soliton, obtained via a double Wick rotation of gravity solutions such as the Schwarzschild–AdS black hole. This setup has been extensively used to model systems dual to pure $\mathrm{QCD}_3$ with additional massive Kaluza–Klein excitations \cite{Witten:1998zw,Horowitz:1998ha}. The AdS soliton constitutes a smooth solution of the supergravity equations of motion, characterized by a cigar-like geometry in the infrared, $\mathrm{d}s^2\sim\mathrm{d}r^2+r^2\mathrm{d}\phi^2$ $(\phi\sim\phi+L_{\phi})$, that caps off smoothly at a finite value of the radial coordinate $r=r_{\star}$ for an appropriate choice of $L_{\phi}$. A defining feature of such geometries is that the spacetime ending at $r=r_{\star}>0$ introduces a mass gap in the dual field theory, which corresponds to ${\cal N}=4$ $\mathrm{SYM}$ compactified on $\mathbb{R}^{1,2}\times\mathbb{S}^{1}_{\phi}$. Depending on the boundary conditions imposed on the fermions, globally well-defined spinors can be periodic or antiperiodic along $\mathbb{S}^1_{\phi}$, with the latter case breaking supersymmetry.

\vspace{1em}

To preserve a fraction of supersymmetry, one may employ a \emph{topological twist}, achieved by mixing a global symmetry with a spacetime symmetry. Concretely, this amounts to modifying the covariant derivative so that the antiperiodic contribution of the spin connection is cancelled, allowing spinors to remain periodic and form supersymmetric multiplets in three dimensions. In practice, this is realized by introducing a Wilson line in the geometry—equivalently, a constant background gauge field ${\cal A}={\cal A}_{\phi}\mathrm{d}\phi$ with nonzero holonomy (see \cite{Cassani:2021fyv, Kumar:2024pcz} for a clear review of the mechanism). In a recent advance, Anabalon and Ross \cite{Anabalon:2021tua} constructed a new AdS soliton solution derived from a charged black hole background that is 1/2 BPS. This solution implements the twisted compactification described above and has proven to be an excellent seed for generating higher-dimensional uplifts yielding gapped and confining systems with $\mathrm{AdS}_5$ factors in type II string theory and M-theory \cite{Fatemiabhari:2024aua,Chatzis:2024top,Chatzis:2024kdu,Castellani:2024ial,Macpherson:2024qfi,Macpherson:2025pqi, Barbosa:2024smw, Fatemiabhari:2024lct}. More recently, another supersymmetric AdS soliton was presented in \cite{Anabalon:2024che}, generalizing the construction of \cite{Anabalon:2021tua} by including a non-trivial scalar profile, an additional gauge field, and an extra free parameter. This broader solution has been interpreted as describing Coulomb branch deformations of ${\cal N}=4$ $\mathrm{SYM}$, in the sense of \cite{Freedman:1999gk,Gubser:2000nd}.

\vspace{1em}

The present work complements and extends the results of \cite{Chatzis:2025dnu}, where new infinite families of string-theory backgrounds were constructed by uplifting the $\mathrm{AdS}$ soliton of \cite{Anabalon:2024che} to type II supergravity and M-theory. These geometries provide holographic duals to Coulomb branch deformations of various $\mathrm{SCFT}_4$s compactified on a circle, which flow in the $\mathrm{IR}$ to three-dimensional gapped $\mathrm{QFT}$s preserving four real supercharges. The resulting solutions include a type $\mathrm{IIB}$ background (already discussed in \cite{Anabalon:2024che}), an infinite family of $\mathrm{M}$-theory backgrounds, and a corresponding family of type $\mathrm{IIA}$ backgrounds. Despite originating from distinct $\mathrm{CFT}$s in the $\mathrm{UV}$, these systems exhibit strikingly similar non-perturbative dynamics in the $\mathrm{IR}$, reflecting universal features of supersymmetric confinement and dimensional reduction. The present paper provides a more detailed analysis of these systems, including new computations that elucidate the properties of the dual field theories. In particular, additionally to the content of \cite{Chatzis:2025dnu}, we start by providing a more thorough presentation of the construction of the backgrounds, using the five-dimensional gauged supergravity solution. We then further probe the confining features of the IR effective dual theories by considering more involved F1 embeddings for the Wilson loop, as well as studying the behaviour of 't Hooft loops and the entanglement entropy on a strip. 
We also included a study of a D7-brane embedding and of the Penrose limit of the solution, providing alternative probes of the confining properties.
Moreover, we include a boundary analysis for the type IIB uplift that matches our previous analysis for the VEVs dual to the operators turned on by the deformation. Finally, as stated in \cite{Chatzis:2025dnu}, we perform a stability analysis on the fluctuations of the Wilson loop configuration used in the type IIB calculation (what in this work is dubbed "embedding I") in section \ref{Wilson_loop_section}.\\

The paper is organized as follows: sections \ref{5d_gauged_SUGRA_section} and \ref{New_BG} review the supergravity solutions of \cite{Chatzis:2025dnu}, with the latter also including a discussion of Page charge quantization. Section \ref{Observables_section} contains detailed computations of various field-theory observables exhibiting {\it universality}. 
A discussion on the embedding of a D7-brane probe in the deformed $\mathrm{AdS}_5\times\mathbb{S}^5$ solution of \cite{Anabalon:2021tua}, is also presented.
In section \ref{Holographic_RG_section}, we perform a boundary analysis and apply holographic renormalization to identify the vacuum expectation values (VEVs) of the operators responsible for deforming the $\mathrm{UV}$ $\mathrm{CFTs}$. Finally, sections \ref{WL_stability_section} and \ref{Penrose_limits_section} are devoted to the stability analysis of the Wilson loop configurations in the type $\mathrm{IIB}$ background and to a study of Penrose limits in the deformed $\mathrm{AdS}_5\times\mathbb{S}^5$ solution of \cite{Anabalon:2021tua}.


\section{5d gauged supergravity}\label{5d_gauged_SUGRA_section}
In this section we study the 5d soliton solution which will act as the "seed solution" for our string backgrounds. We will be uplifting it to get solutions in Type IIA, Type IIB and M-theory (11d supergravity). We first present the solution and its bosonic action, and then study important features like smoothness, which will be carried-up to the higher dimensional uplifts.

\subsection{The supersymmetric AdS soliton}
We start by presenting the five-dimensional gauged supergravity soliton solution found in \cite{Anabalon:2024che}, which one can obtain in a truncation of the compactification of type-IIB supergravity on the five-sphere. This is a generalization of the solution found in \cite{Anabalon:2021tua}, which contains an extra charge parameter as well as a scalar profile.\\

The bosonic sector of the 5d theory, containing a metric, three abelian gauge fields $A_i$ and two scalars $\Phi_1,\Phi_2$, can be written in the following way:
\begin{equation}\label{5d_gauged_action}
\begin{split}
    \mathrm{S} &= \frac{1}{2\kappa}\int \mathrm{d}^5x \sqrt{-g}\left[R - \frac{1}{2}(\partial\Phi_1)^2- \frac{1}{2}(\partial\Phi_2)^2  - \frac{1}{4}\sum _{i=1}^3X_i^{-2}F^i_{\mu\nu}F^{i\, \mu\nu} + \frac{1}{4}\epsilon^{\mu\nu\rho\sigma\lambda}A_{\mu}^1F_{\nu\rho}^2F_{\sigma\lambda}^3+ \frac{4}{L^2}\sum_{i=1}^3X_{i}^{-1}\right],\\
    &\qquad  F^i=\mathrm{d}A^i, \quad X_i=e^{-\frac{1}{2}\vec{a}_i\cdot\vec{\Phi}},\quad \vec{\Phi}=(\Phi_1,\Phi_2),\\
    &\vec{a}_1 = \left(\frac{2}{\sqrt{6}},\sqrt{2}\right), \quad \vec{a}_2 = \left( \frac{2}{\sqrt{6}},-\sqrt{2}\right), \quad \vec{a}_3 = \left(-\frac{4}{\sqrt{6}},0\right).
    \end{split}
\end{equation}
{Here $\kappa$ is related to Newton's constant in 5d as $\kappa^2=8\pi G_{5}$ and the coupling of the gauged supergravity has been set to $g=L^{-1}$.}
The equations of motion for this theory are the Bianchi identities and Maxwell equations for the field strengths, Einstein's equations for the metric as well as the equations of motion for the two scalar fields. They read: 

\begin{equation}\label{EOM_gauged_5d_1}
\begin{split}
    & \mathrm{d}\left(\sqrt{-g}X_i^{-2}F^i\right) = 0, \quad 
    \mathrm{d}\star\left(\sqrt{-g}X_i^{-2} F^i\right) = 0,
\end{split}
\end{equation}

\begin{equation}\label{EOM_gauged_5d_2}
\begin{split}
    & G_{\mu\nu} = \frac{1}{2}T^{\Phi}_{\mu\nu} + \sum_{i=1}^3 \frac{1}{2X_i^2}T^i_{\mu\nu}, \quad 
    T^i_{\mu\nu} = F^i_{\mu\rho}F^{i\, \rho}_{\nu} - \frac{1}{4}g_{\mu\nu}F^i_{\rho\sigma}F^{i\, \rho\sigma} \, ,\\
    &T^{\Phi}_{\mu\nu} = \partial_{\mu}\Phi_1 \partial_{\nu}\Phi_1 +\partial_{\mu}\Phi_2 \partial_{\nu}\Phi_2 - g_{\mu\nu}\left[\frac{1}{2}(\partial\Phi_1)^2 + \frac{1}{2}(\partial\Phi_2)^2 - \frac{4}{L^2}\sum_{i=1}^3X_i^{-1}\right],
    \end{split}
\end{equation}

\begin{equation}\label{EOM_gauged_5d_3}
    \begin{split}
        & \Box \Phi_1 = \sum _{i=1}^3 \left[ -\frac{1}{2}X_i^{-3}\left(\frac{\partial X_i}{\partial\Phi_1}\right)F^i_{\mu\nu}F^{i\, \mu\nu} + \frac{4}{L^2}X_i^{-2}\left(\frac{\partial X_i}{\partial\Phi_1}\right)\right],\\
        & \Box \Phi_2 = \sum _{i=1}^3 \left[ -\frac{1}{2}X_i^{-3}\left(\frac{\partial X_i}{\partial\Phi_2}\right)F^i_{\mu\nu}F^{i\, \mu\nu} + \frac{4}{L^2}X_i^{-2}\left(\frac{\partial X_i}{\partial\Phi_2}\right)\right].
    \end{split}
\end{equation}
{Where we defined $\Box \Phi = \frac{1}{\sqrt{-g}}\partial_{\mu} \left(\sqrt{-g}\partial^{\mu}\Phi\right)$.}
We now consider a consistent truncation of \eqref{5d_gauged_action}, in the sense that the solution we present below can be embedded in the above system:

\begin{equation}\label{5d_soliton}
    \begin{split}
        &\mathrm{d}s^2_5=\frac{r^2\lambda^2(r)}{L^2}\left( -\mathrm{d}t^2+\mathrm{d}z^2+\mathrm{d}w^2+L^2F(r)\mathrm{d}\phi^2\right)+\frac{\mathrm{d}r^2}{r^2\lambda^4(r)F(r)},\\
         &\Phi_1 \equiv \Phi =\sqrt{\frac{2}{3}}\mathrm{ln}\lambda^{-6}(r),\quad \Phi_2 = 0,\\
        & A^1 = A^2 = q_1\left[\lambda^6(r)-\lambda^6(r_{\star})\right]L\, \mathrm{d}\phi,\quad A^3 = q_2\left[\frac{1}{\lambda^6(r)} - \frac{1}{\lambda^6(r_{\star})}\right]L\, \mathrm{d}\phi,\\
        &  F(r) =\frac{1}{L^2}-\frac{\varepsilon\ell^2 L^2}{r^4}\left(q_1^2 -\frac{q_2^2}{\lambda^6(r)}\right),\quad \lambda^6(r)=\frac{r^2+\varepsilon\ell^2}{r^2}.
    \end{split}
\end{equation}

The warp function $F(r)$ has a largest root which we denote as $r_{\star}$ and is responsible for ending the space at the finite point $r=r_{\star}$, $\ell$ is a parameter, while according to \cite{Anabalon:2024che} $\varepsilon=\pm 1$ distinguishes between two non-diffeomorphic branches of the supergravity solution. The truncated action from \eqref{5d_gauged_action} reads:

\begin{equation}
\begin{split}
    \mathrm{S} &= \frac{1}{2\kappa}\int d^5x \sqrt{-g}\left[R - \frac{1}{2}(\partial\Phi)^2+ \frac{4}{L^2}\left(\frac{2}{X}+X^2\right) - \frac{1}{2X^2}F_{\mu\nu} ^{1} F^{1\, \mu\nu} - \frac{1}{4}X^4F^3_{\mu\nu} F^{3\, \mu\nu}\right],\\
&    X=e^{-\Phi/\sqrt{6}}.
\end{split}
\end{equation}

 Let us also note that this metric asymptotes to that of $\mathrm{AdS}_5$ as $r\to \infty$, since $F(r)\to L^{-2}$ and $\lambda(r)\to 1$. One can think of the above solution as a double analytic continuation of electrically charged black hole solutions in $\mathrm{U}(1)^3$ truncated 5d supergravity found in \cite{Cvetic:1999xp}. One can also make use of the ansatz presented in the same work and uplift \eqref{5d_soliton} to a solution of the equations of motion of type IIB supergravity, that is Ricci flat\footnote{The dilaton of this solution is zero and therefore its equation of motion yields $R=0$.}. This solution was also included in \cite{Anabalon:2024che} and we study it in section \ref{IIB_background}.

\subsection{Singularity study}\label{singularity_section}
Before continuing to embeddings of this solution in other supergravities and string backgrounds, let us study more closely the singularity structure in 5d. The contents of this subsection will hold for all the uplifted solutions that we will present in section \ref{New_BG}. From here onwards, we will only consider the supersymmetric case for which we set\footnote{According to the Killing spinor analysis performed in \cite{Anabalon:2024che}, in the 5d solution the supersymmetric point happens when $q_1=-q_2$, while when the solution is uplifted to the 10d string background of section \ref{IIB_background} the relation $q_1=q_2$ also provides supersymmetric preservation. We abuse this detail by writing $Q=|q_1|=|q_2|$ in the 5d solution as well.} $|q_1|=|q_2|=Q$ and four real supercharges are preserved. This is a choice, as the various expressions will be valid even for $q_1$, $q_2$ not satisfying this condition, and the limit $Q\to 0$ corresponds to $q_1\to0,q_2\to 0$. We also choose to work with the dimensionless variable $\xi=\frac{r}{r_{\star}}\geq1$ and the parameter $\hat{\nu}=\varepsilon\frac{\ell^2}{r_\star^2}$, in terms of which the gauge fields and functions take the form:

\begin{equation}
    \begin{split}
    &A^1=A^2 = Q\hat{\nu}L\left(1-\frac{1}{\xi^2}\right)\mathrm{d}\phi,\quad A^3 = \frac{Q\hat{\nu}L(\xi^2-1)}{(1+\hat{\nu})(\xi^2+\hat{\nu})}\mathrm{d}\phi,\\
        &\lambda ^6(\xi)=\frac{\xi^2+\hat{\nu}}{\xi^2},\quad F(\xi)=\frac{(\xi^2-1)\left[ \xi^4+(1+\hat{\nu})\xi^2+1+\hat{\nu} \right]}{L^2\xi^4(\hat{\nu}+\xi^2)},
    \end{split}
\end{equation}
where the condition $F(r_\star)=0$ was used to eliminate $Q$ from the last expression\footnote{$F$ can be brought to the following form in the $r$ variable, isolating the root $r_\star$ from the fourth order polynomial: $$F(r)=(r^2-r_\star^2)\left[ \frac{r^4+(r_\star^2+\varepsilon \ell^2)r^2 + r_\star^2 (r_\star^2 + \varepsilon \ell^2)}{L^2 r^4 (r^2 + \varepsilon\ell^2)}\right].$$}. First we fix the periodicity of the cigar-like coordinate $\phi$ in order to avoid conical singularities. The expansions of the various quantities near the end of the space $\xi= 1$ yield\footnote{Here we denote as $A_{i\phi}$ the components of each 1-form $A^i$.}:
\\
\begin{equation}
\begin{split}
 &\lambda(\xi)\approx (1+\hat{\nu})^{1/6}+\mathcal{O}(\xi-1), ~~~
F(\xi)\approx \frac{2(2\hat{\nu}+3)(\xi-1)}{L^2(1+\hat{\nu})}+\mathcal{O}((\xi-1)^2),\\
 & A_{1\phi}\approx 2 Q\hat{\nu}(\xi-1)+ \mathcal{O}((\xi-1)^2), ~~ A_{3\phi}\approx \frac{(6+4\hat{\nu})}{(1+\hat{\nu})^2}(\xi-1)+ \mathcal{O}((\xi-1)^2).
\end{split}
\end{equation}
Then the metric of the subspace spanned by the angle $\phi$ and the radial coordinate takes the form:

\begin{equation}
\begin{split}
    \mathrm{d}s^2_{r,\phi} &\approx \frac{L^2}{(4\hat{\nu}+6)(\xi-1)}\mathrm{d}\xi^2 + \frac{r_\star^2 (\xi -1)(4\hat{\nu} + 6)}{L^2 (1+\hat{\nu})}\mathrm{d}\phi^2= \mathrm{d}\rho^2 + \frac{ r_\star^2 (4\hat{\nu} + 6)^2}{4L^4 (1+\hat{\nu})} \; \rho^2 \;\mathrm{d}\phi^2 ,\\
    \end{split}
\end{equation}
where in the last step we defined the coordinate $\rho=2L\sqrt{\frac{\xi-1}{4\hat{\nu}+6}}$. If we now redefine the angle to be 
\begin{eqnarray}
    \phi = \frac{2L^2 \sqrt{1+\hat{\nu}}}{r_\star(4\hat{\nu}+6)}\hat\phi,
\end{eqnarray}
the metric is free of conical singularities, as it describes a flat space in cylindrical coordinates $(\rho,\hat{\phi})$ with $\rho\geq 0,\quad \hat{\phi}\in [0,2\pi)$. The latter range of values forces the original cigar coordinate to range as:

\begin{equation}
    \phi\in [0,L_\phi),\quad \text{with}\,\,\, L_\phi=\frac{4\pi L^2\sqrt{1+\hat{\nu}}}{r_\star(4\hat{\nu}+6)}.
\end{equation}

We comment that even though requiring a singularity-free metric is not a feature directly related the confining properties of the dual field theory, it can be the case that a conical singularity in the IR may spoil confinement. This is because even if a singular background does lead to an area law for the Wilson loop, small perturbations of the coordinates will not be necessarily well behaved, as their masses are weighted by components of $R_{\mu\nu\rho\sigma}$ \cite{Canneti:2025rsp} which can be divergent. Overall, it is preferred to have a smooth IR geometry in order to gain a better control over the string corrections, as having a singularity would lead to ambiguous descriptions of the dual QFT.

Let us now focus on studying the root structure of the sixth order polynomial $F$ responsible for capping off the space, which can be rewritten as:

\begin{equation}
\begin{split}
&  F(\xi)= \frac{(\xi^2-1)\left(\xi^2-\xi^2_+ \right)\left(\xi^2 - \xi^2_-\right)}{L^2\xi^4(\hat{\nu}+\xi^2)} \quad {\rm with} \quad 
\xi_\pm^2 = \frac{-(1+\hat{\nu})\pm \sqrt{(\hat{\nu}+1)(\hat{\nu}-3)}}{2}.
\end{split}
\end{equation}

We should first determine the range of possible values for the parameter $\hat{\nu}$ to figure if real values for the roots $\xi_\pm$ can occur. For this, we can directly solve for the equation $F(r_\star)=0$ with the form of $F$ appearing in \eqref{5d_soliton} to acquire an expression in terms of the charge $Q$ and the parameters:

\begin{equation}\label{rstar_explicit}
    \begin{split}
        &r_\star=\frac{1}{\sqrt{6}}\sqrt{2^{2/3}\ell^{4/3}\Lambda + 2\ell^{2}\varepsilon\left(-1+\frac{2^{1/3}\ell^{2/3}\varepsilon}{\Lambda}\right)},\\
        &\Lambda:=\left[ -2\varepsilon \ell^2+3Q\left(9L^4Q+L^2\sqrt{81L^4Q^2-12\ell^2\varepsilon}\right) \right]^{1/3} \, .
    \end{split}
\end{equation}

We notice by taking the limit $Q\to 0$ in \eqref{rstar_explicit} we get $r_\star\to 0$ for $\varepsilon=+1$ and $r_\star\to \ell$ for $\varepsilon\to -1$. In fact, substituting \eqref{rstar_explicit} in the definition of $\hat{\nu}$ reveals that in the $\varepsilon=+1$ branch $\hat{\nu}$ is manifestly positive, while in the branch where $\varepsilon=-1$ it is bounded from below by the value $\hat{\nu}=-1$ for all values of $\ell,L$ and $Q$. Using all of the above we deduce the following:\\

For $\hat{\nu}>-1$ there are no real values for $\xi_\pm$ and therefore the only root of $F$ is $\xi=1$, or $r=r_\star$, where the space terminates smoothly. \\
 
 However when $\hat{\nu}=-1$, which can only be reached when $Q=0$, in which case $\varepsilon=-1$ and $r_\star=\ell$, we have $\xi_+ =\xi_-=0$ and therefore $F$ takes the constant value $L^{-2}$. Moreover, the scalar field $\Phi$ diverges at this point, since $\lambda(r)\to 0$, while the metric component along $\phi$ vanishes.

\begin{figure}[t]
\centering
\begin{subfigure}{0.48\linewidth}
\includegraphics[width=\linewidth]{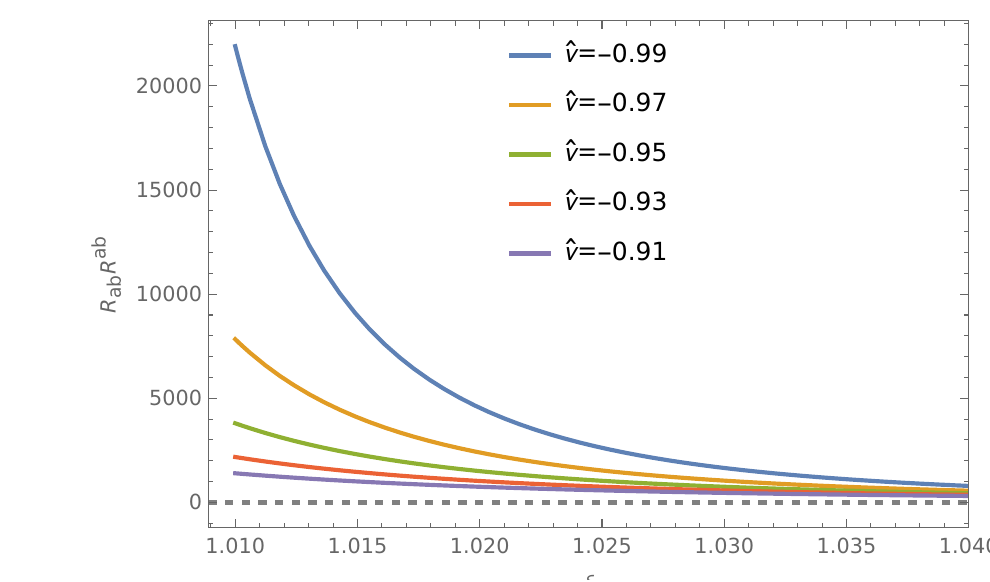}
\caption{Contraction of the Ricci tensor with itself as a function of the dimensionless radial coordinate $\xi$, for various values of $\hat\nu\approx-1$.}
\end{subfigure}
\hfill
\begin{subfigure}{0.48\linewidth}
\includegraphics[width=\linewidth]{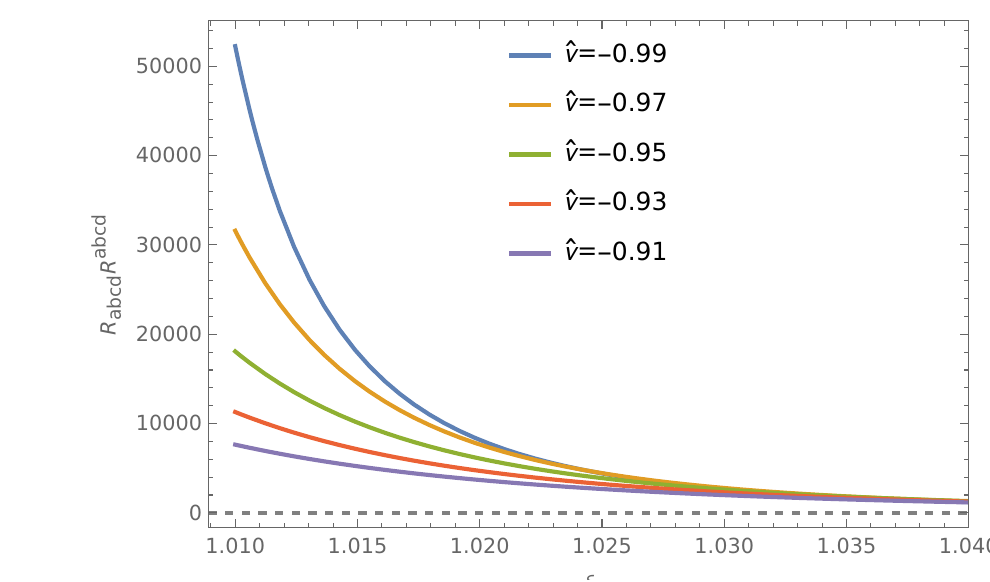}
\caption{Contraction of the Riemann tensor with itself as a function of the dimensionless radial coordinate $\xi$, for various values of $\hat\nu\approx-1$.}
\end{subfigure}
\caption{Invariants of the 5d geometry \eqref{5d_soliton} for values of the parameter $\hat\nu$ close to $-1$. Approximately below the value $-0.95$ they start growing rapidly.}
\label{geometry_invariants_5d}
\end{figure}

We make the claim that at the later value of the parameter $\hat\nu=-1$ and at the point $\xi=1$, the solution is singular (in both Einstein and string frames). To safely deduce this, we plot two invariants of the geometry ($R_{ab}R^{ab}$ and $R_{abcd}R^{abcd}$). These are depicted in figure \ref{geometry_invariants_5d}. What we find is that indeed, their expressions diverge when $\hat\nu=-1 \,\, \& \,\, \xi=1$ which is a confirmation of the metric being singular at this point. Notice however that if we restrict ourselves to study solutions with $\hat\nu>-1$, but still explore the values close to $\hat\nu=-1$, the invariants grow extremely large as $\hat\nu\to-1^+$ but the metric is {\it smooth}.  When the soliton is uplifted to Type IIB--see section \ref{IIB_background}, the point $\hat\nu=-1$ can be thought of as the Coulomb branch singularity in the 10d solution of \cite{Freedman:1999gk}. In fact, many of the gravity systems participating in the holographic modelling of confinement include such naked singularities, see \cite{Girardello:1999hj,Kehagias:1999iy,Brandhuber:1999jr}, which often have 5d origins. The presence of such singularities makes the analysis of observables not trustable.
We wish to emphasize that in our case, we have a {\it smooth geometry} but there is still a region of very large curvatures for specific value of the parameters ($\hat{\nu}\approx -1^+$), at which the supergravity approximation is afflicted by higher curvature corrections that are as important as the leading term. For our purposes, this means that there are some states in the $\mathrm{IR}$ spectrum of the dual $\mathrm{QFT}$ that are not captured by the supergravity approximation.


\section{New deformed Supergravity backgrounds}\label{New_BG}

Having made many of our arguments in the 5d gauged supergravity system, we now present the three types of uplifts of $\eqref{5d_soliton}$ to string theory and eleven dimensional supergravity, in order to apply holography. The first is the solution in type IIB presented in \cite{Anabalon:2024che}, while the rest are new infinite families of solutions of 11d supergravity and type IIA supergravity respectively, introduced in \cite{Chatzis:2025dnu}.


\subsection{Type $\mathrm{IIB}$ background}\label{IIB_background}
In this section we review the ten-dimensional supergravity background constructed in section 5 of \cite{Anabalon:2024che}. This is an uplift to type IIB of the 5d solution we presented earlier, which amounts to further deforming the background of a distribution of D3 branes found in \cite{Freedman:1999gk} (in particular the cases of $n=2$ ($\varepsilon=+1$) and $n=4$ ($\varepsilon=-1$) found in Table 1 of section 2 of \cite{Freedman:1999gk}). These geometries describe a Coulomb branch deformation of $\mathcal{N}=4$ SYM by a nonzero VEV of an operator of dimension $\Delta=2$ proportional to $\varepsilon\ell^2/L^2$ (already presented in \cite{Freedman:1999gk}) as well as a VEV of an operator of dimension $\Delta=3$. The metric takes the form:
\begin{equation}\label{S5-1}
    \begin{split}
        \mathrm{d}s^2&=\frac{\zeta(r,\theta)}{L^2}\left[ r^2(-\mathrm{d}t^2+\mathrm{d}w^2+\mathrm{d}z^2+L^2F(r)\mathrm{d}\phi^2)+\frac{L^2\mathrm{d}r^2}{F(r)r^2\lambda^{6}(r)}+L^4\mathrm{d}\theta^2 \right]\\
        & +\frac{L^2}{\zeta(r,\theta)}\left[ \cos^2\theta\mathrm{d}\psi^2+\cos^2\theta\sin^2\psi \mathrm{D}\phi_1^2 + \cos^2\theta\cos^2\psi \mathrm{D}\phi_2^2 + \lambda^6(r)\sin^2\theta\mathrm{D}\phi_3^2\right],
    \end{split}
\end{equation}
where $\mathrm{D}\phi_i=\mathrm{d}\phi_i+\frac{A^i}{L}$, the function $\zeta$ is defined to be:
\begin{equation}
     \zeta(r , \theta) = \sqrt{1 + \varepsilon \frac{\ell^2}{r^2} \cos^2\theta} ,
\end{equation}
while $\lambda$ and $F$ are defined in \eqref{5d_soliton}. The range of values for the angles parametrizing the deformed $\mathbb{S}^5$ are: $\theta,\psi\in[0,\frac{\pi}{2}]$, $\phi_i\in[0,2\pi)$.
The solution also contains an self-dual RR five-form:
\begin{equation}
 F_5 = (1 + \star) G_5 \, ,
\end{equation}
where
\begin{equation}\label{typeIIB_flux}
 \begin{aligned}
  G_5 & = \frac{2 r^3}{L^4} \big( 1 + \zeta^2 \big) \mathrm{d}t \wedge \mathrm{d}w \wedge \mathrm{d}z \wedge \mathrm{d}\phi \wedge \mathrm{d}r
  - \frac{\varepsilon \ell^2 r^2 F}{L^2} \sin(2\th) \mathrm{d}t \wedge \mathrm{d}w \wedge \mathrm{d}z \wedge \mathrm{d}\phi \wedge \mathrm{d}\theta
  \\[5pt]
  & - \frac{r^3 A'_{1\phi}}{2 L} \cos^2 \theta\, \sin(2 \psi) \mathrm{d}t \wedge \mathrm{d}w \wedge \mathrm{d}z \wedge \mathrm{d}(\phi_1 - \phi_2) \wedge \mathrm{d}\psi
  \\[5pt]
  & - \frac{r^3}{2 L} \sin(2\th) \mathrm{d}t \wedge \mathrm{d}w \wedge \mathrm{d}z \wedge \mathrm{d}\theta\wedge \Big( A'_{1\phi} \big( \sin^2\psi \, \mathrm{d}\phi_1 + \cos^2\psi \, \mathrm{d}\phi_2 \big)
  \\[5pt]
  & - \lambda^{12} A'_{3 \phi} \, \mathrm{d}\phi_3 + \frac{1}{L} \big( A_{1\phi} A'_{1\phi} - \lambda^{12} A_{3\phi} A'_{3\phi} \big) \mathrm{d}\phi \Big) \, .
 \end{aligned}
\end{equation}
Equivalently, we can express things in terms of a frame, in the mostly plus metric convention:
\begin{equation}\label{S5-1-frame}
 \begin{aligned}
  & e^0 = \frac{\sqrt{\zeta}}{L} \, r \, \mathrm{d}t \, , \qquad
     e^1 = \frac{\sqrt{\zeta}}{L} \, r \, \mathrm{d}w \, , \qquad
     e^2 = \frac{\sqrt{\zeta}}{L} \, r \, \mathrm{d}z \, , \qquad
     e^3 = \sqrt{\zeta \, F} \, r \, \mathrm{d}\phi \, ,
 \\[5pt]
  & e^4 = \sqrt{\frac{\zeta}{F}} \frac{\mathrm{d}r}{\lambda^3 r} \, , \qquad
     e^5 = L \sqrt{\zeta} \, \mathrm{d}\theta \, , \qquad
     e^6 = L \frac{\cos\theta}{\sqrt{\zeta}} \, \mathrm{d}\psi \, ,
  \\[5pt]
  & e^7 = L \frac{\cos\theta \, \sin\psi}{\sqrt{\zeta}} \, \Big( \mathrm{d}\phi_1 + \frac{A_1}{L} \Big) \, , \qquad
     e^8 = L \frac{\cos\theta \, \cos\psi}{\sqrt{\zeta}} \, \Big( \mathrm{d}\phi_2 + \frac{A_2}{L} \Big) \, ,
 \\[5pt]
 & e^9 = L \frac{\lambda^3 \sin\theta}{\sqrt{\zeta}} \, \Big( \mathrm{d}\phi_3 + \frac{A_3}{L} \Big) \, ,
 \end{aligned}
\end{equation}
in which case:
\begin{equation}\label{typeIIB_flux_vielbein}
    \begin{split}
        G_5 &=\frac{2\lambda^3(1+\zeta^2)}{L\zeta^{5/2}}e^{01234}-\frac{\varepsilon\ell^2\sqrt{F}\sin(2\theta)}{r^2\zeta^{5/2}}e^{01235}-\frac{\sin\theta\sin\psi A^{\prime}_{1\phi}}{\zeta^{3/2}}e^{01257}\\
        &-\frac{\sin\theta\cos\psi A^{\prime}_{1\phi}}{\zeta^{3/2}}e^{01258}+\frac{\cos\psi A^{\prime}_{1\phi}}{\sqrt{\zeta}}e^{01267}-\frac{\sin\psi A^{\prime}_{1\phi}}{\sqrt{\zeta}}e^{01268}+\frac{\cos\theta A^{\prime}_{3\phi}\lambda^9}{\zeta^{3/2}}e^{01259},
    \end{split}
\end{equation}
where we abbreviate $e^{abcdf}=e^a\wedge e^b\wedge e^c\wedge e^d\wedge e^f$. We denote with a prime the derivative  with respect to the $r$-coordinate.

Alternatively, the five-form $G_5$ can be expressed in terms of a four-form potential $C_4$ as $G_5 = \mathrm{d}C_4$, where
\begin{equation}\label{C4}
 \begin{aligned}
  C_4 & = \Bigg[ \frac{r^4 \zeta^2}{L^4} - \frac{\ell^4}{2} \Big( \frac{q^2_1}{r^2_\star} - \frac{q_2}{L} \frac{A_{3\phi}}{r^2 - r^2_\star} \frac{\lambda^6}{\lambda^6 - 1} \Big) \cos(2 \theta) \Bigg] \mathrm{d}t \wedge \mathrm{d}w \wedge \mathrm{d}z \wedge \mathrm{d}\phi
  \\[5pt]
  & - \frac{\ell^2 \varepsilon}{2} \cos(2 \theta) \mathrm{d}t \wedge \mathrm{d}w \wedge \mathrm{d}z \wedge \Big( q_1 \big( \sin^2 \psi \, \mathrm{d}\phi_1 + \cos^2 \psi \, \mathrm{d}\phi_2 \big) - q_2 \, \mathrm{d}\phi_3 \Big)
  \\[5pt]
  & + \frac{q_1 \ell^2 \varepsilon}{2} \cos^2 \psi \, \mathrm{d}t \wedge \mathrm{d}w \wedge \mathrm{d}z \wedge \mathrm{d}(\phi_1 - \phi_2) \, .
 \end{aligned}
\end{equation}
The Hodge dual of $G_5$ can also be written in terms of a four-form 
$\tilde{C}_4$ (see \eqref{S5-2}) as $\star G_5 = \mathrm{d}\tilde{C}_4$.
The self-dual five-form can then be expressed in terms of $C_4$ and $\tilde{C}_4$ as $F_5 = \mathrm{d}\big( C_4 + \tilde{C}_4 \big)$. Notice that $\lambda_\star$ and $\zeta_\star$ refer to the values of the functions 
$\lambda(r)$ and $\zeta(r , \theta)$ at $r = r_\star$.\\

Lastly, we wish to comment that in this IIB background, the singular locus we described in section \ref{singularity_section} is located, after setting $\hat\nu=-1$, along the curve:
\begin{equation}
    \xi^2-\cos^2\theta=\left( \frac{r}{r_{\star}}\right)^2-\cos^2\theta=0,
\end{equation}
as when this equation is satisfied the geometric invariants diverge. Therefore the singularity appears when $r=r_\star=\ell$ in the direction $\theta=0$. Contrast this with the 5d gauged supergravity where there is no such angle.

Since the soliton we used for the uplift is a consistent truncation of 10d type IIB compactified on $\mathbb{S}^5$, the Killing spinors of the 5d gauged supergravity will uniquelly correspond to Killing spinors of the uplifted solution. The study of preserved supersymmetries was done in detail in \cite{Anabalon:2024che}, where two antiperiodic in $\phi$ complex Killing spinors were identified. Regarding the preserved ${\cal R}$ symmetry, it is a diagonal subgroup $\mathrm{SO}(2)_{\cal R}^3\subset \mathrm{SO}(6)_{\cal R}$. 


\subsection{Embedding in Romans' $\mathrm{SU(2)}\times\mathrm{U}(1)$ 5d gauged Supergravity}

We will now present an embedding of \eqref{5d_soliton} in another five-dimensional gauged supergravity introduced by Romans' in \cite{Romans:1985ps}. The general content of the bosonic sector of this theory consists of a metric, one scalar field $X$, an abelian gauge field $B$, an $\mathrm{SU}(2)$ gauge field $\mathcal{A}^i$ ($i=1,2,3$) and a complex two-form $C$ charged under $B$. The field strengths for these gauge potentials  are:
\begin{equation}
    \begin{split}
        & G = \mathrm{d}B,\\
        &\mathcal{F}^i = \mathrm{d}\mathcal{A}^i - \frac{1}{\sqrt{2}}m\epsilon_{ijk}\mathcal{A}^j\wedge \mathcal{A}^k,\\
        & F = \mathrm{d}C + imB\wedge C,
    \end{split}
\end{equation}
 where the parameter $m$, called the Romans' mass, is related to the $\mathrm{SU}(2)$ coupling. The equations of motion for the theory can be derived from the following five-form Lagrangian density:
\begin{equation}
\begin{split}
     \mathcal{L} =& R\star 1 -3X^{-2}\star \mathrm{d}X\wedge \mathrm{d}X - \frac{1}{2}\star G \wedge G - \frac{1}{2}X^{-2} (\star \mathcal{F}^i\wedge \mathcal{F}^i + \star C \wedge\bar{C})\\
                  &-\frac{i}{2m}C\wedge\bar{F}-\frac{1}{2}\mathcal{F}^i\wedge \mathcal{F}^i\wedge B + 4m^2 (X^2 + 2X^{-1})\star 1,
 \end{split}
 \end{equation}
and they read:
 \begin{equation}\label{Romans_scalar_fluxes_EOM}
     \begin{split}
         & \mathrm{d}(X^{-1}\star \mathrm{d}X) = \frac{1}{3}X^4 \star G\wedge G - \frac{1}{6}X^{-2}\left(\star \mathcal{F}^i\wedge \mathcal{F}^i + \star C\wedge \bar{C}\right) - \frac{4}{3}m^2(X^2 - X^{-1})\star 1,\\
         & \mathrm{d}(X^4\star G) = -\frac{1}{2}\mathcal{F}^i\wedge \mathcal{F}^i - \frac{1}{2}\bar{C}\wedge C \, , \quad  
         D(X^{-2}\star \mathcal{F}^i) = - \mathcal{F}^i\wedge G \, , \quad X^2\star F = im C,
     \end{split}
 \end{equation}
 \begin{equation}\label{Romans_Einstein}
    \begin{split}
        R_{\mu\nu} =& 3X^{-2}\partial_{\mu}X\partial_{\nu}X - \frac{4}{3}m^2(X^2+2X^{-1})g_{\mu\nu} \\
        & + \frac{1}{2}X^4(G_{\mu}^{\rho}G_{\nu\rho}-\frac{1}{6}g_{\mu\nu}G_{\rho\sigma}G^{\rho\sigma}+\frac{1}{2}X^{-2}(\mathcal{F}^{i\, \rho}_{\mu}\mathcal{F}^i_{\nu\rho}-\frac{1}{6}g_{\mu\nu}\mathcal{F}^i_{\rho\sigma}\mathcal{F}^{i\, \rho\sigma})\\
        &+\frac{1}{2}X^{-2}\left[ C_{(\mu}^{\rho}\bar{C}_{\nu)\rho} - \frac{1}{6}C_{\rho\sigma}\bar{C}^{\rho\sigma}\right] \, ,
    \end{split}
\end{equation}
where we defined the covariant derivative $\mathrm{D}(X^{-2}\star \mathcal{F}^i):=\mathrm{d}(X^{-2}\star \mathcal{F}^i) + \sqrt{2}m\epsilon_{ijk}\mathcal{A}^k\wedge (X^{-2}\star \mathcal{F}^j)$.\\

We can embed the solution \eqref{5d_soliton} in this theory by defining the fields in the following way: 
\begin{equation}\label{Romans_configuration}
   X = e^{-\frac{1}{\sqrt{6}}\Phi}=\lambda^2(r), \quad B = A^3, \quad \mathcal{A}^3=\sqrt{2}A^1 \, \quad \mathcal{A}^1 = \mathcal{A}^2 = 0, \quad C = 0.
\end{equation}
After identifying $m^2=L^{-2}$, the equations of motion for the configuration \eqref{Romans_configuration} are satisfied, which coincide with the ones presented in the previous section \eqref{EOM_gauged_5d_1}-\eqref{EOM_gauged_5d_3}.


\subsection{Eleven dimensional Supergravity backgrounds}

In the following, we will present new embeddings of the solution \eqref{Romans_configuration} of Romans' 5d gauged supergravity in the 11 dimensional background of Lin, Lunin and Maldacena \cite{Lin:2004nb}. These are $1/2$-BPS solutions in M-theory preserving 16 supercharges, that are dual to 4 dimensional SCFTs with $\mathcal{N}=2$ supersymmetry. One key feature of these backgrounds, as well as their type IIA reductions we present in \ref{type_IIA_backgrounds}, that differentiates them from the one of section \ref{IIB_background} is that they come with either M5 or D-branes describing flavour degrees of freedom. This means that their dual QFTs contain fundamental matter.\\

We follow the uplift formula of Gauntlett and Varela, see \cite{Gauntlett:2007sm} as well as \cite{Gauntlett_2006} to translate between the notation. The resulting 11d deformed geometries take the form\footnote{Where $\kappa=\frac{\pi}{2}l_p^3$. We will also set $m=1$ for simplicity from here onwards}:
    \begin{align}\label{ds11_LLM1}
        \frac{\mathrm{d}s^2_{11}}{\kappa^{2/3}} = e^{2\hat\lambda}\mathcal{Z}&\left\{ \frac{4}{X}\mathrm{d}s^2_5 + \frac{y^2e^{-6\hat\lambda}}{\mathcal{Z}^3} \mathrm{D}\mu^i\mathrm{D}\mu^i+\frac{4X^3}{\mathcal{Z}^3}\frac{\mathrm{D}\chi^2}{1-y\partial_y D_0}- \frac{\partial_yD_0}{y}\mathrm{d}y^2-\frac{\partial_ye^{D_0}}{y}(\mathrm{d}v_1^2+\mathrm{d}v_2^2)\right\}
    \end{align}
where $\mathrm{d}s^2_5$ stands for the metric in \eqref{5d_soliton} and we define the following quantities:
\begin{equation}\label{ds11_LLM2}
    \begin{split}
    & \mathcal{Z}=\left[ 1+y^2 e^{-6\hat\lambda}(X^3-1)\right]^{1/3},\quad X(r)=\lambda^2(r),\\
    &\mathrm{D}\chi = \mathrm{d}\chi + a_1 + B,\quad  a_1 = \frac{1}{2}(\partial_{v_2}D_0 \mathrm{d}v_1 - \partial_{v_1}D_0 \mathrm{d}v_2),\\
    &B = A^3 =\frac{q_2L\varepsilon\ell^2(r^2-r_{\star}^2)}{(r_{\star}^2+\varepsilon\ell^2)(r^2+\varepsilon\ell^2)}\, \mathrm{d}\phi,\quad \mathcal{A}^{(3)} = \sqrt{2}A^1 =\sqrt{2}q_1L\varepsilon\ell^2\left(\frac{1}{r^2}-\frac{1}{r_{\star}^2}\right)\mathrm{d}\phi,\\
    &\mathrm{D}\mu^i = (\mathrm{d}\mu^1 + \sqrt{2}\mu_2 \mathcal{A}^{(3)})\delta^{i1} + (\mathrm{d}\mu^2-\sqrt{2}\mu_1\mathcal{A}^{(3)})\delta^{i2} + \mathrm{d}\mu^3\delta^{i3},\\[5pt]
    &\mu^1 = \sin\theta\cos\varphi,\quad \mu^2 = \sin\theta\sin\varphi,\quad \mu^3=\cos\theta \, , \\
    \end{split}
\end{equation}
with the $\mu^i$s spanning a two-sphere. The functions $\hat\lambda$ and $D_0$ both have support on $(v_1,v_2,y)$, are related by: 
\begin{equation}
    e^{-6\hat\lambda} = -\frac{\partial_y D_0}{y(1-y\partial_y D_0)},
\end{equation}
and $D_0$, which determines the solution, satisfies the three dimensional Toda equation: 
\begin{equation}
(\partial_{v_1}^2 + \partial_{v_2}^2)D_0 + \partial_y^2 e^{D_0} = 0.
\end{equation}
The 4-form is given by\footnote{With $G = \mathrm{d}B = \mathrm{d}A^3,\quad \mathcal{F}^{(3)} = \mathrm{d}\mathcal{A}^3 = \sqrt{2}\mathrm{d}A^{1},\quad \mathrm{vol}\tilde{\mathbb{S}}^2 = \frac{1}{2}\varepsilon_{ijk}\mu^i\mathrm{D}\mu^j\mathrm{D}\mu^k$.}: 
\begin{equation}
    G_4 = \tilde{G}_4 + G\wedge \beta_2+\mathcal{F}^{(3)}\wedge \beta_2^{(3)}  + \star_5 \mathcal{F}^{(3)}\wedge\beta_1^{(3)},
\end{equation}
where:
\begin{equation}
    \begin{split}
        \tilde{G}_4= -\frac{\kappa}{4}\mathrm{vol}\tilde{\mathbb{S}}^2\wedge&\left\{8 \mathrm{d}\left[\frac{y(1-y^2e^{-6\hat\lambda})}{1+y^2e^{-6\hat\lambda}(X^3-1)}-y\right]\wedge \mathrm{D}\chi-4\partial_y e^{D_0} \mathrm{d}v_1\wedge \mathrm{d}v_2\right. \\
        &\left.+ 8\frac{y(1-y^2e^{-6\hat\lambda})}{1+y^2e^{-6\hat\lambda}(X^3-1)}\mathrm{d} a_1 \right\},
    \end{split}
\end{equation}
\begin{equation}\label{betas_LLM}
\begin{split}
   & \beta_2 = -2\kappa\frac{X^3y^3e^{-6\hat\lambda}}{1+y^2e^{-6\hat\lambda}(X^3-1)}\mathrm{vol}\tilde{\mathbb{S}}^2 \, , \quad 
   \beta_1^{(3)} = -\frac{\kappa\sqrt{8}}{X^2}\mathrm{d}(y\mu^3) \, ,\\
       &\beta_2 ^{(3)} = \sqrt{8}\kappa\left\{ \left[\mu^3\mathrm{d}y + \frac{y(1-y^2e^{-6\hat\lambda})}{1+y^2e^{-6\hat\lambda}(X^3-1)}\mathrm{D}\mu^3\right]\wedge \mathrm{D}\chi+ \frac{1}{2}\mu^3 \partial_y  e^{D_0} \mathrm{d}v_1\wedge \mathrm{d}v_2\right\}.
    \end{split}
\end{equation}

According to \cite{Gauntlett:2007sm}, the equations of motion for this background, which we have verified to be satisfied, are equivalent to the five-dimensional equations of Romans' gauged supergravity \eqref{Romans_scalar_fluxes_EOM} and \eqref{Romans_Einstein}. 
These are the Bianchi identity and Maxwell equation for the flux $G_4$ as well as the eleven dimensional Einstein equations:
\begin{equation}
\begin{split}
& \mathrm{d} G_4 = 0,\quad  \mathrm{d}\star_{11}G_4 =- \frac{1}{2}G_4\wedge G_4,\\
& R_{AB} = \frac{1}{12}G_{4\, A C_1 C_2 C_3}G_{4\, B} ^{C_1 C_2 C_3} - \frac{1}{144}g_{AB}G_{4\, C_1C_2 C_3 C_4}G_4^{C_1 C_2 C_3 C_4}.
\end{split}
\end{equation}

In these families of deformed backgrounds we have two types of extended objects: colour $\mathrm{M}5$ branes sourcing the $\mathrm{AdS}$ geometry, which wrap the Riemann surface $\Sigma_2$ extending in $(v_1,v_2)$ and extend in the Minkowski directions, as well as flavour $\mathrm{M}5$ branes extending in the holographic direction, the Minkowski subspace and the circle parametrised by $\chi$. One can derive the quantization conditions for their charges by integrating the flux $G_4$ on appropriate four-cycles. There are two options available to construct such cycles given in \cite{Lin:2004nb}, which are both isomorphic to four-spheres:
\begin{equation}
    {\cal M}_4\in\{ \mathbb{S}^4_{\text{colour}},\mathbb{S}^4_{\text{flavour}}\}
\end{equation}
For the case of colour M5 branes, we can take an interval $[0,N]$ in the $y$ coordinate and use the circle $\mathbb{S}^1[\chi]$ and the two-sphere $\mathbb{S}^2[\theta,\varphi]$. We take $y=N$ to be a point at which $\mathbb{S}^1[\chi]$ shrinks smoothly\footnote{smoothness holds if $e^{D_0}\sim N-y$ near $y=N$.}, while we also have that $\mathbb{S}^2[\theta,\varphi]$ shrinks to zero at $y=0$. For $0<y<N$ both $\mathbb{S}^1[\chi]$ and $\mathbb{S}^2[\theta,\varphi]$ have finite size. We can therefore fiber the product $\mathbb{S}^1[\chi]\times\mathbb{S}^2[\theta,\varphi]$ over the interval to construct a total space that is a compact cycle isomorphic to a four-sphere: $\mathbb{S}^4_{\text{colour}}\cong\mathbb{S}^1[\chi]\times\mathbb{S}^2[\theta,\varphi]\times[0,N]$.

For the flavour M5s we focus on the $y=0$ slice of the $(y,v_1,v_2)$ subspace and take a closed curve $\Gamma$ in $\Sigma_2[v_1,v_2]$ enclosing a two-surface that is isomorphic to a disc $D^2$ and sits slightly above $y=0$. We can choose this curve such that $\mathbb{S}^1[\chi]$ collapses smoothly on $\partial D^2=\Gamma$. At any point $p\in \text{int}(D^2)$ we can then fiber the $\mathbb{S}^2[\theta,\varphi]$ which has a finite size. Due to our choice of curve\footnote{If one does not make the choice that $\mathbb{S}^1[\chi]$ vanishes at the boundary of the "cup" $D^2$, the total space will have a non trivial boundary isomorphic to $\mathbb{S}^2\times\mathbb{S}^1$ and will not be compact.} the boundary of this four-manifold vanishes and therefore we have a compact cycle that is also isomorphic to a four-sphere: $\mathbb{S}^4_{\text{flavour}}\cong \mathbb{S}^2[\theta,\varphi]\times D^2$. 

Let us perform the calculation for colour branes. In order to do so, we take the limit of the geometry to the boundary $r\to \infty$, which makes $\mathrm{d}s^2_5\to \mathrm{d}s^2_{\mathrm{AdS}_5}$ and $X(r)\to 1$. At the boundary the deformed sphere $\tilde{\mathbb{S}}^2[\theta,\varphi]$ can be recast as a round sphere $\mathbb{S}^2[\theta,\varphi]$, as the fibration in this limit is just a constant which can be absorbed in the definition of the coordinates. The same is true for the fibration in $\mathbb{S}^1[\chi]$. We then have:

\begin{equation}
\begin{split}
    \int _{\mathbb{S}^4} G_4 &= -2\kappa\int _{\mathbb{S}^2}\mathrm{vol}_{\mathbb{S}^2}\int_0^{2\pi}\mathrm{d}\chi\int_0 ^N \mathrm{d}\left[-y^2e^{-6\hat{\lambda}}\right]=(4\pi)^2\kappa N,    
    \end{split}
\end{equation}
where we used that near $y\sim N$ $e^{D_0}\sim(N-y)$ and $e^{3\hat{\lambda}}=N$, which expresses that $\mathbb{S}^1[\chi]$ shrinks in a non-singular fashion. This yields the following condition:
\begin{equation}
    \frac{1}{16\pi^2\kappa}\int_{\mathbb{S}^4}G_4=N\in \mathbb{N},
\end{equation}
 which agrees with the quantization condition in \cite{Gaiotto:2009gz} using their convention which sets $\kappa=\frac{\pi}{2}l_p^3$ where $l_p$ denotes the Planck length. \\

The flavour M5s correspond to punctures on the Riemann surface spanned by $(v_1,v_2)$. If one considers $K_i$ such punctures at different positions $(v_1^i,v_2^i)$, then the integral of $G_4$ over $\mathbb{S}^4_{\text{flavour}}$ surrounding each puncture will yield $K_i$. As $\mathbb{S}^4_{\text{flavour}}$ is difficult to parametrize in the coordinates of the solution \eqref{ds11_LLM1}, the quantized flux can be more easily calculated by focusing near each puncture, where an additional rotational symmetry in $(v_1,v_2)$ is assumed \cite{Gaiotto:2009gz}. We will explore the solution with this additional symmetry, which gives rise to an electrostatic description, in the following subsection.


\subsection{Type $\mathrm{IIA}$ backgrounds}\label{type_IIA_backgrounds}

We will now present a reduction of the former 11d theory on the M-theory circle, which yields new infinite families of deformed type IIA backgrounds of Gaiotto Maldacena type. These are geometries dual to linear quiver SCFTs enjoying $\mathcal{N}=2$ supersymmetry in four dimensions. We start by rewriting the Riemann surface coordinates as $v_1=\rho\cos\beta,\,\, v_1=-\rho\sin\beta$. To make the reduction work, we need to have rotational symmetry in the $[v_1,v_2]$ subspace, meaning that $\partial_\beta$ generates\footnote{The reason for choosing this coordinate and not another available $\mathrm{U}(1)$ being the preservation of $\mathcal{N}=2$ supersymmetry.} a $\mathrm{U}(1)$ isometry and $D_0=D_0(\rho,y)$. We can now make the following change of coordinates first presented in \cite{Gaiotto:2009gz}, where we map the coordinates $(\rho,y)\mapsto (\sigma,\eta)$ and the Toda function $D_0(\rho,y)\mapsto V(\sigma,\eta)$, where:
\begin{equation}\label{coord_change_to_electrostatic}
    \begin{split}
        &\dot{V}=\sigma\partial_{\sigma}V, \quad V^\prime=\partial_{\eta}V,\\
        &\rho^2 e^{D_0}=\sigma^2, \quad y=\dot{V}, \quad \log \rho=V^\prime,
    \end{split}
\end{equation}
and the new function $V$ satisfies the three dimensional cylindrically symmetric Laplace equation:
\begin{equation}\label{laplace}
    \ddot{V}+\sigma^2 V^{\prime\prime}=0,
\end{equation}
supplemented with appropriate boundary conditions dictated by regularity of the metric. With these changes made, the 11d metric \eqref{ds11_LLM1} takes the form: 
\begin{equation}
    \mathrm{d}s^2_{11}= \tilde{f}_1\left[ 4\tilde{f} \mathrm{d}s^2_5 +\tilde{f}_2\mathrm{D}\mu^i\mathrm{D}\mu^i +\tilde{f}_3(\mathrm{d}\chi + B)^2+\tilde{f}_4(\mathrm{d}\sigma^2+\mathrm{d}\eta^2)+\tilde{f}_5(\mathrm{d}\beta + \tilde{f}_6 \mathrm{d}\chi + \tilde{f}_6 B)^2\right],\label{11dGM}
\end{equation}
where we define the various functions to be

\begin{equation}\label{IIA_functions}
    \begin{split}
        & \tilde{f}_1=\kappa^{2/3}\left(\frac{\dot{V}\tilde{\Delta}}{2V^{\prime \prime}}\right)^{1/3},\quad \tilde{f} = X^{-1} Z,\quad \tilde{f}_2=\frac{2\dot{V}V^{\prime\prime}}{Z^2\tilde{\Delta}},\\
        &\quad \tilde{f}_3 = \frac{4X^3\sigma^2V^{\prime\prime}}{2X^3\dot{V}-\ddot{V}}Z,\quad \tilde{f}_4=\frac{2V^{\prime\prime}}{\dot{V}}Z,\quad \tilde{f}_5 = \frac{2(2X^3\dot{V}-\ddot{V})}{Z^2\dot{V}\tilde{\Delta}},\quad \tilde{f}_6 = \frac{2X^3\dot{V}\dot{V}^{\prime}}{2X^3\dot{V}-\ddot{V}}\\
        &\tilde{\Delta}=(\dot{V}^{\prime})^2+V^{\prime\prime}(2\dot{V}-\ddot{V}),\quad Z=\left[ \frac{(\dot{V}^{\prime})^2+V^{\prime\prime}(2X^3\dot{V}-\ddot{V})}{(\dot{V}^{\prime})^2+V^{\prime\prime}(2\dot{V}-\ddot{V})}\right]^{1/3}.
    \end{split}
\end{equation}
Notice that in the absence of a scalar profile, namely $X=1$, we have $Z=1$ and we recover the deformed background presented in \cite{Chatzis:2024kdu} (see equation (5.20) in that paper), while further taking $r\to \infty$  recovers the original Gaiotto-Maldacena background\footnote{The $\mathrm{AdS}_5$ part of the background is recovered when $r\to\infty$, where we find $\lambda(r)\to 1,F(r)\to L^{-2}$ and $\mathrm{d}s^2_5\to \mathrm{d}s^2_{\mathrm{AdS}_5}$.}. After defining the function:
\begin{equation}
    g(\eta,\sigma) = \frac{\dot{V}^\prime}{\ddot{V}V^{\prime\prime} - (\dot{V}^\prime)^2},
\end{equation}
the expressions entering the flux take the following form

\begin{equation}
    \begin{split}
        \tilde{G}_4 =& 2\,\mathrm{vol}\tilde{\mathbb{S}}^2\wedge\left[  \mathrm{d}\tilde{f}_7 \wedge (\mathrm{d}\chi + B -g \mathrm{d}\beta) +\tilde{f}_8\mathrm{d}g\wedge \mathrm{d}\beta +  \tilde{f}_9\left( 
\sigma V^{\prime\prime} \mathrm{d}\eta + \dot{V}^\prime \mathrm{d}\sigma \right)\wedge \mathrm{d} \beta \right],\\     \beta_1^{(3)} =& -\frac{\kappa \sqrt{8}}{X^2} \mathrm{d}(\mu^3 \dot{V}) \, , \quad 
            \beta_2 =\,2 \, \tilde{f}_7 \, \mathrm{vol}\tilde{\mathbb{S}}^2,\\
            \beta_2^{(3)} =& \sqrt{8}\Bigg\{ \left(\tilde{f}_8 \mathrm{D}\mu^3 + \kappa \mu^3 \mathrm{d} \dot{V} \right) \wedge (\mathrm{d}\chi + B -g \mathrm{d}\beta)+ \mu^3 \tilde{f}_9\left( 
\sigma V^{\prime\prime} \mathrm{d}\eta + \dot{V}^\prime \mathrm{d}\sigma \right)\wedge \mathrm{d} \beta \Bigg\},\label{11dGM-2}
    \end{split}
\end{equation}
where
\begin{equation}
        \begin{split}
  \tilde{f}_7=\frac{2\kappa X^3\dot{V}^2V^{\prime\prime}}{\tilde{\Delta}+2V^{\prime\prime}\dot{V}(X^3-1)},\quad \tilde{f}_8=\frac{\kappa \dot{V}\left[(\dot{V}^{\prime})^2-V^{\prime\prime}\ddot{V}\right]}{\tilde{\Delta}+2V^{\prime\prime}\dot{V}(X^3-1)},\quad \tilde{f}_9=\frac{\kappa \sigma V^{\prime\prime}}{(\dot{V}^{\prime})^2-V^{\prime\prime}\ddot{V}},
        \end{split}
\end{equation}
From here, we can apply the reduction following \cite{Gaiotto:2009gz}, as well as \cite{macpherson2024} for a detailed derivation. The ansatz reads:
\begin{equation}
    \begin{split}
        &\mathrm{d}s^2_{11} = e^{-\frac{2}{3}\Phi}\mathrm{d}s^2_{10} + 
        e^{\frac{4}{3}\Phi}(\mathrm{d}\beta + C_1)^2 
        \quad {\rm with} \quad 
        A_3 = C_3 + B_2 \wedge \mathrm{d}\beta \, 
        \quad \text{and} \quad \mathrm{d}A_3=G_4\label{10dGM}
    \end{split}
\end{equation}
where:

\begin{equation}\label{10dGM-2}
    \begin{split}
        &\mathrm{d}s^2_{10}= \tilde{f}_1^{\frac{3}{2}} \tilde{f}_5^{\frac{1}{2}} \left[ 4\tilde{f} \mathrm{d}s^2_5 + \tilde{f}_2 \mathrm{D}\mu_i\mathrm{D}\mu^i + \tilde{f}_3(\mathrm{d}\chi+B)^2 + \tilde{f}_4(\mathrm{d}\sigma^2 + \mathrm{d}\eta^2) \right]\\
        &e^{\frac{4}{3}\Phi} = \tilde{f}_1\tilde{f}_5, \quad C_1=\tilde{f}_6(\mathrm{d}\chi + B),\\
    &H_3=\mathrm{d}B_2=2\mathrm{vol}\tilde{\mathbb{S}}^2\wedge\Big[-g\,\mathrm{d}\tilde{f}_7+\tilde{f}_8\mathrm{d}g+\tilde{f}_9(\sigma V^{\prime\prime}\mathrm{d}\eta+\dot{V}^{\prime}\mathrm{d}\sigma) \Big]\\
    &+\sqrt{8}\mathcal{F}^{(3)}\wedge \left[ \mu^3\tilde{f}_9(\sigma V^{\prime\prime}\mathrm{d}\eta+\dot{V}^{\prime}\mathrm{d}\sigma)-g\left(\tilde{f}_8\mathrm{d}\mu^3 + \kappa \mu^3\mathrm{d}\dot{V} \right)   \right],\\
        &F_4=\mathrm{d}C_3-H_3\wedge C_1= -2\tilde{f}_7\mathrm{vol}\tilde{\mathbb{S}}^2\wedge G -\frac{\kappa \sqrt{8}}{X^2}\star_{5}\mathcal{F}^{(3)}\wedge\mathrm{d}(\mu^3\dot{V})\\
        &+2 \mathrm{vol}\tilde{\mathbb{S}}^2\wedge \Big[\left(1+g \,\tilde{f}_6\right)\mathrm{d}\tilde{f}_7-\tilde{f}_6\tilde{f}_8\mathrm{d}g-\tilde{f}_6\tilde{f}_9(\sigma V^{\prime\prime}\mathrm{d}\eta+\dot{V}^{\prime}\mathrm{d}\sigma)  \Big]\wedge (\mathrm{d}\chi+B)\\
        &+\sqrt{8}\mathcal{F}^{(3)}\wedge\Big[ - \mu^3\tilde{f}_6\tilde{f}_9(\sigma V^{\prime\prime}\mathrm{d}\eta+\dot{V}^{\prime}\mathrm{d}\sigma)+ (1+g\tilde{f}_6)\left(\tilde{f}_8\mathrm{d}\mu^3 +\kappa \mu^3\mathrm{d}\dot{V} \right)\Big]\wedge(\mathrm{d}\chi+B).
    \end{split}
\end{equation}
We have checked that this configuration solves the type $\mathrm{IIA}$ supergravity equations of motion, in units where the $\mathrm{AdS}$ radius is equal to one. Note that this geometry, still contains an $\tilde{\mathbb{S}}^2[\theta,\varphi]$ (expressing the $\mathrm{SU}(2)_{\mathcal{R}}$ symmetry), as well as the $\mathbb{S}^1[\chi]$. Notably, we can rewrite the NS three-form in terms of total derivatives:
\begin{equation}\label{H3_potential}
     \begin{split}
         &\kappa ^{-1} H_3= \mathrm{d}{\cal K} \wedge\mathrm{d}\Omega +\sqrt{8}{\cal F}^{(3)}\wedge\mathrm{d}\tilde{\Omega}, \quad 
         {\cal K}=4\cos\theta A_{1\phi}(r)\mathrm{d}\phi -2\cos\theta\mathrm{d}\varphi\\
         &\Omega=\eta - \frac{\sigma^2\dot{V}\dot{V}^{\prime}}{\ddot{V}^2-2\dot{V}\ddot{V}\lambda^6(r)+\sigma^2(\dot{V}^{\prime})^2},\quad \tilde{\Omega}=\mu^3(\eta-\Omega),
     \end{split}
 \end{equation}
which we can integrate on appropriate cycles to get local expressions for the NS potential $B_2$. The four possible expressions read as follows:
\begin{equation}\label{various_B2s}
    \begin{split}
        &B_2^{(\mathrm{I})}=4A^1\wedge \mathrm{d}(\mu^3(\eta - \Omega)) + 2\Omega(\mathrm{vol}_{\mathbb{S}^2}+2\mu^3\mathrm{d}A^1),\\
        &B_2^{(\mathrm{II})}= 4\mu^3 \eta\, \mathrm{d}A^1 + 2\Omega \mathrm{vol}_{\mathbb{S}^2},\\
        &B_2^{(\mathrm{III})}=4A^1\wedge \mathrm{d}(\mu^3(\eta - \Omega))+2\mu^3(2A^1 - \mathrm{d}\varphi)\wedge\mathrm{d}\Omega,\\
    &B_2^{(\mathrm{IV)}}=4\mu^3(\eta-\Omega)\mathrm{d}A^1 + 2\mu^3(2A^1 - \mathrm{d}\varphi)\wedge\mathrm{d}\Omega,
    \end{split}
\end{equation}
all of which satisfy $\mathrm{d}B_2 =\kappa^{-1} H_3$ and are related by a gauge transformation:
\begin{equation}
\begin{split}
    &B_2^{(\mathrm{II})}=B_2^{(\mathrm{I})}+ \mathrm{d}\Lambda,\quad B_2^{(\mathrm{IV})}=B_2^{(\mathrm{III})}+\mathrm{d}\Lambda \quad {\rm with} \quad \Lambda = 4\mu^3(\eta - \Omega)A^1.
    \end{split}
    \end{equation}
This will be important for the quantization of Page charges associated with $\mathrm{D}4$ branes, as well as some computations for the observables.

Let us briefly comment on the connection to the dual quiver theory. The new compact coordinate $\eta$ together with $\sigma\in[0,\infty)$ constitute a strip on which the function $V$ is supported. Our approach in presenting this will be to consider a fixed superconformal linear quiver on the boundary, whose data is encoded in the boundary conditions of $V(\sigma,\eta)$. After fixing the boundary theory, the geometry will then yield the appropriately quantized Page charges. If we let $\eta\in[0,P]$, where $P$ is the total length of the linear quiver, the boundary conditions for \eqref{laplace} read \cite{Gaiotto:2009gz}:

\begin{equation}\label{boundary_conditions}
    \dot{V}\Big|_{\eta=0,P}=0,\quad \dot{V}\Big|_{\sigma=0}=\mathcal{R}(\eta),\quad V\Big|_{\sigma\to \infty}=0,
\end{equation}
where $\mathcal{R}$ is a piecewise linear and convex function with finitely many discontinuities in its derivative, $\mathcal{R}'(\eta)$, at integer values of $\eta$ which satisfies $\mathcal{R}(0)=\mathcal{R}(P)=0$. This is called the rank function for the quiver and one choice for it which yields linear quivers with $P-1$ nodes containing gauge groups $\mathrm{SU}(N_i)$ is the following \cite{macpherson2024}:
\begin{equation}\label{generic_R}
 {\cal R}(\eta) = \begin{cases} 
       N_1 \eta & \ , \quad \eta\in [0,1]\\
         N_l+ (N_{l+1} - N_l)(\eta-l) & \ , \quad  \eta\in[l,l+1]\ ,\;\;\; l=1,\ldots, P-2\\
         N_{P-1}(P-\eta) & \ , \quad \eta\in[P-1,P],
      \end{cases} \ .
\end{equation}

\begin{figure}[htp]
\begin{center}
	\begin{tikzpicture}
	\node (1) at (-4,0) [circle,draw,thick,minimum size=1.4cm] {N$_1$};
	\node (2) at (-2,0) [circle,draw,thick,minimum size=1.4cm] {N$_2$};
	\node (3) at (0,0)  {$\dots$};
	\node (5) at (4,0) [circle,draw,thick,minimum size=1.4cm] {N$_{P-1}$};
	\node (4) at (2,0) [circle,draw,thick,minimum size=1.4cm] {N$_{P-2}$};
	\draw[thick] (1) -- (2) -- (3) -- (4) -- (5);
	\node (1b) at (-4,-2) [rectangle,draw,thick,minimum size=1.2cm] {F$_1$};
	\node (2b) at (-2,-2) [rectangle,draw,thick,minimum size=1.2cm] {F$_2$};
	\node (3b) at (0,0)  {$\dots$};
	\node (5b) at (4,-2) [rectangle,draw,thick,minimum size=1.2cm] {F$_{P-1}$};
	\node (4b) at (2,-2) [rectangle,draw,thick,minimum size=1.2cm] {F$_{P-2}$};
	\draw[thick] (1) -- (1b);
	\draw[thick] (2) -- (2b);
	\draw[thick] (4) -- (4b);
	\draw[thick] (5) -- (5b);
	\end{tikzpicture}
\end{center}
\caption{\small Quiver diagram for the linear quiver described by the rank function \eqref{generic_R} with $(P-1)$ gauge nodes. The balancing conditions enforce that $F_k= 2N_k - N_{k-1} - N_{k+1}$ for each node.}\label{quiver_fig}
\end{figure}
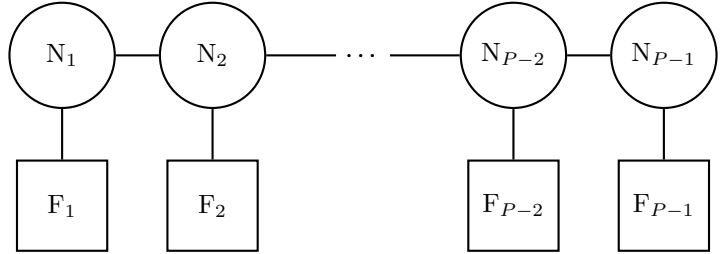

The second derivative of the rank function contains information about the ranks of the flavour groups (we take $N_0=0=N_P$):
\begin{equation}
{\cal R}''(\eta)=\sum_{j=1}^{P-1} (2N_j -N_{j-1} -N_{j+1})\delta(\eta-j)= \sum_{j=1}^{P-1} F_j\delta(\eta-j),
\end{equation}
where $F_j=2N_j - N_{j-1}-N_{j+1}$. We can think of the system \eqref{laplace} with boundary conditions \eqref{boundary_conditions},\eqref{generic_R} as a two-dimensional electrostatic problem of a charge density $\mathcal{R}(\eta)$ located at $\sigma=0$ and two conducting planes at $\eta=0$ and $\eta=P$ with zero potential. One can then proceed using separation of variables to obtain the following solution \cite{Legramandi:2021uds}:
\begin{equation}\label{laplace_potential_generic}
\begin{split}
&V(\sigma,\eta)=-\sum_{k=1}^\infty {\cal R}_k 
\sin \left(\frac{k \pi\eta}{P}\right)
K_0\left( \frac{k \pi \sigma}{P}\right),\quad\mathcal{R}(\eta)=\sum_{k=1}^{\infty}\mathcal{R}_k\sin\left(\frac{k\pi\eta}{P}\right),\\
&{\cal R}_k=\frac{2}{P}\int_0^P {\cal R}(\eta) 
\sin \left( \frac{k \pi\eta}{P}\right) \mathrm{d}\eta=\frac{2P}{(k\pi)^2}\sum_{j=1}^{P}F_j\sin\left(\frac{k\pi j}{P}\right),
\end{split}
\end{equation}
where $K_0$ is the modified Bessel function of the second kind.
%
%
\begin{flushleft}
{\bf Page charges}
\end{flushleft}
In this subsection we will present the quantization of the Page charges. Unlike Maxwell fluxes, Page fluxes provide the correct quantization conditions which express the counting of branes at each stack. These are defined in terms of the RR polyform $F$ as (setting $\alpha^{\prime}=1=g_s$):
\begin{equation}
    \hat{F}=e^{-B_2}\wedge F,\quad F = \sum_{n=1}^5 F_{2n},\quad Q_{\mathrm{D}_p}=\frac{1}{(2\pi)^{7-p}}\int _{\Sigma_{8-p}} \hat{F}_{8-p}
\end{equation}
and for our purposes this gives $\hat{F}_2=F_2$ and $\hat{F}_4=F_4-B_2\wedge F_2$, as we have vanishing Romans mass $F_0=0$. We will follow closely \cite{macpherson2024}, where the reader can find more details. The system described above contains the following extended objects at the following spacetime locations: 
\begin{itemize}
    \item A stack of $\mathrm{NS}5$ branes located at $\sigma\to \infty$ extending in the coordinates of $\mathrm{d}s_5^2$ and $\chi$,\\
    \item A stack of flavour $\mathrm{D}6$ branes located at $\sigma=0$ and at $\eta=k$ (the kinks of the rank function) and extending on $\mathrm{d}s^2_5$ and $\mathbb{S}^2$,\\
    \item A stack of colour $\mathrm{D}4$ branes extending on $\mathbb{R}^{1,2}\times\mathbb{S}^1_{\phi}\times \mathbb{R}_{\eta}$.
\end{itemize}

Let us comment that from the brane configuration explained above, only the $\mathrm{D}6$ flavour branes are made explicit in the solution \eqref{10dGM-2}, as they appear as localized sources, $\mathrm{d}\hat{F}_2\propto \sum_k\delta (\eta - k)$. The rest of the branes have been replaced with units of $\hat{F}_4$ and $H_3$ fluxes.\\
One can consider a flat space $\mathbb{R}^{1,9}$ spanned by coordinates $(x^0,\dots,x^3,y^1,\dots, y^6)$, in which we have the brane setup depicted in Table \ref{table1}.

\begin{table}[H]
\centering
\begin{tabular}{|c|c|c|c|c|}\hline
& $x^0$ $x^1$  $x^2$  $x^3$ & $y^1$  $y^2$  $y^3$  & $y^4$  $y^5$ & $y^6$\\ \hline
    $\mathrm{D}4$ & \textemdash \textemdash \textemdash \textemdash & $\cdot$\quad $\cdot$ \quad $\cdot$ &$\cdot$ \quad $\cdot$ & \textemdash\\ \hline
     $\mathrm{NS}5$  & \textemdash \textemdash \textemdash \textemdash &$\cdot$\quad $\cdot$ \quad $\cdot$ & \textemdash \textemdash &$\cdot$  \\ \hline
     $\mathrm{D}6$  & \textemdash \textemdash \textemdash \textemdash  & \textemdash \textemdash \textemdash &$\cdot$ \quad $\cdot$ & $\cdot$ \\ \hline
\end{tabular}
\caption{Brane configuration for the Gaiotto-Maldacena solution when $G_N\to 0$ in flat space. A line signifies extension of the brane in the corresponding coordinate while a dot that the brane is localized there.}
\label{table1}
\end{table}
This system can be though of as the Hanany-Witten brane set-up at zero Newton constant. As one starts increasing $G_N$, the branes start backreacting, sourcing a curved geometry. There is a nonlinear transformation involving a near horizon limit which then maps the coordinates $(x^i,y^j)$ to the ones appearing in the solution \eqref{10dGM-2}, with $y^6$ giving rise to $\eta$. The precise transformation is not known. The $\mathrm{SU}(2)\times\mathrm{U}(1)\cong\mathrm{SO}(3)\times\mathrm{SO}(2)$ ${\cal R}$-symmetry preserved by the QFT dual of \eqref{10dGM-2} is initially described by isometries of the $(y^1,y^2,y^3)$ and $(y^4,y^5)$ subspaces.

We take again the limit $r\to\infty$ to calculate the fluxes, in which case we have $\mathrm{d}s^2_5\to \mathrm{d}s^2_{\mathrm{AdS}_5}$ as well as: 
\begin{equation}
\begin{split}
&    X\to 1,\quad \tilde{f}\to \kappa^{2/3},\quad \tilde{f}_2\to \frac{2\dot{V}V^{\prime\prime}}{\tilde{\Delta}},\quad \tilde{f}_3\to \frac{4\sigma^2V^{\prime\prime}}{2\dot{V}-\ddot{V}},\\
&\tilde{f}_4\to \frac{2V^{\prime\prime}}{\dot{V}},\quad \tilde{f}_5\to \frac{2(2\dot{V}-\ddot{V})}{\tilde{\Delta}},\quad \tilde{f}_6\to \frac{2\dot{V}\dot{V}^{\prime}}{2\dot{V}-\ddot{V}},\\
&\tilde{f}_7\to \frac{2\kappa \dot{V}^2V^{\prime\prime}}{\tilde{\Delta}},\quad \tilde{f}_8 \to \frac{\kappa \dot{V}\left[ (\dot{V}^{\prime})^2-V^{\prime\prime}\ddot{V}\right]}{\tilde{\Delta}},\quad \tilde{f}_9 \to \frac{\kappa \sigma V^{\prime\prime}}{(\dot{V}^{\prime})^2-\ddot{V}V^{\prime\prime}}.
    \end{split}
\end{equation}
Let us begin with the $\mathrm{NS}5$ branes. In the limit where $\sigma\to \infty$ we can write the approximate form of the Laplace potential from \eqref{laplace_potential_generic} using the asymptotic behaviour of the Bessel function at infinity $K_0(x)\approx \sqrt{\frac{\pi}{2x}}e^{-x}+\mathcal{O}(e^{-x}x^{-3/2})$:
\begin{equation}
    V(\sigma,\eta)\approx -\sqrt{\frac{P}{2\sigma}}\sum_{k=1}^{\infty}\frac{\mathcal{R}_k}{\sqrt{k}}\sin\left(\frac{k\pi}{P}\eta\right)e^{-\frac{k\pi\sigma}{P}}\approx-\sqrt{\frac{P}{2\sigma}}\mathcal{R}_1\sin\left(\frac{\pi\eta}{P}\right)e^{-\frac{\pi\sigma}{P}}+\dots,
\end{equation}
where we keep only the leading term. Using this expression for $V$ the metric takes the form: 
\begin{equation}
 \begin{split}     
  \mathrm{d}s^2_{10}&\to \kappa  \Big\{4\sigma(\mathrm{d}s^2_{\mathrm{AdS}_5}+\mathrm{d}\chi^2)+\frac{2P}{\pi}\sin^2\left(\frac{\pi\eta}{P}\right)\mathrm{d}s^2_{\mathbb{S}^2}+\frac{2\pi}{P}(\mathrm{d}\sigma^2+\mathrm{d}\eta^2)\\
    &-\frac{8LP q_1 \ell^2\varepsilon\sin^2\theta}{\pi r_{\star}^2}\sin^2\left( \frac{\pi \eta}{P}\right)\mathrm{d}\phi\mathrm{d}\varphi + \frac{8 L q_2 \ell^2 \varepsilon\sigma}{r_{\star}^2+\ell^2\varepsilon}\mathrm{d}\phi \mathrm{d}\chi\Big\},
 \end{split}
\end{equation}

The asymptotic expressions for $H_3$ and the dilaton take the form: 
\begin{equation}
    \begin{split}
       \kappa^{-1}  H_3&= 4\sin^2 \left( \frac{\pi\eta}{P}\right)\mathrm{d}\eta \wedge\mathrm{vol}_{\mathbb{S}^2}+\frac{P}{2\pi \sigma}\sin\left( \frac{2\pi \eta }{P}\right)\mathrm{d}\sigma \wedge\mathrm{vol}_{\mathbb{S}^2}\\
        &-\frac{8Lq_1 \ell^2 \varepsilon}{r_{\star}^2}\sin^2\left( \frac{\pi \eta }{P}\right)\sin\theta\,\mathrm{d}\eta \wedge\mathrm{d}\theta\wedge\mathrm{d}\phi-\frac{LPq_1\ell^2 \varepsilon}{\pi \sigma r_{\star}^2}\sin\left(\frac{2\pi \eta}{P}\right)\sin\theta\, \mathrm{d}\sigma\wedge\mathrm{d}\theta\wedge\mathrm{d}\phi,
    \end{split}
\end{equation}
\begin{equation}
    e^{-\Phi}\to\frac{\pi^{3/2}\mathcal{R}_1e^{-\frac{\pi \sigma}{P}}}{2P\sqrt{\sigma}},
\end{equation}
while $C_1\to0$ at leading order. The suitable cycle to integrate $H_3$, see \cite{Gaiotto:2009gz}, is a round unit radius three-sphere which we can construct as $\mathbb{S}^3=\{(\frac{\pi}{P}\eta,\mathbb{S}^2)\,\,\, | \,\,\, \eta\in[0,P]\, \}$ leading to the following quantization\footnote{we remind the reader that we fix $\kappa=\pi/2$ (in units where $\ell _p=1$) in order for the fluxes to be properly quantized.} condition:
\begin{equation}
    Q_{\mathrm{NS}5}=\frac{1}{(4\pi)^2}\int _{\mathbb{S}^3}H_3=\frac{1}{2\pi}\int_{\mathbb{S}^2}\mathrm{vol}_{\mathbb{S}^2}\int_0^P\mathrm{d}\eta\,\sin^2\left(\frac{\pi\eta}{P}\right) = P.
\end{equation}

Moving to the stacks of $\mathrm{D}6$, we will use the following coordinates for the $(\sigma,\eta)$ subspace: 
\begin{equation}
    \eta=k-\rho \cos\alpha,\quad \sigma =\rho\sin\alpha,
\end{equation}
and take $\rho\to 0$, corresponding to the limit $\sigma\to 0$ and $\eta=k\in (0,P)$. By focusing on a specific $k$ we have the relations: 
\begin{equation}
    \dot{V}=N_k,\quad V^{\prime\prime}=\frac{F_k}{2\rho},\quad \dot{V}^{\prime}=\frac{F_k}{2}(1+\cos\alpha)+N_{k+1}-N_k.
\end{equation}

The metric in this limit takes the following form: 
\begin{equation}
     \mathrm{d}s^2_{10}\to 2\kappa\sqrt{N_k}\left[ \sqrt{\frac{\rho}{F_k}}(4\mathrm{d}s^2_{\mathrm{AdS}_5}+\mathrm{d}s^2_{\mathbb{S}^2})+\sqrt{\frac{F_k}{\rho}}(\mathrm{d}\rho^2+\rho^2\mathrm{d}s^2_{\mathbb{S}'^2})\right],
 \end{equation}
where there is second two-sphere spanned by $(\alpha,\chi)$: $\mathrm{d}s^2_{\mathbb{S}'^2}=\mathrm{d}\alpha^2+\sin^2\alpha\,\mathrm{d}\chi^2$. This is indeed, up to a constant factor, the near horizon metric for a stack of $\mathrm{D}6$ branes extending in $\mathrm{AdS}_5$ and $\mathbb{S}^2$. The RR potential and the two-form flux take the following form: 
\begin{equation}\label{C1_and_F2_limit}
  \begin{split}
  &C_1\to \left[ \frac{F_k}{2}(1+\cos\alpha)+N_{k+1}-N_k \right]\mathrm{d}\chi+\frac{Lq_2 \ell^2 \varepsilon \left[ (1+\cos\alpha)F_k - 2N_{k}+2N_{k+1}\right]}{2(r_{\star}^2+\ell^2\varepsilon)}\mathrm{d}\phi,\\
  &F_2=\mathrm{d}C_1\to -\frac{F_k}{2}\mathrm{vol}_{\mathbb{S}'^2}-\frac{L q_2 \ell^2 \varepsilon F_k \sin\alpha\, \mathrm{d}\alpha\wedge\mathrm{d}\phi}{2(r_{\star}^2+\ell^2 \varepsilon)}+\frac{L q_2 \ell^2 \varepsilon F_k \sin^2(2\alpha)\left[ F_k(1+\cos\alpha) - 2N_k + 2N_{k+1}\right]}{32(r_{\star}^2+\ell^2 \varepsilon)N_k}\mathrm{d}\rho \wedge \mathrm{d}\phi\\
  &+\frac{F_k \sin^2(2\alpha)\left[ F_k (1+\cos\alpha) -2N_k +2N_{k+1}\right]}{32N_k}\mathrm{d}\rho \wedge\mathrm{d}\chi,
\end{split}
\end{equation}
which gives us the quantization condition: 
\begin{equation}\label{D6_charge_k}
    Q^k_{\mathrm{D}6}=-\frac{1}{2\pi}\int _{\mathbb{S}'^2}F_2 = F_k=2N_k - N_{k+1}-N_{k-1},\quad k=1,2,\dots,P-1,
\end{equation}
with the total $\mathrm{D}6$ charge being:
\begin{equation}
    Q_{\mathrm{D}6} = \sum_{k=1}^{P-1} Q^k_{\mathrm{D}6} =N_{P-1}+N_1 = \int_0^P\mathrm{d}\eta \, {\cal R}^{\prime\prime}(\eta).
\end{equation}

For the $\mathrm{D}4$ Page charges, we construct the flux $\hat{F}_4=F_4-B_2\wedge F_2$ using the expressions for $B_2$ in \eqref{various_B2s}, where we verify that the last two give a zero Page charge, while the first two yield asymptotically:
\begin{equation}
    \hat{F}_4=\kappa kF_k\sin\alpha\sin\theta\,\mathrm{d}\alpha\wedge\mathrm{d}\theta\wedge\mathrm{d}\varphi\wedge\mathrm{d}\chi =\kappa  kF_k\, \mathrm{vol}_{\mathbb{S}^2}\wedge\mathrm{vol}_{\mathbb{S}^{\prime 2}}.
\end{equation}
This leads to the following expression for the $\mathrm{D}4$ charge in a cell $[k,k+1]$:
\begin{equation}\label{D4_charge_k}
    Q^k_{\mathrm{D}4}=\frac{1}{(2\pi)^3}\int _{\mathbb{S}^2\times\mathbb{S}^{\prime 2}}\hat{F}_4=kF_k,
\end{equation}
while the total Page charge of the $\mathrm{D}4$ branes is given by:
\begin{eqnarray}
    Q_{\mathrm{D}4}=\sum _{k=1}^{P-1} Q^k_{\mathrm{D}4}=\sum_{k=1}^{P-1}k F_k =PN_{P-1}.
\end{eqnarray}
As pointed out in \cite{Macpherson:2024frt}, the calculation of the above quantity includes the induced charges of the $\mathrm{D}4$s on the other stacks of branes. The "true" colour charge of the $\mathrm{D}4$ branes is given by the formula \cite{Nunez:2019gbg}:

\begin{equation}
    Q^{\text{total}}_{\mathrm{D}4} = \int _0^{P}\mathrm{d}\eta \, {\cal R}(\eta).
\end{equation}

Let us comment that the integral \eqref{D4_charge_k} does not yield the rank of the gauge group between $\eta=k$ and $\eta=k+1$, namely $N_k$, as opposed to \eqref{D6_charge_k} which gives off the rank of the flavour group $F_k$. This can be explained in two different ways. For once, the groups $\mathrm{SU}(F_k)$ represent physical symmetries of the dual theory, while the gauge groups $\mathrm{SU}(N_k)$ do not correspond to any physical symmetry and thus it is not necessarily the case that one can extract the various gauge group ranks from quantized charges. Another reason is that in the case of the $\mathrm{D}6$s, one can clearly tell their location in the $\eta$ coordinate (corresponding to a singularity in the metric), which is not the case for $\mathrm{D}4$s, as they are replaced with fluxes. Finally, one might notice that the Page charges have similar expressions as in the undeformed Gaiotto-Maldacena background. This is due to the quantized charges being invariant under bulk deformations which respect the boundary geometry.


\section{Observables of the dual field theories}\label{Observables_section}

In the present section we will calculate various observables in the dual field theories of the backgrounds presented in section \ref{New_BG}. These include:  the expectation value of the Wilson and 't Hooft loop operators, entanglement entropy, flow central charge, holographic complexity as well as the study of $\mathrm{D}7$ brane embeddings in the case of the deformed type IIB background of Anabalon and Ross \cite{Anabalon:2021tua}. We will present the holographic calculation of said observables while emphasizing two key features. The first feature is universality. The resulting expressions are factorized into two parts, one depending on the radial direction which contains the dynamics of each object in gravity, while the other numerical factor containing details of the internal space and expressing information about the UV SCFTs. We attribute this phenomenon to the theorem of Gauntlett and Varela presented as a conjecture in \cite{Gauntlett:2007ma} and later proved in \cite{Cassani:2019vcl} using G-structure techniques. The theorem states that any SUSY solution of supergravity in dimensions D=10 or D=11, which can be written as a warped product $\text{AdS}_{d+1}\times _w {\cal M}_{D-d-1}$ consistently truncates on\footnote{${\cal M}_{D-d-1}$ is required to be Riemannian.} ${\cal M}_{D-d-1}$ resulting in a gauged supergravity in $(d+1)$-dimensions. The field content of the gauged lower dimensional theory is dual to the superconformal current multiplet of the $\text{SCFT}_d$ dual. The fields belonging to this multiplet are the ones responsible for the dynamical/radially dependent part of the observables.  In this sense, the behaviour of these observables, even in the case where they do involve the internal space in their calculation (like the flow central charge and entanglement entropy), depends only on the underlying 5d gauged supergravity. The second feature concerns the singularity study we presented in section \ref{singularity_section}. We notice that even in the case of the solutions being smooth ($\hat\nu\approx-1$ but never $-1$), the Wilson loop observable is affected by the high curvature in the supergravity, signalling that the theory is not a trustable approximation in this region of parameter space. This presents itself as a first order phase transition near this region of parameter space, which we argue is not physically relevant. Additionally, appendices \ref{Appendix_Polyakov_loop} and \ref{appendix_gauge_coupling} are dedicated to the study of Polyakov loops and the effective gauge coupling constant of each QFT.


\subsection{Wilson loop}\label{Wilson_loop_section}
Let us first review the calculation of the Wilson loop in the context of holography \cite{Maldacena:1998im}. Starting from a gauge theory perspective, one can define the Wilson loop as the following non local operator: 
\begin{equation}
    W(\mathcal{C})=\frac{1}{N}\mathrm{tr}\,\mathcal{P}\,\mathrm{exp}\left(i\oint _{\mathcal{C}}A\right),
\end{equation}
where $N$ is the rank of the gauge group and $\mathcal{C}$ is any closed loop, conveniently taken to be a rectangular contour in time and one spatial coordinate. Then the vacuum expectation value of this operator provides a way of investigating the potential between a pair of non dynamical, infinitely massive, quark and anti-quark in the fundamental representation: 
\begin{equation}
    \langle W(\mathcal{C})\rangle \propto e^{-\mathcal{T}\, E(L_{\mathrm{QQ}})},
\end{equation}
where $\mathcal{T}$ is the temporal length of $\mathcal{C}$ and $L_{\mathrm{QQ}}$ the distance between the pair. From here one can deduce if the theory is {\it confining} or if it presents {\it screening}. This calculation is implemented in the dual string theory (in the large $N$ limit, with $g_{\text{YM}}^2N$ fixed) by embedding a probe $\mathrm{F}1$ string with its endpoints fixed on the path $\mathcal{C}$ at the boundary of the spacetime $(r=\infty)$, representing the quark and anti-quark, while the rest of the string enters the bulk in a $\texttt{U}$ shaped fashion. In this regime, the string theory is approximated by supergravity and we can use the Nambu-Goto action of classical string embeddings whose worldsheets, with a fixed boundary $\mathcal{C}$, are minimal surfaces: $\langle W(\mathcal{C})\rangle\sim e^{-S_{\text{NG}}}$, where\footnote{We make the choice to set $\alpha^{\prime}=1$. We also denote all worldvolume coordinates as $\hat\sigma$ instead of $\sigma$ to avoid confusion with the Gaiotto-Maldacena coordinate $\sigma$ in the type IIA backgrounds.}: 
\begin{equation}
    S_{\mathrm{NG}}=\frac{1}{2\pi}\int \mathrm{d}^2\hat\sigma \sqrt{-\mathrm{det}(g_{\text{ind}})},
\end{equation}
with $g_{\text{ind}}$ being the induced metric on the string. \\

In this section we  consider, aside from the usual Wilson loop that was also presented in \cite{Chatzis:2025dnu}, other configurations that wrap or rotate in $\mathbb{S}^1[\phi]$. We will however restrict to embeddings that are not extended along the internal space, resulting in the universal behaviour highlighted earlier in the text. Here are the three types of configurations of interest:\\

$\bullet$\quad {\bf Embedding I.}\quad The first case we will study is the simple embedding in which we give the string a profile in $r$ and its worldsheet coordinates are parametrized as: 
\begin{equation}
    \tau=t,\quad \hat\sigma=w,\quad r=r(\hat\sigma),
\end{equation}
while the rest of the spacetime coordinates take constant values.  In the case of the type IIB background, the consistency of the embedding demands the coordinate $\theta_0$ to be either $0$ or $\frac{\pi}{2}$.

\vspace{10pt}

$\bullet$\quad {\bf Embedding II.}\quad In this embedding we consider a string with the same profile, which now is spinning along $\mathbb{S}^1[\phi]$ with constant angular velocity $\omega$:
\begin{equation}\label{spinning_embedding}
        \tau=t,\quad \hat\sigma=w,\quad r=r(\hat\sigma),\quad \phi(\tau)=\omega \tau.
\end{equation}

\vspace{10pt}

$\bullet$\quad {\bf Embedding III.}\quad For the final embedding we will use, we place a string that wraps the $\mathbb{S}^1[\phi]$, while still having a profile in the radial coordinate: 
\begin{equation}\label{embeddingIII}
            \tau=t,\quad \hat\sigma=w,\quad r=r(\hat\sigma),\quad \phi=\phi(\hat\sigma),
\end{equation}
and in the case of the type $\mathrm{IIB}$ background the embedding is consistent for $\theta_0=0$ and $\psi_0=\frac{\pi}{4}$.\\

In all the above cases, the action of the string reduces to following form:
\begin{equation}\label{WL_action_schematic}
    S_{\mathrm{NG}}\propto  \int_{-L_{\mathrm{QQ}}/2}^{L_{\mathrm{QQ}}/2}\mathrm{d}w\sqrt{\mathcal{F}^2+\mathcal{G}^2r^{\prime 2}},
\end{equation}
with $\mathcal{F}$ and $\mathcal{G}$ being case specific functions of r. One can then derive expressions for the length and energy of the quark-anti-quark pair in terms of $\mathcal{F}$ and $\mathcal{G}$ using the  formulas in \cite{Nunez:2009da,Kol:2014nqa}. Below $r_0$ denotes the turning point of the probe:
\begin{figure}[t]
\centering
\begin{subfigure}{0.44\linewidth}
\includegraphics[width=\linewidth]{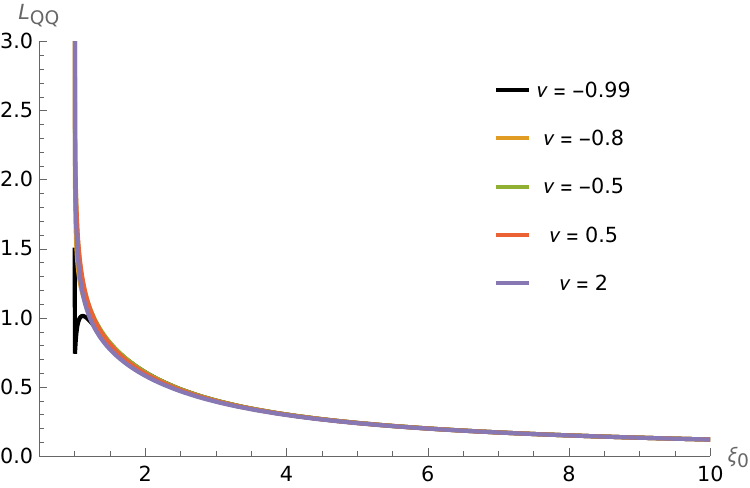}
\caption{Plot of the quark-anti-quark separation for various values of $\hat{\nu}$.}
\end{subfigure}
\hfill
\begin{subfigure}{0.44\linewidth}
\includegraphics[width=\linewidth]{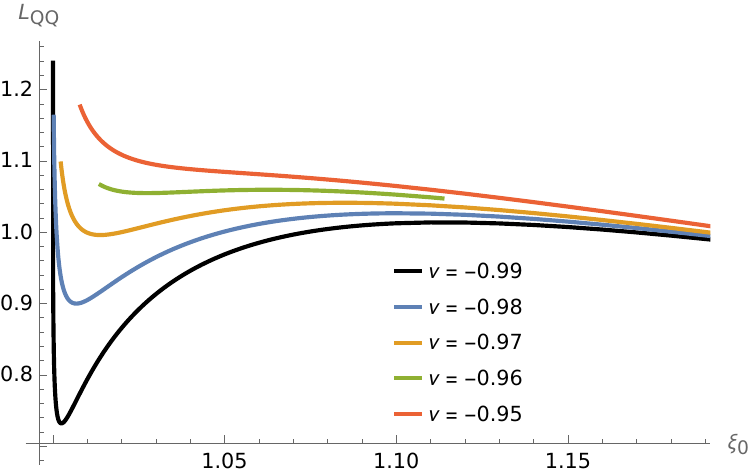}
\caption{Closer look at the plot of the separation length as $\hat{\nu}$ approaches $-1$.}
\end{subfigure}
\hfill
\begin{subfigure}{0.44\linewidth}
    \includegraphics[width=\linewidth]{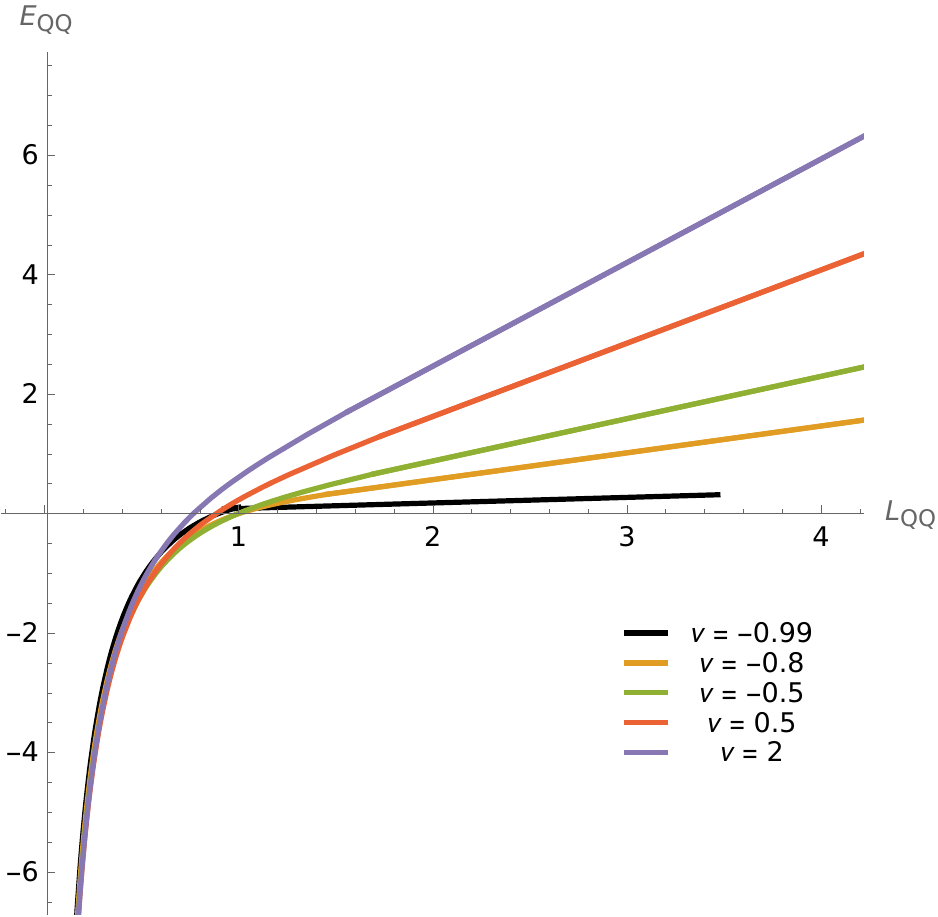}
    \caption{Plot of the potential energy of the quark-anti-quark pair with respect to their separation for various values of $\hat{\nu}$.}
\end{subfigure}
\hfill
\begin{subfigure}{0.46\linewidth}
\includegraphics[width=\linewidth]{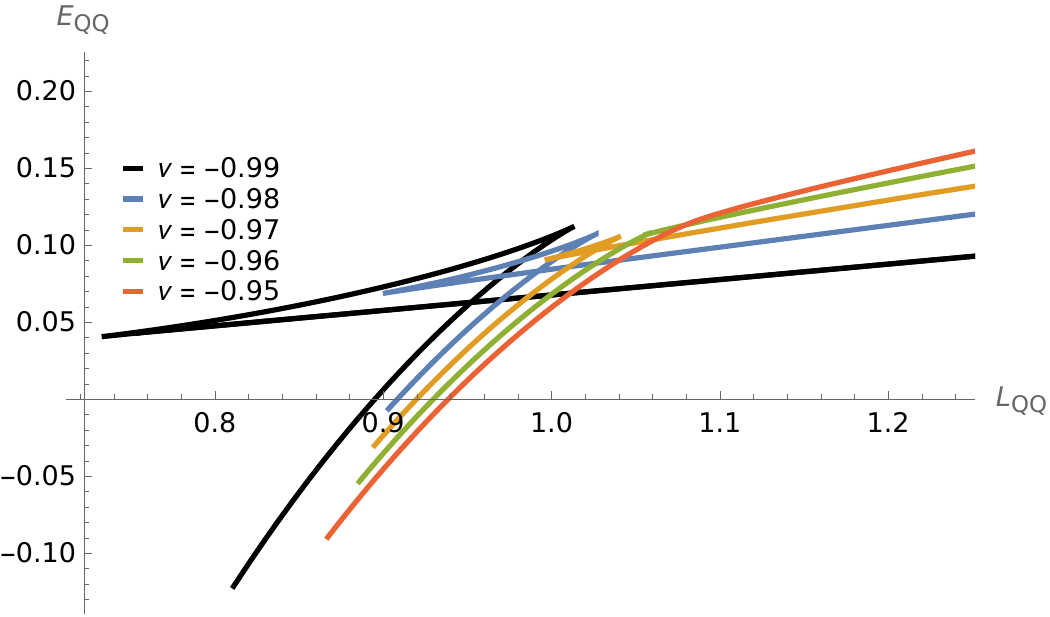}
\caption{Closer look at the plot of the energy as $\hat{\nu}$ approaches $-1$.}
\end{subfigure}
\caption{Resulting plots using numerical integration regarding the WL embedding $\mathrm{I}$ for different values of $\hat\nu$. We notice the length gradually becoming double-valued as $\hat\nu\to-1$. We have set $\theta_0=0$.}
\label{Wilson_plots_embeddingI}
\end{figure}

\begin{equation}\label{WL_integrals}
    \begin{split}
& V_{\mathrm{eff}}(r;r_0)=\frac{\mathcal{F}(r)\sqrt{\mathcal{F}^2(r)-\mathcal{F}^2(r_0)}}{\mathcal{F}(r_0)\mathcal{G}(r)},\quad  L_{\mathrm{QQ}}(r_0)=2\int_{r_0}^{\infty}\frac{\mathrm{d}r}{V_{\mathrm{eff}}(r;r_0)},\\
	& E_{\mathrm{QQ}}(r_0)=\mathcal{F}(r_0)L_{\mathrm{QQ}}(r_0)+2\int_{r_0}^{\infty}\mathrm{d}r \frac{\mathcal{G}(r)}{\mathcal{F}(r)}\sqrt{\mathcal{F}^2(r)-\mathcal{F}^2(r_0)}-2\int_{r_{\star}}^{\infty}\mathrm{d}r\mathcal{G}(r).        
    \end{split}
\end{equation}
These integrals are not analytically solvable, therefore we employ numerical techniques to plot them. However, one can write analytic approximate expressions for the length, developed in \cite{Kol:2014nqa}, and the energy\footnote{This expression was first introduced in the context of entanglement entropy calculations in \cite{Nunez:2025gxq, Nunez:2025puk}.} of the pair:
\begin{equation}\label{approximate_expressions}
    L_{\mathrm{app}}(r_0)=\left.\frac{\pi\mathcal{G}(r)}{\mathcal{F}^{\prime}(r)}\right|_{r=r_0},\quad E_{\text{app}}(r_0)=\int ^{r_0} \mathrm{d}s \,\mathcal{F}(s)\frac{\mathrm{d}L_{\text{app}}(s)}{\mathrm{d}s},
\end{equation}
which serve as a nice way of quickly determining the behaviour of the dynamics of the quark-anti-quark system. Let us now proceed to studying the Wilson loops\footnote{For the remainder of the paper we will use the SUSY condition $|q_1|=|q_2|=Q$.} in each case and highlight the effects of the different embeddings. 

\subsubsection{Wilson loop for the Type IIB background}\label{WL_IIB_subsection}
{\bf Embedding I.}\quad For the simple embedding of case $\mathrm{I}$, we have checked that it is consistent in this background if one chooses the value of the angular coordinate $\theta$ to be $\theta_0=0$ or $\theta_0=\pi/2$. In each case, have the following induced metric: 
\begin{equation}
    \mathrm{d}s^2_{\text{ind}}=-\frac{r^2\zeta(r,\theta_0)}{L^2}\mathrm{d}t^2 + \left[ \frac{r^2\zeta(r,\theta_0)}{L^2}+ \frac{\zeta(r,\theta_0)r^{\prime2}}{r^2F(r)\lambda^6(r)}\right]\mathrm{d}w^2,
\end{equation}
which yields the following action (where $\mathcal{T}=\int\mathrm{d}t$): 
\begin{equation}\label{WL_embedding_I_action_IIB}
\begin{split}
&    \mathrm{S}_{\mathrm{F}1} =\frac{\mathcal{T}}{2\pi}\int_{-L_{\mathrm{QQ}}/2}^{L_{\mathrm{QQ}}/2}\mathrm{d}w\sqrt{\mathcal{F}_{\mathrm{I}}^2+\mathcal{G}_{\mathrm{I}}^2r^{\prime2}} \quad {\rm with} \quad \mathcal{F}_{\mathrm{I}}^2 = \frac{r^4\zeta^2(r,\theta_0)}{L^4},\quad \mathcal{G}_{\mathrm{I}}^2= \frac{\zeta^2(r,\theta_0)}{L^2\lambda^6(r)F(r)}.
\end{split}
\end{equation}
From these functions we can derive the various expressions of interest. It is convenient to use the dimensionless quantity $\hat{\mu}=\left(\frac{r_{\star}}{L}\right)^4$ together with $\hat{\nu}$ and $\xi$ for the various expressions, in terms of which the functions in \eqref{WL_embedding_I_action_IIB} take the following form: 
\begin{equation}\label{F_G_embeddingI_dimles}
    \mathcal{F}_{\mathrm{I}}^2 = \hat{\mu}\xi^2(\xi^2+\hat{\nu}\cos^2\theta_0),\quad \mathcal{G}_{\mathrm{I}}^2 = \frac{\xi^4(\xi^2+\hat{\nu}\cos^2\theta_0)}{\xi^6+\hat{\nu}\xi^4-(1+\hat{\nu})}.
\end{equation}








\begin{figure}[t]
\centering
\begin{subfigure}{0.44\linewidth}
\includegraphics[width=\linewidth]{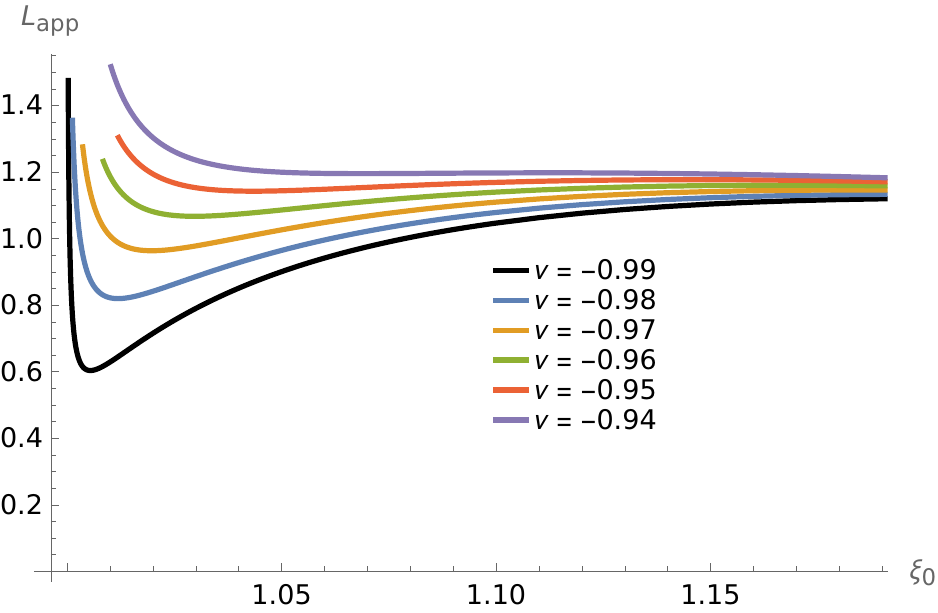}
\caption{Plot of the approximate length function for various values of $\hat{\nu}$ and $\theta_0=0$.}
\end{subfigure}
\hfill
\begin{subfigure}{0.44\linewidth}
\includegraphics[width=\linewidth]{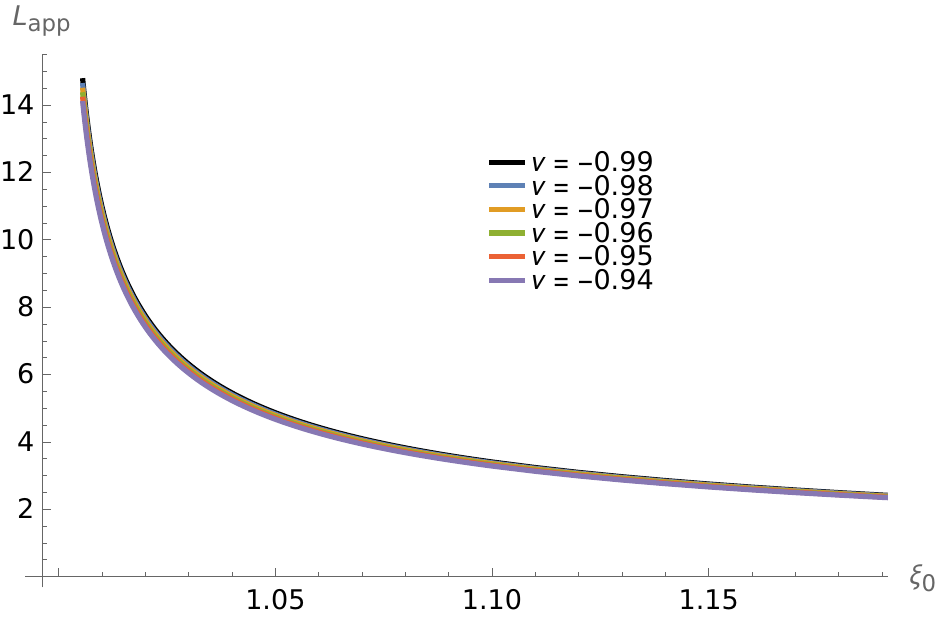}
\caption{Plot of the approximate length function for various values of $\hat{\nu}$ and $\theta_0=\frac{\pi}{2}$.}
\end{subfigure}
\hfill
\begin{subfigure}{0.44\linewidth}
    \includegraphics[width=\linewidth]{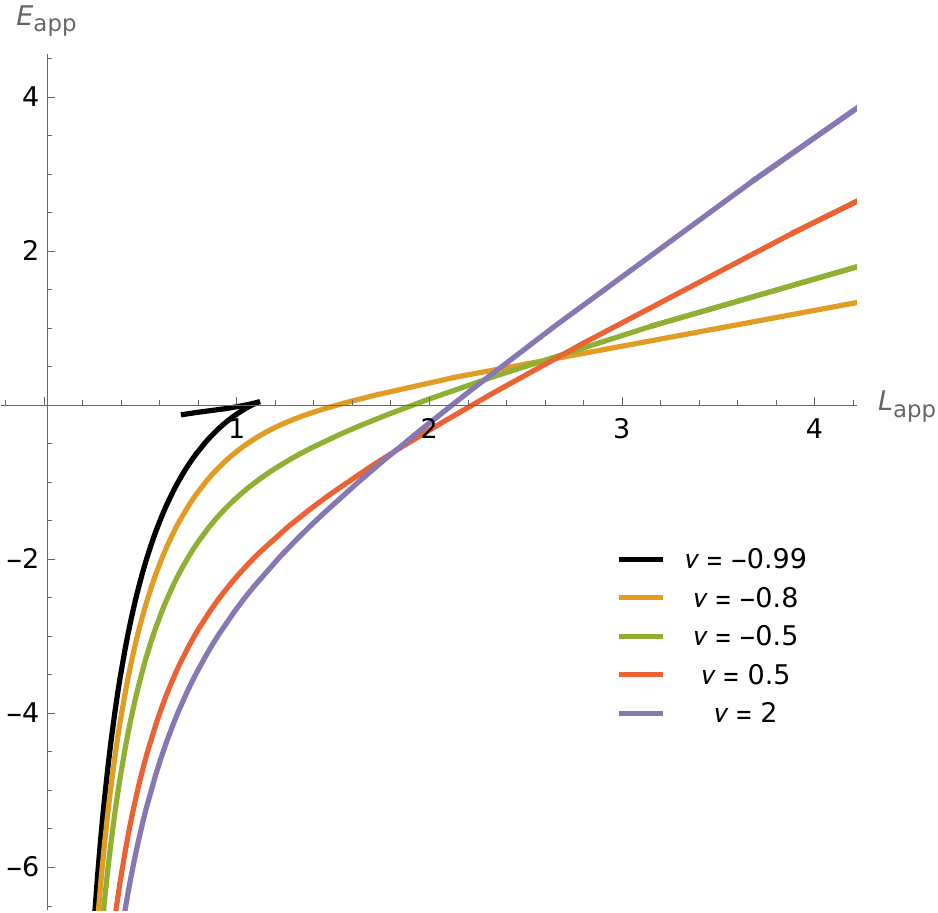}
    \caption{Plot of the approximate potential energy as a function of the approximate length for various values of $\hat\nu$ and $\theta_0=0$.}
\end{subfigure}
\hfill
\begin{subfigure}{0.46\linewidth}
\includegraphics[width=\linewidth]{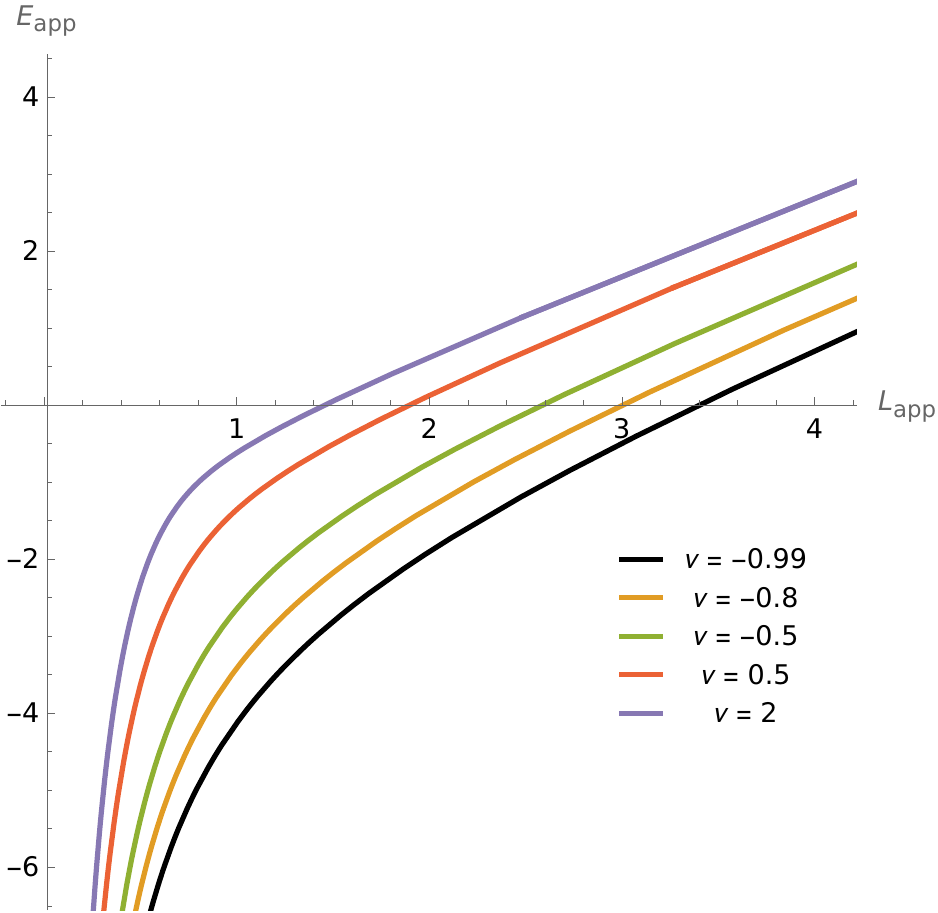}
    \caption{Plot of the approximate potential energy as a function of the approximate length for various values of $\hat\nu$ and $\theta_0=\frac{\pi}{2}$.}
\end{subfigure}
\caption{Wilson loop plots for embedding $\mathrm{I}$ using the approximate length and energy expressions \eqref{approximate_expressions} for various values of $\hat\nu$ and the two options $\theta_0\in\{0,\pi/2\}$.}
\label{Wilson_approximate_plots_embeddingI}
\end{figure}

Using numerical integration, we  plot the length $L_{QQ}$ as a function of the turning point as well as a parametric plot of the potential  energy $E_{QQ}$ between the quark and anti-quark with respect to the length of separation. The results for the embedding with $\theta_0=0$ are depicted in figure \ref{Wilson_plots_embeddingI}. What we discover is that as $\hat\nu$ takes different values away from $-1$ the background displays confining behaviour, that is, $E_{\mathrm{QQ}}$ starts with a Coulombic law for small separations (dictated by conformality in the ultraviolet) while at large $L_{\mathrm{QQ}}$ it is linear, signalling confinement in the infrared. However, when $\hat\nu$ takes values very close  to $-1$ and smaller than $\hat\nu\approx-0.95$,  where the geometry becomes highly curved, the length starts becoming double-valued and there seems to be a first order phase transition appearing in the energy. We claim that what is observed in figure \ref{Wilson_plots_embeddingI} (d) is a manifestation of the supergravity approximation being not trustable in this region in parameter space and for this embedding\footnote{This behaviour does not appear for $\theta_0=\pi/2$, see figure \ref{Wilson_approximate_plots_embeddingI} (b) and (d).}, in other words, this {\it is not} a trustworthy result. Finally, we comment that the approximate expressions \eqref{approximate_expressions} capture very nicely the qualitative behaviour of the system, which can be seen in the plots of figure \ref{Wilson_approximate_plots_embeddingI}.\\

{\bf Embedding II.} The idea of the spinning string embedding is that we might be able to lift the effects observed in figure \ref{Wilson_plots_embeddingI} (d) even for the value $\hat\nu=-0.99$ by introducing the new parameter $\omega$ which can affect the dynamics. More specifically, the angular momentum of the probe $\mathrm{F}1$ introduces restoring effects on the $\texttt{U}$ shaped embedding, much like in the case of a centrifugal force. If we consider for a moment both type the $\mathrm{IIA}$ and type $\mathrm{IIB}$ metrics by writing $\mathrm{d}s^2_{10}=-G_{tt}\mathrm{d}t^2+G_{ww}\mathrm{d}w^2+G_{\phi\phi}\mathrm{d}\phi^2+G_{rr}\mathrm{d}r^2+\dots$, the Nambu-Goto action for embedding \eqref{spinning_embedding} has the form:
\begin{equation}
    \mathcal{L}_{\mathrm{NG}}=\sqrt{(G_{tt}-\omega^2G_{\phi\phi})(G_{ww}+G_{rr}r^{\prime2})},
\end{equation}
from which stems the following bound for the Lagrangian to be real:
\begin{equation}\label{omega_bound}
    \omega^2 G_{\phi\phi}\leq G_{tt}.
\end{equation}
 The range of values for the radial coordinate $r\in[r_{\star},\infty)$ is restricted accordingly such that the above requirement holds, which in turn is interpreted as a $"$restoring force$"$ preventing the string from reaching values arbitrarily close to $r_{\star}$.\\

In the case of the type $\mathrm{IIB}$ solution, we have the induced metric:
\begin{equation}
    \begin{split}
        \mathrm{d}s_{\text{ind}}^2 &= \frac{1}{\zeta(r,\theta_0)}\left\{\frac{\zeta^2(r,\theta_0) r^2\left[L^2\omega^2F(r)-1\right]}{L^2}+\omega^2\cos^2\theta_0A_1^2 + \omega^2\sin^2\theta_0\lambda^6(r)A_3^2\right\}\mathrm{d}t^2\\
        &+\frac{\zeta(r,\theta_0)}{r^2}\left[\frac{r^4}{L^2}+\frac{r^{\prime2}}{F(r)\lambda^6(r)}\right]\mathrm{d}w^2 \, .
    \end{split}
\end{equation}
The Nambu-Goto action then yields:
\begin{equation}
\mathrm{S}_{\mathrm{F}1} =\frac{\mathcal{T}}{2\pi}\int_{-L_{\mathrm{QQ}}/2}^{L_{\mathrm{QQ}}/2}\mathrm{d}w\sqrt{\mathcal{F}_{\mathrm{II}}^2+\mathcal{G}_{\mathrm{II}}^2r^{\prime2}},
\end{equation}
where now,
\begin{equation}\label{F_G_embeddingII_IIB}
\begin{split}
&\mathcal{F}_{\mathrm{II}}^2=\frac{r^4\zeta^2(r,\theta_0)}{L^4}-\frac{r^2\omega^2}{L^2}\left[\cos^2\theta_0A_1^2 + \sin^2\theta_0\lambda^6(r)A_3^2 + r^2F(r)\zeta^2(r,\theta_0)\right],\\
&\mathcal{G}_{\mathrm{II}}^2 = \frac{\zeta^2(r,\theta_0)}{L^2F(r)\lambda^6(r)}-\frac{\omega^2}{r^2F(r)\lambda^6(r)}\left[\cos^2\theta_0A_1^2 + \sin^2\theta_0\lambda^6(r)A_3^2 + r^2F(r)\zeta^2(r,\theta_0)\right].
\end{split}
\end{equation}
or, in terms of the dimensionless quantities:
\begin{equation}
    \begin{split}
        &\mathcal{F}_{\mathrm{II}}^2=        \hat{\mu}\xi^2(\hat{\nu}\cos^2\theta_0+\xi^2)-\frac{\hat{\mu}(\xi^2-1)\omega^2}{2(1+\hat{\nu})}\left[ 4+5\hat{\nu}+2\hat{\nu}^2+2(\hat{\nu}+1)\xi^2 + \hat{\nu}(3+2\hat{\nu})\cos(2\theta_0)\right],\\
        &\mathcal{G}_{\mathrm{II}}^2= \frac{\xi^4\left(\xi^2+\hat\nu\cos^2\theta_0\right)}{\xi^6+\hat{\nu}\xi^4-(1+\hat{\nu})}-\frac{\xi^2\omega^2}{2(1+\hat{\nu})}\frac{\left[ 4+5\hat{\nu}+2\hat{\nu}^2+2(\hat{\nu}+1)\xi^2 + \hat{\nu}(3+2\hat{\nu})\cos(2\theta_0)\right]}{\xi^4+\xi^2(1+\hat{\nu})+1+\hat{\nu}}.
            \end{split}
\end{equation}

In the above expressions we can recognise the same functions as in the previous embedding \eqref{F_G_embeddingI_dimles} in the first terms, while the second terms are controlled by the new $\omega\neq0$ parameter we introduced. The length and energy integral expressions read\footnote{We specialize to the case where $\theta_0=0$ for convenience in writing the expressions explicitly}: 
\begin{figure}[t]
\centering
\begin{subfigure}{0.54\linewidth}
\includegraphics[width=\linewidth]{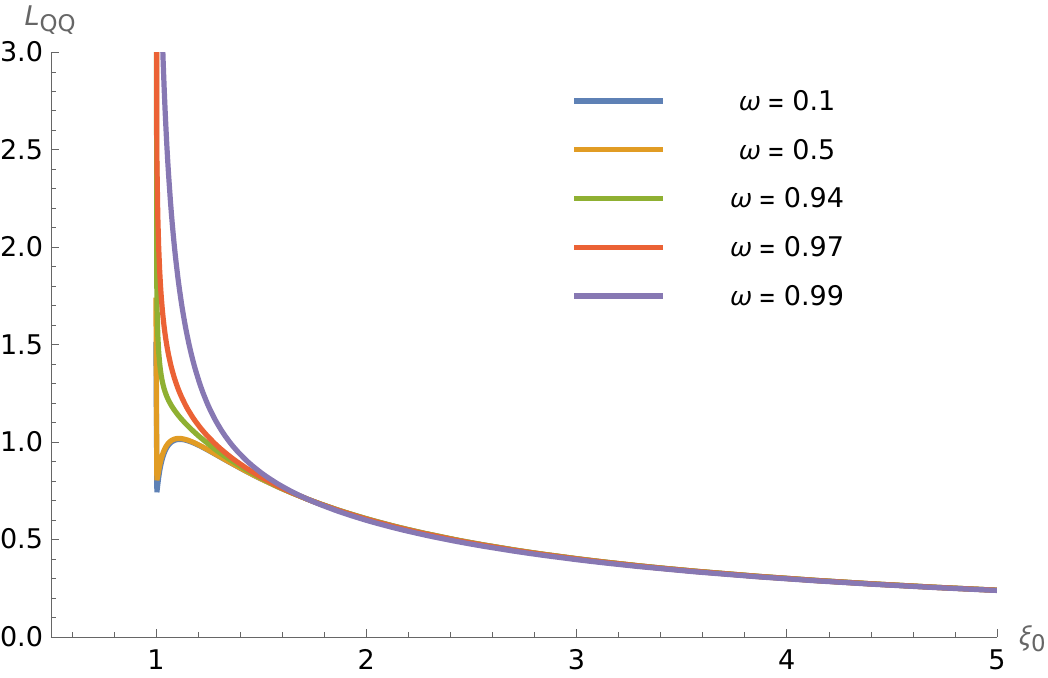}
\caption{Plot of the length of separation \eqref{length_integral_embeddingII} for $\hat\nu=-0.99$, $\theta_0=0$ and various values of $\omega$.}
\end{subfigure}
\hfill
\begin{subfigure}{0.44\linewidth}
\includegraphics[width=\linewidth]{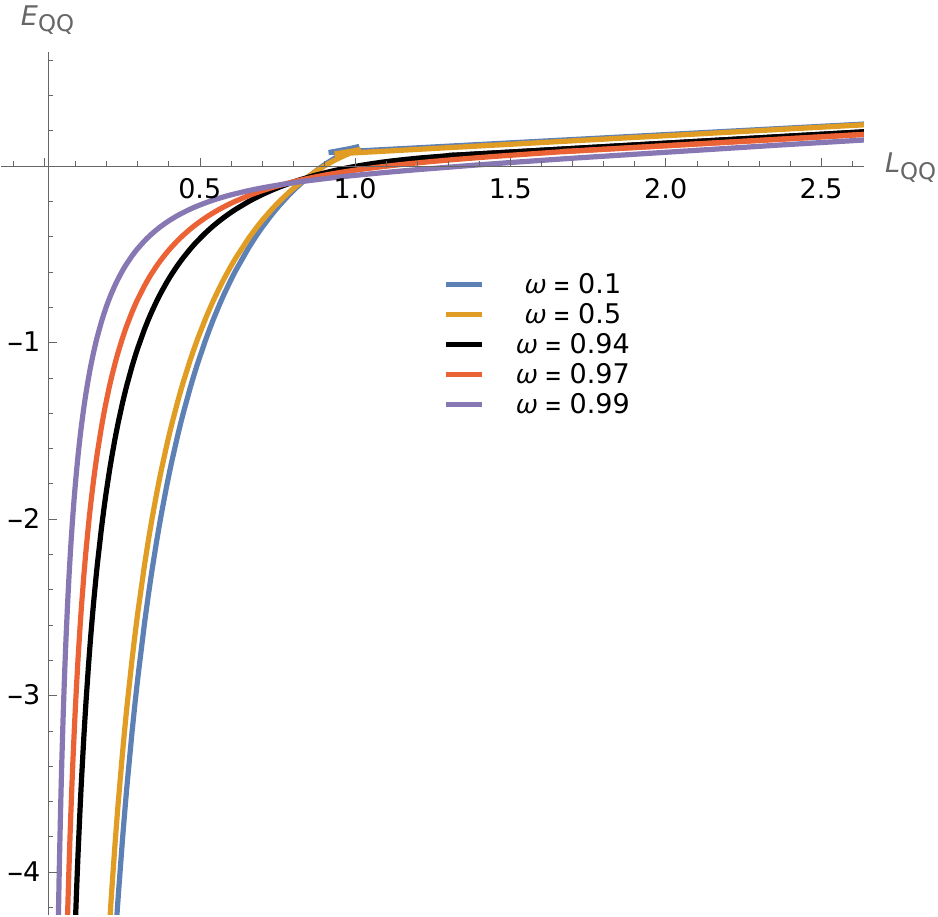}
\caption{Plot of the energy of the pair with respect to the length of separation for $\hat\nu=-0.99$, $\theta_0=0$ and various values of $\omega$.}
\end{subfigure}
\caption{We notice that for values of $\omega$ close to $0.99$ we get a single valued length and no phase transition, even when $\hat\nu=-0.99$ and $\theta_0=0$.}
\label{Wilson_plots_embeddingII}
\end{figure}
\begin{equation}\label{length_integral_embeddingII}
    L_{\mathrm{QQ}}^{(\mathrm{II})}(\xi_0)=2r_\star\sqrt{\hat{\mu}}\int_{\xi_0}^{\infty}\mathrm{d}\xi\xi \sqrt{\frac{\xi_0^2(\hat{\nu}+\xi_0^2)-(\xi_0^2-1)\left[2(1+\hat{\nu})+\xi_0^2\right]\omega^2}{(\xi^2-\xi_0^2)\left[ \xi^6+\hat{\nu}\xi^4-(1+\hat{\nu})\right]\left[ \hat{\nu}+\xi^2+\xi_0^2-(1+2\hat{\nu}+\xi^2+\xi_0^2)\omega^2\right]}}
\end{equation}
and
\begin{equation}\label{Energy_integral_embeddingII}
\begin{split}
    E_{\mathrm{QQ}}^{(\mathrm{II})}(\xi_0)&=\sqrt{\hat{\mu}}\sqrt{\xi_0^2(\hat{\nu}+\xi_0^2)-(\xi_0^2-1)\left[2(1+\hat{\nu})+\xi_0^2\right]\omega^2}L_{\mathrm{QQ}}^{(\mathrm{II})}(\xi_0)\\
    &+2r_\star\int_{\xi_0}^{\infty}\mathrm{d}\xi\xi \sqrt{\frac{(\xi^2-\xi_0^2) \left[\hat{\nu}+\xi^2+\xi_0^2-(1+2\hat{\nu}+\xi^2+\xi_0^2)\omega^2 \right]}{\xi^2+\hat{\nu}\xi^4-(1+\hat{\nu})}}\\
    &-2r_\star\int_1^{\infty}\mathrm{d}\xi\xi\sqrt{\frac{\xi^2(\hat{\nu}+\xi^2)-(\xi^2-1)\left[2(1+\hat{\nu})+\xi^2\right]\omega^2}{\xi^6+\hat{\nu}\xi^4-(1+\hat{\nu})}} \, . 
\end{split}
\end{equation}

We observe in figure \ref{Wilson_plots_embeddingII} that there is indeed a 
range of values for the newly introduced parameter, which is roughly $0.94\lesssim \omega\lesssim0.99$, that lifts the phase transition of 
embedding $\mathrm{I}$ in the case when $\hat\nu=-0.99$ and $\theta_0=0$. One can think of this phenomenon as the spinning motion introducing a $"$restoring force$"$ effect on the string embedding.\\

{\bf Embedding III.}\quad Finally, for the embedding that wraps around the cigar direction $\phi$ we have the following induced metric, where the embedding is constrained by \eqref{embeddingIII}:

\begin{equation}
    \begin{split}
        \mathrm{d}s_{\text{ind}}^2 &= -\frac{r^2\zeta(r,0)}{L^2}\mathrm{d}t^2 + \left\{\frac{r^2\zeta(r,0)}{L^2}+ \frac{\zeta(r,0)r^{\prime 2}}{r^2 F(r)\lambda^6(r)} +\phi^{\prime 2}\left[r^2F(r)\zeta(r,0) + \frac{A_1^2 }{\zeta(r,0)}\right]\right\}\mathrm{d}w^2,
    \end{split}
\end{equation}
with: 
\begin{equation}\label{embeddingIII_IIB_ABC}
\begin{split}
&    \mathrm{S}_{\mathrm{F}1} =\frac{\mathcal{T}}{2\pi}\int _{-L_{\mathrm{QQ}}/2}^{L_{\mathrm{QQ}}/2}\mathrm{d}w\sqrt{\mathcal{A}^2+\mathcal{B}^2r^{\prime2}+\mathcal{C}^2\phi^{\prime2}},\\
    &\mathcal{A}^2 = \frac{r^4\zeta^2(r,0)}{L^4},\quad \mathcal{B}^2 = \frac{\zeta^2(r,0)}{L^2F(r)\lambda^6(r)},\quad\mathcal{C}^2 = \frac{r^2}{L^2}\left[  A_1^2 +r^2F(r)\zeta^2(r,0) \right].
\end{split}
\end{equation}

We can use the equation of motion for $\phi(w)$ to integrate it out, as the corresponding Hamiltonian is conserved:
\begin{equation}
    \frac{\mathrm{d}}{\mathrm{d}w}\frac{\delta\mathcal{L}_{\mathrm{F}1}}{\delta\phi^{\prime}}=\frac{\mathrm{d}}{\mathrm{d}w}\left[\frac{\mathcal{C}^2\phi^{\prime}}{\sqrt{\mathcal{A}^2+\mathcal{B}^2r^{\prime 2}+\mathcal{C}^2\phi^{\prime 2}}}\right]=0,
\end{equation}
and calling the constant in the last expression $c_{\phi}$, we get: 
\begin{equation}
    \phi^{\prime 2}(w) = \frac{c_\phi^2(\mathcal{A}^2+\mathcal{B}^2r^{\prime2})}{\mathcal{C}^2(\mathcal{C}^2-c_\phi^2)}.
\end{equation}

\begin{figure}[t]
\centering
\begin{subfigure}{0.44\linewidth}
\includegraphics[width=\linewidth]{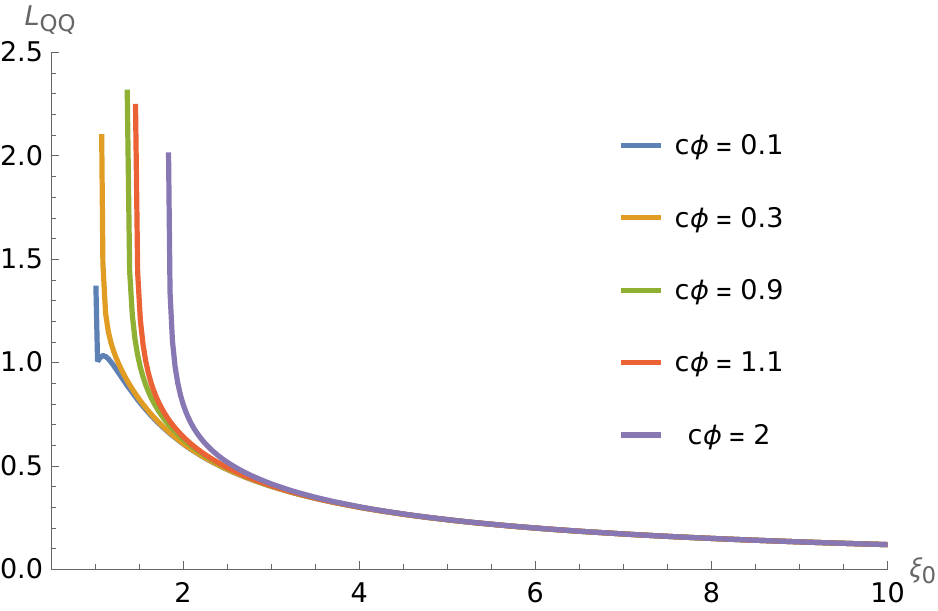}
\caption{Plot of the length of separation as a function of the turning point for various values of $c_{\phi}$ and $\hat\nu=-0.99$.}
\end{subfigure}
\hfill
\begin{subfigure}{0.44\linewidth}
\includegraphics[width=\linewidth]{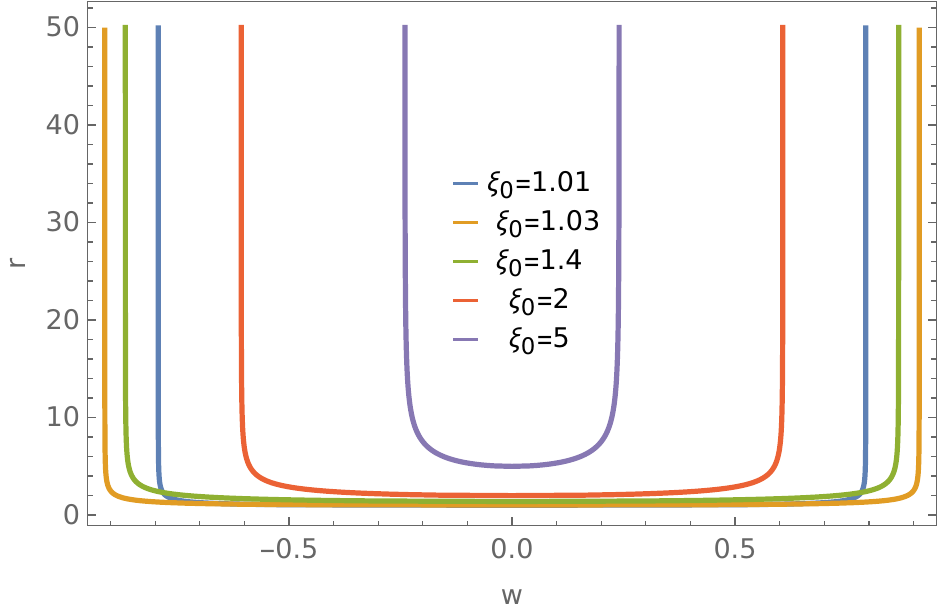}
\caption{Plot of the string profiles in the $(r,w)$ plane for embedding $\mathrm{I}$ ($c_{\phi}=0$), for different turning points and $\hat\nu=-0.99$.}
\end{subfigure}
\hfill
\begin{subfigure}{0.44\linewidth}
    \includegraphics[width=\linewidth]{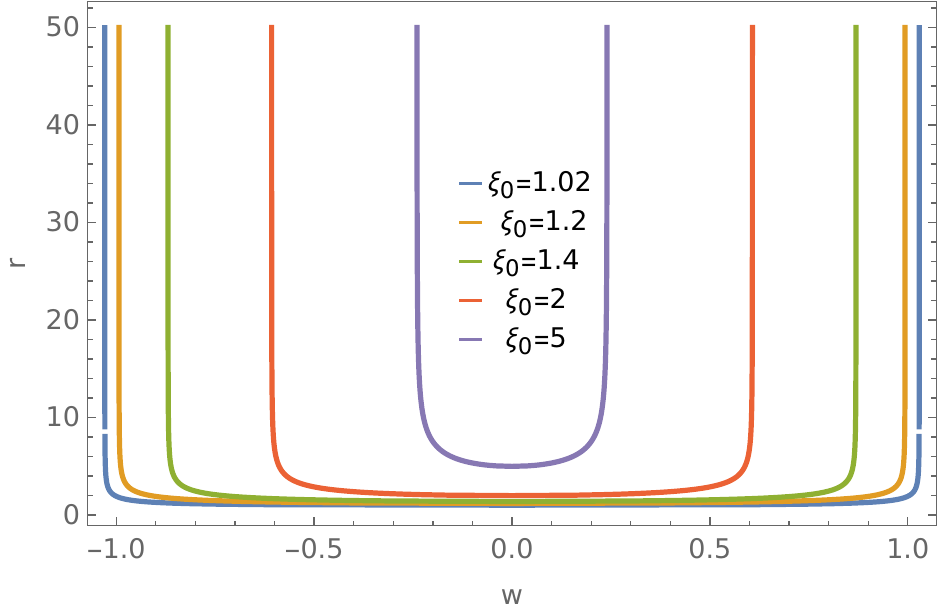}
    \caption{String profiles for embedding $\mathrm{III}$ and various values of $\xi_0$, $\hat\nu=-0.99$ and $c_{\phi}=0.1$.}
\end{subfigure}
\hfill
\begin{subfigure}{0.46\linewidth}
\includegraphics[width=\linewidth]{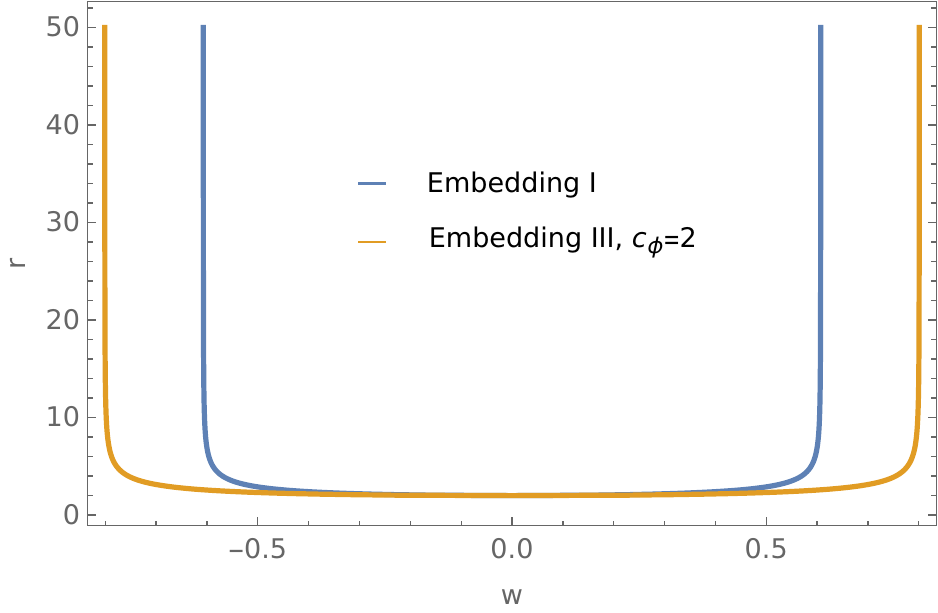}
    \caption{String profiles for both embeddings ($\mathrm{I}$ and $\mathrm{III}$), the same turning point $\xi_0=2$, $\hat\nu=-0.99$ and $c_\phi=2$.}
\end{subfigure}
\caption{We notice that for values of the parameter $c_\phi\gtrapprox 0.3$ the length becomes a single valued function of the turning point when $\hat\nu=-0.99$.}
\label{string_profiles_plots}
\end{figure}

Plugging this back into the action, one arrives at the following expressions:

\begin{equation}
\begin{split}
&\mathrm{S}_{\mathrm{F}1} = \frac{\mathcal{T}}{2\pi}\int_{-L_{\mathrm{QQ}}/2}^{L_{\mathrm{QQ}}/2}\mathrm{d}w\sqrt{\mathcal{F}_{\mathrm{III}}^2+\mathcal{G}_{\mathrm{III}}^2r^{\prime2}},\\
&\mathcal{F}_{\mathrm{III}}^2=\frac{r^6\zeta^2(r,0)\left[A_1^2+r^2F(r)\zeta^2(r,0)\right]}{L^4\left[r^2A_1^2+r^4F(r)\zeta^2(r,0)-c_{\phi}^2L^2\right]},\\
&\mathcal{G}^2_{\mathrm{III}}=\frac{r^2\zeta^2(r,0)\left[A_1^2+r^2F(r)\zeta^2(r,0)\right]}{L^2F(r)\left[r^2A_1^2+r^4F(r)\zeta^2(r,0)-c_{\phi}^2L^2\right]\lambda^6(r)},
\end{split}
\end{equation}
or, in terms of the dimensionless quantities: 
\begin{equation}
    \begin{split}
    &\mathcal{F}^2_{\mathrm{III}}= \frac{\hat{\mu}^2\xi^2(\xi^2-1)(\xi^2+\hat{\nu})\left[ \xi^2+2(1+\hat{\nu})\right]}{\hat{\mu}(\xi^2-1)\left[\xi^2+2(1+\hat{\nu})\right] -c_{\phi}^2},\\    &\mathcal{G}_{\mathrm{III}}^2=\frac{\hat{\mu}\xi^4(\xi^2+\hat{\nu}) \left[\xi^2+2(1+\hat{\nu})\right]}{\left[\xi^4+\xi^2(1+\hat{\nu})+1+\hat{\nu}\right]\left[\hat{\mu}(\xi^2-1)\left(\xi^2+2(1+\hat{\nu})\right)-c_\phi^2 \right]},
    \end{split}
\end{equation}
which agree with \eqref{F_G_embeddingI_dimles} when $c_\phi=0$ and $\theta_0=0$ (the later imposed by the consistency of the embedding). Similarly to the previous embedding of a spinning string, non zero values of the integration constant $c_\phi$ allow the length to become single valued and therefore there is no phase transition in the energy of the configuration, see figure \ref{string_profiles_plots} $\mathrm{(a)}$. The effect of $c_\phi$ is clear when one observes the behaviour of the string profiles: in figure \ref{string_profiles_plots} $\mathrm{(b)}$ we see that as the string approaches the end of space ($\xi_0$ monotonically approaching 1), its $\texttt{U}$ shaped embedding $"$opens$"$ and then $"$closes$"$ again, in other words the profiles are not ordered with the monotonically decreasing $\xi_0$, in contrast with the embedding $\mathrm{III}$ profiles seen in \ref{string_profiles_plots} $\mathrm{(c)}$. This signals the lift of the double-valuedness of $L_{\mathrm{QQ}}(\xi_0)$. The final plot \ref{string_profiles_plots} $\mathrm{(d)}$ shows the profiles for a specific $\xi_0$ in both embeddings, given the value $c_\phi=2$ for embedding $\mathrm{III}$. We note that for the same $\xi_0$ embedding $\mathrm{III}$ gives a smaller length of separation than embedding $\mathrm{I}$, or conversely $c_\phi$ has the effect of obstructing the string from reaching values closer to the end of space compared to embedding $\mathrm{I}$ (for the same fixed length).\\

\subsubsection{Wilson loop for the 11d backgrounds}

In order to study the dynamics of a Wilson loop-like object in the M-theory background \eqref{ds11_LLM1}-\eqref{ds11_LLM2}, the first step we take is making the change of coordinates to write the Riemann surface in polar coordinates: 
\begin{equation}
\mathrm{d}v_1^2+\mathrm{d}v_2^2=\mathrm{d}\rho^2+\rho^2\mathrm{d}\beta^2,
\end{equation}
without making any extra assumptions about a symmetry in $\beta$. We can then use a probe $\mathrm{M}2$ brane whose worldvolume coordinates $\hat{\sigma}$ extend in the submanifold $\Sigma_3[t,w,\beta]$, while it also has a $w$-dependent profile in the radial direction: $r=r(w)$. The rest of the coordinates are kept constant, however one should choose a suitable location for the $\mathrm{M}2$ brane in $y$ such that $y\,\partial_yD_0(y,\rho,\beta)$ approaches infinity. This allows us to compute the $y$ and $D_0$ dependent prefactor that enter in the action of the brane and get an answer in terms of $r$. The induced metric on the brane can be compactly written as: 

 \begin{eqnarray}
& & \mathrm{d}s_{\mathrm{M}2}^2= G_{tt}\mathrm{d}t^2 + (G_{ww} + G_{rr}r'^2)\mathrm{d}w^2+ G_{\beta\beta}\mathrm{d}\beta^2,\nonumber\\
& & G_{tt}= -e^{2\hat{\lambda}} {\cal Z} \frac{4}{X} \frac{r^2\lambda^2(r)}{L^2},~~G_{rr}=e^{2\hat{\lambda}} {\cal Z} \frac{4}{X} \frac{1}{r^2\lambda^4(r) F(r)},\\
 & & G_{\beta\beta}=-e^{2\hat{\lambda}} {\cal Z} \frac{\partial_y e^{D_0}}{y}\rho^2, ~~G_{ww}=-G_{tt}. \nonumber
 \end{eqnarray}

We can use a specific solution for $D_0(y,\rho,\beta)$ in order to illustrate the calculation. Consider the simple solution presented in \cite{Gaiotto:2009gz}:
\begin{equation}
    e^{D_0}=4\frac{N^2-y^2}{(1-\rho^2)^2},\quad y\,\partial_y D_0(y,\rho)=\frac{2y^2}{y^2-N^2},
\end{equation}
suggesting we place the probe $\text{M}2$ at $y=N$. We then have\footnote{we place the brane at $\rho_0$.}
\begin{equation}
\begin{split}
&    -\mathrm{det}\,g_{\text{ind},\mathrm{M2}}=-\frac{16\rho^2e^{6\hat{\lambda}{\cal Z}^3}}{X^2(r)}\frac{\partial_y e^{D_0(y,\rho_0)}}{y}\left[ \frac{r^4\lambda^4(r)}{L^4}+\frac{r^{\prime 2}}{L^2\lambda^2(r)F(r)}\right],\\
&\text{with}\quad -\frac{16\rho_0^2e^{6\hat{\lambda}{\cal Z}^3}}{X^2(r)}\frac{\partial_y e^{D_0(y,\rho_0)}}{y}=\frac{128N^2\rho_0^2}{(1-\rho_0^2)^2}X(r)\left[ \frac{r^4\lambda^4(r)}{L^4}+\frac{r^{\prime 2}}{\lambda^2(r)L^2F(r)}\right]
\end{split}
\end{equation}

and substituting $X(r)=\lambda^2(r)$, the action reads:

\begin{equation}
\begin{split}
   & \mathrm{S}_{\mathrm{M}2}\propto T_{\mathrm{M}2}\int _{\Sigma_3}\mathrm{d}^3\hat\sigma\sqrt{-\mathrm{det}g_{\text{ind,M2}}}\propto 2\pi \mathcal{T}T_{\text{M2}}\int _{-L_{\mathrm{QQ}}/2}^{L_{\mathrm{QQ}}/2}\mathrm{d}w\sqrt{\mathcal{F}^2+\mathcal{G}^2r^{\prime 2}},\\
    &\mathcal{F}^2=\frac{r^4\lambda^6(r)}{L^4},\quad \mathcal{G}^2=\frac{1}{L^2F(r)}.
    \end{split}
\end{equation}

We notice that the functions appearing in the dynamical part of the expression are indeed the same as \eqref{WL_embedding_I_action_IIB} for $\theta_0=0$, therefore the analysis of the previous subsection yields the same results in this case. We conclude that the dynamics of the effective string described by the $\mathrm{M}2$ in these backgrounds is the same as the $\mathrm{F}1$ string of the Type IIB solution and it captures confinement.

\subsubsection{Wilson loop for the Type IIA backgrounds}

We now wish to apply the same analysis for the family of backgrounds presented in \eqref{10dGM-2}. This calculation proves to be more subtle, however it too contains the same dynamical part as the previous two cases. The complication with this backgrounds enters because of the various functions $\tilde{f}_i$ which after our deformation have acquired a non-trivial dependence on $r$. To circumvent this and extract the $r$-dependent part out of the expression, which holds the information of the dynamics, we choose to localise the probe in the $(\sigma,\eta)$ subspace by taking $\eta=\eta_\star$ to be an integer in $(0,P)$ and $\sigma\to 0$. The later will offer us a clearer interpretation of the dual quiver data, as we have $\dot{V}(\sigma,\eta  )\Big|_{\sigma=0}=\mathcal{R}(\eta)$, where $\mathcal{R}$ contains information about the location of each gauge group on the quiver. To calculate the various expressions involving the $\tilde{f}_i$s we can first expand the Laplace potential\footnote{With boundary conditions in the $\eta$ direction: $\dot{V}(\sigma,\eta)\Big|_{\eta=0}=0=\dot{V}(\sigma,\eta)\Big|_{\eta=P}$. The connection with \eqref{laplace_potential_generic} can be made by setting $\mathcal{R}_k = \frac{P a_k}{k \pi}$.} $V$ and its derivatives in Fourier-Bessel series as:
\begin{equation}\label{V_expansions}
    \begin{split}
        &V(\sigma,\eta)=- \sum_{k=1}^\infty\left(\frac{P}{k\pi}\right) a_k\sin\left(\frac{k\pi}{P}\eta\right)  K_0\left(\frac{k\pi}{P}\sigma\right),\\
        &\dot{V}(\sigma,\eta)=  \sigma \sum_{k=1}^\infty a_k\sin\left(\frac{k\pi}{P}\eta\right)K_1\left(\frac{k\pi}{P}\sigma\right) ,\\
        &V^{\prime\prime}(\sigma,\eta)=\sum_{k=1}^\infty\left(\frac{k\pi}{P}\right) a_k\sin\left(\frac{k\pi}{P}\eta\right)K_0\left( \frac{k\pi}{P}\sigma\right),\\
        &\ddot{V}(\sigma,\eta) = -\sigma^2\sum_{k=1}^\infty\left(\frac{k\pi}{P}\right) a_k\sin\left(\frac{k\pi}{P}\eta\right)K_0\left(\frac{k\pi}{P}\sigma\right) ,\\
      & \dot{V}^{\prime}(\sigma,\eta) = \sigma \sum_{k=1}^\infty \left(\frac{k\pi}{P}\right)a_k\cos\left(\frac{k\pi}{P}\eta\right)K_1\left(\frac{k\pi}{P}\sigma\right),
    \end{split}
\end{equation}
and use the known asymptotic behaviour of the modified Bessel functions: 
\begin{equation}
  \text{As}\,\, x\to 0:\quad  xK_1(x)\to1, \,\, K_0(x) \approx -\gamma_{\mathrm{EM}}+\log\frac{2}{x}+\mathcal{O}(x^2).
\end{equation}

One can notice that all of the above expressions are finite as $\sigma\to 0$, except $V^{\prime\prime}$. Let us write $\sigma=\epsilon$ with $\epsilon\to 0$, which then gives the following asymptotic behaviour:
\begin{equation}
    \begin{split}
        &\lim_{\epsilon\to0}\dot{V}(\epsilon,\eta)=\mathcal{R}(\eta), \quad \lim_{\epsilon\to0}\dot{V}^{\prime}(\epsilon,\eta)=\mathcal{R}^{\prime}(\eta),\\
        &\lim_{\epsilon\to0}\ddot{V}(\epsilon,\eta)=0, \quad V^{\prime\prime}(\epsilon,\eta)\approx \hat{M}(\epsilon),
    \end{split}
\end{equation}
where we denoted as $\hat M(\epsilon)$ the limiting expression of $V^{\prime\prime}$ which diverges as $\log \epsilon^{-1}$. This in turn, allows us to write:
\begin{equation}\label{typeIIA_limits}
    \begin{split}
    \text{As}\,\,\sigma\to 0:\quad     &\tilde{\Delta}\approx 2\hat{M}(\epsilon)\mathcal{R}(\eta),\quad Z^3\to X^3,\quad \tilde{f}_1\to \mathcal{R}^{2/3}(\eta)\\[5pt]
        &\tilde{f}\to 1,\quad \tilde{f}_1^3\tilde{f}_5\approx 2X\mathcal{R}(\eta)\hat{M}^{-1}(\epsilon)\to 0,\quad \tilde{f}_2\to X^{-2},\quad \tilde{f}_3\to0,
    \end{split}
\end{equation}
which we will make use of in our calculations. 

In the following we introduce probe fundamental strings and consider all the different embeddings that were studied in subsection \ref{WL_IIB_subsection}.

{\bf Embedding I.}\quad Let us start by the simplest $\mathrm{F1}$ embedding with a profile $r=r(w)$, $t=\tau$ and $w=\hat\sigma$. We find the induced metric to be: 
\begin{equation}
 \mathrm{d}s^2_{\text{ind}}= 4\tilde{f}_1^{3/2}\tilde{f}_5^{1/2}\tilde{f}\frac{r^2\lambda^2(r)}{L^2}\left\{ -\mathrm{d}t^2 + \left[1 + \frac{L^2r^{\prime 2}}{r^4F(r)\lambda^6(r)} \right]\mathrm{d}w^2\right\},
\end{equation}
and the Nambu-Goto action\footnote{We use the shorthand notation $\tilde{f}_i\equiv\tilde{f}_i(\sigma,\eta_{\star},r)$.}: 
\begin{equation}
\begin{split}
&\mathrm{S}_{\mathrm{F}1} =\frac{\mathcal{T}}{2\pi}\int _{-L_{\mathrm{QQ}}/2}^{L_{\mathrm{QQ}}/2}\mathrm{d}w\sqrt{\hat{\mathcal{F}}_{\text{I}}^2 + \hat{\mathcal{G}}_{\text{I}}^2r^{\prime 2}} \quad {\rm with} \quad  \hat{\mathcal{F}}_{\text{I}}^2= 16\tilde{f}_1^3\tilde{f}_5\tilde{f}^2\frac{r^4\lambda^4(r)}{L^4},\quad \hat{\mathcal{G}}_{\text{I}}^2 =\frac{16\tilde{f}_1^3\tilde{f}_5\tilde{f}^2}{L^2F(r)\lambda^2(r)}.
\end{split}
\end{equation}
By employing \eqref{typeIIA_limits} we get as $\sigma=\epsilon\to 0$: 
\begin{equation}\label{F_and_G_limits_WLI_IIA}
    \hat{\mathcal{F}}_{\text{I}}^2\approx\frac{32\mathcal{R}(\eta_\star)}{\hat{M}(\epsilon)}X(r)\frac{r^4\lambda^4(r)}{L^4},\quad \hat{\mathcal{G}}_{\text{I}}^2\approx \frac{32\mathcal{R}(\eta_\star) }{\hat{M}(\epsilon)}\frac{X(r)}{L^2F(r)\lambda^2(r)}.
\end{equation}

We notice again the radial part of the expressions being the same as the one found in the 11d and type IIB backgrounds and therefore the behaviour of the Wilson loops in this case is also described by figure \ref{Wilson_plots_embeddingI}. Moreover, The numerical factor describing the conformal quiver in the UV, contains the rank of the gauge group in which we choose to compute the Wilson loop, $N_{\star}={\cal R}(\eta_{\star})$.\\

\textbf{Embedding II.} Here we repeat the study of the spinning string embedding in this background by taking: 
\begin{equation}
\begin{split}
  &  t=\tau,\quad \hat{\sigma}=w,\quad \phi=\omega\tau,\quad r=r(w),\quad z=z_0,\\[5pt]
  &\theta=\theta_0,\quad \varphi=\varphi_0,\quad \chi=\chi_0,\quad \sigma=\epsilon\to0,\quad \eta=\eta_\star \, .
    \end{split}
\end{equation}
Then the induced metric on the string is: 
\begin{equation}
    \begin{split}
        \mathrm{d}s^2_{\mathrm{ind}}&=(\tilde{f}_1^3\tilde{f}_5)^{1/2}\left[ -4\tilde{f}\frac{r^2\lambda^2(r)}{L^2}+\omega^2\left(\tilde{f}_3A_3^2+2\tilde{f}_2\sin^2\theta_0A_1^2+4L^2\tilde{f}F(r)\frac{r^2\lambda^2(r)}{L^2} \right) \right]\mathrm{d}t^2\\
        &+4(\tilde{f}_2^3\tilde{f}_5)^{1/2}\tilde{f}\left[ \frac{r^2\lambda^2(r)}{L^2}+\frac{r^{\prime 2}}{r^2F(r)\lambda^4(r)}\right]\mathrm{d}w^2,
    \end{split}
\end{equation}
and the Nambu-Goto action reads: 
\begin{equation}
    S_{\mathrm{F}1}=\frac{1}{2\pi}\int\mathrm{d}t\mathrm{d}w\sqrt{-\mathrm{det}g_{\mathrm{ind}}}=\frac{\mathcal{T}}{2\pi} \int_{-L_{\mathrm{QQ}}/2}^{L_{\mathrm{QQ}}/2}\mathrm{d}w\sqrt{\hat{\mathcal{F}}_{\mathrm{II}}^2+\hat{\mathcal{G}}^2_{\mathrm{II}}r^{\prime2}},
\end{equation}
where:
\begin{equation}
    \begin{split}
        &\hat{\mathcal{F}}_{\mathrm{II}}^2=4\tilde{f}_1^3\tilde{f}_5\tilde{f}\frac{r^2\lambda^2}{L^2}\left[ 4\tilde{f}\frac{r^2\lambda^2}{L^2} -\omega^2\left( \tilde{f}_3A_3^2+2\tilde{f}_2\sin^2\theta_0A_1^2+4\tilde{f}Fr^2\lambda^2\right)\right],\\
        &\hat{\mathcal{G}}_{\mathrm{II}}^2=4\tilde{f}_1^3\tilde{f}_5\tilde{f}\left[ \frac{4\tilde{f}}{L^2\lambda^2F} -\omega^2\frac{\tilde{f}_3A_3^2+2\tilde{f}_2\sin^2\theta_0A_1^2+4F\tilde{f}r^2\lambda^2}{r^2\lambda^4F}\right].
    \end{split}
\end{equation}
We can now take the limit $\epsilon\to 0$ using \eqref{typeIIA_limits} to obtain the following expressions: 
\begin{equation}
    \begin{split}
        &\hat{\mathcal{F}}_{\mathrm{II}}^2\approx\frac{32\mathcal{R}(\eta_\star)}{\hat{M}(\epsilon)}\left[ X(r)\frac{r^4\lambda^4}{L^4} - \omega^2\frac{r^2\lambda^2X(r)}{4L^2}\left(2X^{-2}(r)\sin^2\theta_0A_1^2+4F(r)r^2\lambda^2(r) \right) \right],\\
        &\hat{\mathcal{G}}_{\mathrm{II}}^2\approx \frac{32\mathcal{R}(\eta_\star) }{\hat{M}(\epsilon)}\left[ \frac{X(r)}{L^2\lambda^2(r)F(r)}- \frac{\omega^2}{2}\frac{\sin^2\theta_0A_1^2+2F(r)X^2(r)r^2\lambda^2(r)}{r^2X(r)\lambda^4(r)F(r)}\right].
    \end{split}
\end{equation}

We first notice that when $\omega=0$ the above agree with the case of embedding $\mathrm{I}$ \eqref{F_and_G_limits_WLI_IIA}, which in turn agrees with the type $\mathrm{IIB}$ results of embedding $\mathrm{I}$ \eqref{WL_embedding_I_action_IIB}. The expressions appearing in the $\omega^2$ terms also match the ones in \eqref{F_G_embeddingII_IIB} when one imposes $\theta_0=0$ for the $\mathrm{IIB}$ case and $\theta_0=\pi/2$ for the $\mathrm{IIA}$ (notice that the angle $\theta_0$ is not the same coordinate for the two backgrounds). Consequently, one gets the same behaviour as the one depicted in the figure \ref{Wilson_plots_embeddingII} for the case of the type $\mathrm{IIB}$ version of this embedding.

\textbf{Embedding III.} We will now study the third and final embedding for the type IIA backgrounds, that is a string wrapping the shrinking circle. In this case \eqref{embeddingIII} gives the following induced metric on the string: 
\begin{equation}
    \mathrm{d}s^2_{\mathrm{ind}}=\tilde{f}_1^{3/2}\tilde{f}_5^{1/2}4\tilde{f}\frac{r^2\lambda^2(r)}{L^2}\left\{ -\mathrm{d}t^2+\left[1+\frac{L^2r^{\prime 2}}{r^4\lambda^6(r)F(r)} +\left(F(r)+ \frac{\tilde{f}_2L^2\cos^2\theta_0 (\mathcal{A}^{(3)})^2}{2\tilde{f}r^2\lambda^2(r)} \right)\phi^{\prime 2}\right]\mathrm{d}w^2 \right\}.
\end{equation}
After taking limits of the various functions as $\sigma=\epsilon\to 0$ we obtain the Nambu-Goto action: 
\begin{equation}
    \mathrm{S}_{\mathrm{F}1} = \frac{1}{2\pi}\int \mathrm{d}t\mathrm{d}w\sqrt{-\mathrm{det}g_{\mathrm{ind}}}\approx\frac{\mathcal{T}}{2\pi}\sqrt{\frac{32\mathcal{R}(\eta_\star)}{\hat{M}(\epsilon)}}\int _{-L_{\mathrm{QQ}}/2}^{L_{\mathrm{QQ}}/2}\mathrm{d}w\sqrt{\hat{\mathcal{A}}^2+\hat{\mathcal{B}}^2r^{\prime2}+\hat{\mathcal{C}}^2\phi^{\prime2}},
\end{equation}
with 
\begin{equation}
    \hat{\mathcal{A}}^2=\frac{r^4 \lambda^6(r)}{L^4},\quad \hat{\mathcal{B}}^2=\frac{X(r)}{\lambda^2(r)F(r)L^2},\quad \hat{\mathcal{C}}^2= \frac{r^4\lambda^4(r)}{L^4}X(r)\left[F(r)+\frac{L^2\cos^2\theta_0A_1^2}{X^2(r)r^2\lambda^2(r)}\right].
\end{equation}

Once again, these functions are the same as the ones found in the corresponding embedding in the type $\mathrm{IIB}$ case \eqref{embeddingIII_IIB_ABC} when we set $\theta_0=0$. One can then follow the same procedure of using the conserved momentum of $\phi$ to integrate it out and introduce the new parameter $c_{\phi}$ to the problem. The results are the same as in the type $\mathrm{IIB}$ case, modulo the divergent prefactor in the action, and can be seen in figure \ref{string_profiles_plots}.\\

Let us summarize our findings regarding the Wilson loop observable. Our previous study in section \ref{singularity_section} revealed that the supergravity background in the case where $\varepsilon=-1$ has a singularity when $\hat{\nu}=-1$ at $r=r_\star$. We deduced that this happens when $q_1=q_2=0$, in which case the gauge fields $A_1,A_2$ and $A_3$ are set to zero and there is no fibration. We therefore decided to accept the {\it smooth} supergravity solutions in the regime $\hat{\nu}>-1$ with nonzero gauge fields.\\

There is however the following interesting phenomenon: for values of $\hat\nu$ very close to $-1$ Wilson loop embeddings start "feeling" the need for higher curvature corrections in the action, which results in a first order phase transition making its appearance, much like the one in the singular background of \cite{Brandhuber:1999jr}. This happens for the case of embedding $\mathrm{I}$ in the type $\mathrm{IIB}$ background in a specific direction $\theta=0$.\\

Our analysis however, shows that the backgrounds are dual to confining field theories\footnote{as long as one does not allow the string embedding to explore the $\eta$ direction, where there are localized sources describing fundamental degrees of freedom. In this case the Wilson loop will probably undergo screening, as was shown in a different system in \cite{Giliberti:2024eii}.}, since the energy as a function of the length of separation for a probe quark-anti-quark pair interpolates from a Coulombic behaviour in the $\mathrm{UV}$, due to the solutions being asymptotically $\mathrm{AdS}$, to a linear one in the $\mathrm{IR}$. This holds for all values of the parameters of the theory as well as all values of $\hat{\nu}>-1$. The later is supported by the study of different types of embeddings, which introduce new parameters allowing us to reveal the confining behaviour and lift the apparent phase transition for $\hat\nu\approx-1$.


\subsection{'t Hooft loop}

In this section we will study the 't Hooft loop, which is the non local operator associated with magnetic charges, in analogy with its electric equivalent, the Wilson loop. The idea is to probe the background with some suitable brane\footnote{One usually aims to wrap the shrinking circle $\mathbb{S}^1[\phi]$ which plays an important role in the $\mathrm{IR}$ dynamics, as well as some closed submanifold of the internal space.} that effectively looks like a magnetic string in the submanifold spanned by the field theory coordinates $\mathbb{R}^{1,2}[t,w,\phi]$, which in turn captures the dynamics of monopole-anti-monopole pairs. One can then deduce if screening takes place, that is, if the pair moves freely at no energy cost, or equivalently, the tension of the $"$chromomagnetic string$"$ is zero. In the gravity picture this amounts to the system energetically preferring a disconnected embedding of the probe, unlike a $\mathrm{U}$ shaped one, which can happen when two different values of the parameter $r_0$ correspond to the same value for the length of separation $L_{\mathrm{MM}}(r_0)$. \\

The action for the 't Hooft loop will have a similar form as the one for the Wilson loop in the effective two dimensional worldsheet part:
\begin{equation}
S_{\mathrm{D}_p}=T_{\mathrm{D}_p}\int_{\Sigma_{p+1}}\mathrm{d}^{p+1}\hat\sigma\sqrt{-e^{-2\Phi}\mathrm{det}(g_{\mathrm{ind}})}\propto \int_{-L_{\mathrm{MM}}/2}^{L_{\mathrm{MM}}/2}\mathrm{d}w\sqrt{\mathcal{F}_t^2+\mathcal{G}_t^2r^{\prime 2}},
\end{equation}
where one can relate the value of the function $\mathcal{F}_t$ at the end of space with the tension of the string:
\begin{equation}
T_{\mathrm{eff}}\propto\mathcal{F}_t(r_0).
\end{equation}
The inclusion of the shrinking coordinate in the induced metric will typically contribute a multiplicative factor of the warping function $F(r)$ in $\mathcal{F}_t$, which vanishes smoothly as $r\to r_{\star}$. This is the case when a theory is in the confining phase, where the potential for Wilson loop obeys the area law, while the one for the 't Hooft loop the perimeter law \cite{tHooft:1977nqb}.\\

In the following, we study the dynamics of the effective string using various type IIB brane embeddings. Up to constant prefactors, we find that all the different probes give the same 't Hooft loop observable.

\subsubsection{'t Hooft loop for the Type IIB background}

{\bf D3 probe.} First, we look at a D3 probe extended in $\Sigma_4=[t,w,\phi,\phi_3]$ with the following embedding ansatz:
\begin{equation}
    r=r(w), \quad \theta=\theta_0=\frac{\pi}{2}, \quad \phi_1 =\phi_2 =\text{constant}, \quad z=z_0, \quad \psi=\psi_0.
\end{equation}
The equations of motion are satisfied and the embedding is consistent as long as $\theta_0=\frac{\pi}{2}.$ 

We have the induced metric
\begin{equation}
\begin{split}
    \mathrm{d}s^2_{\mathrm{ind, D3}} =& \frac{\zeta(r,\theta_0)}{L^2}\left[ -r^2 \mathrm{d}t^2 + F(r)r^2L^2\mathrm{d}\phi^2 + \mathrm{d}w^2\left(r^2 + \frac{L^2 r^{\prime 2}}{F(r)\lambda(r)^6 r^2}\right)  \right] \\
    &+ \frac{L^2}{\zeta(r,\theta_0)}\lambda^6(r) \left(\mathrm{d}\phi_3 + \frac{A_3}{L} \right)^2.
\end{split}
\end{equation}
Here we have $\zeta(r,\theta_0=\frac{\pi}{2})=1$, and the dilaton is trivial. The Dirac-Born-Infeld action for the $\mathrm{D}3$ yields the following:
\begin{equation}
    S_{\mathrm{D}3} = \mathrm{T}_{\mathrm{D}3}\int_{\Sigma_4} \mathrm{d}^4\hat\sigma \sqrt{-e^{-2\Phi}\mathrm{det}(g_{\mathrm{ind},\mathrm{D}3})} = \hat{{\cal N}}_{\text{D}3} \int_{-L_{\mathrm{MM}}}^{L_{\mathrm{MM}}} \mathrm{d}w \; \sqrt{\mathcal{F}^2_t + \mathcal{G}^2_t r^{\prime 2}},
\end{equation}
where we once again notice the universal nature of the observable. The functions appearing in the dynamical part are: 
\begin{equation}\label{thooft_F_and_G}
\mathcal{F}^2_t=r^6 \lambda(r)^6 F(r),\quad \mathcal{G}_t^2 =L^2r^{ 2}
\end{equation}
and the constant $\text{UV}$ prefactor reads:
\begin{equation}
   \hat{{\cal N}}_{\text{D}3} = \frac{T_{D3}}{L}\int \; \mathrm{d}t \; \mathrm{d}\phi \; \mathrm{d}\phi_3=\frac{\mathcal{T}T_{\mathrm{D}3}}{L}2\pi L_{\phi},
\end{equation}
where $\mathcal{T}$ is the extent of the time coordinate. In terms of the dimensionless parameters we have:

\begin{figure}[t]
\centering
\begin{subfigure}{0.44\linewidth}
\includegraphics[width=\linewidth]{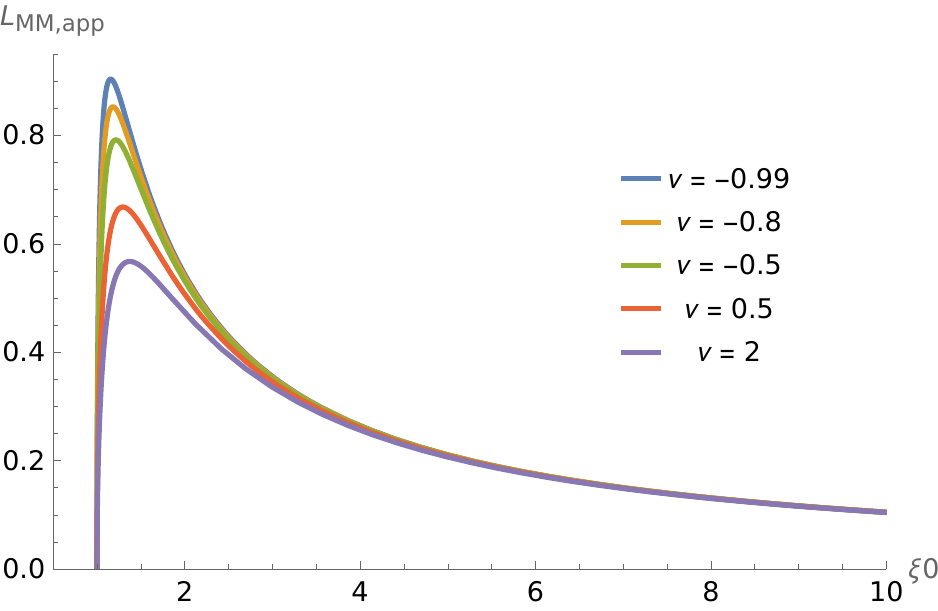}
\caption{Plot of the approximate length of separation of a monopole-anti-monopole pair as a function of the the turning point for different values of $\hat{\nu}$.}
\end{subfigure}
\hfill
\begin{subfigure}{0.44\linewidth}
\includegraphics[width=\linewidth]{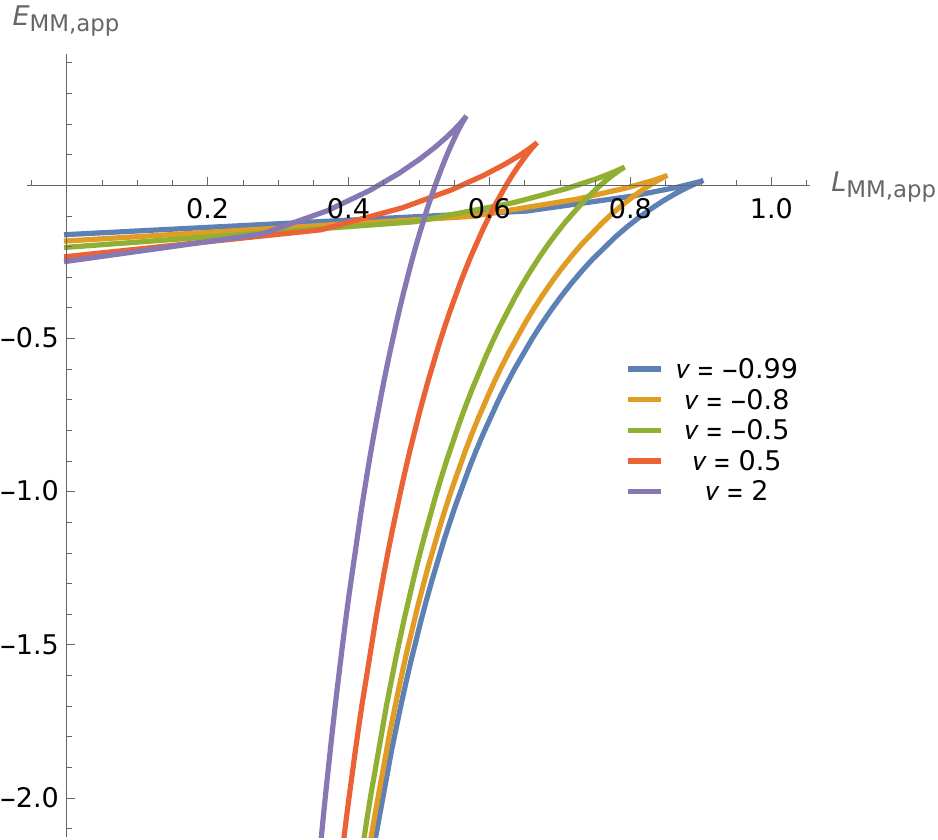}
\caption{Plot of the approximate energy of the pair as a function of the approximate length for different values of $\hat{\nu}$.}
\end{subfigure}
\hfill
\begin{subfigure}{0.44\linewidth}
    \includegraphics[width=\linewidth]{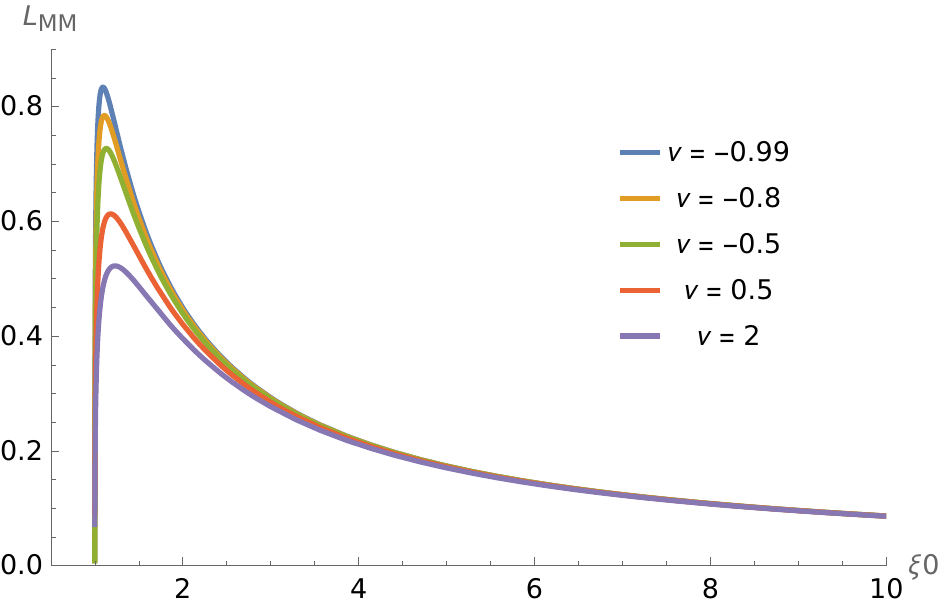}
    \caption{Plot of the length as a function of the the turning point for different values of $\hat{\nu}$.}
\end{subfigure}
\hfill
\begin{subfigure}{0.46\linewidth}
\includegraphics[width=\linewidth]{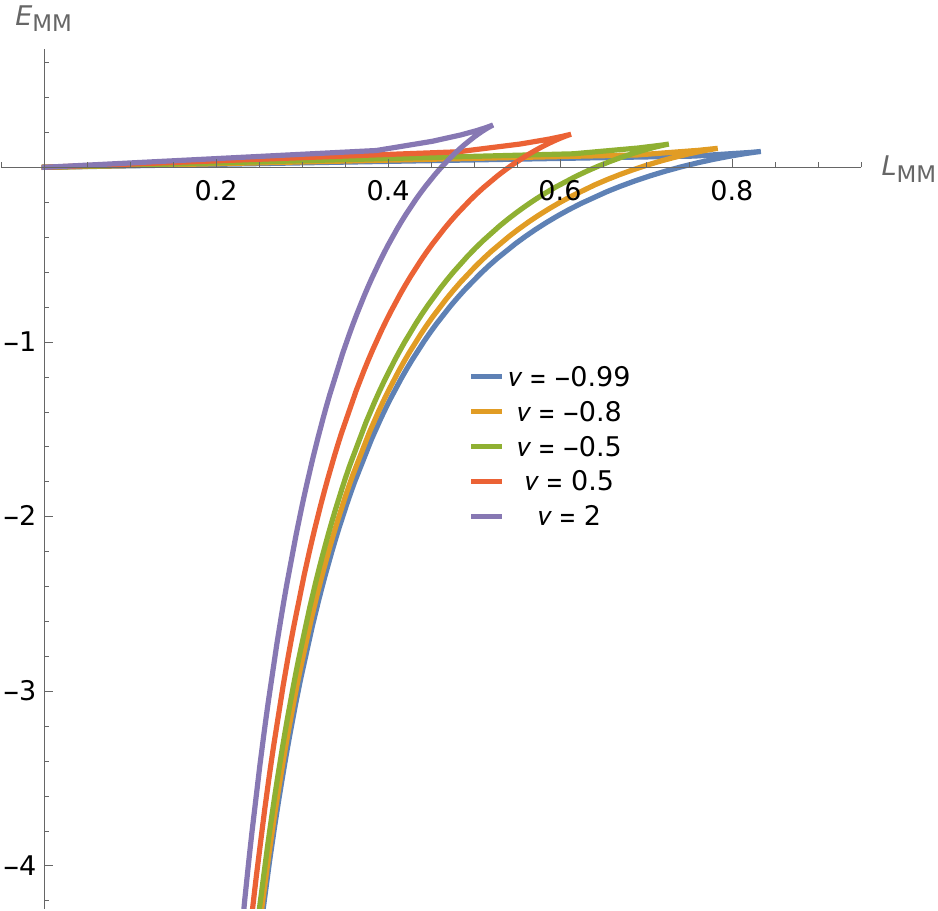}
    \caption{Plot of the energy with respect to the lenth of separation for different values of $\hat{\nu}$.}
\end{subfigure}
\caption{Resulting plots for the 't Hooft loop length and energy expressions, given in \eqref{length_integral_thooft_IIB}-\eqref{approximate_expressions_thooft_IIB}.}
\label{thooft_plots_IIB}
\end{figure}

\begin{equation}
    \mathcal{F}_t^2 = \frac{r_\star^6}{L^2}\left[ \xi^4(\hat{\nu}+\xi^2)-(1+\hat{\nu})\right],\quad \mathcal{G}_t^2 = L^2r_\star^2\xi^2,
\end{equation}
from which we can form the length and energy integrals using the same equations as in the case of the Wilson loop \eqref{WL_integrals}, \eqref{approximate_expressions}. The resulting expressions are listed in appendix \ref{Details_computations}, in equations \eqref{length_integral_thooft_IIB}, \eqref{energy_integral_thooft_IIB} and 
\eqref{approximate_expressions_thooft_IIB}.

The resulting plots of both the approximate and exact expressions are depicted in figure \ref{thooft_plots_IIB}, where we once more notice how well the approximate expressions capture the dynamics of the system. The double-valuedness of the length for any value of $\hat\nu$ leads to a phase transition and monopole screening taking place. The system prefers energetically the disconnected embedding in which the monopole-anti-monopole pair moves freely. This phase transition in the 't Hooft loop (that can be interpreted as the screening of non-dynamical monopoles) further supports our claims for quark confinement in the dual field theory \cite{tHooft:1977nqb}.\\

{\bf D5 probe.}\quad Next we consider a D5 brane extended in the submanifold $\Sigma_6=[t,w,\phi,\phi_1,\phi_2,\phi_3]$, with\footnote{The full equations of motion enforce $\cos(2\psi_0)=0$ and $-1+3\cos(2\theta_0)=0$, in order for the embedding to be consistent.}:
\begin{equation}
    r=r(w),\quad z=z_0,\quad \theta=\theta_0=\frac{1}{2}\arccos\left( \frac{1}{3} \right),\quad \psi_0=\frac{\pi}{4}.
\end{equation}
The induced metric on the brane is:
\begin{equation}
    \begin{split}
        \mathrm{d}s_{\text{ind}}^2 &= \frac{\zeta(r,\theta_0)r^2}{L^2}\left( -\mathrm{d}t^2+F(r)L^2\mathrm{d}\phi^2 \right) + \frac{L^2}{\zeta(r,\theta_0)}\cos^2\theta_0\sin^2\psi_0\left( \mathrm{d}\phi_1 + \frac{A_1}{L}\mathrm{d}\phi \right)^2 \\
        &+\frac{L^2}{\zeta(r,\theta_0)}\cos^2\theta_0\cos^2\psi_0\left( \mathrm{d}\phi_2 + \frac{A_2}{L}\mathrm{d}\phi \right)^2 + \frac{L^2\lambda^6(r)}{\zeta(r,\theta_0)}\sin^2\theta_0 \left( \mathrm{d}\phi_3 + \frac{A_3}{L}\mathrm{d}\phi \right)^2\\
        &+\left[\frac{r^2\zeta(r,\theta_0)}{L^2}+\frac{\zeta(r,\theta_0)r^{\prime 2}}{r^2F(r)\lambda^6(r)}\right]\mathrm{d}w^2,
    \end{split}
\end{equation}
and its action gives the following expression:
\begin{equation}
\begin{split}
&    S_{\mathrm{D}5} =\mathrm{T}_{\mathrm{D}5}\int_{\Sigma_6}\mathrm{d}^6\hat\sigma \sqrt{-e^{-2\Phi}\mathrm{det}(g_{\mathrm{ind},\mathrm{D}5})}=\hat{{\cal N}}_{\mathrm{D}5}\int_{-L_{\mathrm{MM}}/2}^{L_{\mathrm{MM}}/2}\mathrm{d}w\sqrt{\mathcal{F}_t^2+\mathcal{G}_t^2r^{\prime 2}},\\
&\mathcal{F}_t=r^3\lambda^3(r)\sqrt{F(r)},\quad \mathcal{G}_t=Lr,\quad \hat{{\cal N}}_{\mathrm{D}5}= \frac{8\pi^3}{3\sqrt{3}}L\mathcal{T} T_{\mathrm{D}5} L_\phi.
\end{split}
\end{equation}
We see that the dynamical part, comprising of the $w$-integral, is the same as the one in the previous embedding as we find the same functions as in \eqref{thooft_F_and_G}, and therefore the same plots of figure \ref{thooft_plots_IIB} apply here.\\

{\bf D7 probe.}\quad Let us consider a D7 that extends in the following coordinates $\Sigma_8=[t,w,\phi,\theta, \psi, \phi_1,\phi_2,\phi_3]$ with $r=r(w)$ and $z=z_0$ kept fixed. The embedding is consistent and the equations of motion are satisfied. The induced metric on the $\mathrm{D}7$ is:
\begin{equation}
\begin{split}
    \mathrm{d}s^2_{\mathrm{ind,D7}} =& \frac{\zeta(r,\theta)}{L^2}\left[ -r^2 \mathrm{d}t^2 + F(r)L^2r^2\mathrm{d}\phi^2 + \mathrm{d}w^2\left(r^2 + \frac{L^2 r^{\prime 2}}{F(r) \lambda(r)^6r^2} \right) + L^4 \mathrm{d}\theta^2 \right]\\
    &+\frac{L^2}{\zeta(r,\theta)}\left[\cos^2\theta\mathrm{d}\psi^2 + \cos^2\theta\sin^2\psi\left( \mathrm{d}\phi_1 + \frac{A_1}{L}\right)^2 + \cos^2\theta \cos^2\psi \left( \mathrm{d}\phi_2 + \frac{A_2}{L}\right)^2\right.\\
    &\left. + \lambda(r)^6\sin^2\theta \left( \mathrm{d}\phi_3 + \frac{A_3}{L}\right)^2\right],
\end{split}
\end{equation}
while the $\mathrm{DBI}$ yields:
\begin{equation}
\begin{split}
&     S_{\mathrm{D}7}=T_{\mathrm{D}7}\int_{\Sigma_8}\mathrm{d}^8\hat\sigma \sqrt{-e^{-2\Phi}\mathrm{det}(g_{\mathrm{ind},\mathrm{D}7})} = \hat{{\cal N}}_{\mathrm{D}7} \int_{-L_{\mathrm{MM}}}^{L_{\mathrm{MM}}} \mathrm{d}w \; \sqrt{\mathcal{F}_t^2 + 
   \mathcal{G}_t^2 r^{\prime 2}},\\
&\mathcal{F}_t=r^3\lambda^3(r)\sqrt{F(r)},\quad \mathcal{G}_t=Lr,\quad \hat{{\cal N}}_{\mathrm{D}7} = \frac{4}{3}\pi^3L^3\mathcal{T} T_{\mathrm{D}7} L_{\phi}.
\end{split}
\end{equation}

Once again, we obtain the same dynamics and therefore this probe captures the same effects of the 't Hooft loop as the previous ones.

\subsubsection{'t Hooft loop for the Type IIA backgrounds}
Here, we study a probe in the Type IIA family of backgrounds that can be associated with the 't Hooft loop and displays screening of the monopole-anti-monopole pair. The dynamics as we argue is analogous to the one in the Type IIB system.
\\

{\bf D2 probe.} We will first use a $\mathrm{D}2$ probe brane extended in $\Sigma_3[t,w,\phi]$ with the direction $r=r(w)$ and the rest of the coordinates being fixed to constant values: 
\begin{equation}\label{thooft_ansatz_GM}
    z=z_0,\quad \theta=\theta_0,\quad \varphi = \varphi_0,\quad \chi=\chi_0,\quad \eta=\eta_\star,\quad \sigma = \epsilon\to 0,
\end{equation}
with $0<\eta_\star<P$. This is the simplest choice one can have for describing the 't Hooft loop, as the pullback of the NS potential $B_2$ on $\Sigma_3$ is zero and therefore the DBI action is easy to handle. The induced metric on the brane reads:
\begin{equation}
    \begin{split}
\mathrm{d}s_{\mathrm{ind}}^2&=-4\tilde{f}_1^{3/2}\tilde{f}_5^{1/2}\tilde{f}\frac{r^2\lambda^2}{L^2}\, \mathrm{d}t^2+4\tilde{f}_1^{3/2}\tilde{f}_5^{1/2}\tilde{f}\left(\frac{r^2\lambda^2}{L^2}+\frac{r^{\prime 2}}{r^2F\lambda^4}\right)\mathrm{d}w^2\\
&+\tilde{f}_1^{3/2}\tilde{f}_5^{1/2}\left[\tilde{f}_3A_{3\phi}^2+2\left(\tilde{f}_2\sin^2\theta _0A_{1\phi}^2+2F\tilde{f}r^2\lambda^2\right)\right]\mathrm{d}\phi^2.
\end{split}
\end{equation}
where in the above expression all the functions are to be taken at the values of the ansatz \eqref{thooft_ansatz_GM}: $\tilde{f}_i\equiv\tilde{f}_i(r,\eta_\star,\sigma_0)$, etc.  We can now choose the value for $\theta_0=0$ and take $\sigma_0\to 0$. Using coordinates $\hat{\sigma}\in\{t,w,\phi\}$ where $t\in[0,\mathcal{T}]$ on the worldvolume $\Sigma_3$, we have: 
\begin{equation}
    \begin{split}
S_{\mathrm{D}2}&=T_{\mathrm{D}2}\int_{\Sigma_3}\mathrm{d}^3\hat{\sigma}\sqrt{-e^{-2\Phi}\mathrm{det}(g_{\mathrm{ind},\mathrm{D2}})}\\
&=8 LT_{\mathrm{D}2}L_{\phi}\mathcal{T}\lim_{\epsilon\to 0}\tilde{f}_1^{3/2}(\epsilon,\eta_{\star})\tilde{f}(\epsilon,\eta_{\star})\int_{-L_{\mathrm{MM}}}^{L_{\mathrm{MM}}}\mathrm{d}w \sqrt{F\frac{r^4\lambda^4}{L^4}\left(\frac{r^2\lambda^2}{L^2}+\frac{r^{\prime 2}}{r^2F\lambda^4}\right)}\\
&=\hat{{\cal N}}_{\text{D}2}\int_{-L_{\mathrm{MM}}}^{L_{\mathrm{MM}}}\mathrm{d}w\sqrt{\mathcal{F}_t^2+\mathcal{G}_t^2r^{\prime 2}},\qquad
\hat{{\cal N}}_{\text{D}2}=\frac{8T_{\mathrm{D}2}L_{\phi}\mathcal{T}\mathcal{R}(\eta_\star)}{L^2}
    \end{split}
\end{equation}
where the functions $\mathcal{F}_t$ and $\mathcal{G}_t$ are the same as the ones found in the Type IIB background \eqref{thooft_F_and_G}. Consequently, the plots are similar to figure \ref{thooft_plots_IIB}, indicating the screening of monopole-anti-monopole pairs in this background as well.\\

One could also study the 't Hooft loops obtained by suitably probing the geometry with D4 and D6 branes. In these cases, the non-zero pull back of the NS two form makes the calculation more cumbersome. We do not discuss these further in this paper. Instead, we focus on another relevant observable.


\subsection{Entanglement entropy}

In this section, we  study the entanglement entropy of the  QFTs dual to our backgrounds. We consider codimension two submanifolds which explore the bulk, while their boundary at $r=\infty$ divides the dual theory in two entangled regions. Following the proposal of Ryu and Takayanagi \cite{Ryu_2006}, the action for the minimization of the hypersurface for the $10\mathrm{d}$ theories is given by
\begin{equation}\label{EE_formula}
  S_{\mathrm{EE}}=\frac{1}{4G_N}\int_{\Sigma_8}\mathrm{d}^8\hat{\sigma}\sqrt{e^{-4\Phi}\mathrm{det}(g_{\Sigma_8})},
\end{equation}
where $G_N$ is the ten dimensional Newton constant and $\Phi$ the dilaton of the theory.
In $11\mathrm{d}$ we propose the following formula\footnote{To the best of our knowledge this definition for 11d geometries has not been written down explicitly.}
\begin{equation}\label{EE_11d_formula}
    S_{\mathrm{EE}}=\frac{1}{4G_N^{(11)}}\int_{\Sigma_9}\mathrm{d}^9\hat{\sigma}\sqrt{\mathrm{det}(g_{\Sigma_{9}})},
\end{equation}
where $G_N^{(11)}$ denotes the eleven dimensional Newton constant and $\Sigma_9$ a codimension two manifold in eleven dimensions. The above reduces to \eqref{EE_formula} in the case of a $\mathrm{U}(1)$ isometry in the M-theory solution. In that case, using \eqref{10dGM} one can indeed verify that\footnote{one can choose to write the expression of the metric in type II theories either in the string or Einstein frame, the difference being $\mathrm{d}s^2_{E}=e^{-\Phi/2}\mathrm{d}s^2_{S}$.}

\begin{equation}
   \sqrt{\mathrm{det}(g_{\Sigma_9})}= \sqrt{e^{-4\Phi}\mathrm{det}(g_{\Sigma_8})}.
\end{equation}

In the absence of a $\mathrm{U}(1)$ isometry we expect that \eqref{EE_11d_formula} will yield a similar dynamical integral as \eqref{EE_formula}, which if we consider giving the RT surface a profile in the radial coordinate will have the same form as the integral found in the previous observables \eqref{WL_action_schematic}, with the constant of proportionality in this case being related to the central charge.

In the following, we study this observable in the three different backgrounds presented in this work.

\subsubsection{Entanglement Entropy for the Type IIB background}

For the case of the type $\mathrm{IIB}$ background, we choose an eight-manifold $\Sigma_8$ that extends in the directions $[w,z,\phi,\theta,\psi,\phi_1,\phi_2,\phi_3]$ with the profile of the radial coordinate being $r=r(w)$. At $r=\infty$ we take this manifold to have a boundary which splits the field theory in two regions: one being a strip of length $L_{\mathrm{EE}}$, and the other the rest of the space.

The induced metric on $\Sigma_8$ is then
\begin{equation}
     \begin{split}
         \mathrm{d}s^2_{\mathrm{ind}}&=\frac{\zeta(r,\theta)r^2}{L^2}\left[\mathrm{d}z^2+L^2F(r)\mathrm{d}\phi^2 + \left( 1 + \frac{L^2r^{\prime 2}}{F(r)\lambda^6(r)r^4}\right)\mathrm{d}w^2\right]\\
         &+L^2\left[ \zeta(r,\theta)\mathrm{d}\theta^2 + \frac{\cos^2\theta}{\zeta(r,\theta)}\mathrm{d}\psi^2 + \frac{\cos^2\theta\sin^2\psi}{\zeta(r,\theta)}\left(\mathrm{d}\phi_1 +\frac{A_1}{L}\right)^2\right.\\
         &\left. +\frac{\cos^2\theta\cos^2\psi}{\zeta(r,\theta)}\left(\mathrm{d}\phi_2 +\frac{A_2}{L}\right)^2 + \frac{\lambda^6(r)\sin^2\theta}{\zeta(r,\theta)}\left(\mathrm{d}\phi_3 +\frac{A_3}{L}\right)^2 \right],
     \end{split}
 \end{equation}
and the determinant of this yields
\begin{equation}
    \mathrm{det}(g_{\Sigma_8}) = L^6r^6F(r)\lambda^6(r)\cos^6\theta\cos^2\psi\sin^2\theta\sin^2\psi\left[1+\frac{L^2r^{\prime 2}}{\lambda^6(r)F(r)r^4}\right].
\end{equation}
Given that the dilaton for this background is trivial, we arrive at the following expression for the entanglement entropy action 
\begin{equation}\label{SEE_IIB}
    \begin{split}
&S_{\mathrm{EE}}=\hat{\cal N}_{\text{IIB}}^{\text{EE}}\int_{-L_{\mathrm{EE}}/2}^{L_{\mathrm{EE}}/2}\mathrm{d}w\sqrt{\mathcal{F}_{\mathrm{EE}}^2+\mathcal{G}_{\mathrm{EE}}^2r^{\prime 2}},\\
& \mathcal{F}_{\mathrm{EE}}^2 = r^6F(r)\lambda^6(r),\quad \mathcal{G}_{\mathrm{EE}}^2 = L^2r^2,\quad \hat{\cal N}^{\text{EE}}_{\text{IIB}}=\frac{L^3L_{\phi}L_z\pi^3}{4G_N}.
    \end{split}
    \end{equation}
We  find the same functions as the ones appearing in the integrand on the 't Hooft loop observable in \eqref{thooft_F_and_G}. From this it follows that the entanglement entropy undergoes a phase transition as well, since the plots for the length of the strip $L_{\mathrm{EE}}$ and the corresponding energy are identical to the ones presented in figure \ref{thooft_plots_IIB}, where a phase transition in the energy $E(L_{\mathrm{EE}})$ is apparent. There is a critical length after which the system prefers to have zero entanglement entropy, corresponding to two disconnected Ryu-Takayanagi surfaces that are attached at $w=\pm L_{\text{EE}}/2$ at the boundary ($r=\infty$) and extend deep in the bulk, reaching the end of space. This behaviour of the entanglement entropy further supports the confinement-deconfinement phase transition \cite{Klebanov:2007ws, Kol:2014nqa, Georgiou:2015pia}.

\subsubsection{Entanglement Entropy for the Type IIA background}

For this background, we will choose an eight manifold $\Sigma_8$ extended in the coordinates $[w,z,\phi,\theta,\varphi,\chi,\sigma,\eta]$ with $r=r(w)$. The induced metric on this manifold is
\begin{equation}
\begin{split}
    \mathrm{d}s_{\Sigma_8}^2=\tilde{f}_1^{3/2}\tilde{f}_5^{1/2}&\left[  4\tilde{f}\frac{r^2 \lambda(r)^2}{L^2} (\mathrm{d}z^2+L^2F\mathrm{d}\phi^2)\right.    +4\tilde{f}\tilde{f}_1^{3/2}\tilde{f}_5^{1/2}\left(\frac{r^{\prime 2}}{r^2\lambda^4F}+\frac{r^2 \lambda(r)^2}{L^2} \right)\mathrm{d}w^2\\ &\left.+\tilde{f}_2\mathrm{D}\mu^i\mathrm{D}\mu^i+\tilde{f}_3(\mathrm{d}\chi+B)^2+\tilde{f}_4(\mathrm{d}\sigma^2+\mathrm{d}\eta^2) \right].
\end{split}
\end{equation}
Taking the determinant of the above metric, we arrive at
\begin{equation}
   \mathrm{det}(g_{\Sigma_8})=4^3F(r)\frac{r^6 \lambda(r)^6}{L^4}\sin^2\theta\tilde{f}_1^{12}\tilde{f}_2^2\tilde{f}_3\tilde{f}_4^2\tilde{f}_5^4\tilde{f}^3\left[1+\frac{L^2 r^{\prime 2}}{r^4 F(r)\lambda^6(r)}\right],
    \end{equation}
which in turn yields the following action for the surface (setting $\kappa=1$):
\begin{equation}
\begin{split}\label{SEE_IIA}
   &  S_{\mathrm{EE}}=\hat{\cal N}^{\text{EE}}_{\text{IIA}}\int_{-L_{\mathrm{EE}}/2}^{L_{\mathrm{EE}}/2}\mathrm{d}w\sqrt{\mathcal{F}_{\mathrm{EE}}^2+\mathcal{G}_{\mathrm{EE}} ^2 r^{\prime 2}},\\
&\mathcal{F}_{\mathrm{EE}}^2=r^6F(r)\lambda^6(r),\quad \mathcal{G}_{\mathrm{EE}}^2=L^2r^2, \quad \hat{\cal N}^{\text{EE}}_{\text{IIA}}=\frac{64\pi^2 \kappa^{3} L_{\phi}L_z}{L^2G_N}\int_0^{\infty}\mathrm{d}\sigma\int_0^P\mathrm{d}\eta \,\sigma\dot{V}V^{\prime\prime},
    \end{split}
\end{equation}
Once again we arrive at the same dynamical part as in the type $\mathrm{IIB}$ background and the 't Hooft loops, hence this systems also exhibits a phase transition in the entanglement entropy. In addition to the constant numerical prefactor, we also obtain an integral in the subspace spanned by $\sigma$ and $\eta$, which makes its appearance in many of these systems when one calculates the entanglement entropy or other observables that encode the degrees of freedom of the dual field theory, like the c-function \cite{macpherson2024, Bea_2015, Chatzis:2024top}. We further comment on this in the next section where we calculate the flow central charge. 

\subsubsection{Entanglement Entropy in $11\mathrm{d}$ supergravity}

To calculate the entanglement entropy directly in the $11\mathrm{d}$ system \eqref{11dGM}, we take the codimension 2 manifold $\Sigma_9$ with $t$ constant and $r=r(w)$. The induced metric on this subspace is then
\begin{equation}
\begin{split}
    \mathrm{d}s^2_9 = e^{2\lambda}& \left\{ \frac{4Z}{X^3}\left[ \frac{r^2 \lambda(r)^2}{L^2} \left( \mathrm{d}z^2 + L^2 F(r) \mathrm{d}\phi^2 \right) +\left( \frac{r^2 \lambda(r)^2}{L^2} + \frac{r^{\prime 2}}{r^2 \lambda(r)^4 F(r)} \right)\mathrm{d}w^2  \right] + \frac{y^2 e^{-6\lambda}}{Z^2}(D\mu^i)^2 \right. \\
    &\left. + \frac{4X^3}{Z^2}\frac{(D\chi^2)}{1-y\partial_y D_0} - \frac{Z}{y}\partial_y D_0\mathrm{d}y^2 -\frac{Z^2}{y}\partial_y e^{D_0}(\mathrm{d}v_1^2 + \mathrm{d}v_2^2)\right\}.
\end{split}
\end{equation}
where we have again set $\kappa=1=m$. The determinant of this metric is
\begin{equation}
    \operatorname{det}g_{\Sigma_9} = 256 L^2 e^{2D_0}y^2 (\partial_y D_0)^2\sin^2\theta \left[F(r)\frac{r^6 \lambda(r)^6}{L^6}+\frac{r^2 r^{\prime 2}}{L^2} \right]
\end{equation}
and the action is written as
\begin{align}
    &S_{\mathrm{EE}}= \hat{{\cal N}}^{\text{EE}}_{\text{LLM}} \int_{-L_{\mathrm{EE}}/2}^{L_{\mathrm{EE}}/2} \mathrm{d}w \; \sqrt{\mathcal{F}_{\mathrm{EE}}^2 + \mathcal{G}^2_{\mathrm{EE}} r^{\prime 2}},\\
    &\mathcal{F}_{\mathrm{EE}}^2=r^6 F(r) \lambda^6(r),\quad \mathcal{G}^2_{\mathrm{EE}}=r^2L^2,\quad\hat{{\cal N}}_{\text{LLM}}^{\text{EE}} =\frac{32\pi^2L_zL_\phi}{G_N^{(11)}}\int\mathrm{d}y\mathrm{d}v_1\mathrm{d}v_2 \, y\partial_y e^{D_0}.\label{calN11d}
\end{align}
Once again, the observable factorizes itself in a universal manner and the $r$-dependent factor is the same as in the previous cases. An interesting comment here is that the integral in the subspace spanned by $y,v_1$ and $v_2$ appearing in the $\mathrm{UV}$ factor above, is the $11\mathrm{d}$ version of the integral found in the electrostatic case \eqref{SEE_IIA}. To see this in detail, we make use of the change of coordinates \eqref{coord_change_to_electrostatic} to translate this subspace to the "electrostatic variables": $(y,v_1,v_2)\mapsto(\sigma,\eta,\beta)$. Writting the $(v_1,v_2)$ plane in polar coordinates $(\rho,\beta)$ and differentiating these relations, we find:  
\begin{equation}
\begin{split}
    &\mathrm{d}y = \dot{V}^\prime \mathrm{d}\eta  - \sigma V^{\prime\prime}\mathrm{d}\sigma, \quad \mathrm{d}v_1 \wedge \mathrm{d}v_2 = - \rho \; \mathrm{d}\rho \wedge \mathrm{d}\beta 
    \quad {\rm and } \quad e^{\frac{D_0}{2}}\mathrm{d}\rho = \sigma V^{\prime\prime} \mathrm{d}\eta + \dot{V}^\prime \mathrm{d}\sigma,
\end{split}
\end{equation}
and then we write the integrand as:
\begin{equation}
    y \partial_y e^{D_0} = \frac{2V^{\prime\prime} \dot{V} \sigma^2}{\rho^2 \left[ \ddot{V}V^{\prime\prime} - (\dot{V^{\prime}})^2 \right]}.
\end{equation}
This maps the integral in $(y,v_1,v_2)$ to the following: 
\begin{equation}
    \int \mathrm{d}y \; \mathrm{d}v_1 \; \mathrm{d}v_2 \; y \partial_y e^{D_0} = -2 \int \mathrm{d}\beta \; \mathrm{d}\eta \; \mathrm{d}\sigma \; \dot{V}V^{\prime\prime}\sigma,\quad \text{with}\,\,V=V(\beta,\eta,\sigma),
\end{equation}
which is indeed proportional to the integral in \eqref{SEE_IIB} in the electrostatic case, that is when the extra $\mathrm{SO}(2)$ symmetry in the $v_1,v_2$ plane renders $V$ independent of $\beta$. The same integral appears also in flow central charge calculation, carried out below.


\subsection{Flow central charge}\label{cflow_section}

We now study an observable called  flow central charge. The calculation details are included in \cite{Chatzis:2024kdu,Chatzis:2025dnu,Bea_2015, merrikin2023compactification6dcaln10}. This is a monotonic quantity related to the number of degrees of freedom of the flow and asymptotes to the UV central charge value. It provides an effective number of degrees of freedom for every value of the holographic coordinate along the flow from the $\mathrm{CFT}_4$ to the strongly coupled $3\mathrm{d}$ systems in the $\mathrm{IR}$.

Interestingly, notice that this quantity is monotonic for the backgrounds here studied, in which Lorentz invariance $SO(1,3)$ is broken to $SO(1,2)\times SO(2)$ under the RG flow. In summary, we write a quantity that detects the four dimensional fixed point, the gapped character of the three-dimensional QFT (giving a vanishing result) and is monotonic along the flow.
\\

Consider a dilaton and metric of the form
\begin{equation}
  \mathrm{d} s^2=-a_0 \mathrm{d} t_0^2+\sum_{i=1}^{d} a_i \mathrm{d} x_i^2+\prod_{i=1}^{d}\left(a_i \right)^{\frac{1}{d}} \tilde{b}(r) \mathrm{d} r^2+h_{a b} (\mathrm{d}y^a-A^a) (\mathrm{d}y^b-A^b), ~~~\Phi(\vec{y}, r).\label{backaniso}
\end{equation}
To calculate the holographic flow central charge we use \cite{Macpherson_2015, Bea_2015, merrikin2023compactification6dcaln10}
\begin{align}
& G_{i j} \mathrm{d} X^i \mathrm{d} X^j=\sum_{i=1}^{d} a_i \mathrm{d} x_i^2+h_{a b} (\mathrm{d} y^a-A^a) (\mathrm{d} y^b-A^b) \\
& \hat{H}=\left(\int \prod_{a=1}^{8} \mathrm{d} X^a \sqrt{e^{-4 \Phi} \operatorname{det}\left(G_{i j}\right)}\right)^2, \quad c_{\mathrm{flow}} =\text{Vol}_d \,d^d \frac{\tilde{b}(r)^{\frac{d}{2}} \hat{H}^{\frac{2 d+1}{2}}}{G_N^{(10)}\left(\hat{H}^{\prime}\right)^d}.
\end{align}
For $G_{ij}$ the 8-dimensional internal space and field theory directions and $\mathrm{Vol}_d$ the volume of the field theory subspace spanned by $\{x_i\}$. 
\subsubsection{Flow central charge in Type IIB Background}

Calculating the flow central charge in the IIB background, we find with $d=3$,
\begin{equation}
    \tilde{b}(r) = \frac{L^{\frac{4}{3}}}{F(r)^{\frac{4}{3}}\lambda(r)^{6}r^4}.\label{btilde}
\end{equation}
The 8-dimensional space is
\begin{align}
    G_{ij}\mathrm{d}x^i \mathrm{d}x^j &=\frac{\zeta(r,\theta)r^2}{L^2}\left[\mathrm{d}w^2+\mathrm{d}z^2+L^2F(r)\mathrm{d}\phi^2
    \right]\nonumber\\
    & +\frac{L^2}{\zeta(r,\theta)}\left[ \zeta(r,\theta)^2 \mathrm{d}\theta^2 +\cos^2\theta\mathrm{d}\psi^2+\cos^2\theta\sin^2\psi \mathrm{D}\phi_1^2 + \cos^2\theta\cos^2\psi \mathrm{D}\phi_2^2 + \lambda^6(r)\sin^2\theta\mathrm{D}\phi_3^2\right].\nonumber
\end{align}
Taking the determinant we find:
\begin{align}
&\mathrm{det}G_{ij}=r^6 L^6 F(r) \lambda(r)^6 \sin^2\theta\cos^6\theta \cos^2\psi \sin^2\psi ,\nonumber \\
&\int \mathrm{d}^8x \; \sqrt{e^{-4\Phi}\operatorname{det}G_{ij}} =\pi^3 L^3 r^3 \lambda(r)^3\sqrt{F(r)} L_wL_zL_{\phi}=\hat{H}^{1/2},\nonumber\\
\end{align}
and the flow central charge takes the form
\begin{equation}
     c_\mathrm{{flow}} = \frac{3^3 \;\tilde{b}(r)^\frac{3}{2} \hat{H}^{\frac{7}{2}} L_\phi L_z L_w}{G_{N}^{(10)}(\hat{H}')^3}.
\end{equation}
We write the radially dependent part in terms of dimensionless variables $\xi, \hat{\nu}$ defined in \eqref{singularity_section} and the final expression takes the universal form:

\begin{equation}
c_\mathrm{flow}=27\hat{\mathcal{N}}_{\text{IIB}}^c \;\frac{ \; \xi^6 \left(1+\frac{\hat{\nu}}{\xi^2}\right)^2\left(\frac{\xi^6-1+\hat{\nu}(\xi^4-1)}{\xi^4(\hat{\nu}+\xi^2)}\right)^\frac{3}{2}}{(2\hat{\nu}+3\xi^2)^3} 
\quad {\rm with} \quad 
\hat{\mathcal{N}}_{\text{IIB}}^c = 
\frac{\pi^3 L^8 L_w L_z L_\phi}{8 G_N^{(10)}} .\label{chols5}
\end{equation}

\begin{figure}[H]
    \centering
    \includegraphics[width=0.7\linewidth]{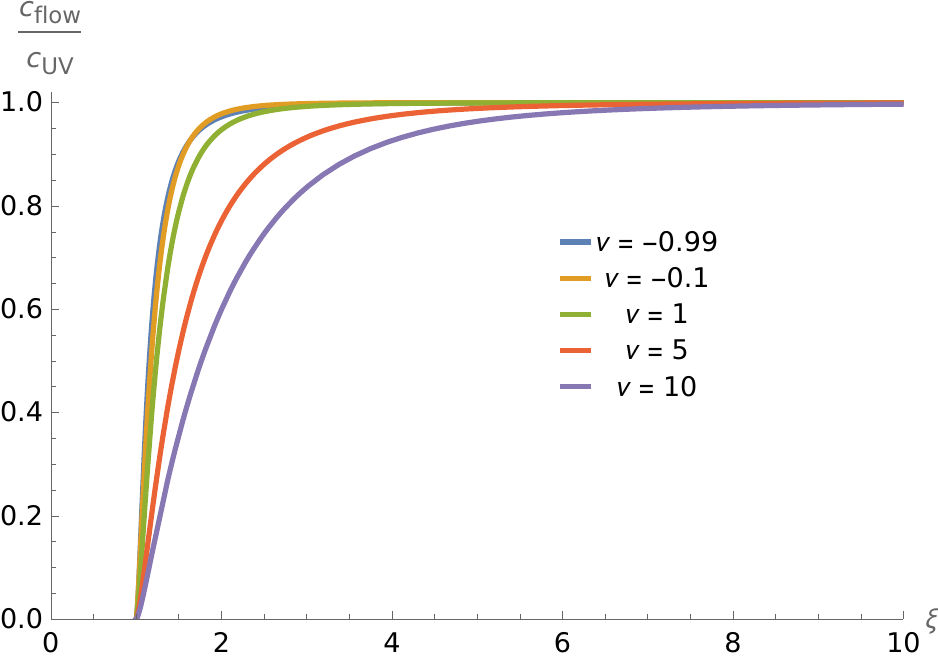}
    \caption{Plot of the normalised flow central charge in eq.\eqref{chols5} as a function of $\xi$ for different values of $\hat{\nu}$ and $L=1$. The $c_{\text{flow}}$ goes to the UV central charge $ c_{\text{UV}}= {\hat{\mathcal{N}}}^c_{\text{IIB}}$ for large values of $\xi$. This plot is the same for all of the three backgrounds.}
    \label{cflow_figure}
\end{figure}

In figure \ref{cflow_figure} we observe that the point $r=r_{\star}$ corresponds to an energy scale in the dual field theory below which there are no dynamical degrees of freedom. This indicates that the system has a gapped spectrum and $r_{\star}$ is associated with the gap scale.\\

Moreover, we notice that the non-universal part $\hat{\cal N}^c_{\text{IIB}}$ of the observable which equals the central charge of the $\text{UV}$ theory, is also contained in the $\text{UV}$ factor of the entanglement entropy calculation \eqref{SEE_IIB}. The two are related as: 
\begin{equation}
    \hat{\cal N}^{c}_{\text{IIB}}=c_{\text{UV}}=\frac{L^2L_w^2L_zL_{\phi}}{2}\hat{{\cal N}}^{\text{EE}}_{\text{IIB}}.
\end{equation}
Similar relations hold for the rest of the backgrounds under study. The interested reader can consult \cite{Jokela:2025cyz} for more on the relation between c-functions and entanglement entropy.

\subsubsection{Flow central charge for Type $\mathrm{IIA}$ background}

In the same way as above, we calculate the flow central charge for the 11d and IIA backgrounds. We have the same $\tilde{b}(r)$ as in eq. (\ref{btilde}). The 8-manifold metric reads
\begin{align}
    G_{ij}\mathrm{d}x^i \mathrm{d}x^j &=\tilde{f}_1^{3/2}\tilde{f}_5^{1/2}\left[\frac{4\tilde{f}\; r^2 \lambda(r)^2}{L^2} (\mathrm{d}w^2+ \mathrm{d}z^2+L^2 F(r)\mathrm{d}\phi^2)+\tilde{f}_2\mathrm{D}\mu^i\mathrm{D}\mu^i+\tilde{f}_3(\mathrm{d}\chi+B)^2+\tilde{f}_4(\mathrm{d}\sigma^2+\mathrm{d}\eta^2) \right],
\end{align}
and we find
\begin{equation}
    e^{-4\Phi}\mathrm{det}(G_{ij})= 64 \tilde{f}_1^9 \tilde{f}_2^2 \tilde{f}_3 \tilde{f}_4^2 \tilde{f}_5 \tilde{f}^3 \frac{r^6 \lambda(r)^6}{L^4} F(r)
\end{equation}
Setting $\kappa=1$ and utilising 
\begin{equation}
    \tilde{f}_1^9 \tilde{f}_2^2 \tilde{f}_3 \tilde{f}_4^2 \tilde{f}_5 \tilde{f}^3 = 16 \dot{V}^2 V^{\prime\prime\; 2} \sigma^2,
\end{equation}
 we get:
\begin{eqnarray}
    \hat{H}^{\frac{1}{2}}= \frac{265\pi^2 L_\phi}{L^2} \int \mathrm{d}\sigma \mathrm{d}\eta \; \dot{V}V^{\prime\prime}\sigma \; r^3 \lambda(r)^3 \sqrt{F(r)}.
\end{eqnarray}
This yields the result: 

\begin{equation}
    c_{\text{flow}}=\hat{\mathcal{N}}_{\text{IIA}}^c \;27\frac{ \; \xi^6 \left(1+\frac{\hat{\nu}}{\xi^2}\right)^2\left(\frac{\xi^6-1+\hat{\nu}(\xi^4-1)}{\xi^4(\hat{\nu}+\xi^2)}\right)^\frac{3}{2}}{(2\hat{\nu}+3\xi^2)^3},\qquad     \hat{\mathcal{N}}_{\text{IIA}}^c = \frac{32\pi^2L_{\phi}^2L_z^2L_w^2L^2}{G_N}\int_0^{\infty}\mathrm{d}\sigma\int_0^P \mathrm{d}\eta \dot{V}V^{\prime\prime}\sigma,
\end{equation}
where the flow factor matches that of the type $\mathrm{IIB}$ calculation above. The normalized flow central charge for this solution will also be depicted by figure \ref{cflow_figure}.
Again, one observes that the expression $\int \mathrm{d}\sigma\mathrm{d}\eta\, \dot{V}\sigma $ makes its appearance in the corresponding constant in the entanglement entropy calculation. More specifically,

\begin{equation}
    c_{\text{UV}}=\hat{{\cal N}}^c_{\text{IIA}}=\frac{L^4L_{\phi}L_wL_z}{2}\hat{{\cal N}}^{\text{EE}}_{\text{IIA}}.
\end{equation}

\subsubsection{Flow central charge in $11\mathrm{d}$ supergravity}

For the calculation in $11\mathrm{d}$, we compute a 9-volume and there is no dilaton. Here we find:
\begin{equation}
     \hat{H}^{\frac{1}{2}}= \frac{256\pi^2 L_\phi}{L^2} \int \mathrm{d}y \; \mathrm{d}v_1  \mathrm{d}v_2 \; y \partial_y e^{D_0}\; r^3 \lambda(r)^3 \sqrt{F(r)},\label{VintII}
\end{equation}
and therefore
\begin{equation}
    c_{\text{flow}}=\hat{\cal N}^c_{\text{LLM}}\;27\frac{ \; \xi^6 \left(1+\frac{\hat{\nu}}{\xi^2}\right)^2\left(\frac{\xi^6-1+\hat{\nu}(\xi^4-1)}{\xi^4(\hat{\nu}+\xi^2)}\right)^\frac{3}{2}}{(2\hat{\nu}+3\xi^2)^3},\qquad    \hat{\mathcal{N}}_{\text{IIA}}^c = \frac{32\pi^2 L_w^2L_z^2L_\phi^2}{G_N^{(11)}} \int \mathrm{d}y \; \mathrm{d}v_1  \mathrm{d}v_2  y \partial_y e^{D_0},
\end{equation}
in which case we have the connection with the entanglement entropy $\text{UV}$ factor: 
\begin{equation}
    c_{\text{UV}}=\hat{{\cal N}}^c_{\text{LLM}}=L_zL_\phi L_w^2\hat{\cal N}^{\text{EE}}_{\text{LLM}}.
\end{equation}


\subsection{D7 brane embeddings in the Anabalon-Ross deformed $\mathrm{AdS}_5\times\mathbb{S}^5$ solution}

\begin{figure}[H] 
   \centering
   \includegraphics[width=7cm,height=7cm]{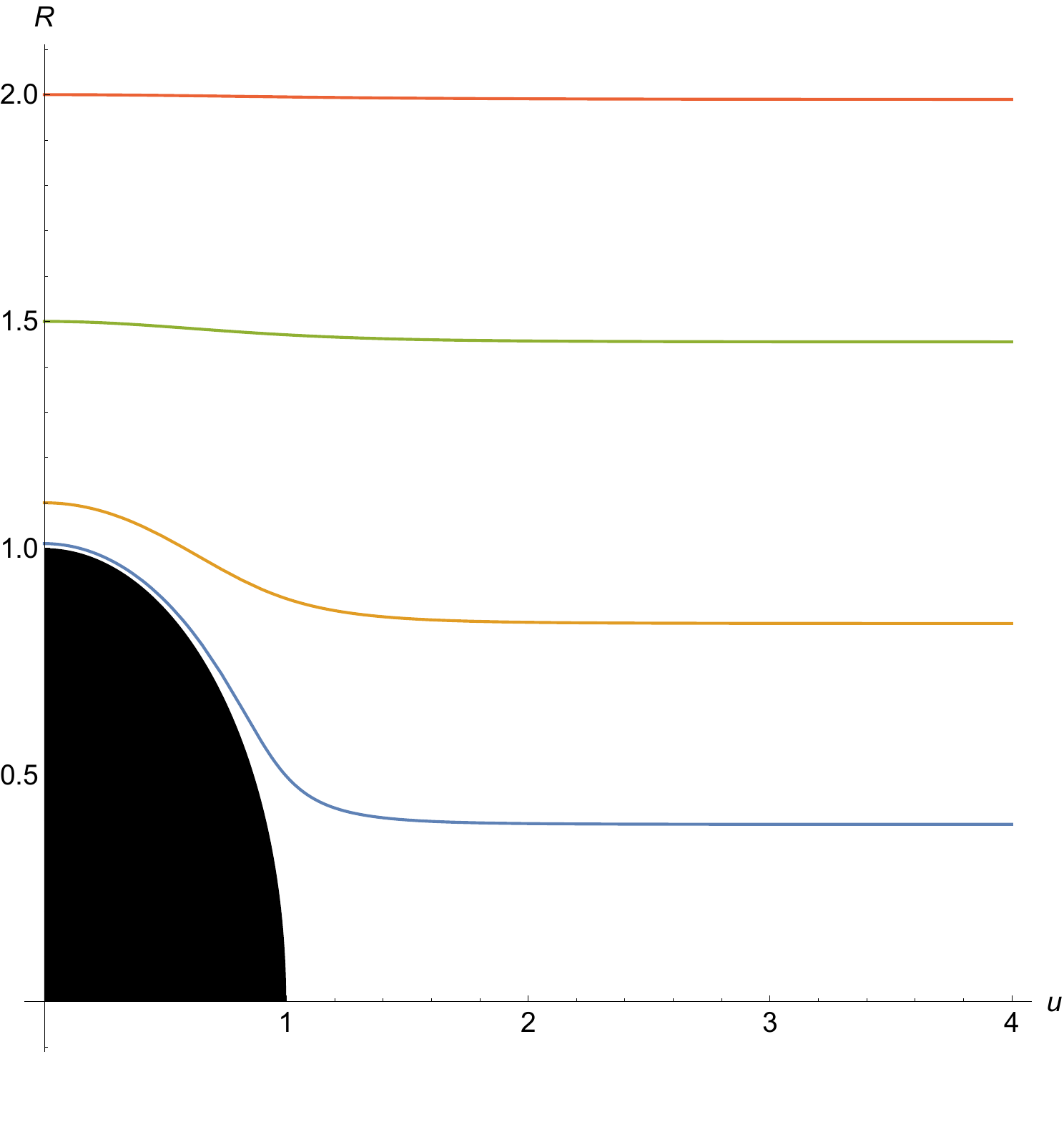}
    \includegraphics[width=7cm,height=7cm]{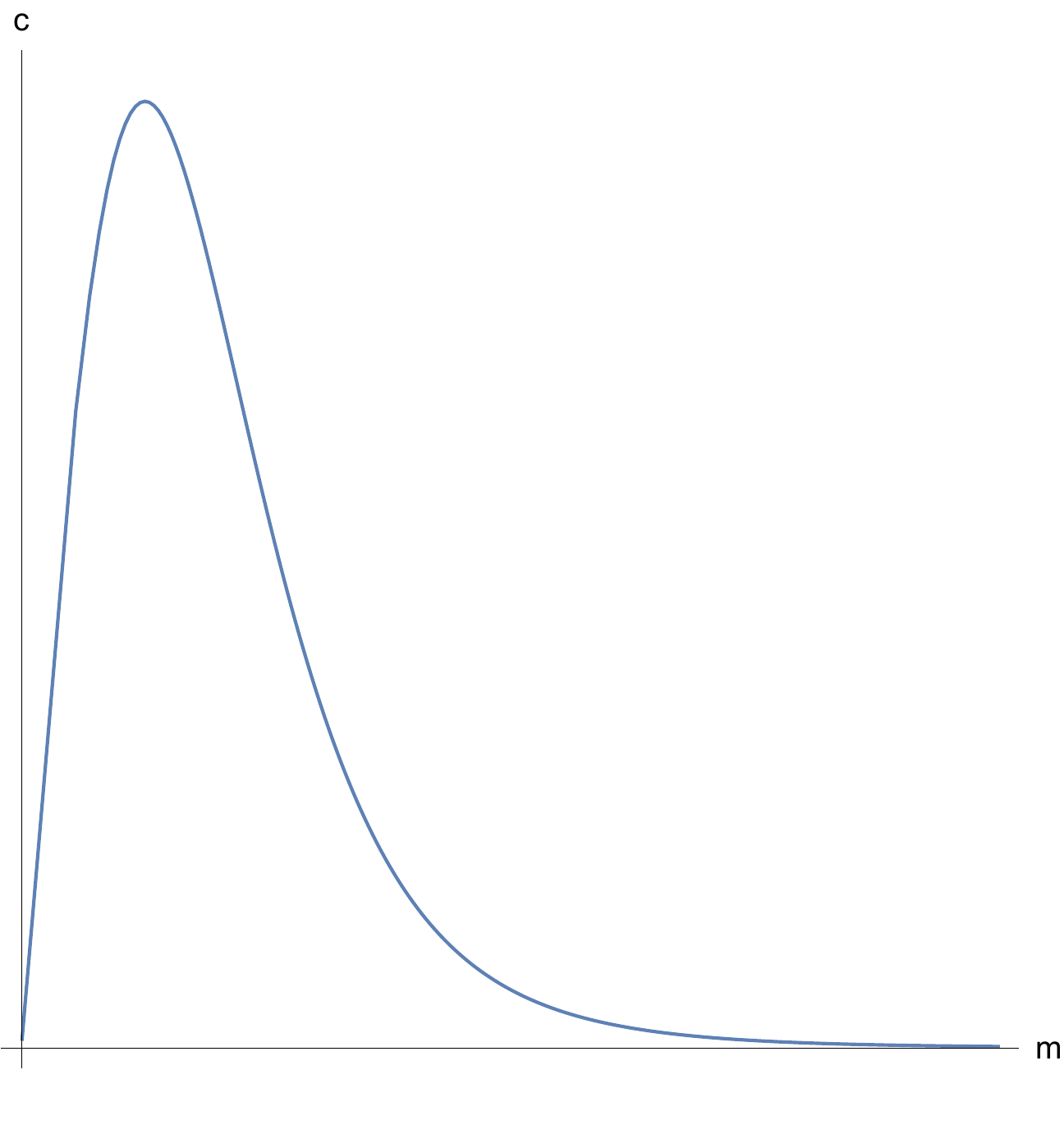}
     \caption{In the left panel, plots of the D7-brane embeddings for different values of the constant $R_0$. 
     The black region indicates the end of space which is an inaccessible region of the geometry. In the right panel, qualitative plot of the parameter $c$ 
     versus the parameter $m$, as they are read off from the boundary behaviour of the solution. }
   \label{D7_embeddings-c_vs_m}
\end{figure}

The main objective of this subsection is to study the properties of a D7-brane probe embedded in a confined background. 
To facilitate the analysis we will focus on the type IIB confined solution and set to zero the Coulomb branch parameter $\ell$. 
In this way we essentially analyze the addition of D7-brane probes to the Anabalon-Ross deformed 
$\mathrm{AdS}_5 \times \mathbb{S}^5$ background \cite{Anabalon:2021tua}. The background in this case is 
\begin{eqnarray} \label{metric_D7_embedding}
\mathrm{d}s_{10}^2 &=& \frac{r^2}{L^2} \Bigg[\mathrm{d}x_{1,2}^2 + f(r) \, L^2 \, \mathrm{d}\phi^2 + \frac{L^2 \, \mathrm{d}r^2}{r^4 \, f(r)}\Bigg] +
L^2 \Bigg\{\mathrm{d}\theta^2 + \sin^2 \theta \, 
\left(\mathrm{d}\phi_3 + \frac{A_3}{L}\right)^2 
\nonumber \\[5pt]
&+& \cos^2 \theta \Bigg[ \mathrm{d}\psi^2 + \sin^2 \psi \left(\mathrm{d}\phi_1 + \frac{A_1}{L}\right)^2+ 
\cos^2 \psi \left(\mathrm{d}\phi_2 + \frac{A_2}{L}\right)^2\Bigg] \Bigg\} \, ,
\end{eqnarray}
with\footnote{The metric in \eqref{metric_D7_embedding} is the 
$\ell\rightarrow 0$ limit of \eqref{metric_S5_WL}. 
Due to the redefinition of equation \eqref{def_Q}, the parameter $Q$ has dimensions of length.}
\begin{equation} \label{metric_D7_embedding_functions}
  f(r) = \frac{1}{L^2} - \frac{L^2\, Q^2}{r^6}  \quad {\rm and} \quad  
 A_1 = A_2 =A_3 = - \frac{Q^{1/3}}{L^{1/3}}\mathrm{d}\phi + \frac{L Q}{r^2}\mathrm{d}\phi,
\end{equation}
where we have explicitly used the expression for $r_{\star}=L^{2/3}Q^{1/3}$ coming from the condition $F(r_{\star})=0$.
A consistent ansatz for a D7-brane probe with the following worldvolume coordinates 
\begin{equation}
\zeta^{\alpha} = (t,y,z,\phi, r, \psi, \phi_1,\phi_2)
\quad {\rm is} \quad 
\theta = \theta(r) \quad {\rm and} \quad \phi_3 = \frac{L^{2/3}Q^{1/3}}{L^2} \, \phi \, . 
\end{equation}
Notice that in the limit of $Q\to0$, 
in which case we obtain the familiar $\mathrm{AdS}_5 \times \mathbb{S}^5$ solution,
%
%
the ansatz for the embedding function $\phi_3$ becomes zero, and we obtain the D7-brane embedding of  \cite{Kruczenski:2003be}. 
To simplify the study of the embeddings, it is convenient to introduce an isotropic system of 
coordinates, along the lines of \cite{Mateos:2007vn, Jokela:2012dw}. To identify these coordinates, 
we have to isolate the $(r, \theta)$ part of the induced metric
 \begin{equation}
\frac{\mathrm{d}r^2}{r^2 \, f(r)} + L^2 \mathrm{d}\theta^2  = L^2 \Bigg[\frac{\mathrm{d}r^2}{r^2 \, L^2 \, f(r)} + \mathrm{d}\theta^2 \Bigg]
\end{equation}
and exchange the radial coordinate $r$ with a new radial coordinate $\rho$, such that such 
the first term inside the brackets becomes $\mathrm{d}\rho^2/\rho^2$. This means 
\begin{equation}
\frac{\mathrm{d}r}{r \, \sqrt{1 - \frac{L^4 \, Q^2}{r^6}}} = \frac{\mathrm{d}\rho}{\rho} \quad \Rightarrow \quad 
r = \frac{L^{2 \over 3}\, Q^{\frac{1}{3}}}{2^{\frac{1}{3}} \, \rho} \left(1+ \rho^6\right)^{\frac{1}{3}} \quad {\rm with} \quad \rho \in [1,\infty) \, . 
\end{equation}
Describing the embedding in terms of the function $\chi(\rho) =\sin \theta(\rho) $ simplifies the analysis, 
and the Euclidean D7-brane action becomes
\begin{equation}
\frac{I}{{\cal N}} = \int \frac{\mathrm{d} \rho}{\rho^5} \left(1+ \rho^6\right)^{\frac{1}{3}} 
\sqrt{\left(1-\rho^6\right)^2 + 4 \, \rho^6 \, \chi^2}
\left(1- \chi^2\right) \sqrt{\big.1- \chi^2 + \rho^2 \, \chi'^2}
\end{equation}
with
\begin{equation}
{\cal N} = \frac{\pi ^2 L^{8 \over 3} Q^{4 \over 3}}{\sqrt[3]{2}} \, T_{D_7}
\end{equation}
The equation of motion for $\chi (\rho)$ can be obtained from the Euclidean D7-brane action and 
implies that the behaviour of the field at infinity is 
\begin{equation}
\chi = \frac{m}{\rho} + \frac{c}{\rho^3} + \cdots
\end{equation}
where the dimensionless constants $m$ and $c$ are related to the quark mass and condensate.
Let us now introduce a system of Cartesian-like coordinates $(u, R)$, defined as follows
\begin{equation}
R = \rho \, \chi \quad {\rm and} \quad u = \rho \, \sqrt{1- \chi^2} \, . 
\end{equation}
The Euclidean D7-brane action becomes
\begin{equation} \label{action_R_u}
\frac{I}{{\cal N}'} = \int \mathrm{d}u \, \frac{u^3 \left[1 +\left(u^2+ R^2\right)^{3}\right]^{\frac{1}{3}}}{\left(u^2+R^2\right)^{4}}
\sqrt{4 R^2 \left(u^2+R^2\right)^2+\left[1-\left(u^2+R^2\right)^3\right]^2}
\sqrt{1+R'^2} \, 
\end{equation}
where the embedding of the D7-brane is given by $R = R(u)$. For large values of $u$ we have
\begin{equation}
R(u) \, = \, m + \frac{c}{u^2} + \cdots \, .  
\end{equation}
Even if there are two free parameters in solving the equation of motion for the embedding function $R(u)$, 
the dynamics in the IR of the theory allows that for a given value of $m$ a finite number of values for $c$ exist, 
in order to have solutions that are well behaved. In the current case,  there is only one value of $c$ for every $m$. 
To solve numerically the equation of motion for $R(u)$ from \eqref{action_R_u}, we integrate from the IR with boundary 
conditions $R(0)=R_0$ and $R'(0)=0$. In the left panel of figure \ref{D7_embeddings-c_vs_m}, 
some of these solutions are presented for different values of the constant $R_0$. The values of the parameters $m$ and $c$ 
are read off at the boundary and are presented in the right panel of figure \ref{D7_embeddings-c_vs_m}.

In other confining backgrounds (see e.g. \cite{Seo:2009kg}), the interior singularity triggers repulsion, causing the D7-branes to avoid the central region. We observe similar avoidance behavior for the probe brane in our case (see the left panel of Figure \ref{D7_embeddings-c_vs_m}), even without an interior singularity.
Moreover, the plot of the condensate with respect to the quark mass 
(in the right panel of figure \ref{D7_embeddings-c_vs_m}), shows an interesting non monotonic behaviour. 
The condensate flows to zero both when the quark mass is large 
(something that is expected since away from the singularity the embeddings are flat), but also for very small quark masses. 
And all this without ``jumping'' from Minkowski to Black hole embeddings, 
as it happens in the presence of the temperature \cite{Mateos:2007vn}.
Notice here that in the embeddings that are studied in \cite{Seo:2009kg},
in a Gubser dilaton flow geometry \cite{Gubser:1999pk}  that is not supersymmetric, the condensate 
is finite for zero quark mass indicating spontaneous chiral symmetry.
In our case, there is a finite value of the quark mass that the condensate obtains a maximal value. 
This behaviour has to be investigated further. First we have to determine the exact relation between 
the parameters $m$, $c$ with the quark mass and the condensate. 
This should be followed by a holographic renormalization, in order to calculate the 
``on-shell'' action, and by the computation of the meson spectrum.


\section{Holographic renormalization}\label{Holographic_RG_section}

\subsection{Anabalon-Ross solution}

In this section we will study the holographic renormalization of the solution presented by Anabalon and Ross in \cite{Anabalon:2021tua}, using the methods developed by Skenderis et al in \cite{Henningson:1998gx,deHaro:2000vlm,Bianchi:2001de,Papadimitriou:2004ap}. The configuration we wish to study is a solution of five-dimensional gauged ${\cal N}=2$ supergravity with the following action:
\begin{equation}
     S=\frac{1}{16\pi G}\int \mathrm{d}^5x \sqrt{-g}\left( R + \frac{12}{L^2}-\frac{3}{4}\mathcal{F}_{\mu\nu}\mathcal{F}^{\mu\nu} \right) + \frac{1}{16\pi G}\int \sqrt{-g}\mathcal{F}\wedge\mathcal{F}\wedge\mathcal{A},
 \end{equation}
with a gauge field 
\begin{equation}
     \mathcal{A}=Q\left( 
    \frac{1}{r^2}-\frac{1}{r_{\star}^2}\right)\mathrm{d}\phi,\quad \mathcal{F}=\mathrm{d}\mathcal{A}.
 \end{equation}
The equations of motion for the gauge field and the metric read: 
\begin{equation}
     \begin{split}
         &\mathrm{d}\star_5{\cal F}+{\cal F}\wedge {\cal F}=0,
         \,\,\,\, G_{\mu\nu}=\tilde{T}_{\mu\nu}({\cal{A}}) 
         \,\,\,\, {\rm with} \,\,\,\, 
         \tilde{T}_{\mu\nu}({\cal A})=\frac{3}{2}\left( \mathcal{F}_{\mu\rho}\mathcal{F}^{\rho}_{\nu}-\frac{1}{6}g_{\mu\nu}\mathcal{F}_{\rho\sigma}\mathcal{F}^{\rho\sigma} \right) +\frac{6}{L^2}g_{\mu\nu}.
     \end{split}
 \end{equation}
The metric is given by
\begin{equation}\label{ds5_AR}
 \begin{split}
  \mathrm{d}s^2&=\frac{r^2}{L^2}\Big(-\mathrm{d}t^2+\mathrm{d}x_1^2+\mathrm{d}x_2^2+f(r)\mathrm{d}\phi^2\Big)+\frac{L^2\mathrm{d}r^2}{r^2f(r)} 
  \quad {\rm with} \quad f(r)=1-\frac{\mu \, l^2}{r^{4}}-\frac{Q^{2} L^2}{r^{6}} .
 \end{split}
\end{equation}
The configuration preserves four real supercharges when $\mu=0$, or equivalently when $r_{\star}=(QL)^{1/3}$. Following the steps leading to holographic renormalization \cite{Bianchi:2001de,deHaro:2000vlm,Papadimitriou:2004ap}, we first transform our coordinate system to Fefferman-Graham coordinates:
\begin{equation}\label{FG_metric}
    \mathrm{d}s^2_5=\frac{L^2}{\rho^2}\Big( \mathrm{d}\rho^2+g_{ij}(x,\rho)\mathrm{d}x^i\mathrm{d}x^j\Big),
\end{equation}
where the variable $\rho$ is defined such that the boundary is now at $\rho=0$. To achieve this we will treat the coordinate change asymptotically near the boundary. First we use $r\mapsto z^{-1}$ and expand around $z=0$. The radial part of the metric then becomes (notice that we have to go up to fifth order for the parameter $Q$ to appear):
\begin{equation}
    \frac{L\mathrm{d}r}{r\sqrt{f(r)}}\approx-\left( \frac{L}{z}+\frac{L^3\mu}{2}z^3+\frac{L^3Q^2}{2}z^5+\mathcal{O}(z^7)\right)\mathrm{d}z=\frac{L\mathrm{d}\rho}{\rho},
\end{equation}
which we can integrate to find $\rho(z)$ asymptotically:
\begin{equation}
\begin{split}
        &\Rightarrow\rho=\exp\left\{ \frac{1}{L}\int\mathrm{d}z\left( \frac{L}{z}+\frac{L^3\mu}{2}z^3+\frac{L^3Q^2}{2}z^5+\mathcal{O}(z^7) \right)\right\} \\
    &\Rightarrow\rho \approx z + \frac{L^2\mu}{8}z^5 + \frac{L^2Q^2}{12}z^7+\mathcal{O}(z^9),
    \end{split}
\end{equation}
and inverting this we get the relation for $z(\rho)$:
\begin{equation}
    z\approx \rho - \frac{L^2\mu}{8}\rho^5-\frac{L^2Q^2}{12}\rho^7+\mathcal{O}(\rho^9).
\end{equation}
We can now expand the expressions of the boundary metric $g_{ij}$ around $\rho=0$:
\begin{equation}
    \begin{split}
        &|g_{tt}|=g_{x_1x_1}=g_{x_2x_2}\approx \frac{1}{L^2}+\frac{\mu}{4}\rho^4+\frac{Q^2}{6}\rho^6+\mathcal{O}(\rho^8) \, , \quad  
        g_{\phi\phi}\approx \frac{1}{L^2}-\frac{3\mu}{4}\rho^4 - \frac{5Q^2}{6}\rho^6+\mathcal{O}(\rho^8).
    \end{split}
\end{equation}
One can then use the following formula to calculate the vacuum expectation value of the stress energy momentum tensor in the four-dimensional boundary field theory \cite{deHaro:2000vlm}:
\begin{equation}
    \langle T_{ij}\rangle = \frac{4L^3}{16\pi G_{\text{N}}}g_{(4)ij},
\end{equation}
where $g_{(4)ij}$ denotes the coefficient of the fourth order in $\rho$ term in the expansion. The result reads: 

\begin{equation}
 -  \langle T_{tt}\rangle= \langle T_{x_1x_1}\rangle=\langle T_{x_2x_2}\rangle=-\frac{L^3\mu}{16\pi G_{\text{N}} },\quad \langle T_{\phi\phi}\rangle =\frac{3L^3\mu}{16\pi G_{\text{N}}} .
\end{equation}

As expected, the SUSY point $\mu=0$ has vanishing vev for the energy momentum tensor. As for the gauge field, its near boundary expansion in these variables is:
\begin{equation}
    \mathcal{A}\approx\left( - \frac{Q}{r_{\star}^2}+Q \rho^2 - \frac{L^2 Q\mu}{4}\rho^6+\mathcal{O}(\rho^8)\right)\mathrm{d}\phi,
\end{equation}
yielding the vev for the current: 
\begin{equation}
    \langle J_{\phi}\rangle = Q,
\end{equation}
which couples to a global current expressing the ${\cal R}$-symmetry of the dual theory. We comment that the preserved ${\cal R}$ symmetry is still realized by internal cycles inside the deformed $\mathbb{S}^5$, namely the $\mathrm{U}(1)^3\cong\mathbb{S}^1\times\mathbb{S}^1\times\mathbb{S}^1\subset\mathbb{S}^5$.

\subsection{Anabalon-Nastase-Oyarzo}
The same procedure can be carried for the solution of \cite{Anabalon:2024che}. In this case the change of variables bringing the metric of \eqref{5d_soliton} asymptotically to the Fefferman-Graham form \eqref{FG_metric}, as well as its inverse, read\footnote{again, after letting $r=z^{-1}$ to bring the boundary at $z=0$}: 
\begin{equation}
\begin{split}
    &
    \rho \approx z - \frac{\ell^2\varepsilon}{6}z^3 + \frac{9L^4(q_1^2-q_2^2)\ell^2\varepsilon + {\color{teal}5\ell^4}}{72}z^5+\mathcal{O}(z^7),\\
    &z\approx \rho + \frac{\ell^2\varepsilon}{6}\rho^3 - \frac{9L^4(q_1^2-q_2^2)-{\color{teal}5\ell^4}}{72}\rho^5+\mathcal{O}(\rho^7).
\end{split}
\end{equation}

Let us now look at the expansions of the 4d metric coefficients near the boundary. We have: 
\begin{equation}\label{expansions_1}
\begin{split}
|g_{tt}|=g_{ww}=g_{zz} =\frac{r^2\lambda^2(r)}{L^2}& \approx \frac{1}{L^2}+\left[ -\frac{L^4\ell^2\varepsilon(q_1^2-q_2^2)}{4L^2}-{\color{teal}{\frac{\ell^4}{18L^2}}}\right]\rho^4
    \\
& + \left[ -\frac{L^2\ell^4(q_1^2-q_2^2)}{8}+{\color{teal}\frac{13\ell^6\varepsilon}{648L^2}}\right]\rho^6 + \mathcal{O}(\rho^8),
\\
g_{\phi\phi} =r^2\lambda^2(r)F(r) & \approx \frac{1}{L^2}+\left[ -\frac{3L^4\ell^2(q_1^2-q_2^2)\varepsilon}{4L^2}-{\color{teal}\frac{\ell^4}{18L^2}} \right]\rho^4 
\\
& + \frac{-27L^4\ell^4 (19q_1^2+5q_2^2)+ {\color{teal}13\ell^6\varepsilon}}{648L^2}\rho^6+\mathcal{O}(\rho^8).
\end{split}
\end{equation}

We notice that the boundary expansions contain constant isotropic terms, highlighted in blue. In order to ensure that the VEV of the stress-energy-momentum tensor vanishes, as it should for a SUSY theory, we can remove these terms by adding the following boundary expression to the supergravity action: 
\begin{equation}\label{Sct}
   S_{\text{c.t.}}= \frac{1}{8\pi G_N} \Big(
   \frac{L\ell^4}{9}\Big)\int_{\mathbb{R}^{1,3}}\mathrm{d}^4x\sqrt{\gamma},
\end{equation}
where $\gamma$ is the induced metric on the boundary of $\mathrm{AdS}_5$. Such a term respects all the symmetries of the theory, is covariant and does not interfere with the equations of motion. It only results in a shifting of the VEV of $T_{ij}$.
The final expression reads:
\begin{equation}
    \langle T_{ij}\rangle =\frac{4L^3}{16\pi G_{N}}\Big( g_{(4)ij}+\eta_{ij}\frac{\ell^4}{18L^2} \Big)=\frac{L^3\mu}{16\pi G_N} \,\mathrm{diag}(-1,1,1,-3),
\end{equation}
where $\mu=-\eta L^4(q_1^2-q_2^2)$, $g_{(4)ij}$ refers to the fourth order coefficients in \eqref{expansions_1} and the second term is the effect of \eqref{Sct}. Clearly, in the supersymmetric case where $q_1=q_2$ we get $\langle T_{ij}\rangle =0$. Let us note that counterterms of this form were first introduced in \cite{Balasubramanian:1999re} for the purpose of removing divergences from physical observables like the $\mathrm{AdS}$ mass density. As noted in \cite{Balasubramanian:1999re} as well as \cite{deHaro:2000vlm}, it is often necessary to fix such terms, expressing ambiguities, in order to match the ground state energy. As for the scalar and gauge fields, their near boundary behaviour is: 
\begin{equation}
    \begin{split}
        &\Phi(r)=\sqrt{\frac{2}{3}}\ln\lambda^{-6}(r)\approx -\sqrt{\frac{2}{3}}\ell^2\varepsilon\rho^4 + \frac{\ell^4}{3\sqrt{6}}\rho^6+\mathcal{O}(\rho^8),\\
        &A^1_{\phi}=A^2_{\phi}\approx  \frac{L\ell^2\varepsilon q_1}{r_{\star}^2}-L\ell^2 \varepsilon q_1\rho^2-\frac{L\ell^4 q_1}{3}\rho^4 + \frac{L\ell^4 q_1}{36}\left[ 9L^4 (q_1^2-q_2^2)-2\ell^2\varepsilon \right]\rho^6 +\mathcal{O}(\rho^8),\\
&A^3_{\phi}\approx \frac{L\ell^2\varepsilon q_2}{r_{\star}^2+\ell^2\varepsilon}-L\ell^2\varepsilon q_2\rho^2+\frac{2L\ell^4q_2}{3}\rho^4 - \frac{L q_2}{36}\left[ -9L^4(q_1^2-q_2^2)\ell^4+14\ell^6\varepsilon\right]\rho^6 + \mathcal{O}(\rho^8).
\end{split}
\end{equation}
From these we can read the VEV of the dimension two\footnote{the prefactor of 2 in the VEV expression enters because in this special case where $\Delta =2$ we have $(2\Delta -d)=0$, as explained in \cite{deHaro:2000vlm}} operator dual to the scalar field\footnote{we remind the reader that $\eta=-\varepsilon\ell^2/L^2$.}: 
\begin{equation}
    \langle O_2\rangle =2 \times\left(  -\sqrt{\frac{2}{3}}\ell^2\varepsilon\right)=\frac{2\sqrt{2}L^2}{\sqrt{3}}\eta,
\end{equation}
as well as the VEVs of the currents contained in the $\mathcal{O}(\rho^2)$ coefficients of the gauge fields: 
\begin{equation}
    \begin{split}
        & \langle J_1\rangle =-L\ell^2 \varepsilon q_1,\quad \langle J_3\rangle = -L\ell^2\varepsilon q_2,
    \end{split}
\end{equation}
which are equal at the SUSY point where $q_1=q_2=q$. We see that one can recover the results of the previous subsection by setting $Q=\ell ^2 Lq$ and taking the limit $\ell\to 0$ which maps the Anabalon-Nastase-Oyarzo solution to the Anabalon-Ross one as explained in \cite{Chatzis:2025dnu}.



\section{The stability analysis of the classical solution for the Wilson Loops}\label{WL_stability_section}

In this subsection we will examine the stability of some of the Wilson loop configurations that were calculated in \cite{Chatzis:2025dnu}.
The analysis will be for Wilson loops in the Type IIB case, however we expect that the conclusions of the subsequent analysis will be valid also for the other cases.


\subsection{The classical solution}

We consider the ten-dimensional supergravity background constructed in section 5 of \cite{Anabalon:2024che}, presented here in \ref{IIB_background} (we are especially interested in the supersymmetric case, 
in which the  parameters $q_1$ and $q_2$ are equal) and we perform the following redefinition (see also \cite{Chatzis:2025dnu})  
\begin{equation} \label{def_Q}
q_1 = q_2 = \frac{Q}{\ell^2 } \, , 
\end{equation}
where $Q$ has dimensions of length. In this way the background functions and the gauge fields become\footnote{For simplicity we have chosen  
$\varepsilon=-1$. Notice that we can reverse this choice at any time by $\ell \rightarrow i \, \ell$.}
\begin{eqnarray}
 \zeta(r , \theta)^2 &=& 1- \frac{\ell^2}{r^2} \cos^2\theta \, , \quad 
 \lambda(r)^6 = 1 - \frac{\ell^2}{r^2} \, , \quad 
  F(r) = \frac{1}{L^2}\left[ 1 - \frac{L^4\, Q^2}{r^4 (r^2-  \ell^2)}\right]
  \nonumber\\[5pt]
 A_1 &=& A_2= \,  - \, L \, Q \, \frac{r^2 - r^2_{\star}}{r^2_{\star} \, r^2} \, d\phi \, , \quad
 A_3 = \, - \, L \, Q  \,
 \frac{r^2 - r^2_{\star}}{ (r^2_{\star} - \ell^2) \, (r^2 -\ell^2) } \,  d\phi \, .
\end{eqnarray}
The background metric is given by \eqref{S5-1}, see \cite{Anabalon:2024che,Chatzis:2025dnu}, which we repeat here for convenience
\begin{eqnarray} \label{metric_S5_WL}
\mathrm{d}s^2 &=&  \frac{r^2}{L^2}  \, \zeta(r,\theta)
\left[\mathrm{d}x_{1,2}^2+ L^2 F(r) \mathrm{d}\phi^2 +\frac{L^2 \mathrm{d}r^2}{F(r)\,  r^4 \, \lambda^3(r)} \right] + L^2 \,  \zeta(r,\theta) \,  \mathrm{d}\theta^2
\nonumber \\
&  +& \frac{L^2}{\zeta(r,\theta)}\Bigg\{\cos^2\theta \left[\mathrm{d}\psi^2+ \sin^2\psi \left(\mathrm{d}\phi_1 +\frac{A_1}{L}\right)^2 +
\cos^2\psi \left(\mathrm{d}\phi_2 +\frac{A_2}{L}\right)^2 \right]
\\ 
& +&\lambda^6(r) \sin^2\theta \left( \mathrm{d}\phi_3 +\frac{A_3}{L}\right)\Bigg\}.
\nonumber
\end{eqnarray}
with $C_4$ given in \eqref{C4} which does not participate in the calculation of the Wilson loops and their stability analysis. 

To obtain generic expressions for the string configuration and for the equations of the fluctuation modes, we consider the following general form for the metric 
\begin{equation} \label{metric_general}
\mathrm{d}s^2 = G_{tt} \mathrm{d}t^2 + G_{ww} \mathrm{d}w^2 + G_{zz} \mathrm{d}z^2 + G_{\phi \phi} \mathrm{d}\phi^2 + G_{rr} \mathrm{d}r^2+ G_{\theta \theta} \mathrm{d}\theta^2 + \ldots 
\end{equation}
Notice that in the cases we will consider $G_{tt} $ will be negative and $y$ denotes the spatial side of the Wilson loop along which the Wilson loop extends. 
To write down the expression for the string trajectory in a compact form, we introduce the following functions
\begin{equation}
g(r,\theta) = - G_{tt} G_{rr}\, , \quad  f_y(r,\theta) = - G_{tt} G_{yy} \, . 
\end{equation}
The Nambu-Goto action for a string that propagates in the gravity background becomes 
\begin{equation} \label{string_NG}
S = - {1 \over 2 \pi} \int \mathrm{d} \tau \mathrm{d} \sigma \sqrt{- \det g_{\a \b} } 
\quad {\rm with} \quad  
g_{\alpha \beta} = G_{\mu\nu} \partial_\alpha x^\mu \partial_\b x^\nu \, . 
\end{equation}
To fix reparametrization invariance, we make the choice:  $t=\tau$ and $r=\sigma$, 
and the ansatz for the string embedding corresponds to that of embedding I (with $\phi_0=0$) studied in section \ref{WL_IIB_subsection}
\begin{equation} \label{string_embedding}
y = y(\sigma) \, , \quad 
\phi =0 \, , \quad 
\theta  = \theta_0 = \Big\{0 \,\, {\rm or} \,\, \frac{\pi}{2}\Big\} \quad {\rm and} \quad {\rm rest  = constant} \, . 
\end{equation}
Notice here that the two values for the angle $\theta$ came from the substitution of the ansatz  \eqref{string_embedding} into the equations of motion, that are derived from \eqref{string_NG}.
The equation that the embedding obeys becomes
\begin{equation} \label{yclassical}
{f_y \, y_{\rm cl}^\prime \over \sqrt{ g + f_y \, y_{\rm cl}^{\prime 2}}} = \pm \left(f_y^0\right)^{1/2} 
\quad  \Rightarrow \quad 
y_{\rm cl}^\prime = \pm {\sqrt{f_y^0 \, F}\over f_y} 
\quad  {\rm with} \quad 
F = {g \, f_y \over f_y - f_y^0} \, .
\end{equation}
All the functions are calculated at $\theta=\theta_0$ and $r_0$ is the value of the holographic coordinate at the turning point. 
Moreover, $f_y^0 \equiv f_y(r_0,\theta_0)$, $y_{\rm cl}$ is the classical solution and $y_{\rm cl}^\prime$ is the derivative of the classical solution with respect to $\sigma$ at $\theta=\theta_0$.
The two signs denote the two symmetric branches of the string trajectory around the turning point.
Integrating  \eqref{yclassical} we obtain the expression for the separation length and inserting the solution for $y_{\rm cl}^{\prime}$ into the Nambu-Goto action 
(and properly subtracting the contribution from two straight strings extending until the end of the geometry), we obtain the expression for the energy. This is the same procedure we followed to obtain equations \eqref{WL_integrals} and in this context they read
\begin{equation} \label{WL-length_energy}
L = 2\sqrt{ f_{y}^0} \int_{r_0}^{\infty} \mathrm{d} r {\sqrt{F} \over f_y}\ .
\quad {\rm and } \quad
E = {1 \over \pi} \int_{r_0}^\infty \mathrm{d} r \sqrt{F}-\frac{1}{\pi}\int_{r_{\star}}^{\infty}\mathrm{d}r \sqrt{g} \, . 
\end{equation}
In principle, we would like to analytically evaluate the preceding integrals, express the parameter $r_0$ as a function of the separation length $L$ and then insert that expression into the expression for the energy. 
In that way, we would obtain an expression of the energy in terms of the separation length, $E(L)$. 
In practice this can be done only for a limited amount of ``cases" (i.e. backgrounds), and the confined background is not one of those cases. 
Alternatively, we consider the expressions of \eqref{WL-length_energy} as parametric equations of the length and of the energy with parameter $r_0$.  

In figure \ref{Wilson_plots_embeddingI} $(\mathrm{c})$ and $(\mathrm{d})$ we have plotted the energy of the Wilson loop as a function of the separation length for different values of $\hat{\nu}$, or equivalently, different values of $Q$ (the case $\hat{\nu}=-1$ corresponding to $Q=0$). 
For large values of the separation length, which is equivalent to small values of the parameter $r_0$ (i.e. the turning point of the string approaches the confinement scale $Q$), the energy has a confining behaviour. 
This means that it increases linearly with the separation length. 
In this section we are going to investigate the perturbative stability of the string configuration, paying special attention to the region of small  $r_0$. 
We are going to analyze the case with $\theta_0=0$, but the main conclusion will be also valid for the other cases. 
Notice that in the $\theta_0=0$ case and for very small values of $Q$ (or $\hat{\nu}\approx-1^{+}$), there is a region in the $E(L)$ plot (figure \ref{Wilson_plots_embeddingI} $(\mathrm{d})$) in which a ``swallow tail" is observed. 
In \cite{Chatzis:2025dnu} we have plotted the quantities $R_{ab}R^{ab}$ and $R_{abcd}R^{abcd}$ for different values of $\theta_0$ and $Q$ and we have observed that when both $\theta_0$ and  $Q$ approach zero, those quantities become very large,  almost infinite. As a result, background corrections are needed since the supergravity approximation is not valid. It is in this regime that the 
``swallow tail" appears, and for this reason we argue that the phase transition that is seen in the behaviour of the Wilson loop does not correspond to an actual physical situation. Based on these arguments we are not considering this region of the parametric space for $Q$ in the following analysis.


\subsection{Expansion of the action and equations of motion for the fluctuations}

To investigate the stability of the string configuration that is described in \eqref{yclassical}, we consider small fluctuations about the classical solution. 
We are going to fluctuate all the coordinates, either transverse or longitudinal to the quark-antiquark axis.
We perturb the embedding ansatz as follows
\begin{eqnarray} \label{expansion_ansatz}
&& z = z_0 + \delta z (\tau, \sigma)\, , \quad 
y = y_{\rm cl}(\sigma) + \delta y (\tau, \sigma) \, , \quad  
\theta = \theta_0 + \delta \theta (\tau, \sigma)  \, , \quad
\phi = \delta \phi (\tau, \sigma)  
\nonumber \\[5pt]
&& \varphi_1 =  \delta \varphi_1 (\tau, \sigma) \, , \quad 
\varphi_2 =  \delta \varphi_2 (\tau, \sigma) \, , \quad  
\varphi_3 =  \delta \varphi_3 (\tau, \sigma)\, , \quad
\psi = \delta \psi (\tau, \sigma) \, . 
\end{eqnarray}
Notice that we keep the expansion ansatz for the angle $\theta$ general. 
In this way we will obtain a system of equations for the fluctuations that could be potentially applied for both values of $\theta_0$. 
The next step is to substitute the ansatz \eqref{expansion_ansatz} into the Nambu-Goto action and expand in powers of the fluctuations.
In this way we obtain an expression of the form
\begin{equation}
S = S_0 + S_1 + S_2 + \ldots
\end{equation}
where the zeroth order term is the classical action and the first order contribution is
\begin{equation}
S_1 = - {1 \over 2\pi} \int \mathrm{d}\tau \mathrm{d}\sigma 
\Bigg\{ \sqrt{f_{y}^0}\,
\delta y^\prime + \left[ {1 \over 2 F^{1/2}} \partial_\theta g + {f_{y}^0 \, F^{1/2}
 \over 2 f_y^2} \partial_\theta f_y \right] \delta \theta \Bigg\}
\end{equation}
where $\partial_\theta g$ and $\partial_\theta f_y$ denote derivatives of the functions $g$ and
$f_y$ with respect to $\theta$ evaluated at the value $\theta=\theta_0$, while $\delta y^\prime$ denotes derivative with respect to $\sigma$.
In the cases that we consider the derivatives of the functions $g$ and $f_y$ vanish and the first term is a surface contribution that will not affect the equations of motion. 
The second order contribution can split to a sum of terms. Some of them contain contribution from just one fluctuation
\begin{eqnarray}
S_2^{\delta z} &=&  {f_z \over 2 \, F^{1/2}}
\delta z^{\prime 2} - {h \, f_z \, F^{1/2} \over 2 \, g \, f_y} \delta  \dot{z}^2
\quad \& \quad 
S_2^{\delta y} = {g f_y \over 2 \, F^{3/2}} \delta y^{\prime 2} - {h \over 2 \, F^{1/2}} \delta \dot{y}^2
\nonumber \\  
S_2^{\delta \theta} &=& {f_\theta \over 2 \, F^{1/2}} \delta \theta^{\prime 2} - 
{h \, f_\theta \,  F^{1/2} \over 2 \, g \, f_y} \delta \dot{\theta}^2 + \left( {1 \over 4 \, F^{1/2}} \partial_\theta^2 g
+ {f_{y}^0 \, F^{1/2} \over 4 \, f_y^2} \partial_\theta^2 f_y \right) \delta \theta^2
\nonumber \\ 
S_2^{\delta \psi} &=&  {f_{\psi} \over 2 \, F^{1/2}} \delta \psi^{\prime 2} - {h \, f_{\psi} \, F^{1/2} \over 2 \, g \, f_y} \delta  \dot{\psi}^2
\quad \& \quad 
S_2^{\delta \phi} = {f_{\phi} \over 2 \, F^{1/2}} \delta \phi^{\prime 2} - {h \, f_{\phi} \, F^{1/2} \over 2 \, g \, f_y} \delta  \dot{\phi}^2
\nonumber \\ 
S_2^{\delta \varphi_i} & = &  {f_{\varphi_i} \over 2 \, F^{1/2}} \delta \varphi_i^{\prime 2} - {h \, f_{\varphi_i} \, F^{1/2} \over 2 \, g \, f_y} \delta  \dot{\varphi_i}^2 
\quad {\rm with} \quad i=1,2,3
\end{eqnarray}
and there are three more that contribute to the coupling between the modes $\delta \phi $ and $\delta \varphi_i$, for $i=1,2,3$
\begin{eqnarray}
S_2^{\delta \phi \, \delta \varphi_i} &=& f_{12} \Bigg[\frac{ f_{\varphi_i}}{F^{1/2}} \,  \delta \phi^{\prime } \, \delta \varphi_i^{\prime} - 
{h \, f_{\varphi_i} \, F^{1/2} \over  g \, f_y} \,  \delta  \dot{\phi} \, \delta  \dot{\varphi_i} \Bigg] 
\,\,\, {\rm for} \,\,\, i=1,2
\,\,\, {\rm with} \,\,\,  f_{12} = \,  -  \, Q \, \frac{r^2 - r^2_{\star}}{r^2_{\star} \, r^2}  \qquad 
\nonumber \\ 
S_2^{\delta \phi \, \delta \varphi_3} &=& f_{3} \Bigg[\frac{ f_{\varphi_3}}{F^{1/2}} \,  \delta \phi^{\prime } \, \delta \varphi_3^{\prime} - 
{h \, f_{\varphi_3} \, F^{1/2} \over  g \, f_y} \,  \delta  \dot{\phi} \, \delta  \dot{\varphi_3} \Bigg] 
\,\,\, {\rm with} \,\,\,  f_3 = \, -  \, Q  \, \frac{r^2 - r^2_{\star}}{ (r^2_{\star} - \ell^2) \, (r^2 -\ell^2) } \, . \qquad 
\end{eqnarray}
The definitions of the functions $f_z$,  $f_\theta$, $f_\phi$, $f_\psi$, $h$ and $f_{\varphi_i}$ with $i=1,2,3$,  
are listed in \eqref{auxiliary_functions}.
The equations for the fluctuations $\delta z$, $\d \psi$, $\delta y$ and $\delta \theta $ are decoupled, 
while the equations for $\delta \phi $, $\delta \varphi_{12}$ and $\delta \varphi_{3}$ are coupled. 
After introducing a time dependence of the form  $e^{-{\rm i} \, \omega \, \tau}$ (see \eqref{fluctuation_time_dependence}), 
they take the form that is listed in \eqref{eom-decoupled}, \eqref{eom-coupled_1} and \eqref{eom-coupled_23}. 
Therefore, we have reduced the problem of examining the stability of the string trajectory to an eigenvalue problem 
for the differential operators that refer to the different types of fluctuation modes. The differential equations are of the general 
Sturm-Liouville type
\begin{equation} \label{Sturm-Liouville}
\left\{ - {\mathrm{d} \over \mathrm{d}\sigma} \left[ p(\sigma) {\mathrm{d} \over \mathrm{d}\sigma} \right] - r(\sigma) \right\} \Phi(\sigma) = \omega^2 \, q(\sigma) \Phi(\sigma) \quad {\rm with} \quad  
\sigma_{\rm min} \le \sigma < \infty
\end{equation}
where the functions $p(\sigma)$, $q(\sigma)$ and $r(\sigma)$ are read off from the corresponding fluctuation equation 
and depend on $r_0$ and the confinement parameter $Q$. 
Summarizing, the task is to solve the Sturm-Liouville problem and determine the range of values of $r_0$ and $Q$, 
for which $\omega^2$ is negative. For this range of values the classical solution will be perturbatively unstable.


\subsection{Transformation to a Schr\"{o}dinger potential}

\begin{figure}[ht] 
   \centering
   \includegraphics[width=7cm,height=7cm]{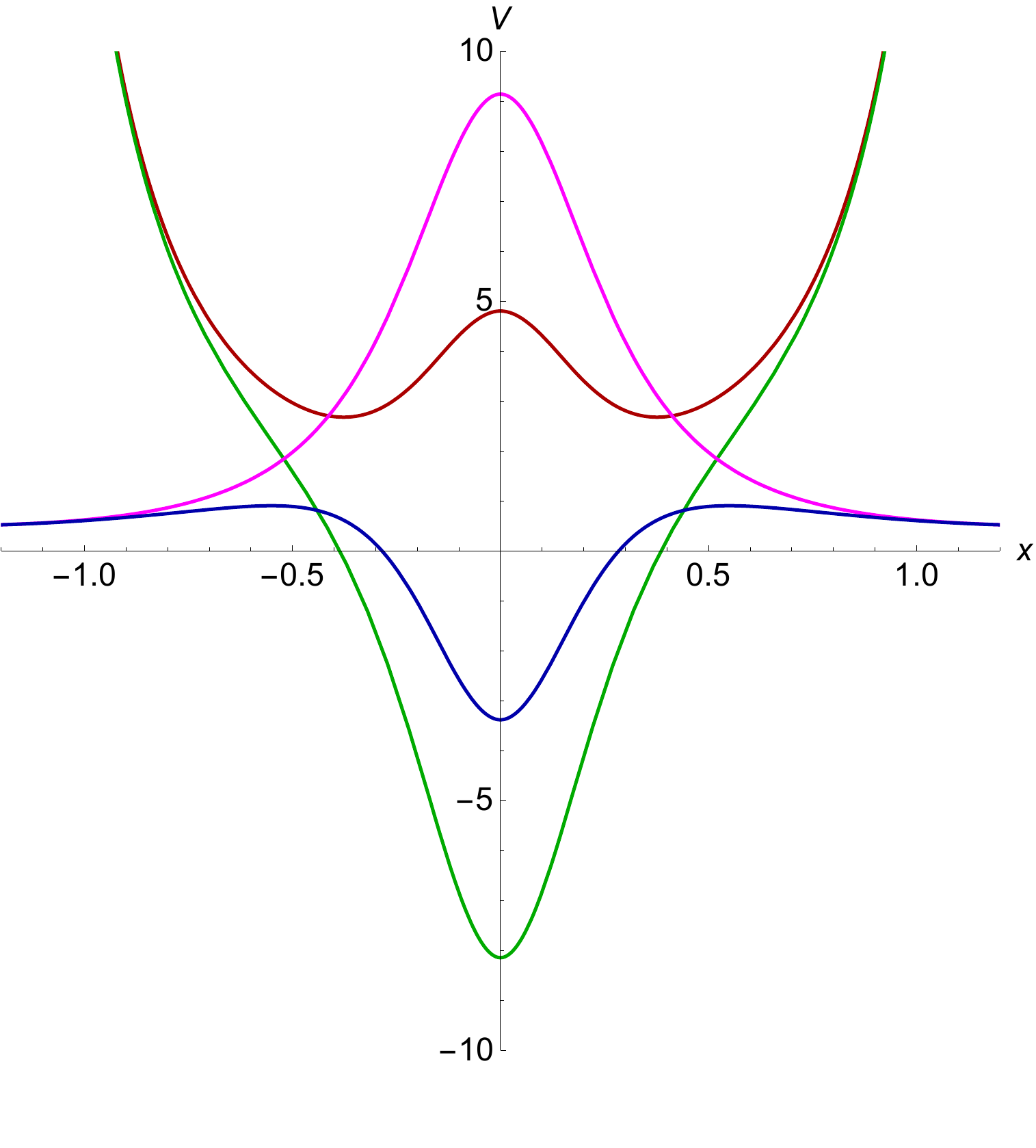}
     \caption{Plots of the Schr\"{o}dinger potentials from \eqref{Schr_potentials_Q_zero} as a function of  $x$ from \eqref{def_x_Q_zero} 
     for $Q=0$ and $r_0=1.2$.  The red, green, magenta and blue solid lines correspond to $V_{\delta z}$, 
     $V_{\delta y}$, $V_{\delta \theta}$ and $V_{\delta \psi}$.} 
   \label{Potentials-Zero-Q}
\end{figure}

To perform the stability analysis it is often more convenient to transform the problem to a Schr\"{o}dinger one. 
For this reason we perform the following change of variable and change of function
\begin{equation} \label{change_variables_general}
x =\pm  \int_{r_0}^{\sigma} \mathrm{d}\sigma ^{\prime} \sqrt{q \over p} \quad {\rm with} \quad  \Phi = (p q)^{-1/4} \Psi
\end{equation}
where the two different signs for the new variable $x$, correspond to the two different branches of the U-shaped string embedding.
Now equation \eqref{Sturm-Liouville} becomes
\begin{equation} \label{Schrodinger}
\left[ -{\mathrm{d}^2 \over \mathrm{d}x^2} + V(x) \right] \Psi(x) = \omega^2 \Psi(x)
\end{equation}
and the expression for the potential is
\begin{equation}  \label{Schrodinger_potential}
V = -{r \over q} + {p^{1/4} \over q^{3/4}} {\mathrm{d} \over \mathrm{d}\sigma} \left[ \left( {p \over q} \right)^{1/2} {\mathrm{d} \over \mathrm{d}\sigma} (p q)^{1/4} \right] 
=  -{r \over q} + (p q)^{-1/4} {\mathrm{d}^2 \over \mathrm{d}x^2} (p q)^{1/4} \, . 
\end{equation}
In the first part of the above relation the potential is a function of the ``old" variable $\sigma$, 
while in the second part the potential depends on the `new" variable $x$. 
\begin{flushleft}
{\bf Toy model - $Q=0$ case - reproducing the results of \cite{Avramis:2006nv}}
\end{flushleft}
To set-up the conventions and compare with existing results in the literature, we begin the analysis with the $Q=0$ case
and consider $\theta_0=0$. The expressions 
for the Schr\"{o}dinger potentials in each fluctuation mode are presented in \eqref{Schr_potentials_Q_zero}.
Notice that the analysis of the ${\delta \psi}$ mode was absent in \cite{Avramis:2006nv}. For all the cases that we consider, the 
change of variables \eqref{change_variables_general} in this case becomes\footnote{Notice that in the following analysis 
we have set $\ell=1$. This means that all the values of $r$ that we consider are with respect to $\ell$: i.e. $r_0=1.2$ for  $\ell=1$, 
transforms to $r_0=1.2 \,  \ell$ for generic $\ell$. }
\begin{equation} \label{def_x_Q_zero}
x(\sigma, r_0) = \pm \int_{r_0}^{\sigma} \frac{\mathrm{d}\xi}{\sqrt{\left(\xi ^2-r_0^2\right) \left(\xi ^2+r_0^2-1\right)}} \, . 
\end{equation}
To decide about the stability of the solution against perturbations, in figure \ref{Potentials-Zero-Q}, we are plotting the expressions of the different potentials from \eqref{Schr_potentials_Q_zero} as a function of  $x$ from \eqref{def_x_Q_zero}. 
We have chosen to plot for $r_0=1.2$, but the conclusions we draw are valid for any value of $r_0>1$. 
Notice that the $Q=0$ case corresponds to the Coulomb branch (i.e. without confinement) and there is a singularity at $r=1$.  
Since $V_{\delta z}$ and $V_{\delta \theta}$ are positive for every value of $r_0$, the solution is stable under fluctuations along 
$z$ and $\theta$. 
Contrary to the above behaviour, $V_{\delta y}$ and $V_{\delta \psi}$ are positive for large values of $x$ and turn negative 
as $x$ approaches zero. Because of those negative values for the potential, a numerical computation of the spectrum is needed, 
in order  to decide whether tachyonic modes are produced or not. As we will see from the analysis of the next subsection, 
the small negative part of the potential $V_{\delta \psi}$ is not sufficient to produce tachyons for that mode, 
however we will identify tachyons, below a critical value of $r_0$, for the mode $\delta y$.  
\begin{flushleft}
{\bf Finite  $Q$ - confining geometry}
\end{flushleft}
For finite $Q$, as can be seen from equations \eqref{eom-coupled_1} and \eqref{eom-coupled_23}, two of the modes 
(namely $ \delta \phi$ and $\delta \varphi_{12}$ for $\theta=0$) are coupled and all the rest are decoupled. 

\begin{figure}[ht] 
   \centering
   \includegraphics[width=7cm,height=7cm]{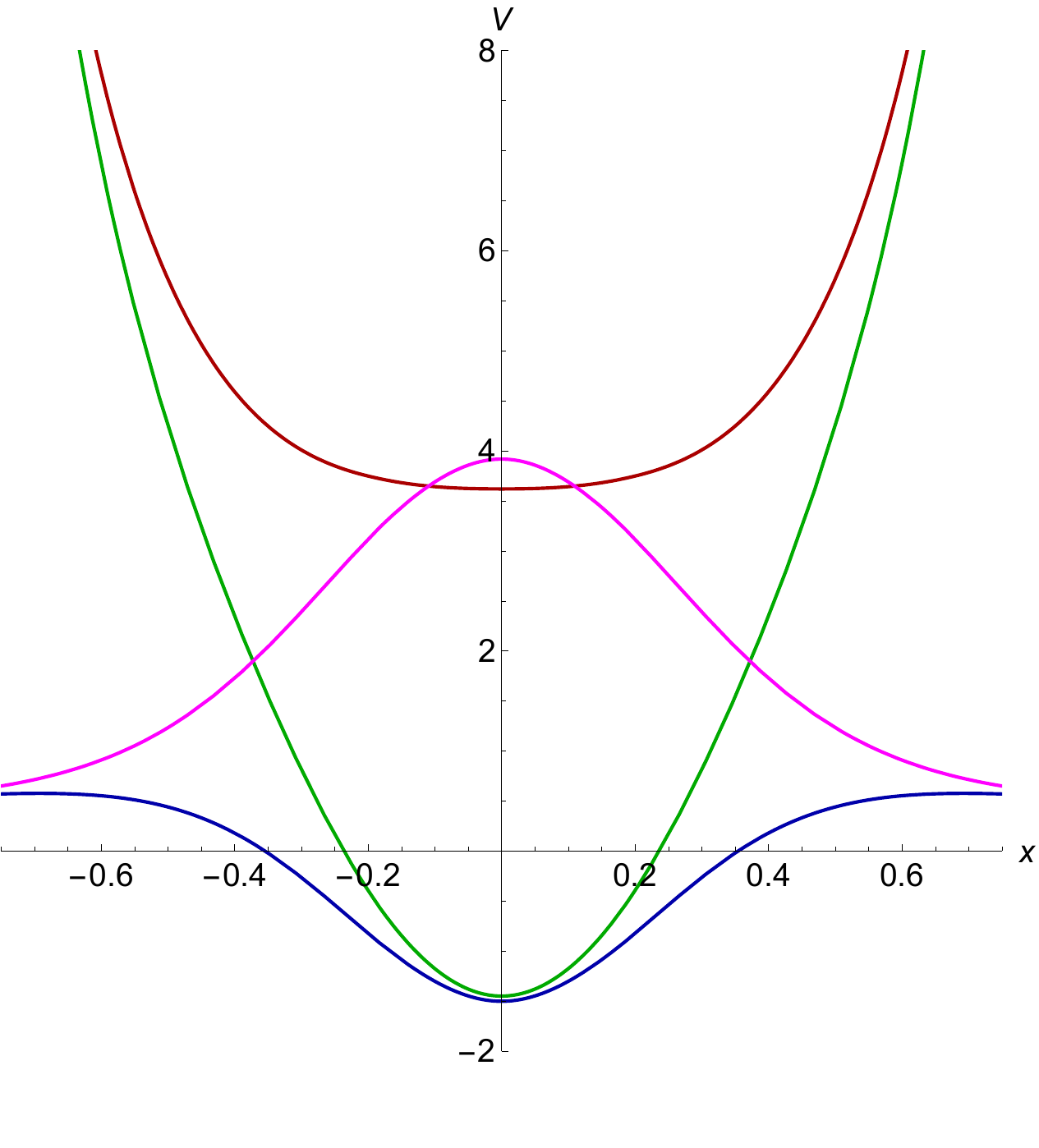}
    \includegraphics[width=7cm,height=7cm]{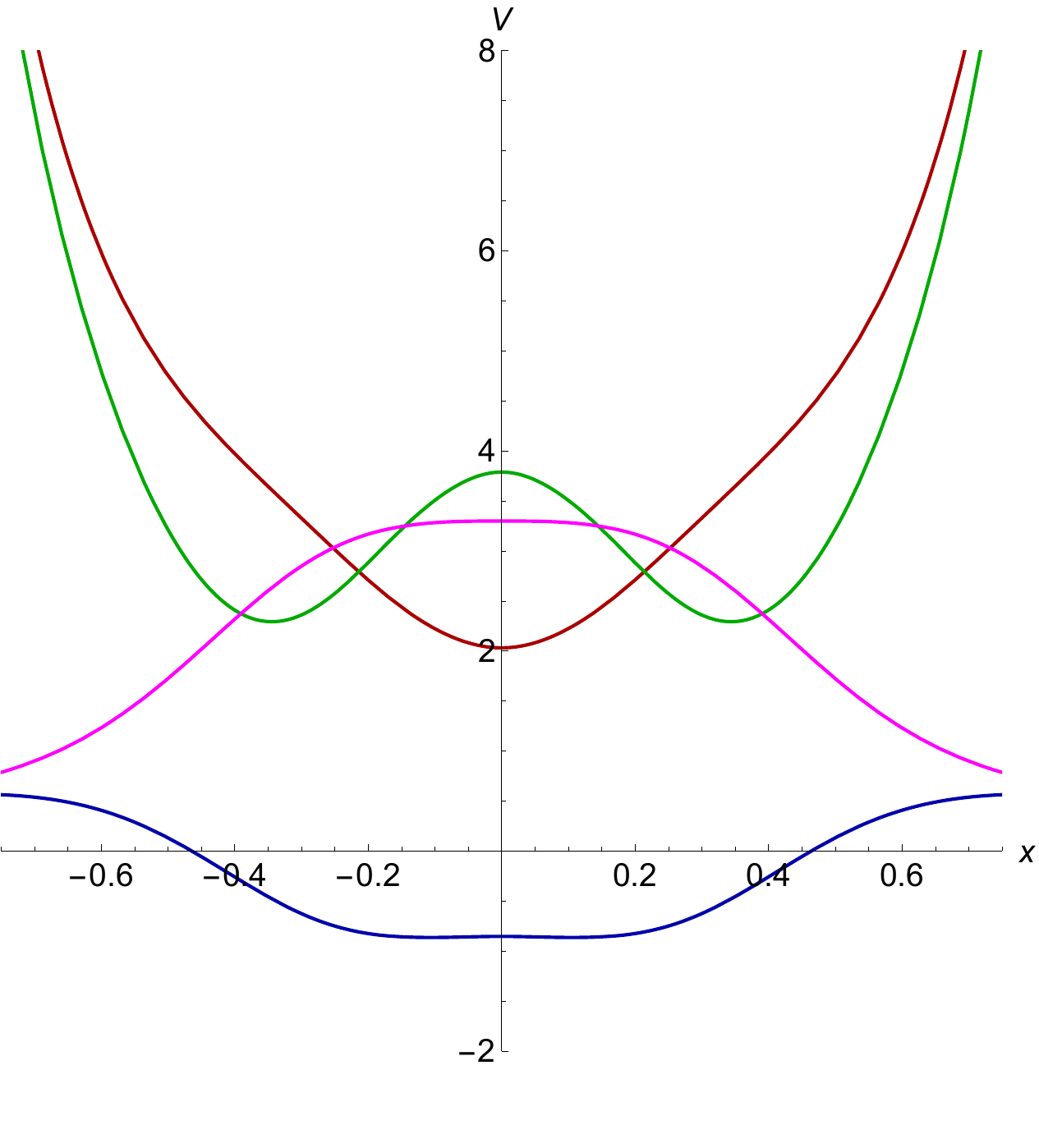}
     \caption{Plots of the decoupled Schr\"{o}dinger potentials for finite values of $Q$ as a function of  $x$ from \eqref{def_x_Q_finite}. 
     In the left panel $Q=1/2$ and in the right panel $Q=1$. The value of the parameter $r_0$ is fixed to $r_0=1.3$, for all the plots. 
     The red, green, magenta and blue solid lines correspond to $V_{\delta z}$, $V_{\delta y}$, $V_{\delta \theta}$ and $V_{\delta \psi}$. 
     Notice that increasing $Q$, decreases the region of potential instability. If we keep increasing the value of $Q$, 
     all the potentials will become positive for any value of $x$. 
     To produce the plots of this figure, we have the used the full expression for each potential and not the perturbative expansions of 
     \eqref{Schr_potentials_Q_finite_perturbative}.} 
   \label{Potentials-finite-Q}
\end{figure}

The expressions for the potentials of the decoupled modes are complicated and in  \eqref{Schr_potentials_Q_finite_perturbative} we are presenting their expansions for small value of $Q$. The change of variables for finite $Q$ becomes
\begin{equation} \label{def_x_Q_finite}
x(\sigma, r_0,Q) =  \pm \int_{r_0}^{\sigma} \frac{\xi ^2 \sqrt{\xi ^2-1} \, \mathrm{d}\xi }{\sqrt{\left(\xi ^6-\xi ^4-Q^2\right) 
\left(\xi^2-r_0^2\right) \left(\xi ^2+r_0^2-1\right)}} \, . 
\end{equation}
Notice that this expression holds both for the coupled and the decoupled modes and in the limit of $Q\rightarrow 0$, 
it reduces to \eqref{def_x_Q_zero}. In figure \ref{Potentials-finite-Q} we are plotting the decoupled potentials for small values of $Q$. 
From this plot it is clear that the presence of the confinement contributes to the stability of the system, 
since increasing the value of $Q$ is decreasing the region within which some of the potentials have negative values. 
Moreover the potentials that were positive for $Q=0$ remain positive for finite $Q$. 
In the next subsection we will solve numerically the fluctuation equations for ${\delta y}$ and ${\delta \psi}$ 
(that have regions of $x$ with negative values or the corresponding potentials) to confirm the absence of tachyonic modes for finite $Q$.

\begin{figure}[ht] 
   \centering
   \includegraphics[width=7cm,height=7cm]{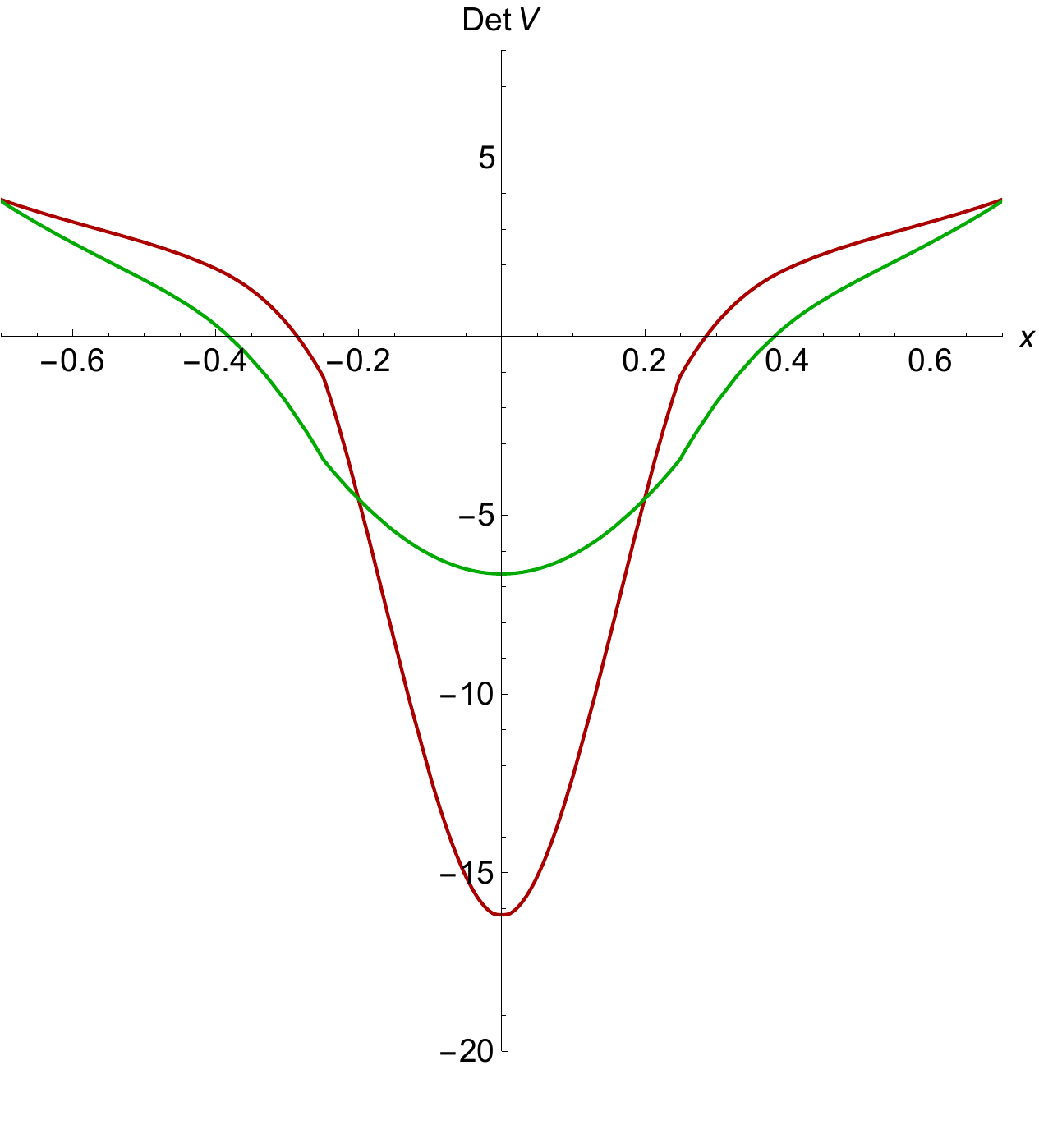}
    \includegraphics[width=7cm,height=7cm]{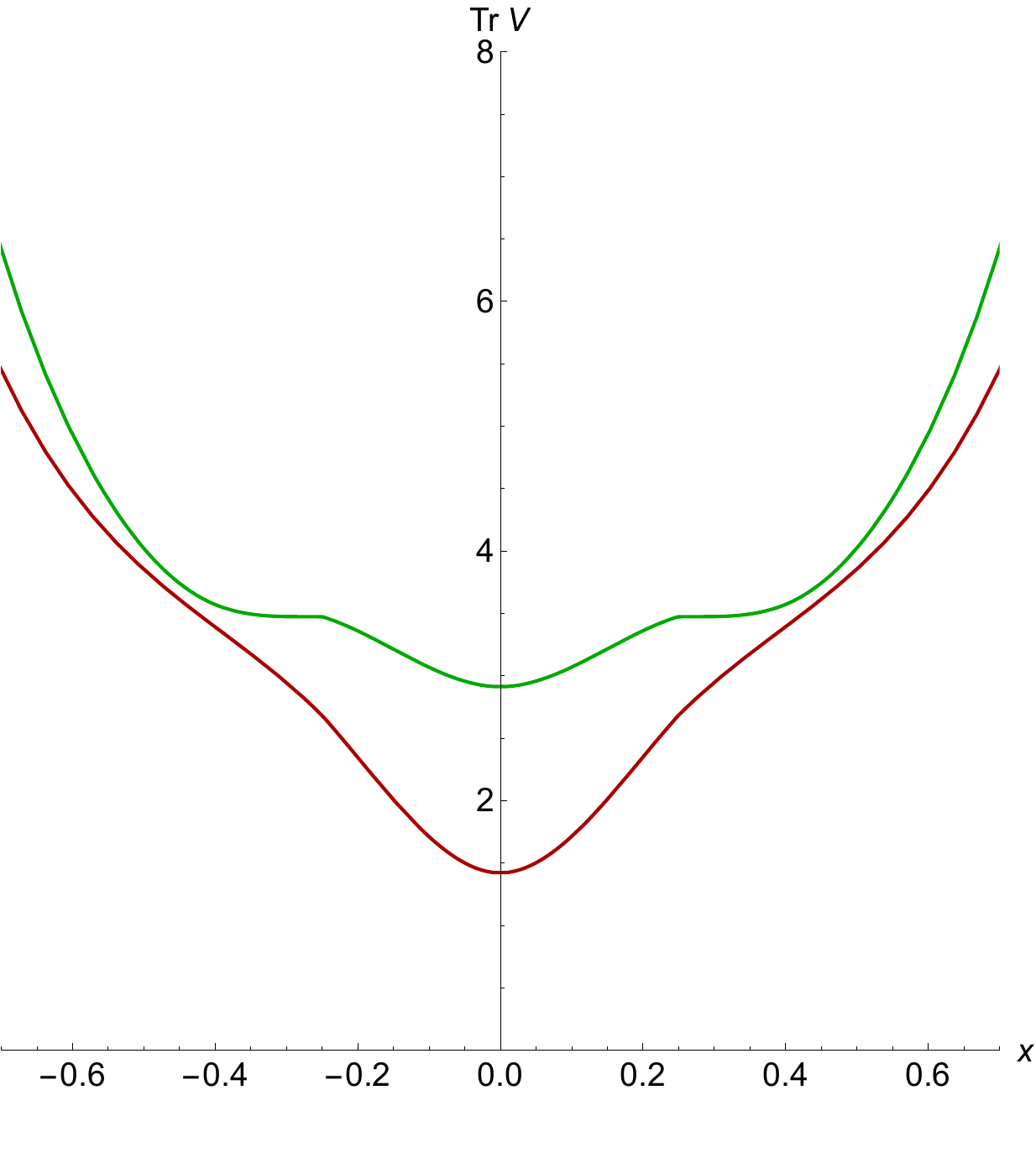}
     \caption{Plots of the determinant (left panel) and of the trace (right panel) of the matrix potential $V$ for two values of the parameter $Q$. 
      The red and green solid lines correspond to $Q=0$ and $Q=1/2$. 
      While the trace is always positive, the determinant is initially negative (i.e. for small values of $Q$), but as the value of 
      $Q$ increases the region that could potentially lead to an instability is shrinking. To produce the plots of this figure, 
      we have used a perturbative expansion for the trace and the determinant that includes more terms than those that appear in 
      \eqref{V_matrix}. In particular we have included terms of order $Q^4$. } 
   \label{finite_Q_det_tr}
\end{figure}

The coupled system of $\delta \phi$ and $\delta \varphi_{12}$ can be brought into a Schr\"{o}dinger form with the following transformation 
\begin{equation}
\begin{pmatrix} 
      \delta \phi \\
      \delta \varphi_{12}\\
   \end{pmatrix} 
   \,=\,
      \begin{pmatrix} 
      \Omega_1 & \Omega_2\\
      \Omega_3& \Omega_4\\
   \end{pmatrix} .
   \begin{pmatrix} 
      \Delta_1 \\
      \Delta_2\\
   \end{pmatrix} 
\end{equation}
together with the change of variables of \eqref{def_x_Q_finite}. The differential equations to determine the functions $ \Omega_i$, 
with $i=1, \cdots, 4$ are coupled and cannot be solved analytically. 
Alternatively, one can expand in powers of $Q$ and solve the coupled system order by order in $Q$. 
Until the linear in $Q$ term, the expressions of the functions  $ \Omega_i$, with $i=1, \cdots, 4$ are listed in  \eqref{Omega1234}. 
The coupled system becomes
\begin{equation}\label{Matrix-Schrodinger} 
\partial_{x}^2\,\begin{pmatrix} \Delta_1 \\  \Delta_2\\  \end{pmatrix}+\left(\omega^2\,\hat 1- V \right) . \begin{pmatrix} \Delta_1 \\  \Delta_2\\  \end{pmatrix}=0 \quad {\rm with} \quad V=\begin{pmatrix} V_{11} & V_{12} \\  V_{21} & V_{22}\\  \end{pmatrix}
\end{equation}
where the expressions for $V_{11} $, $V_{12}$, $V_{21}$ and $V_{22}$ are listed in \eqref{V_matrix}. 
For normalizable solutions of  \eqref{Matrix-Schrodinger} that vanish at infinity, $\omega^2$ is positive 
if the matrix potential $V$ is positively definite. This means that ${\rm Tr} \,  V >0$ and ${\rm det}\,V >0$. In figure \ref{finite_Q_det_tr} we have plotted the determinant and the trace of the 
potential $V$, including terms of order $Q^4$. While the trace is positive even for $Q=0$, the negative values of the determinant, in the left panel of figure \ref{finite_Q_det_tr}, reduce and 
eventually disappear when the value of $Q$ increases. As a result it is plausible to expect that the spectrum is tachyon free. Solving numerically the coupled system of equations 
for $\delta \phi$ and $\delta \varphi_{12}$, we will verify this expectation. 


\subsection{Solving for the fluctuations}

The final step in the investigation of the stability of the Wilson loop configuration is to solve numerically the equations of motion for the fluctuations.  
We will focus to fluctuation modes for which the corresponding Schr\"{o}dinger potential had regions of potential instability, 
i.e. regions of negative values. 
\begin{flushleft}
{\bf Toy model - $Q=0$ case - reproducing the results of \cite{Avramis:2006nv}}
\end{flushleft}
To describe the steps we have to follow in order to solve numerically the fluctuation equations, we start the analysis 
from the simpler $Q=0$ case. 
In this way we will reproduce the results of the analysis of \cite{Avramis:2006nv}.

Solving from \eqref{eom-decoupled} the equation of motion for $\delta \psi $ approximately around $\sigma = r_0$ 
reveals two classes of solutions, namely 
\begin{equation} \label{delta_psi_approx_r0}
\delta \psi  = A + B \, \sqrt{\big.\sigma - r_0} \, . 
\end{equation}
In a parallel approach, it is possible to solve perturbatively the Schr\"odinger equation for $\delta \psi$ with the potential $V_{\delta \psi}$ 
from \eqref{Schr_potentials_Q_zero} around $x=0$ (as it is defined in \eqref{def_x_Q_zero}). From this analysis two types of modes arise, 
namely even and odd ones \cite{Kuperstein:2008cq,Filev:2014bna}. 
Inverting \eqref{def_x_Q_zero} to express $\sigma$ as a function of $x$ when $\sigma$ approaches $r_0$, 
we can relate the even and the odd types of modes with the two classes of solutions of \eqref{delta_psi_approx_r0}.
From this identification, we have
\begin{eqnarray}
&& A \, \neq \, 0 \,\,\,\, \mathrm{and} \,\,\,\, B \, = \, 0 \quad  \Rightarrow \quad  \Psi^{'}(0)\, = \, 0 
\,\,\,\, {\rm and} \,\,\,\, \Psi(0) \, \neq \, 0 \quad  \Rightarrow \; \text{even modes} \label{even_psi}
\nonumber \\[5pt]
&& A \, =\, 0 \,\,\,\,  \mathrm{and} \,\,\,\, B \, \neq \, 0 \quad  \Rightarrow \quad  \Psi^{'}(0)\, \neq \, 0 
\,\,\,\, {\rm and} \,\,\,\, \Psi(0) \, = \, 0 \quad  \Rightarrow \; \text{odd modes} \label{odd_psi} \, . 
\end{eqnarray}
Solving numerically the equation of motion for $\delta \psi$, and imposing either odd or even boundary conditions, 
we find a spectrum that is free of instabilities for any value of $r_0$. 
This happens even if the Schr\"odinger potential for $\delta \psi$, as can be seen in figure \ref{Potentials-Zero-Q}, 
becomes negative for a range of values of $r_0$. For $r_0=1.1$ the first several excited states are,
\begin{eqnarray}
&& \omega_{\rm even} \, = \, 0.928,  \, 3.440, \, 5.774, \, 8.084, \, 10.390, \,  \dots
\nonumber \\
&& \omega_{\rm odd} \, = \, 2.442, \, 4.650, \, 6.936, \, 9.238, \, 11.544,  \, \dots 
\end{eqnarray}
From the analysis of the previous subsection we know that the other mode that could potentially lead to an instability is $\delta y $.
Analysing the behaviour of $\delta y $ around $\sigma = r_0$, we obtain the following behaviour
\begin{equation} \label{delta_y_approx_r0}
\delta y  = A + \frac{B}{\sqrt{\big.\sigma - r_0}}
\end{equation}
and performing a similar analysis as before, we identify odd and even modes in the following fashion
\begin{eqnarray}
&& A \, = \, 0 \,\,\,\, \mathrm{and} \,\,\,\, B \, \neq \, 0 \quad  \Rightarrow \quad  \Psi^{'}(0)\, = \, 0 
\,\,\,\, {\rm and} \,\,\,\, \Psi(0) \, \neq \, 0 \quad  \Rightarrow \; \text{even modes} \label{even_y}
\nonumber \\[5pt]
&& A \, \neq\, 0 \,\,\,\,  \mathrm{and} \,\,\,\, B \, = \, 0 \quad  \Rightarrow \quad  \Psi^{'}(0)\, \neq \, 0 
\,\,\,\, {\rm and} \,\,\,\, \Psi(0) \, = \, 0 \quad  \Rightarrow \; \text{odd modes} \label{odd_y} \, . 
\end{eqnarray}
Computation of the spectrum reveals a critical value for the parameter $r_0$, namely $r_0^{\rm crit} = 1.125$, 
below which the spectrum is tachyonic. This is in agreement with the findings from  \cite{Avramis:2006nv}. 
Notice that exactly for this critical value of $r_0$, the spectrum has a zero mode
\begin{eqnarray}
&& \omega^{\rm crit}_{\rm even} \, = \, 0.000,  \, 4.391, \, 6.881, \, 9.299, \, 11.692, \,  \dots
\nonumber \\
&& \omega^{\rm crit}_{\rm odd} \, = \, 3.154, \, 5.665, \, 8.096, \, 10.497, \, 12.885,  \, \dots 
\end{eqnarray}
%
\begin{flushleft}
{\bf Finite  $Q$ - confining geometry}
\end{flushleft}
%
Now we move to the case that we are interested in, namely of finite $Q$,  and start solving numerically from the mode $\delta \psi$. 
Similarly to the $Q=0$ case, we find no sign of instability and the first several excited states 
for $Q=1$ and $r_0=r_0^{\rm min}(Q=1)+0.1=1.31$ are,
\begin{eqnarray}
&& \omega_{\rm even} \, = \, 1.192, \, 4.014, \, 6.678,\,  9.346, \, 12.015, \,  \dots
\nonumber \\
&& \omega_{\rm odd} \, = \, 2.704, \, 5.340, \,  8.011, \, 10.680, \, 13.350  \, \dots 
\end{eqnarray}
For the $\delta y$ modes, one has to separate the analysis is two cases. When $Q<0.22$, in agreement with the analysis of the $Q=0$ case, there is always a critical value of $r_0$ 
(that depends on $Q$) where a tachyon appears and for this value there is a zero mode. However, as we extensively discussed in \cite{Chatzis:2025dnu}, these WL trajectories are 
pathological since the string is "touching" the Coulomb branch singularity. The reason is that the values of the Coulomb scale and of the confining scale are close to each other. 
Remarkably when $Q>0.22$ the tachyons disappear from the spectrum for any value above the $r_0^{\rm min}(Q)$. 
For $Q=1$ and $r_0=r_0^{\rm min}(Q=1)+0.1=1.31$ the first several excited states that we find are,
\begin{eqnarray}
&& \omega_{\rm even} \, = \, 2.388, \, 5.163, \, 7.851, \, 10.521, \, 13.183, \,  \dots
\nonumber \\
&& \omega_{\rm odd} \, = \, 3.727, \, 6.505, \, 9.188, \, 11.853, \, 14.511  \, \dots 
\end{eqnarray}
Finally we solve numerically the coupled system of equations  \eqref{eom-coupled_1} and the first of  
\eqref{eom-coupled_23} (this is the one that survives when $\theta=0$). 
As previously the modes are either even or odd depending on the boundary conditions and it turns out that even/odd modes of $\delta \phi$ couple to even/odd modes of $\delta\varphi_{12}$, respectively.  From the numerical solution, no presence of tachyon appears and
the first several excited states that we find, for $Q=1$, $r_0=r_0^{\rm min}(Q=1)+0.1=1.31$, are
\begin{eqnarray}
&& \omega_{\rm even-even} \, = \, 1.177, \, 2.687, \,  4.002, \,  5.146 \, , 6.652, \,  \dots
\nonumber \\
&& \omega_{\rm odd-odd} \, = \, 2.680,  \, 3.780, \, 5.278, \, 6.270, \, 7.814  \, \dots 
\end{eqnarray}


\section{Penrose limits of the Anabalon-Ross deformed $\mathrm{AdS}_5\times \mathbb{S}^5$ solution}\label{Penrose_limits_section}

Here we review the uplift of the supersymmetric Anabalon-Ross solution \cite{Anabalon:2021tua} in type-IIB supergravity, following the notation of \cite{Chatzis:2024kdu}. This consists of a $10D$ metric and a self-dual RR $5$-form. As mentioned in \cite{Chatzis:2025dnu} this solution can be reached by taking the limit $\ell\to 0$ with $\ell^2 Q$ fixed in the solution of section \ref{IIB_background} (considering the SUSY case where $Q=|q_1|=|q_2|$). For the metric we have
\footnote{Notice that the internal coordinates $(\th, \, \varphi, \varphi_i)$ here correspond to $(\th - \frac{\pi}{2}, \, \psi, \, \phi_i)$ in section \ref{IIB_background}.}
\begin{equation}
 \label{metricAR}
 \begin{aligned}
  L^{-2} \mathrm{d}s^2 & = r^2 \big(- \mathrm{d}t^2 + \mathrm{d}x^2_1 + \mathrm{d}x^2_2\big) + \frac{\mathrm{d}r^2}{r^2 f(r)} + r^2 f(r) \mathrm{d}\phi^2
  \\[5pt]
  & + \mathrm{d}\th^2 + \sin^2\th \mathrm{d}\vphi^2 + \sin^2\th \sin^2\vphi \Big( \mathrm{d}\vphi_1 + \frac{1}{L} \cA \Big)^2
  \\[5pt]
  & + \sin^2\th \cos^2\vphi \Big( \mathrm{d}\vphi_2 + \frac{1}{L} \cA \Big)^2
  + \cos^2\th \Big( \mathrm{d}\vphi_3 + \frac{1}{L} \cA \Big)^2 \, ,
 \end{aligned}
\end{equation}
where
\begin{equation}
 f(r) = 1 - \left( \frac{Q}{r} \right)^6 \, , \qquad \cA = L Q \left( 1 - \left( \frac{Q}{r} \right)^2 \right) \mathrm{d} \phi = L A(r)\mathrm{d} \phi\, .
\end{equation}
The periodicities of the various angular coordinates are
\begin{equation}
 \label{periodicitiesthphi}
 0 \leq \th \leq \frac{\pi}{2} \, , \qquad 0 \leq \vphi \leq \frac{\pi}{2} \, , \qquad 0 \leq \vphi_i \leq 2 \pi \quad (i = 1, 2, 3) \, .
\end{equation}
The RR $5$-form is given by the formula below
\begin{equation}
 F_5 = (1 + \star) G_5 \, , \qquad G_5 = 4 L^4 \, \textrm{vol}_5 - L^3 \sum\limits_{i = 1}^3 \m_i \mathrm{d}\m_i \wedge \Big( \mathrm{d}\vphi_i + \frac{1}{L} \cA \Big) \wedge \star_5 \mathrm{d}\cA \, .
\end{equation}

In the above expression, the operator $\star_5$ stands for Hodge duality in the $5D$ space with line element given by the first line in \eqref{metricAR}. Also, with $\text{vol}_5$ we refer to the volume form of the aforementioned $5D$ space. Therefore
\begin{equation}
 \text{vol}_5 = r^3 \, \mathrm{d}t \wedge \mathrm{d}x_1 \wedge \mathrm{d}x_2 \wedge \mathrm{d}r \wedge \mathrm{d}\phi \, , \qquad
 \star_5 \mathrm{d}\cA = 2 L Q^3 \, \mathrm{d}t \wedge \mathrm{d}x_1 \wedge \mathrm{d}x_2 \, .
\end{equation}
Moreover, $\m_i \, (i = 1, 2, 3)$ are the embedding coordinates in $\mathbb{R}^3$ of a unit two-sphere parametrised by $(\th, \vphi)$. In particular
\begin{equation}
 \m_1 = \sin\th \sin\vphi \, , \qquad \m_2 = \sin\th \cos\vphi \, , \qquad \m_3 = \cos\th \, .
\end{equation}
Given the ranges of the coordinates $(\th, \vphi)$ in \eqref{periodicitiesthphi}, the embedding coordinates $\m_i \, (i = 1, 2, 3)$ cover only the $\frac{1}{8}$ of the sphere surface.

\noindent{Notice} that we factor out the scale $L^2$ in the line element \eqref{metricAR}. This is done by redefining the radial coordinate $r$ of \cite{Chatzis:2024kdu} as $r \to L^2 r$ and the parameter $Q$ as $Q \to L Q$. The reason behind this redefinition is that we will use the scale $L$ later to compute Penrose limits in the geometry \eqref{metricAR}.\\

Besides the time-like Killing vector, the $10D$ geometry \eqref{metricAR} admits four more Killing vectors representing the symmetry under shifts in the directions $\phi$ and $\vphi_i \, (i = 1, 2, 3)$. In this section, we will consider the Penrose limits that correspond to null geodesics that are associated with the motion of a point particle in $t$ and the Killing directions $\phi$ and $\vphi_i \, (i = 1, 2, 3)$.

\subsection{Motion along $(t , \, \phi , \, \vphi_1)$}

Let us start with the case where a particle moves in the spacetime \eqref{metricAR} with fixed $x_1 , \, x_2 , \, r , \, \th , \, \vphi$. In particular, if we take $\th = \vphi = \frac{\pi}{2}$ then the particle does not "feel" the presence of $\vphi_2$ and $\vphi_3$. The geodesic equations of motion, except that for $r$, require
\begin{equation}
 t = \cE u \, , \qquad
 \phi = \Big( \cJ + \frac{r^2_0}{Q^2 - r^2_0} \cJ_1 \Big) u \, , \qquad
 \vphi_1 = Q \cJ_1 u \, ,
\end{equation}
where $r_0$ is the fixed value of the $r$ direction and $\cE , \, \cJ , \, \cJ_1$ are integration constants. The geodesic equation for the $r$ direction provides an algebraic constraint for the constants $\cE , \, \cJ$ and $\cJ_1$, which is
\begin{equation}
 \cE^2 = \left(1 + 2 \frac{Q^4}{r^4_0} \right) \cJ^2 + \frac{2}{r^4_0} \frac{Q^6 + Q^4 r^2_0 + r^6_0}{Q^2 - r^2_0} \cJ \cJ_1 + \frac{2 Q^6 + r^6_0}{\big( Q^2 - r^2_0 \big)^2} \frac{\cJ^2_1}{r^2_0} \, .
\end{equation}
Moreover, the null condition implies that $\cJ$ and $\cJ_1$ are related as
\begin{equation}
 \label{valuesJphi1}
 \cJ = - \frac{Q^2}{Q^2 - r^2_0} \, \cJ_1 \qquad \text{or} \qquad \cJ = - \frac{3 Q^2 r^2_0}{4 Q^4 - 5 Q^2 r^2_0 + r^4_0} \, \cJ_1 \, .
\end{equation}
The above values for $\cJ$ correspond to
\begin{equation}
 \label{valuesEphi1}
 \cE^2 = \cJ^2_1 \qquad \text{and} \qquad \cE^2 = \frac{8 Q^6 + r^6_0}{\big( 4 Q^2 - r^2_0 \big)^2} \, \frac{\cJ^2_1}{r^2_0}
\end{equation}
respectively.

For the Penrose limit we will adopt the rescalings
\begin{equation}
 x_1 \to \frac{x_1}{L} \, , \qquad
 x_2 \to \frac{x_2}{L} \, , \qquad
 r \to r_0 + \frac{\r}{L} \, , \qquad
 \th \to \frac{\pi}{2} + \frac{y_1}{L} \, , \qquad
 \vphi \to \frac{\pi}{2} + \frac{y_2}{L}
\end{equation}
together with
\begin{equation}
 t \to \cE u \, , \qquad \phi \to \Big( \cJ + \frac{r^2_0}{Q^2 - r^2_0} \cJ_1 \Big) u + \frac{w}{L} \, , \qquad \vphi_1 \to Q \cJ_1 u + c \frac{w}{L} + \frac{v}{L^2} \, .
\end{equation}
Here $c$ is a constant which will be specified later, while $\cE$ and $\cJ$ take the values \eqref{valuesJphi1} and \eqref{valuesEphi1}. Therefore, we have the following two cases
\footnote{
Of course one can also consider the case where $\cJ + \frac{r^2_0}{Q^2 - r^2_0} \cJ_1 = 0$, which amounts to rescaling $\phi$ as $\phi \to \frac{\phi}{L}$. This would just correspond to motion along $t$ and $\vphi_1$ only.
}
\begin{equation}
 \begin{aligned}
  & \text{\textbf{Case A:}} \qquad \cE^2 = \cJ^2_1 \, , \qquad\qquad\qquad\qquad\,\, \cJ = - \frac{Q^2}{Q^2 - r^2_0} \, \cJ_1 \, ,
  \\[5pt]
  & \text{\textbf{Case B:}} \qquad \cE^2 = \frac{8 Q^6 + r^6_0}{\big( 4 Q^2 - r^2_0 \big)^2} \, \frac{\cJ^2_1}{r^2_0} \, , \qquad \cJ = - \frac{3 Q^2 r^2_0}{4 Q^4 - 5 Q^2 r^2_0 + r^4_0} \, \cJ_1 \, .
 \end{aligned}
\end{equation}
Let us examine each case separately.

\noindent\textbf{\underline{Case A}}

\vskip 10pt

In order for the Penrose limit to be valid the constant $c$ must take the value
\footnote{
This value ensures the vanishing of the term that is proportional to $L$ in the expansion of the line element for large $L$.
}
\begin{equation}
 \label{cExample1A}
 c = \frac{r^4_0 - Q^4}{Q^3} \, .
\end{equation}
Sending $L$ to infinity, one derives a pp-wave geometry, which can be cast in Brinkman form after the following sequence of transformations
\begin{enumerate}[i.]
 \item Rescalings
 \begin{equation}
  \begin{aligned}
   & x_1 \to \frac{x_1}{r_0} \, , \qquad 
  x_2 \to \frac{x_2}{r_0} \, , \qquad
  w \to \frac{Q^3}{r_0 \sqrt{r^6_0 - Q^6}} \, w \, ,
  \\[5pt]
  &  \r \to \frac{\sqrt{r^6_0 - Q^6}}{r^2_0} \, \r \, , \qquad
  v \to \om_1 \, v + \om_2 \, \r \, w \, ,
  \end{aligned}
 \end{equation}
 with
 \begin{equation}
  \label{om12Case1}
  \om_1 = \frac{r^2_0}{Q^3 \cJ_1} \, , \qquad \om_2 = 2 \, .
 \end{equation}
 \item Rotation
 \begin{equation}
  \label{RtationwrhoCase1}
  \begin{pmatrix}
   w
   \\[5pt]
   \r
  \end{pmatrix} \to
  \begin{pmatrix}
   \cos\left( \frac{\om_2}{\om_1} u \right) & \sin\left( \frac{\om_2}{\om_1} u \right)
   \\[5pt]
   -\sin\left( \frac{\om_2}{\om_1} u \right) & \cos\left( \frac{\om_2}{\om_1} u \right)
  \end{pmatrix} \cdot
  \begin{pmatrix}
   w
   \\[5pt]
   \r
  \end{pmatrix} \, .
 \end{equation}
 \item Shifts
 \begin{equation}
  \vphi_2 \to \vphi_2 + Q \cJ_1 \Big( 1 - \frac{Q^2}{r^2_0} \Big) u \, , \qquad \vphi_3 \to \vphi_3 + Q \cJ_1 \Big( 1 - \frac{Q^2}{r^2_0} \Big) u \, .
 \end{equation}
\end{enumerate}

 As a result one finds
 \begin{equation}
  \label{metricpp1A}
  \begin{aligned}
   \mathrm{d}s^2 & = 2 \mathrm{d}u \mathrm{d}v + \mathrm{d}x^2_1 + \mathrm{d}x^2_2 + \mathrm{d}y^2_1 + y^2_1 \, \mathrm{d}\vphi^2_3 + \mathrm{d}y^2_2 + y^2_2 \, \mathrm{d}\vphi^2_2 + \mathrm{d}w^2 + \mathrm{d}\r^2
   \\[5pt]
   & - \frac{Q^6 \cJ^2_1}{r^4_0} \big( y^2_1 + y^2_2 + 4 w^2 + 4 \r^2 \big) \mathrm{d}u^2 \, .
  \end{aligned}
 \end{equation}
 Notice that each of the pairs $(y_1 , \, \vphi_3)$ and $(y_2 , \, \vphi_2)$ parametrise an $\mathbb{R}^2$.
 
 The above transformations, when applied to the self-dual five-form give
 \begin{equation}
  \label{F5pp1A}
  \begin{aligned}
   F_5 & = \frac{2 Q^3 \cJ_1}{r^2_0} \, \Big( 2 \, \mathrm{d}u \wedge \mathrm{d}x_1 \wedge \mathrm{d}x_2 \wedge \mathrm{d}\r \wedge \mathrm{d}w
   + y_1 \, \mathrm{d}u \wedge \mathrm{d}y_1 \wedge \mathrm{d}\vphi_3 \wedge \mathrm{d}\r \wedge \mathrm{d}w
   \\[5pt]
   & + y_2 \, \mathrm{d}u \wedge \mathrm{d}y_2 \wedge \mathrm{d}\vphi_2 \wedge \mathrm{d}\r \wedge \mathrm{d}w
   - y_1 \, \mathrm{d}u \wedge \mathrm{d}x_1 \wedge \mathrm{d}x_2 \wedge \mathrm{d}y_1 \wedge \mathrm{d}\vphi_3
   \\[5pt]
   & - y_2 \, \mathrm{d}u \wedge \mathrm{d}x_1 \wedge \mathrm{d}x_2 \wedge \mathrm{d}y_2 \wedge \mathrm{d}\vphi_2
   - 2 \, y_1 y_2 \, \mathrm{d}u \wedge \mathrm{d}y_1 \wedge \mathrm{d}y_2 \wedge \mathrm{d}\vphi_2 \wedge \mathrm{d}\vphi_3 \Big) \, .
  \end{aligned}
 \end{equation}
 
\noindent\textbf{\underline{Case B}}
 
\vskip 10pt

Now the Penrose limit makes sense when
\begin{equation}
 \label{cExample1B}
 c = - Q + \frac{4 Q^3}{3 r^2_0} - \frac{r^4_0}{3 Q^3} \, .
\end{equation}
The associated pp-wave geometry can be expressed in Brinkman coordinates after the following operations
\begin{enumerate}[i.]
 \item Rescalings
 \begin{equation}
  \begin{aligned}
   & x_1 \to \frac{x_1}{r_0} \, , \qquad 
  x_2 \to \frac{x_2}{r_0} \, , \qquad
  w \to \frac{3 Q^3 r^2_0}{\sqrt{r^{12}_0 + 7 Q^6 r^6_0 - 8 Q^{12}}} \, w \, ,
  \\[5pt]
  &  \r \to \frac{\sqrt{r^6_0 - Q^6}}{r^2_0} \, \r \, , \qquad
  v \to \om_1 \, v + \om_2 \r \, w \, ,
  \end{aligned}
 \end{equation}
 where
 \begin{equation}
  \label{om12Case2}
  \om_1 = \frac{4Q^2 - r^2_0}{3 Q^3 \cJ_1} \, , \qquad \om_2 = - \frac{2}{3 r^3_0} \sqrt{\frac{r^{12}_0 + 7 Q^6 r^6_0 - 8 Q^{12}}{r^6_0 - Q^6}} \, .
 \end{equation}

 \item Rotation in the $(\r , \, w)$-plane as in \eqref{RtationwrhoCase1}, where now the parameters $\om_1$ and $\om_2$ are given in \eqref{om12Case2}.
 
 \item Shifts
 \begin{equation}
  \vphi_2 \to \vphi_2 + Q \, \cJ_1 \, \frac{Q^2 - r^2_0}{4 Q^2 - r^2_0} u \, , \qquad \vphi_3 \to \vphi_3 + Q \, \cJ_1 \, \frac{Q^2 - r^2_0}{4 Q^2 - r^2_0} u \, .
 \end{equation}
\end{enumerate}
The outcome of this is
 \begin{equation}
  \label{metricpp1B}
  \begin{aligned}
   \mathrm{d}s^2 & = 2 \mathrm{d}u \mathrm{d}v + \mathrm{d}x^2_1 + \mathrm{d}x^2_2 + \mathrm{d}y^2_1 + y^2_1 \, \mathrm{d}\vphi^2_3 + \mathrm{d}y^2_2 + y^2_2 \, \mathrm{d}\vphi^2_2 + \mathrm{d}w^2 + \mathrm{d}\r^2
   \\[5pt]
   & - \frac{Q^6 \cJ^2_1}{r^6_0 \big( r^2_0 - 4 Q^2 \big)^2} \left( 9 r^6_0 \big( y^2_1 + y^2_2 \big) + 4 \big( 4 Q^6 + 5 r^6_0 \big) \big( w^2 + \r^2 \big) \right.
   \\[5pt]
   & \left. + 16 \big( Q^6 - r^6_0 \big) \big( \tilde{w}^2 - \tilde{\r}^2 \big) \right) \mathrm{d}u^2 \, ,
  \end{aligned}
 \end{equation}
 where
 \begin{equation}
  \label{etrtpp1B}
  \tilde{w} = \sin\Big( \frac{\om_2}{\om_1} u \Big) \r + \cos\Big( \frac{\om_2}{\om_1} u \Big) w \, , \qquad \tilde{\r} = \cos\Big( \frac{\om_2}{\om_1} u \Big) \r - \sin\Big( \frac{\om_2}{\om_1} u \Big) w \, .
 \end{equation}
 Like in the previous case, each of the pairs $(y_1 , \, \vphi_3)$ and $(y_2 , \, \vphi_2)$ parametrise a Euclidean plane.
 
 For the self-dual five-form we find
 \begin{equation}
  \label{F5pp1B}
  \begin{aligned}
   F_5 & = \frac{8}{3} \frac{\sqrt{r^6_0 + 8 Q^6}}{r^3_0 \om_1 \om_2} \Big( \mathrm{d}u \wedge \mathrm{d}x_1 \wedge \mathrm{d}x_2 \wedge \mathrm{d}\r \wedge \mathrm{d}w
   - \frac{\sqrt{r^6_0 + 8 Q^6}}{6 r^3_0} y_1 \mathrm{d}u \wedge \mathrm{d}y_1 \wedge \mathrm{d}\vphi_3 \wedge \mathrm{d}\r \wedge \mathrm{d}w
   \\[5pt]
   & - \frac{\sqrt{r^6_0 + 8 Q^6}}{6 r^3_0} y_2 \mathrm{d}u \wedge \mathrm{d}y_2 \wedge \mathrm{d}\vphi_2 \wedge \mathrm{d}\r \wedge \mathrm{d}w
   - \frac{\om_2}{4} y_1 \mathrm{d}u \wedge \mathrm{d}x_1 \wedge \mathrm{d}x_2 \wedge \mathrm{d}y_1 \wedge \mathrm{d}\vphi_3\
   \\[5pt]
   & - \frac{\om_2}{4} y_2 \mathrm{d}u \wedge \mathrm{d}x_1 \wedge \mathrm{d}x_2 \wedge \mathrm{d}y_2 \wedge \mathrm{d}\vphi_2
   - \frac{3 r^3_0 \om_2}{2 \sqrt{r^6_0 + 8 Q^6}} y_1 y_2 \mathrm{d}u \wedge \mathrm{d}y_1 \wedge \mathrm{d}y_2 \wedge \mathrm{d}\vphi_2 \wedge \mathrm{d}\vphi_3\Big) \, .
  \end{aligned}
 \end{equation}
 
 Notice that when $Q = 0$, both pp-wave backgrounds that we discussed above reduce to flat spaces with vanishing RR fields.

 \subsection{Motion along $(t , \, \phi , \, \vphi_2)$}
 
 We will now consider the case where the motion of the particle is realised in the $t , \, \phi$ and $\vphi_2$ directions. Therefore, we will take $x_1 , \, x_2 , \, r , \, \th , \, \vphi$ to be fixed and $\th = \frac{\pi}{2}$, $\vphi = 0$. One can easily see from \eqref{metricAR}, that for this values of $\th$ and $\vphi$, the terms of the line element involving $\mathrm{d}\vphi_1$ and $\mathrm{d}\vphi_3$ vanish. Therefore, the particle does not "feel" the presence of $\vphi_2$ and $\vphi_3$. From the geodesic equations of motion we find that
 \begin{equation}
 t = \cE u \, , \qquad
 \phi = \Big( \cJ + \frac{r^2_0}{Q^2 - r^2_0} \cJ_2 \Big) u \, , \qquad
 \vphi_2 = Q \cJ_2 u \, ,
\end{equation}
where $r_0$ is the fixed value of the $r$ direction and $\cE , \, \cJ , \, \cJ_2$ are integration constants. On top of that, the equation for the radial direction $r$ reduces into the algebraic constraint
\begin{equation}
 \cE^2 = \left(1 + 2 \frac{Q^4}{r^4_0} \right) \cJ^2 + \frac{2}{r^4_0} \frac{Q^6 + Q^4 r^2_0 + r^6_0}{Q^2 - r^2_0} \cJ \cJ_2 + \frac{2 Q^6 + r^6_0}{\big( Q^2 - r^2_0 \big)^2} \frac{\cJ^2_2}{r^2_0} \, .
\end{equation}
The null condition relates $\cJ$ with $\cJ_2$ and it turns out that one has to consider the following two cases
\footnote{
Again, one can also consider the case where $\cJ + \frac{r^2_0}{Q^2 - r^2_0} \cJ_2 = 0$, which corresponds to particle motion only along $t$ and $\vphi_2$.
}
\begin{equation}
 \begin{aligned}
  & \text{\textbf{Case A:}} \qquad \cE^2 = \cJ^2_2 \, , \qquad\qquad\qquad\qquad\,\, \cJ = - \frac{Q^2}{Q^2 - r^2_0} \, \cJ_2 \, ,
  \\[5pt]
  & \text{\textbf{Case B:}} \qquad \cE^2 = \frac{8 Q^6 + r^6_0}{\big( 4 Q^2 - r^2_0 \big)^2} \, \frac{\cJ^2_2}{r^2_0} \, , \qquad \cJ = - \frac{3 Q^2 r^2_0}{4 Q^4 - 5 Q^2 r^2_0 + r^4_0} \, \cJ_2 \, .
 \end{aligned}
\end{equation}

Like in the previous example, in order to take the Penrose limit we will adopt the rescalings
\begin{equation}
 x_1 \to \frac{x_1}{L} \, , \qquad
 x_2 \to \frac{x_2}{L} \, , \qquad
 r \to r_0 + \frac{\r}{L} \, , \qquad
 \th \to \frac{\pi}{2} + \frac{y_1}{L} \, , \qquad
 \vphi \to \frac{y_2}{L}
\end{equation}
together with
\begin{equation}
 t \to \cE u \, , \qquad \phi \to \Big( \cJ + \frac{r^2_0}{Q^2 - r^2_0} \cJ_2 \Big) u + \frac{w}{L} \, , \qquad \vphi_2 \to Q \cJ_2 u + c \frac{w}{L} + \frac{v}{L^2} \, .
\end{equation}
Here again $c$ is a constant which must be fixed to the value \eqref{cExample1A} (\textbf{case A}) or \eqref{cExample1B} (\textbf{case B}), so that the Penrose limit makes sense
\footnote{
These values ensure the vanishing of the term that is proportional to $L$ in the expansion of the line element for large $L$.
}
.

It turns out that if we compute the limit in the \textbf{case A}, one derives the pp-wave background \eqref{metricpp1A} and \eqref{F5pp1A} with $\vphi_2 \to \vphi_1$ and $\cJ_1 \to \cJ_2$. Similarly, in the \textbf{case B} one finds \eqref{metricpp1B}, \eqref{etrtpp1B} and \eqref{F5pp1B}, where again $\vphi_2 \to \vphi_1$ and $\cJ_1 \to \cJ_2$.

\subsection{Motion along $(t , \, \phi , \, \vphi_3)$}

In the last example, we will consider the case where the particle moves in the $t , \, \phi$ and $\vphi_3$ directions. For this reason, we will fix $x_1 , \, x_2 , \, r \, , \th$ and set $\th = 0$ while keeping $\vphi$ arbitrary. As it can be confirmed from \eqref{metricAR}, the terms involving $\mathrm{d}\vphi_1$ and $\mathrm{d}\vphi_2$ shrink to zero, and therefore the particle does not "feel" these directions. The geodesic equations of motion now imply
\begin{equation}
 t = \cE u \, , \qquad
 \phi = \Big( \cJ + \frac{r^2_0}{Q^2 - r^2_0} \cJ_3 \Big) u \, , \qquad
 \vphi_3 = Q \cJ_3 u \, ,
\end{equation}
with $r_0$ being the fixed value of the $r$ direction and $\cE , \, \cJ , \, \cJ_3$ are integration constants. Like in the previous two examples, the equation of motion for the radial direction $r$ is equivalent to an algebraic constraint, where now
\begin{equation}
 \cE^2 = \left(1 + 2 \frac{Q^4}{r^4_0} \right) \cJ^2 + \frac{2}{r^4_0} \frac{Q^6 + Q^4 r^2_0 + r^6_0}{Q^2 - r^2_0} \cJ \cJ_3 + \frac{2 Q^6 + r^6_0}{\big( Q^2 - r^2_0 \big)^2} \frac{\cJ^2_3}{r^2_0} \, .
\end{equation}
On top of that, we need to take into account the null condition of the geodesic which provides a relation between $\cJ$ and $\cJ_3$, namely
\begin{equation}
 \label{valuesJphi3}
 \cJ = - \frac{Q^2}{Q^2 - r^2_0} \, \cJ_3 \qquad \text{or} \qquad \cJ = - \frac{3 Q^2 r^2_0}{4 Q^4 - 5 Q^2 r^2_0 + r^4_0} \, \cJ_3 \, .
\end{equation}
This suggests that we need to consider the following two cases
\begin{equation}
 \begin{aligned}
  & \textbf{\text{Case A:}} \qquad \cE^2 = \cJ^2_3 \, , \qquad\qquad\qquad\qquad\,\, \cJ = - \frac{Q^2}{Q^2 - r^2_0} \, \cJ_3 \, ,
  \\[5pt]
  & \textbf{\text{Case B:}} \qquad \cE^2 = \frac{8 Q^6 + r^6_0}{\big( 4 Q^2 - r^2_0 \big)^2} \, \frac{\cJ^2_3}{r^2_0} \, , \qquad \cJ = - \frac{3 Q^2 r^2_0}{4 Q^4 - 5 Q^2 r^2_0 + r^4_0} \, \cJ_3 \, .
 \end{aligned}
\end{equation}

For the Penrose limit we will adopt the rescalings
\begin{equation}
 x_1 \to \frac{x_1}{L} \, , \qquad
 x_2 \to \frac{x_2}{L} \, , \qquad
 r \to r_0 + \frac{\r}{L} \, , \qquad
 \th \to \frac{y}{L}
\end{equation}
together with
\footnote{
Like before, one can also consider the case where $\cJ + \frac{r^2_0}{Q^2 - r^2_0} \cJ_3 = 0$, which corresponds to particle motion only along $t$ and $\vphi_3$.
}

\begin{equation}
 t \to \cE u \, , \qquad \phi \to \Big( \cJ + \frac{r^2_0}{Q^2 - r^2_0} \cJ_3 \Big) u + \frac{w}{L} \, , \qquad \vphi_3 \to Q \cJ_3 u + c \frac{w}{L} + \frac{v}{L^2} \, .
\end{equation}
Like in the previous two examples, $c$ is a constant which must be fixed to the value \eqref{cExample1A} (\textbf{case A}) or \eqref{cExample1B} (\textbf{case B}), to ensure the vanishing of the term that is linear in $L$ when taking the Penrose limit.

At the end of the day, one obtains a pp-wave background for each of the two cases A and B. However, it turns out that both pp-wave backgrounds are equivalent to the ones found in the previous two examples. In particular, if we want to match the two solutions with the ones that correspond to motion along $\vphi_1$ all we have to do is to apply the following change of coordinates
\begin{equation}
 y \to \sqrt{y^2_1 + y^2_2} \, , \qquad \vphi \to \tan^{-1} \frac{y_1}{y_2} \, \qquad \vphi_1 \to \vphi_3 \, ,\label{change-coord-pp}
\end{equation}
together with $\cJ_3 \to \cJ_1$.
\\

We can make the following comment: as we are fibering the three $\mathrm{U(1)}$ isometries $\varphi_i$ inside the $\mathbb{S}^5$ in an identical way, these directions are indistinguishable from each other which explains the same behaviour found for the three geodesic motions studied above (cases A and B respectively). Even though the embedding coordinates $\mu_i$ are different with respect to each other, one can chose any one of the expressions to correspond to $\mu_1$ etc. In other words, there is a freedom to exchange $\mu_i$s among themselves. We conclude that the pp-wave analysis showcases an isotropy between the three $\mathrm{U}(1)$ coordinates, encoded in the coordinate change \eqref{change-coord-pp}. 


\section{Conclusions and discussion}

In this work, we studied in detail the solutions presented in \cite{Chatzis:2025dnu} which describe supersymmetric RG flows between various $\mathrm{UV}$ $\mathrm{SCFT}s$ in 4d and $(2+1)$ dimensional $\mathrm{SQFT}s$ in the $\mathrm{IR}$. The construction of these systems was done by implementing SUSY preserving Coulomb branch deformations, realized as a twisted compactification in the dual supergravity backgrounds. More specifically,

\begin{itemize}
    \item{} We have constructed smooth supersymmetric asymptotically AdS solutions in type~II and eleven-dimensional supergravity, realizing holographic RG flows from four-dimensional SCFTs to three-dimensional SQFTs which have a mass gap and in some cases, confine external quarks. We extend and complement the results of \cite{Chatzis:2025dnu}, showing that distinct UV fixed points exhibit universal IR behaviour.

    \item{} For the uplifts we used the five-dimensional gauged supergravity soliton of \cite{Anabalon:2024che} to construct new infinite families of solutions in type~IIA, type~IIB, and M-theory. In all cases, the geometries are smooth and free of conical singularities for appropriate parameter ranges, ensuring physical regularity of the dual theories.

    \item{} We computed a variety of holographic observables not included in \cite{Chatzis:2025dnu}, namely new embeddings for the Wilson loops, 't~Hooft loops and entanglement entropy. These quantities display a universal factorization property: the dependence on the internal space is separated from the dynamics along the holographic radial direction. This universality points towards a deeper geometric structure underlying the supersymmetric confinement mechanism. The new results showcasing phase transitions in the 't Hooft loops and entanglement entropy, complementing the Wilson loop calculation performed in \cite{Chatzis:2025dnu}, further hint towards a confining behaviour for the dual. That confining behaviour is also reflected in the form of the D7-brane embeddings that avoid the central region of the geometry. The plot of the condensate with respect to the quark mass shows an interesting non monotonic behaviour, where the condensate flows to zero both for large and small quark mass.

    \item{} The holographic renormalization of the type~IIB background revealed the operator VEVs driving the flow, offering a clear dictionary between the bulk deformation parameters and boundary field-theory operators. The boundary analysis confirmed the consistency of the UV and IR asymptotics and provided finite counterterms for the renormalized action.

\item{} We investigated in detail the stability of the Wilson loop embedding used to study the type IIB background, under linear fluctuations of the coordinates. We found that the configuration is stable, however, when $\hat{\nu}\approx-1^{+}$ (or $Q\ll1$) we see tachyonic modes appearing, further supporting our claims about this parametric region being untrustworthy, as discussed in \cite{Chatzis:2025dnu}.

    \item{} The study of Penrose limits for the Anabalon--Ross deformed AdS$_5 \times S^5$ solution offers insight into the spectrum of excitations and possible integrable subsectors in the dual theory, paving the way for future studies of string dynamics on these backgrounds.
    \end{itemize}

From a broader perspective, our results provide a unified and systematic framework for studying supersymmetric compactifications of SCFT$_4$s on a circle that flow to confining three-dimensional theories. Is is important to emphasize that the dual $\mathrm{QFT}s$ described by these systems are all strongly coupled. Our deformation procedure should be applicable to any four-dimensional superconformal theory with a known holographic dual, as long as the later admits a consistent truncation to five-dimensional gauged supergravity. The interplay between topological twisting, smooth IR geometries, and holographic universality could shed light on nonperturbative dynamics beyond the present models.\\

    This work opens many interesting research directions for future exploration, some of which include:

    \begin{itemize}

        \item Extension of the holographic renormalization analysis to the 11d and type~IIA uplifts, to extract the corresponding VEVs and compare the operator maps across dimensions.
        \item Investigation of the stability of the (non-SUSY) soliton configurations under perturbations and exploration of possible phase transitions associated with varying the compactification radius or the holonomy parameters.
        \item Study the supersymmetric defect and domain-wall configurations supported by these geometries, which could model interfaces between different confining phases.
        \item Analysis of the Penrose limits and pp-wave sectors in more detail to identify potential integrable subsectors and compute semiclassical string spectra.
        \item We should explore the inclusion of flavour branes and their impact on the universal structure of observables, particularly in the 11d embeddings with M5 flavour branes.
        \item{It would be interesting if more elaborated solitons than the ones studied in \cite{Anabalon:2024che} could cure the singularities of the flavoured backgrounds in \cite{Casero:2006pt}, along the lines discussed in \cite{Macpherson:2025pqi}.} 
    \end{itemize}

  Finally, we wish to examine the effects of higher-curvature corrections near the large-curvature region ($\hat{\nu} \approx -1$) to assess their impact on the holographic predictions for IR observables. This will potentially provide information about the light modes appearing near the Coulomb branch VEV, which were not accounted for in the supergravity solutions presented here. There are two possible corrections one can include: corrections to the metric and corrections to the observable itself (in the case of the Wilson loop, which is able to detect the high curvature effects). We believe that if the needed corrections in next-to-leading order in $\alpha^{\prime}$ are found and included in the calculation, the Wilson loop phase transition observed near the region where $\hat{\nu}=-1$ will potentially disappear. This will be a fascinating phenomenon to observe. We hope to address these interesting points and many others in future work.

\section*{Acknowledgments}
The authors would like to thank the following colleagues for the useful discussions, their interesting comments and for sharing their knowledge and ideas with us: Andres Anabal\'on, Alexandre Mathieu Frederic Belin, Francesco Bigazzi, Nicol\'o Bragagnolo, Federico Castellani, Aldo Lorenzo Cotrone, Anton Faedo, Ali Fatemiabhari, Prem Kumar, Yolanda Lozano, Noppadol Mekareeya, Ren\'e Meyer, Alfonso Ramallo, Ricardo Stuardo, Daniel C. Thompson,  Alessandro Tomasiello. The research of D.C. has been supported by the STFC consolidated grand ST/Y509644-1. D.C. would also like to thank the universities of Santiago de Compostela, Oviedo, as well as the Galileo Galilei Institute for theoretical physics in Florence, Milano Bicocca and INFN for their hospitality while in the last stages of this work.  M.H. has been supported by the STFC consolidated grant ST/Y509644/1. M.H. would like to thank Humboldt-Universit\"{a}t zu Berlin during the KMPB school. G. I. is supported by the Einstein Stiftung Berlin via the Einstein International Postdoctoral Fellowship program “Generalised dualities and their holographic applications to condensed matter physics” (project number IPF- 2020-604). C. N. is supported by STFC’s grants ST/Y509644-1, ST/X000648/1 and ST/T000813/1.
This paper has been financed by the funding programme ``MEDICUS", 
of the University of Patras (D.Z. with grant number: 83800).


\appendix

\section{Polyakov loop embedding}\label{Appendix_Polyakov_loop}
Here we explore two cases of embeddings of $\mathrm{D}1$ branes in the type IIB solution of section \ref{IIB_background}, which will \textit{not} give the same dynamics as the 't Hooft loop. The reason being that they are not extended enough to include time, a profile in $r(w)$, as well as the shrinking circle $\phi$. The inclusion of the later in the various probes is responsible for the vanishing of the tension of the effective string at $r=r_\star$, as it conspires to the appearance of a multiplicative factor of $\sqrt{F(r)}$ in $\mathcal{F}_t$ through the determinant. This suggests to us that the minimal probe brane which is capable of capturing the 't Hooft loop in this background is a $\mathrm{D}3$.

\textbf{Case I:}
We first consider a $\mathrm{D}1$ on $\Sigma_2=[t,w]$ with $r=r(w)$ and all the other coordinates set to constant values. We have the induced metric on the $\mathrm{D}1$:
\begin{equation}
    \mathrm{d}s^2_{\mathrm{ind, D1}} = \frac{\zeta(r,\theta)}{L^2}\left[ -r^2 \mathrm{d}t^2 + \mathrm{d}w^2\left(r^2 + \frac{L^2 r^{\prime 2}}{F(r) \lambda(r)^6r^2} \right)  \right],
\end{equation}
and its action which reads:
\begin{equation}
\begin{split}
&      S_{\mathrm{D}1}=T_{\mathrm{D}1}\int_{\Sigma_2}\mathrm{d}^2\sigma \sqrt{-e^{-2\Phi}\mathrm{det}(g_{\mathrm{ind},\mathrm{D}1})} = T_{\mathrm{D}1}\mathcal{T} \int\mathrm{d}w \,\sqrt{\mathcal{F}^2_{\mathrm{D}1}+\mathcal{G}_{\mathrm{D}1}^2r^{\prime 2}},\\
&\mathcal{F}_{\mathrm{D}1}=\frac{\zeta(r,\theta_0)r^2}{L^2},\quad \mathcal{G}_{\mathrm{D}1}=\frac{\zeta(r,\theta_0)}{L\sqrt{F(r)}\lambda^3(r)}.
\end{split}
\end{equation}

We notice that one does not get the same functions as is the case for the rest of the probes, furthermore, the function $\zeta(r,\theta_0)$ is strictly nonzero for all values of $r$ and $\theta_0$ from which it follows that the object we are calculating does not have a vanishing tension at the end of the space.\\

\textbf{Case II:}
Another interesting embedding one can calculate is that of a Eucledian $\mathrm{D}1$ that extends on $\Sigma_2=[w,\phi]$, is localized in time and the rest of the coordinates are constant. The induced metric in this case is:
\begin{equation}
\begin{split}
    \mathrm{d}s^2_{\mathrm{ind,ED1}} &= \frac{1}{\zeta(r,\theta_0)}\left[ \cos^2\theta_0A_1^2+\sin^2\theta_0\lambda^6(r)A_3^2 +r^2F(r)\zeta^2(r,\theta_0)\right]\mathrm{d}\phi^2\\
    &+\zeta(r,\theta_0)\left[ \frac{r^2}{L^2}+\frac{r^{\prime 2}}{r^2F(r)\lambda^6(r)}\right]\mathrm{d}w^2,
\end{split}
\end{equation}
and its action is given by:
\begin{equation}
\begin{split}
&S_{\mathrm{ED}1} =T_{\mathrm{ED}1} \int_{\Sigma_2}\mathrm{d}^2\sigma \sqrt{e^{-2\Phi}\mathrm{det}(g_{\mathrm{ind},\mathrm{ED}1})}=\mathrm{T}_{\mathrm{ED}1}L_{\phi}\int\mathrm{d}w\sqrt{\mathcal{F}^2_{\mathrm{ED}1}+\mathcal{G}^2_{\mathrm{ED}1}r^{\prime 2}},\\
&\mathcal{F}^2_{\mathrm{ED}1}=\frac{r^2}{L^2}\left[ \cos^2\theta_0A_1^2+\sin^2\theta_0\lambda^6(r)A_3^2 + r^2F(r)\zeta^2(r,\theta_0)\right],\\
&\mathcal{G}^2_{\mathrm{ED}1}=\frac{\cos^2\theta_0 A_1^2+\sin^2\theta_0\lambda^6(r)A_3^2+r^2F(r)\zeta^2(r,\theta_0)}{r^2F(r)\lambda^6(r)}.
\end{split}
\end{equation}

Again, the functions appearing in the dynamical part of this observable disagree with the ones in the 't Hooft loop.

\section{Gauge coupling}\label{appendix_gauge_coupling}

We calculate the gauge coupling for the effective (2+1)-dimensional theories dual to the type IIB and IIA backgrounds. The procedure is as written in \cite{nunez2023,Chatzis:2024kdu}. We begin by writing the DBI action for a probe Dp-brane and perform an expansion of the field strength
\begin{equation}
    \mathrm{S}_{\mathrm{D}p,\mathrm{DBI}}= T_{\mathrm{D}p}\int_{\Sigma_{p+1}} \mathrm{d}^{p+1}\hat\sigma\sqrt{-e^{2\Phi}\mathrm{det}(g_{\text{ind}}+F)}.
\end{equation}
From the following we can read off the gauge coupling,
\begin{equation}
    S_{\mathrm{D}_p, \mathrm{DBI}}=T_{\mathrm{D}_p} \int_{\Sigma_{p+1}} \mathrm{d}^{p+1} \sigma \sqrt{-e^{-2 \Phi} \operatorname{det}(g_{\mathrm{ind}})}\left(1+\frac{1}{4} F_{\mu \nu} F^{\mu \nu}+\mathcal{O}\left(F^4\right)\right).
\end{equation}
\vspace{10pt}

{\bf Gauge coupling for the Type IIB background}\\

We consider a probe $\mathrm{D}3$ that extends in $[t,w,z,\phi]$, with the shrinking circle $\phi$ being wrapped. The worldvolume field strength is turned on which we take without loss of generality to have only the $F_{tw}$ component be non-zero. The rest of the coordinates are kept fixed. The induced metric on the $\mathrm{D}3$ is 
\begin{equation}
    \begin{split}
    \mathrm{d}s^2_{\text{ind}}&=\frac{r^2\zeta(r,\theta_0)}{L^2}(-\mathrm{d}t^2+\mathrm{d}w^2+\mathrm{d}z^2)\\
        &+ \left\{ r^2F(r)\zeta(r,\theta_0)+\frac{1}{\zeta(r,\theta_0)}\left[\cos^2\theta _0A_1^2 + \sin^2\theta_0 \lambda^6(r)A_3^2\right]\right\}\mathrm{d}\phi^2.
    \end{split}
\end{equation}

We will now compute the DBI action for the $\mathrm{D}3$ over the manifold $\Sigma_4$ spanned by $[t,w,z,\phi]$, which takes the form\footnote{Notice that the dilaton factor in this case is trivial}
\begin{equation}
    \begin{split}
      \mathrm{S}_{\mathrm{D}3} &= T_{\mathrm{D}3}\int_{\Sigma_4} \mathrm{d}^4\hat\sigma\sqrt{-e^{2\Phi}\mathrm{det}(g_{\text{ind}}+F)} =T_{\mathrm{D}3}\int_{\Sigma_4}\mathrm{d}^4\hat\sigma \sqrt{\alpha - \beta F_{tw}^2},
    \end{split}
\end{equation}
with the expressions:
\begin{equation}
    \begin{split}
        &\alpha:=r^4\zeta^2(r,\theta_0)\left[\cos^2\theta_0 A_1^2 + r^2F(r)\zeta^2(r,\theta_0)+\sin^2\theta\lambda^6(r)A_3^2\right],\\
       & \beta:=L^4\left[\cos^2\theta_0 A_1^2+r^2F(r)\zeta^2(r,\theta)+\sin^2\theta_0\lambda^6(r)A_3^2\right].
    \end{split}
\end{equation}
We can now expand for small values of the field strength ($|F_{tw}|\ll1$) up to $\mathcal{O}(F^2)$ to get
\begin{equation}
\mathrm{S}_{\mathrm{D}3}\approx T_{\mathrm{D}3}\int_{\Sigma_4}\mathrm{d}^4\sigma \frac{2r^4\zeta^2(r,\theta)-L^4F_{tw}^2}{2L^3r\zeta(r,\theta)}\sqrt{\cos^2\theta _0A_1^2+r^2F(r)\zeta^2(r,\theta_0)+\sin^2\theta_0\lambda^6(r)A_3^2},
\end{equation}

By expanding the DBI action to quadratic order in $F_{tw}$ and extracting the prefactor of the kinetic term, one obtains an expression for the gauge coupling. Its dependence on the radial coordinate $r$ reflects how the effective coupling evolves across energy scales in the dual QFT. Hence we can study the gauge coupling's behaviour in strongly coupled regimes that are otherwise difficult to access.
\begin{equation}
\begin{split}
    \frac{1}{g_{\mathrm{YM}}^2}&=\frac{T_{\mathrm{D}3}L L_{\phi}}{r\zeta(r,\theta_0)}\sqrt{\cos^2\theta _0A_1^2+r^2F(r)\zeta^2(r,\theta_0)+\sin^2\theta_0\lambda^6(r)A_3^2}\\
    &=\frac{2T_{\mathrm{D}3}LL_\phi}{\xi \sqrt{\nu+2\xi^2+\nu\cos(2\theta_0)}}\sqrt{\frac{(\xi^2-1)\left[4+5\nu+2\nu^2+2(1+\nu)\xi^2+\nu(3+2\nu)\cos(2\theta_0)\right]}{1+\nu}},
    \end{split}
\end{equation}
where in the last line we expressed everything in terms of the dimensionless variable $\xi=r/r_{\star}$ for the case where $q_1=q_2=q$.

 \begin{figure}[H]
    \centering
    \begin{subfigure}{0.44\linewidth}
    \includegraphics[width=\linewidth]{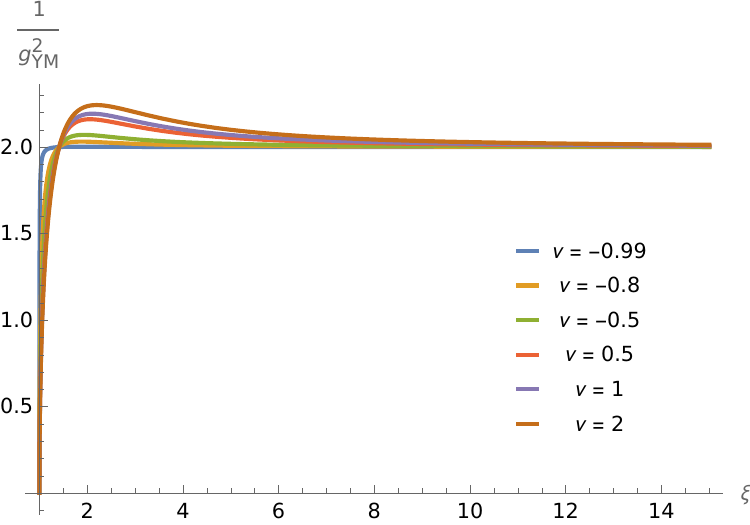}
    \caption{Plot of the square of the inverse gauge coupling for $\theta_0=0$.}
    \end{subfigure}
    \hfill
    \begin{subfigure}{0.44\linewidth}
    \includegraphics[width=\linewidth]{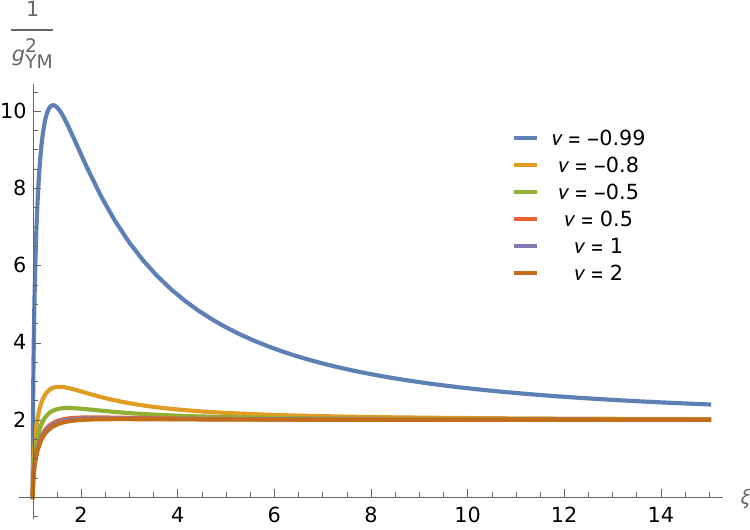}
    \caption{Plot of the square of the inverse gauge coupling for $\theta_0=\pi/2$.}
    \end{subfigure}
    \caption{Plots of the gauge coupling constant of the dual QFT with respect to the radial coordinate for the type $\mathrm{IIB}$ background.}
    \label{gauge_coupling_plots}
    \end{figure}

We can plot the above function for given values of $\theta_0$. The result can be seen in Figure \ref{gauge_coupling_plots}. We see that as $r\to r_{\star}$ the gauge coupling constant diverges, reflecting the fact that the dual QFTs are strongly coupled in the deep IR. In the UV the coupling constant takes a fixed value, which is a consequence of the conformal symmetry that is present as $r\to\infty$.\\

{\bf Gauge Coupling for Gaiotto-Maldacena Background} \\

We consider a probe D4 extending along $\Sigma_5 = [t, w, z, \phi, \eta]$ between $\eta_i, \eta_{i+1}$ and has a field strength which we take to have only the $F_{tz}$ component being non-zero. The other coordinates are kept fixed. 

The induced metric is 
\begin{equation}
    \mathrm{d}s^2_{\mathrm{ind}}=(\tilde{f}_1^3 \tilde{f}_5)^{\frac{1}{2}} \left\{ 4\tilde{f} \left[ \frac{r^2 \lambda(r)^2}{L^2}\left( -\mathrm{d}t^2 + \mathrm{d}w^2 + \mathrm{d}z^2 + L^2 F(r)\mathrm{d}\phi^2 \right)\right] + \tilde{f}_4\mathrm{d}\eta^2 \right\}.
\end{equation}
We utilise the formula as given in \cite{nunez2023,Chatzis:2024kdu},
\begin{equation}
    S_{\mathrm{D4}} = T_{\mathrm{D4}} \int_{\Sigma_5} \mathrm{d}^5 \hat{\sigma} \sqrt{-e^{-2\Phi}\operatorname{det}(g_{\mathrm{ind}}+F)}.
\end{equation}
The expansion gives
\begin{equation}
    \sqrt{-e^{-2\Phi}\operatorname{det}(g_{\mathrm{ind}}+F)}  = 32 \tilde{f}_1^3 \tilde{f}_4^{\frac{1}{2}} \tilde{f}_5^{\frac{1}{2}} \tilde{f}^2 \frac{r^4 \lambda(r)^4 \sqrt{F(r)}}{L^3} \left[ 1 - \frac{L^4 F_{tz}^2}{32 \tilde{f}^2 \tilde{f}_1^3\tilde{f}_5 \; r^4 \lambda(r)^4} \right].
\end{equation}
Hence we have
\begin{equation}
    S_{\mathrm{D4}} = 32 \; T_{\mathrm{D4}} L_{\phi} \int_{\eta_i}^{\eta_{i+1}} \mathrm{d}\eta \int \mathrm{d}t\mathrm{d}w\mathrm{d}z \tilde{f}_1^3 \tilde{f}_4^{\frac{1}{2}} \tilde{f}_5^{\frac{1}{2}} \tilde{f}^2 \frac{r^4 \lambda(r)^4 \sqrt{F(r)}}{L^3} \left[ 1 - \frac{L^4 F_{tz}^2}{32 \tilde{f}^2 \tilde{f}_1^3\tilde{f}_5 \; r^4 \lambda(r)^4} \right].
\end{equation}
We find the gauge coupling as
\begin{equation}
    \frac{1}{4g^2_{\mathrm{YM}}} = T_{\mathrm{D4}} L_{\phi} L \sqrt{F(r)}  \int_{\eta_i}^{\eta_{i+1}} \mathrm{d}\eta \; \tilde{f}_5^{-\frac{1}{2}}\tilde{f}_4^{\frac{1}{2}}.
\end{equation}


\section{Details of the computations}
\label{Details_computations}

In this Appendix, we have gathered lengthy expressions and useful details of the calculations that are discussed in the main text.

\no The Hodge dual of $G_5$ can also be written in terms of a four-form $\tilde{C}_4$ as $\star G_5 = \mathrm{d}\tilde{C}_4$, where
\begin{equation}
 \begin{aligned}
  \tilde{C}_4 & = \frac{L^4 q_1}{2} \big( \lambda^6 - 1 \big) \sin(2 \theta) \, \mathrm{d}\theta\wedge \mathrm{d}\phi \wedge \big( \sin^2\psi \, \mathrm{d}\phi_1 + \cos^2 \psi \, \mathrm{d}\phi_2 \big) \wedge \mathrm{d}\phi_3
  \\[5pt]
  & + \frac{L^4 q_2}{2} \frac{\big( \lambda^6 - 1 \big) \zeta^2_\star}{\lambda^6_\star \zeta^2} \cos^4\theta\sin(2 \psi) \, \mathrm{d}\phi \wedge \mathrm{d}\phi_1 \wedge \mathrm{d}\phi_2 \wedge \mathrm{d}\psi
  \\[5pt]
  & + \frac{L^4 q_1}{8} \frac{\big( \lambda^6 - 1 \big) \zeta^2_\star}{\zeta^2} \sin^2(2 \theta) \sin(2 \psi) \, \mathrm{d}\phi \wedge \mathrm{d}(\phi_1 - \phi_2) \wedge \mathrm{d}\phi_3 \wedge \mathrm{d}\psi
  \\[5pt]
  & - \frac{L^4}{8} \frac{\zeta^2 - 1}{\zeta^2} \sin^2(2 \theta) \sin(2 \psi) \, \mathrm{d}\phi_1 \wedge \mathrm{d}\phi_2 \wedge \mathrm{d}\phi_3 \wedge \mathrm{d}\psi
  \\[5pt]
  & - \frac{L^4 q_2}{\lambda^6_\star} \big( \zeta^2_\star - 1 \big) \sin(2 \theta) \cos^2\psi \, \mathrm{d}\theta\wedge \mathrm{d}\phi \wedge \mathrm{d}\phi_1 \wedge \mathrm{d}\phi_2
  \\[5pt]
  & + L^4 q_1 \big( \zeta^2_\star - 1 \big) \sin(2 \theta) \cos^2\psi \, \mathrm{d}\theta\wedge \mathrm{d}\phi \wedge \mathrm{d}(\phi_1 - \phi_2) \wedge \mathrm{d}\phi_3
  \\[5pt]
  & - L^4 \sin(2 \theta) \cos^2\theta\cos^2\psi \, \mathrm{d}\theta\wedge \mathrm{d}\phi_1 \wedge \mathrm{d}\phi_2 \wedge \mathrm{d}\phi_3 \, .
  \label{S5-2}
 \end{aligned}
\end{equation}

\no The expressions for the length and the energy in the case 
of the 't Hooft loops read:
\begin{equation}\label{length_integral_thooft_IIB}
    \begin{split}
        L_{\mathrm{MM}}(\xi_0)=\frac{\sqrt{2}L^2}{r_\star}\sqrt{-(1+\hat\nu)+\xi_0^4(\xi_0^2+\hat\nu)}\int_{\xi_0}^\infty \mathrm{d}\xi \sqrt{\frac{\xi}{\left[-(1+\hat\nu)+\xi^4(\xi^2+\hat\nu)\right]\left[\xi^6-\xi_0^6+\hat\nu(\xi^4-\xi_0^4)\right]}},
    \end{split}
\end{equation}

\begin{equation} \label{energy_integral_thooft_IIB}
    \begin{split}
        E_{\mathrm{MM}}(\xi_0)&=\frac{r_\star^3}{L}\sqrt{-(1+\hat\nu)+\xi_0^4(\xi_0^2+\hat\nu)}L_{\mathrm{MM}(\xi_0)}
        \\[5pt] & +2Lr_\star^2\int_{\xi_0}^\infty\frac{\mathrm{d}\xi \, \xi\sqrt{\left[\xi^6-\xi_0^6+\hat\nu(\xi^4-\xi_0^4)\right]}}{\sqrt{-(1+\hat\nu)+\xi_0^4(\xi_0^2+\hat\nu)}}-2Lr_\star^2\int_{1}^\infty \mathrm{d}\xi \, \xi,
    \end{split}
\end{equation}
\begin{equation}\label{approximate_expressions_thooft_IIB}
    \begin{split}
        &L_{\mathrm{MM,app}}(\xi_0)=\frac{L^2\pi\sqrt{-(1+\hat\nu)+\xi_0^4(\xi_0^2+\hat\nu)}}{r_\star^2\xi^2(2\hat\nu+3\xi_0)},\\[5pt]
       & E_{\mathrm{MM,app}}(\xi_0)=Lr_\star\pi\int^{\xi_0}\mathrm{d}s\, \frac{3\hat\nu(1+\hat\nu)+12(1+\hat\nu)s^2-4\hat\nu s^6-3s^8}{s^3(2\hat\nu+3s^3)^2},
    \end{split}
\end{equation}

\no Now moving to the calculations for the stability analysis of the Wilson loop, to write the equations of the fluctuations in a compact form, we introduce the following functions
\begin{eqnarray} \label{auxiliary_functions}
&& f_z(r,\theta) = - \, G_{tt} G_{zz} \, , \quad  
f_\theta(r,\theta) = - \, G_{tt} G_{\theta \theta} \, , \quad
f_\phi(r,\theta) = - \, G_{tt} G_{\phi \phi} 
\nonumber \\[5pt]
&& 
f_{\varphi_1}(r,\theta,\psi) = - \, G_{tt} G_{\varphi_1 \varphi_1} \, , \quad 
f_{\varphi_2}(r,\theta,\psi) = - \, G_{tt} G_{\varphi_2 \varphi_2} 
\\[5pt]
&& 
f_{\psi}(r,\theta) = - \, G_{tt} G_{\psi \psi} \, , \quad
f_{\varphi_3}(r,\theta) = - \, G_{tt} G_{\varphi_3 \varphi_3}  \, , \quad
h(r,\theta) = G_{rr} G_{yy} \, . 
\nonumber 
\end{eqnarray}

\no Introducing a time dependence in the fluctuations of the form $ e^{-{\rm i} \, \omega \, \tau}$ , we have 
\begin{eqnarray} \label{fluctuation_time_dependence}
&& \delta z (\tau,\sigma) = \delta z (\sigma) \, e^{-{\rm i} \, \omega \, \tau} \, , 
\quad 
\delta y (\tau,\sigma) = \delta y (\sigma) \, e^{-{\rm i} \, \omega \, \tau} \, , 
\quad 
\delta \theta (\tau,\sigma) = \delta \theta (\sigma) \, e^{-{\rm i} \, \omega \, \tau} 
\nonumber \\[5pt]
&& 
\delta \phi (\tau,\sigma) = \delta \phi (\sigma) \, e^{-{\rm i} \, \omega \, \tau} \, , 
\quad
\delta \varphi_1 (\tau,\sigma) =\delta \varphi_2 (\tau,\sigma) = \delta  \varphi_{12} (\sigma) \, e^{-{\rm i} \, \omega \, \tau}
\\[5pt]
&&
\delta \varphi_3 (\tau,\sigma) = \delta  \varphi_3 (\sigma) \, e^{-{\rm i} \, \omega \, \tau} \, ,  
\quad
\delta \psi (\tau,\sigma) = \delta \psi (\sigma) \, e^{-{\rm i} \, \omega \, \tau} \, . 
\nonumber
\end{eqnarray}
Notice that we have used the same function for the $\sigma$ dependence of the modes of $\delta \varphi_1$ and $\delta \varphi_2$, 
something that is consistent with the equations of motion for the corresponding fluctuations. 

\no The equations for the fluctuations $\delta z$, $\d \psi$, $\delta y$ and $\delta \theta $ are
\begin{eqnarray} \label{eom-decoupled}
&&\Bigg[ {\mathrm{d} \over \mathrm{d}\sigma} \left({f_z \over  F^{1/2}} {\mathrm{d} \over \mathrm{d}\sigma} \right)
+ \omega^2 \, {h \, f_z   \, F^{1/2} \over g \, f_y} \Bigg] \delta z = 0 
\\
&&
\Bigg[ {\mathrm{d} \over \mathrm{d}\sigma} \left({f_{\psi} \over  F^{1/2}} {\mathrm{d} \over \mathrm{d}\sigma} \right)
+ \omega^2 \, {h f_{\psi} \,  F^{1/2} \over g \, f_y} \Bigg] \delta \psi = 0  \, , \quad 
\Bigg[ {\mathrm{d} \over \mathrm{d}\sigma } \left( {g \, f_y \over F^{3/2}} {\mathrm{d} \over \mathrm{d}\sigma} \right)
+ \omega^2 \, {h \over F^{1/2}} \Bigg] \d y = 0
\nonumber \\
&&\Bigg[ {\mathrm{d} \over \mathrm{d}u} \left({f_\theta \over F^{1/2}} {\mathrm{d} \over \mathrm{d}\sigma} \right)
 +  \omega^2 \, {h f_\theta F^{1/2} \over g \, f_y}
- {1 \over 2 \, F^{1/2}} \, \partial_\theta^2 g - 
{f^0_{y} \, F^{1/2} \over 2 \, f_y^2} \, \partial_\theta^2 f_y  \Bigg] \delta \theta = 0
\nonumber
\end{eqnarray}
while the equations for the modes $\delta \phi $, $\delta \varphi_{12}$ and $\delta \varphi_{3}$ are
\begin{eqnarray}  \label{eom-coupled_1}
&& {\mathrm{d} \over \mathrm{d}\sigma} \left[{f_{\phi} \over  F^{1/2}} \,  \delta \phi' + f_{12} \, {f_{\varphi_1}+f_{\varphi_2} \over  F^{1/2}} \,  \delta \varphi_{12}' 
+ f_{3} \, {f_{\varphi_3} \over  F^{1/2}} \,  \delta \varphi_{3}'  \right]
\nonumber \\ 
&& 
+ \, \omega^2 \, {h  \, F^{1/2} \over g \, f_y}  \Bigg[ f_{\phi} \,  \delta \phi  + f_{12} \, \left(f_{\varphi_1}+f_{\varphi_2} \right) \,  \delta \varphi_{12}
+ f_{3} \, f_{\varphi_3}\,  \delta \varphi_{3} \Bigg]= 0 
\end{eqnarray}
\begin{eqnarray}  \label{eom-coupled_23}
&& {\mathrm{d} \over \mathrm{d}\sigma} \left[ {f_{\varphi_1}+f_{\varphi_2} \over  F^{1/2}} \,  \left(\delta \varphi_{12}' + f_{12} \, \delta \phi' \right) \right]
+ \, \omega^2 \, {h  \, F^{1/2} \over g \, f_y}\,  \left(f_{\varphi_1}+f_{\varphi_2} \right)   \Bigg[ \delta \varphi_{12} +  f_{12} \, \delta \phi \Bigg]= 0 
\nonumber \\ 
&& 
 {\mathrm{d} \over \mathrm{d}\sigma} \left[ {f_{\varphi_3} \over  F^{1/2}} \,  \left(\delta \varphi_{3}' + f_{3} \, \delta \phi' \right) \right]
+ \, \omega^2 \, {h  \, F^{1/2} \over g \, f_y}\, f_{\varphi_3}  \Bigg[ \delta \varphi_{3} +  f_{3} \, \delta \phi \Bigg]= 0 \, .
\end{eqnarray}

 \no The expressions for the Schr\"{o}dinger potentials in each fluctuation mode, in the $Q=0$ case, are 
\begin{eqnarray} \label{Schr_potentials_Q_zero}
V_{\delta z}(\sigma, r_0) & = & V_{\delta \phi}(\sigma, r_0)  = \frac{\left[2 r_0^2 \left(r_0^2-1\right)-1\right] \sigma ^2+r_0^2 \left(r_0^2-1\right)+8
   \sigma ^8-18 \sigma ^6+11 \sigma ^4}{4 \sigma ^2 \left(\sigma ^2-1\right)^2}
 \nonumber \\
V_{\delta y}(\sigma, r_0) & = &\frac{-\left[8 r_0^2 \left(r_0^2-1\right)-11\right] \sigma ^4
+\left[6 r_0^2 \left(r_0^2-1\right)-1\right] \sigma ^2
-3 r_0^2 \left(r_0^2-1\right)+8 \sigma ^8-18 \sigma ^6}{4 \sigma ^2 \left(\sigma ^2-1\right)^2}
 \nonumber \\
V_{\delta \theta}(\sigma, r_0) & = &\frac{\left[6 r_0^2 \left(r_0^2-1\right)-1\right] \sigma ^2-3 r_0^2 \left(r_0^2-1\right)+2
\sigma ^6-\sigma ^4}{4 \sigma ^2 \left(\sigma ^2-1\right)^2}
\nonumber \\
V_{\delta \psi}(\sigma, r_0) & = & V_{\delta \varphi_{12}}(\sigma, r_0) = \frac{-\left[ \left(r_0^2-1\right) r_0^2+1\right] \sigma ^2+r_0^2 \left(r_0^2-1\right)+
2 \sigma ^6-\sigma ^4}{4 \sigma ^2 \left(\sigma ^2-1\right)^2} \, . 
\end{eqnarray}
The expressions for the Schr\"{o}dinger potentials of the decoupled modes in the case of finite $Q$, when expanded around the $Q=0$ value, are
\begin{eqnarray} \label{Schr_potentials_Q_finite_perturbative}
V_{\delta z}& = & V^{Q=0} _{\delta z} + \frac{r_0^4 \left(-12 \sigma ^4+12 \sigma ^2-5\right)+r_0^2 \left(12 \sigma ^4-12
   \sigma ^2+5\right)+4 \sigma ^8-8 \sigma ^6+7 \sigma ^4-3 \sigma ^2}{4 \sigma ^6
   \left(\sigma ^2-1\right)^3} \, Q^2 + {\cal O}(Q^4)
 \nonumber \\
V_{\delta y}& = & V^{Q=0}_{\delta y} + \frac{r_0^4 \left(20 \sigma ^4-20 \sigma ^2+7\right)+r_0^2 \left(-20 \sigma ^4+20
   \sigma ^2-7\right)+4 \sigma ^8-8 \sigma ^6+7 \sigma ^4-3 \sigma ^2}{4 \sigma ^6
   \left(\sigma ^2-1\right)^3} \, Q^2 + {\cal O}(Q^4)
 \nonumber \\
V_{\delta \theta} & = &V^{Q=0}_{\delta \theta} + \frac{\left(-12 r_0^4+12 r_0^2+5\right) \sigma ^2+7 r_0^2 \left(r_0^2-1\right)+8
   \sigma ^6-13 \sigma ^4}{4 \sigma ^6 \left(\sigma ^2-1\right)^3}\, Q^2 + {\cal O}(Q^4)
\nonumber \\
V_{\delta \psi} & = &V^{Q=0}_{\delta \psi} + \frac{3 \left(4 r_0^4-4 r_0^2-1\right) \sigma ^2-5 r_0^2 \left(r_0^2-1\right)-8 \sigma
   ^6+11 \sigma ^4}{4 \sigma ^6 \left(\sigma ^2-1\right)^3} \, Q^2 + {\cal O}(Q^4) \, . 
\end{eqnarray}
From the transformation of the coupled system of  $\delta \phi$ and $\delta \varphi_{12}$ to a  Schr\"{o}dinger form, the expressions of the 
functions $ \Omega_i$, with $i=1, \cdots, 4$ are
\begin{eqnarray} \label{Omega1234}
&& \Omega_1 = \frac{1}{\sqrt{\sigma } \sqrt[4]{\sigma ^2-1}} + {\cal O} (Q^2) \,, \quad  
\Omega_3 =\frac{Q}{2 \, \sigma^{5/2} \sqrt[4]{\sigma ^2-1}} 
\Bigg[\sqrt{\sigma ^2-1}+\sigma ^2 \arctan \left(\sqrt{\sigma^2-1}\right)\Bigg]+ {\cal O} (Q^3)
\nonumber \\ 
&& \Omega_3 =Q\, \frac{\sqrt[4]{\sigma ^2-1}}{2 \, \sigma^{5/2 }} \Bigg[\sqrt{\sigma ^2-1}-\sigma ^2 \arctan \left(\sqrt{\sigma^2-1}\right)\Bigg]+ {\cal O} (Q^3) \,, \quad  
\Omega_4 =\frac{\sqrt[4]{\sigma ^2-1}}{\sqrt{\sigma}}+ {\cal O} (Q^2) \, . 
 \end{eqnarray}
 while the insertions of the matrix potential $V$ from \eqref{Matrix-Schrodinger} are
 \begin{eqnarray} \label{V_matrix}
&& V_{11} = V^{Q=0} _{\delta z} \, , \quad 
V_{12} = V_{21} =  \frac{Q}{2} \Bigg\{\frac{\left(2 r_0^4-2 r_0^2+2 \sigma ^6-5
   \sigma ^4+3 \sigma ^2\right) \arctan\left(\sqrt{\sigma
   ^2-1}\right)}{\left(\sigma ^2-1\right)^2}  \qquad  \qquad 
   \nonumber \\
   &&  \quad  + \,  \frac{r_0^4 \left(4-6 \sigma ^2\right)+r_0^2
   \left(6 \sigma ^2-4\right)+(\sigma -1) \sigma ^2 (\sigma +1)
   \left(2 \sigma ^4+\sigma ^2-2\right)}{\sigma ^4 \left(\sigma
   ^2-1\right)^{3/2}} \Bigg\}
   \, , \quad V_{22} = V^{Q=0} _{\delta \psi}\, . 
\end{eqnarray}

\bibliography{main.bib}

@article{Ryu_2006,
   title={Holographic Derivation of Entanglement Entropy from the anti–de Sitter Space/Conformal Field Theory Correspondence},
   volume={96},
   ISSN={1079-7114},
   url={http://dx.doi.org/10.1103/PhysRevLett.96.181602},
   DOI={10.1103/physrevlett.96.181602},
   number={18},
   journal={Physical Review Letters},
   publisher={American Physical Society (APS)},
   author={Ryu, Shinsei and Takayanagi, Tadashi},
   year={2006},
   month=may }

@article{Klebanov:2007ws,
    author = "Klebanov, Igor R. and Kutasov, David and Murugan, Arvind",
    title = "{Entanglement as a probe of confinement}",
    eprint = "0709.2140",
    archivePrefix = "arXiv",
    primaryClass = "hep-th",
    reportNumber = "PUPT-2241",
    doi = "10.1016/j.nuclphysb.2007.12.017",
    journal = "Nucl. Phys. B",
    volume = "796",
    pages = "274--293",
    year = "2008"
}

@misc{macpherson2024,
      title={Marginally deformed AdS$_5$/CFT$_4$ and spindle-like orbifolds}, 
      author={Niall T. Macpherson and Paul Merrikin and Carlos Nunez},
      year={2024},
      eprint={2403.02380},
      archivePrefix={arXiv},
      primaryClass={hep-th},
      url={https://arxiv.org/abs/2403.02380}, 
}

@article{Macpherson_2015,
   title={Type IIB supergravity solutions with AdS5 from Abelian and non-Abelian T dualities},
   volume={2015},
   ISSN={1029-8479},
   url={http://dx.doi.org/10.1007/JHEP02(2015)040},
   DOI={10.1007/jhep02(2015)040},
   number={2},
   journal={Journal of High Energy Physics},
   publisher={Springer Science and Business Media LLC},
   author={Macpherson, Niall T. and Núñez, Carlos and Zayas, Leopoldo A. Pando and Rodgers, Vincent G. J. and Whiting, Catherine A.},
   year={2015},
   month=feb }

@article{Bea_2015,
   title={Compactifications of the Klebanov-Witten CFT and new $AdS_3$ backgrounds},
   volume={2015},
   ISSN={1029-8479},
   url={http://dx.doi.org/10.1007/JHEP05(2015)062},
   DOI={10.1007/jhep05(2015)062},
   number={5},
   journal={Journal of High Energy Physics},
   publisher={Springer Science and Business Media LLC},
   author={Bea, Yago and Edelstein, José D. and Itsios, Georgios and Kooner, Karta S. and Nunez, Carlos and Schofield, Daniel and Sierra-García, J. Aníbal},
   year={2015},
   month=may }

@misc{merrikin2023compactification6dcaln10,
      title={Compactification of 6d $ \mathcal{N}=(1,0) $ quivers, 4d SCFTs and their holographic dual Massive IIA backgrounds}, 
      author={Paul Merrikin and Carlos Nunez and Ricardo Stuardo},
      year={2023},
      eprint={2210.02458},
      archivePrefix={arXiv},
      primaryClass={hep-th},
      url={https://arxiv.org/abs/2210.02458}, 
}

@misc{nunez2023,
      title={Confinement in $(1+1)$ dimensions: a holographic perspective from I-branes}, 
      author={Carlos Nunez and Marcelo Oyarzo and Ricardo Stuardo},
      year={2023},
      eprint={2307.04783},
      archivePrefix={arXiv},
      primaryClass={hep-th},
      url={https://arxiv.org/abs/2307.04783}, 
}

@article{Chatzis:2024kdu,
    author = "Chatzis, Dimitrios and Fatemiabhari, Ali and Nunez, Carlos and Weck, Peter",
    title = "{SCFT deformations via uplifted solitons}",
    eprint = "2406.01685",
    archivePrefix = "arXiv",
    primaryClass = "hep-th",
    doi = "10.1016/j.nuclphysb.2024.116659",
    journal = "Nucl. Phys. B",
    volume = "1006",
    pages = "116659",
    year = "2024"
}

@article{Cvetic:1999xp,
    author = "Cvetic, Mirjam and Duff, M. J. and Hoxha, P. and Liu, James T. and Lu, Hong and Lu, J. X. and Martinez-Acosta, R. and Pope, C. N. and Sati, H. and Tran, Tuan A.",
    title = "{Embedding AdS black holes in ten-dimensions and eleven-dimensions}",
    eprint = "hep-th/9903214",
    archivePrefix = "arXiv",
    reportNumber = "UPR-0840-T, CTP-TAMU-11-99, RU-99-4-B",
    doi = "10.1016/S0550-3213(99)00419-8",
    journal = "Nucl. Phys. B",
    volume = "558",
    pages = "96--126",
    year = "1999"
}

@article{Anabalon:2024che,
    author = "Anabal\'on, Andr\'es and Nastase, Horatiu and Oyarzo, Marcelo",
    title = "{Supersymmetric AdS solitons and the interconnection of different vacua of $ \mathcal{N} $ = 4 Super Yang-Mills}",
    eprint = "2402.18482",
    archivePrefix = "arXiv",
    primaryClass = "hep-th",
    doi = "10.1007/JHEP05(2024)217",
    journal = "JHEP",
    volume = "05",
    pages = "217",
    year = "2024"
}

@article{Freedman:1999gk,
    author = "Freedman, D. Z. and Gubser, S. S. and Pilch, K. and Warner, N. P.",
    title = "{Continuous distributions of D3-branes and gauged supergravity}",
    eprint = "hep-th/9906194",
    archivePrefix = "arXiv",
    reportNumber = "CERN-TH-99-189, HUTP-99-A029, MIT-CTP-2877, USC-99-03",
    doi = "10.1088/1126-6708/2000/07/038",
    journal = "JHEP",
    volume = "07",
    pages = "038",
    year = "2000"
}

@article{Romans:1985ps,
    author = "Romans, L. J.",
    title = "{Gauged $N=4$ Supergravities in Five-dimensions and Their Magnetovac Backgrounds}",
    reportNumber = "NSF-ITP-85-113",
    doi = "10.1016/0550-3213(86)90398-6",
    journal = "Nucl. Phys. B",
    volume = "267",
    pages = "433--447",
    year = "1986"
}

@article{Lin:2004nb,
    author = "Lin, Hai and Lunin, Oleg and Maldacena, Juan Martin",
    title = "{Bubbling AdS space and 1/2 BPS geometries}",
    eprint = "hep-th/0409174",
    archivePrefix = "arXiv",
    reportNumber = "PUPT-2136",
    doi = "10.1088/1126-6708/2004/10/025",
    journal = "JHEP",
    volume = "10",
    pages = "025",
    year = "2004"
}

@article{Gauntlett:2007sm,
    author = "Gauntlett, Jerome P. and Varela, Oscar",
    title = "{D=5 SU(2) x U(1) Gauged Supergravity from D=11 Supergravity}",
    eprint = "0712.3560",
    archivePrefix = "arXiv",
    primaryClass = "hep-th",
    reportNumber = "IMPERIAL-TP-2007-JG-04",
    doi = "10.1088/1126-6708/2008/02/083",
    journal = "JHEP",
    volume = "02",
    pages = "083",
    year = "2008"
}

@article{Gauntlett_2006,
   title={AdS spacetimes from wrapped M5 branes},
   volume={2006},
   ISSN={1029-8479},
   url={http://dx.doi.org/10.1088/1126-6708/2006/11/053},
   DOI={10.1088/1126-6708/2006/11/053},
   number={11},
   journal={Journal of High Energy Physics},
   publisher={Springer Science and Business Media LLC},
   author={Gauntlett, Jerome P and Conamhna, Oisín A.P. Mac and Mateos, Toni and Waldram, Daniel},
   year={2006},
   month=nov, pages={053–053} }

@article{Gaiotto:2009gz,
    author = "Gaiotto, Davide and Maldacena, Juan",
    title = "{The Gravity duals of N=2 superconformal field theories}",
    eprint = "0904.4466",
    archivePrefix = "arXiv",
    primaryClass = "hep-th",
    doi = "10.1007/JHEP10(2012)189",
    journal = "JHEP",
    volume = "10",
    pages = "189",
    year = "2012"
}

@article{Anabalon:2021tua,
    author = "Anabalon, Andres and Ross, Simon F.",
    title = "{Supersymmetric solitons and a degeneracy of solutions in AdS/CFT}",
    eprint = "2104.14572",
    archivePrefix = "arXiv",
    primaryClass = "hep-th",
    doi = "10.1007/JHEP07(2021)015",
    journal = "JHEP",
    volume = "07",
    pages = "015",
    year = "2021"
}

@article{Cassani:2019vcl,
    author = "Cassani, Davide and Josse, Gr\'egoire and Petrini, Michela and Waldram, Daniel",
    title = "{Systematics of consistent truncations from generalised geometry}",
    eprint = "1907.06730",
    archivePrefix = "arXiv",
    primaryClass = "hep-th",
    doi = "10.1007/JHEP11(2019)017",
    journal = "JHEP",
    volume = "11",
    pages = "017",
    year = "2019"
}

@article{Gauntlett:2007ma,
    author = "Gauntlett, Jerome P. and Varela, Oscar",
    title = "{Consistent Kaluza-Klein reductions for general supersymmetric AdS solutions}",
    eprint = "0707.2315",
    archivePrefix = "arXiv",
    primaryClass = "hep-th",
    doi = "10.1103/PhysRevD.76.126007",
    journal = "Phys. Rev. D",
    volume = "76",
    pages = "126007",
    year = "2007"
}

@article{Fatemiabhari:2024aua,
    author = "Fatemiabhari, Ali and Nunez, Carlos",
    title = "{From conformal to confining field theories using holography}",
    eprint = "2401.04158",
    archivePrefix = "arXiv",
    primaryClass = "hep-th",
    doi = "10.1007/JHEP03(2024)160",
    journal = "JHEP",
    volume = "03",
    pages = "160",
    year = "2024"
}

@article{Chatzis:2025dnu,
    author = "Chatzis, Dimitrios and Hammond, Madison and Itsios, Georgios and Nunez, Carlos and Zoakos, Dimitrios",
    title = "{Universal observables, SUSY RG-flows and holography}",
    eprint = "2506.10062",
    archivePrefix = "arXiv",
    primaryClass = "hep-th",
    reportNumber = "HU-EP-25/19",
    doi = "10.1007/JHEP08(2025)134",
    journal = "JHEP",
    volume = "08",
    pages = "134",
    year = "2025"
}

@article{Kehagias:1999iy,
    author = "Kehagias, A. and Sfetsos, K.",
    title = "{On asymptotic freedom and confinement from type IIB supergravity}",
    eprint = "hep-th/9903109",
    archivePrefix = "arXiv",
    reportNumber = "CERN-TH-99-63",
    doi = "10.1016/S0370-2693(99)00431-1",
    journal = "Phys. Lett. B",
    volume = "456",
    pages = "22--27",
    year = "1999"
}

@article{Girardello:1999hj,
    author = "Girardello, L. and Petrini, M. and Porrati, M. and Zaffaroni, A.",
    title = "{Confinement and condensates without fine tuning in supergravity duals of gauge theories}",
    eprint = "hep-th/9903026",
    archivePrefix = "arXiv",
    reportNumber = "CERN-TH-99-46, BICOCCA-FT-99-05, IMPERIAL-TP-98-99-42, NYU-TH-98-2-03",
    doi = "10.1088/1126-6708/1999/05/026",
    journal = "JHEP",
    volume = "05",
    pages = "026",
    year = "1999"
}

@article{Brandhuber:1999jr,
    author = "Brandhuber, A. and Sfetsos, K.",
    title = "{Wilson loops from multicenter and rotating branes, mass gaps and phase structure in gauge theories}",
    eprint = "hep-th/9906201",
    archivePrefix = "arXiv",
    reportNumber = "CERN-TH-99-191",
    doi = "10.4310/ATMP.1999.v3.n4.a4",
    journal = "Adv. Theor. Math. Phys.",
    volume = "3",
    pages = "851--887",
    year = "1999"
}

@article{Gubser:2000nd,
    author = "Gubser, Steven S.",
    title = "{Curvature singularities: The Good, the bad, and the naked}",
    eprint = "hep-th/0002160",
    archivePrefix = "arXiv",
    reportNumber = "PUPT-1916",
    doi = "10.4310/ATMP.2000.v4.n3.a6",
    journal = "Adv. Theor. Math. Phys.",
    volume = "4",
    pages = "679--745",
    year = "2000"
}

@article{Nunez:2009da,
    author = "Nunez, Carlos and Piai, Maurizio and Rago, Antonio",
    title = "{Wilson Loops in string duals of Walking and Flavored Systems}",
    eprint = "0909.0748",
    archivePrefix = "arXiv",
    primaryClass = "hep-th",
    doi = "10.1103/PhysRevD.81.086001",
    journal = "Phys. Rev. D",
    volume = "81",
    pages = "086001",
    year = "2010"
}

@article{Kol:2014nqa,
    author = "Kol, Uri and Nunez, Carlos and Schofield, Daniel and Sonnenschein, Jacob and Warschawski, Michael",
    title = "{Confinement, Phase Transitions and non-Locality in the Entanglement Entropy}",
    eprint = "1403.2721",
    archivePrefix = "arXiv",
    primaryClass = "hep-th",
    doi = "10.1007/JHEP06(2014)005",
    journal = "JHEP",
    volume = "06",
    pages = "005",
    year = "2014"
}

@article{Nunez:2025gxq,
    author = "Nunez, Carlos and Roychowdhury, Dibakar",
    title = "{Time-like Entanglement Entropy: a top-down approach}",
    eprint = "2505.20388",
    archivePrefix = "arXiv",
    primaryClass = "hep-th",
    month = "5",
    year = "2025"
}

@article{Nunez:2019gbg,
    author = "N\'u\~nez, Carlos and Roychowdhury, Dibakar and Speziali, Stefano and Zacar\'\i{}as, Salom\'on",
    title = "{Holographic aspects of four dimensional ${\cal N }=2$ SCFTs and their marginal deformations}",
    eprint = "1901.02888",
    archivePrefix = "arXiv",
    primaryClass = "hep-th",
    doi = "10.1016/j.nuclphysb.2019.114617",
    journal = "Nucl. Phys. B",
    volume = "943",
    pages = "114617",
    year = "2019"
}

@article{Macpherson:2024frt,
    author = "Macpherson, Niall T. and Merrikin, Paul and Nunez, Carlos",
    title = "{Marginally deformed AdS$_{5}$/CFT$_{4}$ and spindle-like orbifolds}",
    eprint = "2403.02380",
    archivePrefix = "arXiv",
    primaryClass = "hep-th",
    doi = "10.1007/JHEP07(2024)042",
    journal = "JHEP",
    volume = "07",
    pages = "042",
    year = "2024"
}

@article{Maldacena:1997re,
    author = "Maldacena, Juan Martin",
    title = "{The Large $N$ limit of superconformal field theories and supergravity}",
    eprint = "hep-th/9711200",
    archivePrefix = "arXiv",
    reportNumber = "HUTP-97-A097, HUTP-98-A097",
    doi = "10.4310/ATMP.1998.v2.n2.a1",
    journal = "Adv. Theor. Math. Phys.",
    volume = "2",
    pages = "231--252",
    year = "1998"
}

@article{Maldacena:1998im,
    author = "Maldacena, Juan Martin",
    title = "{Wilson loops in large N field theories}",
    eprint = "hep-th/9803002",
    archivePrefix = "arXiv",
    reportNumber = "HUTP-98-A014",
    doi = "10.1103/PhysRevLett.80.4859",
    journal = "Phys. Rev. Lett.",
    volume = "80",
    pages = "4859--4862",
    year = "1998"
}

@article{tHooft:1977nqb,
    author = "'t Hooft, Gerard",
    title = "{On the Phase Transition Towards Permanent Quark Confinement}",
    reportNumber = "Print-78-0099 (UTRECHT)",
    doi = "10.1016/0550-3213(78)90153-0",
    journal = "Nucl. Phys. B",
    volume = "138",
    pages = "1--25",
    year = "1978"
}

@article{Henningson:1998gx,
    author = "Henningson, M. and Skenderis, K.",
    title = "{The Holographic Weyl anomaly}",
    eprint = "hep-th/9806087",
    archivePrefix = "arXiv",
    reportNumber = "CERN-TH-98-188, KUL-TF-98-21",
    doi = "10.1088/1126-6708/1998/07/023",
    journal = "JHEP",
    volume = "07",
    pages = "023",
    year = "1998"
}

@article{deHaro:2000vlm,
    author = "de Haro, Sebastian and Solodukhin, Sergey N. and Skenderis, Kostas",
    title = "{Holographic reconstruction of space-time and renormalization in the AdS / CFT correspondence}",
    eprint = "hep-th/0002230",
    archivePrefix = "arXiv",
    reportNumber = "SPIN-2000-05, ITP-UU-00-03, PUTP-1921",
    doi = "10.1007/s002200100381",
    journal = "Commun. Math. Phys.",
    volume = "217",
    pages = "595--622",
    year = "2001"
}

@article{Bianchi:2001de,
    author = "Bianchi, Massimo and Freedman, Daniel Z. and Skenderis, Kostas",
    title = "{How to go with an RG flow}",
    eprint = "hep-th/0105276",
    archivePrefix = "arXiv",
    reportNumber = "MIT-CTP-3143, DAMTP-2001-41, ITP-01-41, ROM2F-2001-15, PUTP-1987",
    doi = "10.1088/1126-6708/2001/08/041",
    journal = "JHEP",
    volume = "08",
    pages = "041",
    year = "2001"
}

@article{Papadimitriou:2004ap,
    author = "Papadimitriou, Ioannis and Skenderis, Kostas",
    editor = "Biquard, O.",
    title = "{AdS / CFT correspondence and geometry}",
    eprint = "hep-th/0404176",
    archivePrefix = "arXiv",
    reportNumber = "ITFA-2004-17",
    doi = "10.4171/013-1/4",
    journal = "IRMA Lect. Math. Theor. Phys.",
    volume = "8",
    pages = "73--101",
    year = "2005"
}

@article{Kumar:2024pcz,
    author = "Kumar, S. Prem and Stuardo, Ricardo",
    title = "{Twisted circle compactification of $ \mathcal{N} $ = 4 SYM and its holographic dual}",
    eprint = "2405.03739",
    archivePrefix = "arXiv",
    primaryClass = "hep-th",
    doi = "10.1007/JHEP08(2024)089",
    journal = "JHEP",
    volume = "08",
    pages = "089",
    year = "2024"
}

@article{Aramini:2025twg,
    author = "Aramini, Fabrizio and Argurio, Riccardo and Bertolini, Matteo and Garc{\'\i}a-Valdecasas, Eduardo and Moroni, Pietro",
    title = "{Gravity, finite duality cascades and confinement}",
    eprint = "2506.18988",
    archivePrefix = "arXiv",
    primaryClass = "hep-th",
    month = "6",
    year = "2025"
}

@article{Klebanov:1998hh,
    author = "Klebanov, Igor R. and Witten, Edward",
    title = "{Superconformal field theory on three-branes at a Calabi-Yau singularity}",
    eprint = "hep-th/9807080",
    archivePrefix = "arXiv",
    reportNumber = "IASSNS-HEP-98-64, PUPT-1804",
    doi = "10.1016/S0550-3213(98)00654-3",
    journal = "Nucl. Phys. B",
    volume = "536",
    pages = "199--218",
    year = "1998"
}

@article{Klebanov:2000nc,
    author = "Klebanov, Igor R. and Tseytlin, Arkady A.",
    title = "{Gravity duals of supersymmetric SU(N) x SU(N+M) gauge theories}",
    eprint = "hep-th/0002159",
    archivePrefix = "arXiv",
    reportNumber = "PUPT-1919, OHSTPY-HEP-T-00-002",
    doi = "10.1016/S0550-3213(00)00206-6",
    journal = "Nucl. Phys. B",
    volume = "578",
    pages = "123--138",
    year = "2000"
}

@article{Klebanov:2000hb,
    author = "Klebanov, Igor R. and Strassler, Matthew J.",
    title = "{Supergravity and a confining gauge theory: Duality cascades and chi SB resolution of naked singularities}",
    eprint = "hep-th/0007191",
    archivePrefix = "arXiv",
    reportNumber = "IASSNS-HEP-00-56, PUPT-1944",
    doi = "10.1088/1126-6708/2000/08/052",
    journal = "JHEP",
    volume = "08",
    pages = "052",
    year = "2000"
}

@article{Maldacena:2000yy,
    author = "Maldacena, Juan Martin and Nunez, Carlos",
    title = "{Towards the large N limit of pure N=1 superYang-Mills}",
    eprint = "hep-th/0008001",
    archivePrefix = "arXiv",
    doi = "10.1103/PhysRevLett.86.588",
    journal = "Phys. Rev. Lett.",
    volume = "86",
    pages = "588--591",
    year = "2001"
}

@article{Atiyah:2000zz,
    author = "Atiyah, Michael and Maldacena, Juan Martin and Vafa, Cumrun",
    title = "{An M theory flop as a large N duality}",
    eprint = "hep-th/0011256",
    archivePrefix = "arXiv",
    reportNumber = "HUTP-00-A045",
    doi = "10.1063/1.1376159",
    journal = "J. Math. Phys.",
    volume = "42",
    pages = "3209--3220",
    year = "2001"
}

@article{Edelstein:2001pu,
    author = "Edelstein, Jose D. and Nunez, Carlos",
    title = "{D6-branes and M theory geometrical transitions from gauged supergravity}",
    eprint = "hep-th/0103167",
    archivePrefix = "arXiv",
    reportNumber = "HUTP-01-A014",
    doi = "10.1088/1126-6708/2001/04/028",
    journal = "JHEP",
    volume = "04",
    pages = "028",
    year = "2001"
}

@article{Castellani:2024ial,
    author = "Castellani, Federico and Nunez, Carlos",
    title = "{Holography for confined and deformed theories: TsT-generated solutions in type IIB supergravity}",
    eprint = "2410.00094",
    archivePrefix = "arXiv",
    primaryClass = "hep-th",
    doi = "10.1007/JHEP12(2024)155",
    journal = "JHEP",
    volume = "12",
    pages = "155",
    year = "2024"
}

@article{Macpherson:2025pqi,
    author = "Macpherson, Niall T. and Merrikin, Paul and Nunez, Carlos and Stuardo, Ricardo",
    title = "{Twisted-circle compactifications of SQCD-like theories and holography}",
    eprint = "2506.15778",
    archivePrefix = "arXiv",
    primaryClass = "hep-th",
    doi = "10.1007/JHEP08(2025)146",
    journal = "JHEP",
    volume = "08",
    pages = "146",
    year = "2025"
}

@article{Chatzis:2024top,
    author = "Chatzis, Dimitrios and Fatemiabhari, Ali and Nunez, Carlos and Weck, Peter",
    title = "{Conformal to confining SQFTs from holography}",
    eprint = "2405.05563",
    archivePrefix = "arXiv",
    primaryClass = "hep-th",
    doi = "10.1007/JHEP08(2024)041",
    journal = "JHEP",
    volume = "08",
    pages = "041",
    year = "2024"
}

@article{Horowitz:1998ha,
    author = "Horowitz, Gary T. and Myers, Robert C.",
    title = "{The AdS / CFT correspondence and a new positive energy conjecture for general relativity}",
    eprint = "hep-th/9808079",
    archivePrefix = "arXiv",
    reportNumber = "NSF-ITP-98-076, MCGILL-98-13",
    doi = "10.1103/PhysRevD.59.026005",
    journal = "Phys. Rev. D",
    volume = "59",
    pages = "026005",
    year = "1998"
}

@article{Witten:1998zw,
    author = "Witten, Edward",
    editor = "Bergstrom, L. and Lindstrom, U.",
    title = "{Anti-de Sitter space, thermal phase transition, and confinement in gauge theories}",
    eprint = "hep-th/9803131",
    archivePrefix = "arXiv",
    reportNumber = "IASSNS-HEP-98-21",
    doi = "10.4310/ATMP.1998.v2.n3.a3",
    journal = "Adv. Theor. Math. Phys.",
    volume = "2",
    pages = "505--532",
    year = "1998"
}

@article{Legramandi:2021uds,
    author = "Legramandi, Andrea and Nunez, Carlos",
    title = "{Electrostatic description of five-dimensional SCFTs}",
    eprint = "2104.11240",
    archivePrefix = "arXiv",
    primaryClass = "hep-th",
    doi = "10.1016/j.nuclphysb.2021.115630",
    journal = "Nucl. Phys. B",
    volume = "974",
    pages = "115630",
    year = "2022"
}

@article{Macpherson:2024qfi,
    author = "Macpherson, Niall T. and Merrikin, Paul and Stuardo, Ricardo",
    title = "{Circle compactifications of Minkowski$_{D}$ solutions, flux vacua and solitonic branes}",
    eprint = "2412.15102",
    archivePrefix = "arXiv",
    primaryClass = "hep-th",
    doi = "10.1007/JHEP08(2025)143",
    journal = "JHEP",
    volume = "08",
    pages = "143",
    year = "2025"
}

@article{Giliberti:2024eii,
    author = "Giliberti, Mauro and Fatemiabhari, Ali and Nunez, Carlos",
    title = "{Confinement and screening via holographic Wilson loops}",
    eprint = "2409.04539",
    archivePrefix = "arXiv",
    primaryClass = "hep-th",
    doi = "10.1007/JHEP11(2024)068",
    journal = "JHEP",
    volume = "11",
    pages = "068",
    year = "2024"
}

@article{Nunez:2025puk,
    author = "Nunez, Carlos and Roychowdhury, Dibakar",
    title = "{Holographic Timelike Entanglement Across Dimensions}",
    eprint = "2508.13266",
    archivePrefix = "arXiv",
    primaryClass = "hep-th",
    month = "8",
    year = "2025"
}

@article{Jokela:2025cyz,
    author = "Jokela, Niko and Kastikainen, Jani and Nunez, Carlos and Pen{\'\i}n, Jos{\'e} Manuel and Ruotsalainen, Helime and Subils, Javier G.",
    title = "{On entanglement c-functions in confining gauge field theories}",
    eprint = "2505.14397",
    archivePrefix = "arXiv",
    primaryClass = "hep-th",
    reportNumber = "HIP-2025-16/TH",
    month = "5",
    year = "2025"
}

@article{Polchinski:2000uf,
    author = "Polchinski, Joseph and Strassler, Matthew J.",
    title = "{The String dual of a confining four-dimensional gauge theory}",
    eprint = "hep-th/0003136",
    archivePrefix = "arXiv",
    reportNumber = "IAS-TH-00-18, NSF-ITP-00-16",
    month = "3",
    year = "2000"
}

@article{Nunez:2001pt,
    author = "Nunez, Carlos and Park, I. Y. and Schvellinger, Martin and Tran, Tuan A.",
    title = "{Supergravity duals of gauge theories from F(4) gauged supergravity in six-dimensions}",
    eprint = "hep-th/0103080",
    archivePrefix = "arXiv",
    reportNumber = "HUTP-01-A012, CTP-MIT-3094, CTP-TAMU-09-01",
    doi = "10.1088/1126-6708/2001/04/025",
    journal = "JHEP",
    volume = "04",
    pages = "025",
    year = "2001"
}

@article{Bigazzi:2002gyi,
    author = "Bigazzi, F. and Cotrone, A. L. and Petrini, M. and Zaffaroni, A.",
    title = "{Supergravity duals of supersymmetric four-dimensional gauge theories}",
    eprint = "hep-th/0303191",
    archivePrefix = "arXiv",
    reportNumber = "IC-2003-11, CPHT-RR-007-0203, BICOCCA-FT-03-3",
    journal = "Riv. Nuovo Cim.",
    volume = "25N12",
    pages = "1--70",
    year = "2002"
}

@article{Bertolini:2003iv,
    author = "Bertolini, Matteo",
    title = "{Four lectures on the gauge / gravity correspondence}",
    eprint = "hep-th/0303160",
    archivePrefix = "arXiv",
    reportNumber = "NORDITA-2003-10-HE",
    doi = "10.1142/S0217751X03016811",
    journal = "Int. J. Mod. Phys. A",
    volume = "18",
    pages = "5647--5712",
    year = "2003"
}

@article{Gauntlett:2001ps,
    author = "Gauntlett, Jerome P. and Kim, Nakwoo and Martelli, Dario and Waldram, Daniel",
    title = "{Wrapped five-branes and N=2 superYang-Mills theory}",
    eprint = "hep-th/0106117",
    archivePrefix = "arXiv",
    reportNumber = "QMUL-PH-01-07",
    doi = "10.1103/PhysRevD.64.106008",
    journal = "Phys. Rev. D",
    volume = "64",
    pages = "106008",
    year = "2001"
}

@article{Bigazzi:2001aj,
    author = "Bigazzi, F. and Cotrone, A. L. and Zaffaroni, A.",
    title = "{N=2 gauge theories from wrapped five-branes}",
    eprint = "hep-th/0106160",
    archivePrefix = "arXiv",
    reportNumber = "BICOCCA-IFT-01-15",
    doi = "10.1016/S0370-2693(01)01100-5",
    journal = "Phys. Lett. B",
    volume = "519",
    pages = "269--276",
    year = "2001"
}

@article{Cassani:2021fyv,
    author = "Cassani, Davide and Komargodski, Zohar",
    title = "{EFT and the SUSY Index on the 2nd Sheet}",
    eprint = "2104.01464",
    archivePrefix = "arXiv",
    primaryClass = "hep-th",
    doi = "10.21468/SciPostPhys.11.1.004",
    journal = "SciPost Phys.",
    volume = "11",
    pages = "004",
    year = "2021"
}

@article{Barbosa:2024smw,
    author = "Barbosa, Marcelo and Nastase, Horatiu and Nunez, Carlos and Stuardo, Ricardo",
    title = "{Penrose limits of I-branes, twist-compactified D5-branes, and spin chains}",
    eprint = "2405.08767",
    archivePrefix = "arXiv",
    primaryClass = "hep-th",
    doi = "10.1103/PhysRevD.110.046015",
    journal = "Phys. Rev. D",
    volume = "110",
    number = "4",
    pages = "046015",
    year = "2024"
}

@article{Fatemiabhari:2024lct,
    author = "Fatemiabhari, Ali and Nunez, Carlos and Piai, Maurizio and Rucinski, James",
    title = "{Stability of holographic confinement with magnetic fluxes}",
    eprint = "2411.16854",
    archivePrefix = "arXiv",
    primaryClass = "hep-th",
    doi = "10.1103/PhysRevD.111.066009",
    journal = "Phys. Rev. D",
    volume = "111",
    number = "6",
    pages = "066009",
    year = "2025"
}

@article{Casero:2006pt,
    author = "Casero, Roberto and Nunez, Carlos and Paredes, Angel",
    title = "{Towards the string dual of N=1 SQCD-like theories}",
    eprint = "hep-th/0602027",
    archivePrefix = "arXiv",
    reportNumber = "CPHT-RR-010-0106, SWAT-06-454",
    doi = "10.1103/PhysRevD.73.086005",
    journal = "Phys. Rev. D",
    volume = "73",
    pages = "086005",
    year = "2006"
}

@article{Kuperstein:2008cq,
    author = "Kuperstein, Stanislav and Sonnenschein, Jacob",
    title = "{A New Holographic Model of Chiral Symmetry Breaking}",
    eprint = "0807.2897",
    archivePrefix = "arXiv",
    primaryClass = "hep-th",
    reportNumber = "ULB-TH-08-23, TAUP-2879-08",
    doi = "10.1088/1126-6708/2008/09/012",
    journal = "JHEP",
    volume = "09",
    pages = "012",
    year = "2008"
}

@article{Filev:2014bna,
    author = "Filev, Veselin G. and Ihl, Matthias and Zoakos, Dimitrios",
    title = "{Holographic Bilayer/Monolayer Phase Transitions}",
    eprint = "1404.3159",
    archivePrefix = "arXiv",
    primaryClass = "hep-th",
    doi = "10.1007/JHEP07(2014)043",
    journal = "JHEP",
    volume = "07",
    pages = "043",
    year = "2014"
}

@article{Avramis:2006nv,
    author = "Avramis, Spyros D. and Sfetsos, Konstadinos and Siampos, Konstadinos",
    title = "{Stability of strings dual to flux tubes between static quarks in N = 4 SYM}",
    eprint = "hep-th/0612139",
    archivePrefix = "arXiv",
    doi = "10.1016/j.nuclphysb.2007.01.026",
    journal = "Nucl. Phys. B",
    volume = "769",
    pages = "44--78",
    year = "2007"
}

@article{Balasubramanian:1999re,
    author = "Balasubramanian, Vijay and Kraus, Per",
    title = "{A Stress tensor for Anti-de Sitter gravity}",
    eprint = "hep-th/9902121",
    archivePrefix = "arXiv",
    reportNumber = "HUTP-99-A002, EFI-99-6, NSF-ITP-98-132",
    doi = "10.1007/s002200050764",
    journal = "Commun. Math. Phys.",
    volume = "208",
    pages = "413--428",
    year = "1999"
}

@article{Seo:2009kg,
    author = "Seo, Yunseok and Shock, Jonathan P. and Sin, Sang-Jin and Zoakos, Dimitrios",
    title = "{Holographic Hadrons in a Confining Finite Density Medium}",
    eprint = "0912.4013",
    archivePrefix = "arXiv",
    primaryClass = "hep-th",
    doi = "10.1007/JHEP03(2010)115",
    journal = "JHEP",
    volume = "03",
    pages = "115",
    year = "2010"
}

@article{Gubser:1999pk,
    author = "Gubser, Steven S.",
    title = "{Dilaton driven confinement}",
    eprint = "hep-th/9902155",
    archivePrefix = "arXiv",
    reportNumber = "HUTP-99-A010",
    month = "2",
    year = "1999"
}

@article{Jokela:2012dw,
    author = "Jokela, Niko and Mas, Javier and Ramallo, Alfonso V. and Zoakos, Dimitrios",
    title = "{Thermodynamics of the brane in Chern-Simons matter theories with flavor}",
    eprint = "1211.0630",
    archivePrefix = "arXiv",
    primaryClass = "hep-th",
    doi = "10.1007/JHEP02(2013)144",
    journal = "JHEP",
    volume = "02",
    pages = "144",
    year = "2013"
}

@article{Mateos:2007vn,
    author = "Mateos, David and Myers, Robert C. and Thomson, Rowan M.",
    title = "{Thermodynamics of the brane}",
    eprint = "hep-th/0701132",
    archivePrefix = "arXiv",
    doi = "10.1088/1126-6708/2007/05/067",
    journal = "JHEP",
    volume = "05",
    pages = "067",
    year = "2007"
}

@article{Kruczenski:2003be,
    author = "Kruczenski, Martin and Mateos, David and Myers, Robert C. and Winters, David J.",
    title = "{Meson spectroscopy in AdS / CFT with flavor}",
    eprint = "hep-th/0304032",
    archivePrefix = "arXiv",
    doi = "10.1088/1126-6708/2003/07/049",
    journal = "JHEP",
    volume = "07",
    pages = "049",
    year = "2003"
}

@article{Georgiou:2015pia,
    author = "Georgiou, George and Zoakos, Dimitrios",
    title = "{Entanglement entropy of the Klebanov-Strassler model with dynamical flavors}",
    eprint = "1505.01453",
    archivePrefix = "arXiv",
    primaryClass = "hep-th",
    doi = "10.1007/JHEP07(2015)003",
    journal = "JHEP",
    volume = "07",
    pages = "003",
    year = "2015"
}

@article{Canneti:2025rsp,
    author = {Canneti, Tommaso and Castellani, Federico and M{\"u}ck, Wolfgang},
    title = "{Vacuum configuration of winding superstrings from non-standard semiclassical quantization}",
    eprint = "2501.14532",
    archivePrefix = "arXiv",
    primaryClass = "hep-th",
    doi = "10.1007/JHEP06(2025)172",
    journal = "JHEP",
    volume = "06",
    pages = "172",
    year = "2025"
}
\bibliographystyle{utphys}

\end{document}